\documentclass[aps,prx,twocolumn,secnumarabic,nobalancelastpage,amsmath,amssymb,nofootinbib,floatfix]{revtex4-2}

\usepackage{overpic}
\usepackage{graphicx}
\usepackage{dcolumn}
\usepackage{bm}
\usepackage[dvipsnames]{xcolor}
\usepackage{tikz}
\usetikzlibrary{%
    decorations.pathreplacing,%
    decorations.pathmorphing,%
    arrows
}
\usetikzlibrary{positioning, calc,intersections,shapes,matrix,fit}

\usepackage{algorithm}
\usepackage{algpseudocode}
\usepackage{amssymb}
\usepackage{enumerate}
\usepackage{comment}
\usepackage{listings}
\usepackage{mathrsfs}

\usepackage[colorlinks=true,linkcolor=MidnightBlue,citecolor=blue,filecolor=green,urlcolor=PineGreen]{hyperref}

\usepackage{blochsphere}
\usepackage{tikz}
\usetikzlibrary{positioning,arrows,calc,math,angles,quotes}

\def\be{\begin{equation}}
\def\ee{\end{equation}}

\newcommand{\ket}[1]{\left\vert #1 \right\rangle}
\newcommand{\bra}[1]{\left\langle #1 \right\vert}
\newcommand{\ensavg}[1]{\left\langle #1 \right\rangle}
\newcommand{\ip}[2]{\langle #1 \vert #2 \rangle}

\newcommand{\tr}[1]{\mathrm{tr}\left( #1 \right)}
\newcommand{\longstate}[1]{\State \hangindent=1.5em \hangafter=1 #1}

\definecolor{xkcd_purple}{RGB}{126, 30, 156}
\definecolor{xkcd_lilac}{RGB}{206, 162, 253}
\definecolor{xkcd_sky}{RGB}{117, 187, 253}
\definecolor{xkcd_cobalt}{RGB}{3, 10, 167}
\definecolor{xkcd_pine}{RGB}{10, 72, 30}
\definecolor{xkcd_teal}{RGB}{2, 147, 134}
\definecolor{xkcd_aquamarine}{RGB}{4, 216, 178}
\definecolor{xkcd_chartreuse}{RGB}{193, 248, 10}
\definecolor{xkcd_goldenrod}{RGB}{250, 194, 5}
\definecolor{xkcd_pumpkin}{RGB}{225, 119, 1}
\definecolor{xkcd_brightred}{RGB}{255, 0, 13}
\definecolor{xkcd_deeppink}{RGB}{203, 1, 98}
\definecolor{xkcd_wine}{RGB}{128, 1, 63}
\definecolor{xkcd_charcoal}{RGB}{52, 56, 55}
\definecolor{xkcd_greypurple}{RGB}{130, 109, 140}

\definecolor{codegreen}{rgb}{0,0.6,0}
\definecolor{codegray}{rgb}{0.5,0.5,0.5}
\definecolor{codepurple}{rgb}{0.58,0,0.82}
\definecolor{backcolour}{rgb}{0.95,0.95,0.95}
\definecolor{huntergreen}{rgb}{0.21, 0.37, 0.23}
\definecolor{lavenderpurple}{rgb}{0.59, 0.48, 0.71}
\definecolor{patriarch}{rgb}{0.7, 0.0, 1.0}
\definecolor{coquelicot}{rgb}{1.0, 0.22, 0.0}
\definecolor{crimsonglory}{rgb}{0.75, 0.0, 0.2}
\definecolor{deeppink}{rgb}{1.0, 0.08, 0.58}
\definecolor{electricviolet}{rgb}{0.56, 0.0, 1.0}
\definecolor{electricgreen}{rgb}{0.0, 1.0, 0.0}
\definecolor{mint}{rgb}{0.24, 0.71, 0.54}
\definecolor{dodgerblue}{rgb}{0.12, 0.56, 1.0}
\definecolor{lincolngreen}{rgb}{0.11, 0.35, 0.02}
\definecolor{persianblue}{rgb}{0.11, 0.22, 0.73}
\definecolor{pl}{rgb}{0.5, 0.0, 0.5}
\definecolor{amaranth}{rgb}{0.9, 0.17, 0.31}
\definecolor{candyapplered}{rgb}{1.0, 0.03, 0.0}
\definecolor{blueryb}{rgb}{0.01, 0.28, 1.0}

\newcommand{\bw}[1]{{\color{blue} #1}}

\begin{document}
\title{Solving $k$--SAT problems with generalized quantum measurement}

\author{Yipei Zhang} \email{zyp123@berkeley.edu}
\affiliation{Department of Chemistry, University of California, Berkeley, CA 94720, USA}
\affiliation{Berkeley Center for Quantum Information and Computation, Berkeley, CA 94720, USA}
\author{Philippe Lewalle} \email{plewalle@berkeley.edu}
\affiliation{Department of Chemistry, University of California, Berkeley, CA 94720, USA}
\affiliation{Berkeley Center for Quantum Information and Computation, Berkeley, CA 94720, USA}
\author{K.~Birgitta Whaley} 
\affiliation{Department of Chemistry, University of California, Berkeley, CA 94720, USA}
\affiliation{Berkeley Center for Quantum Information and Computation, Berkeley, CA 94720, USA}

\date{\today}

\begin{abstract}
We generalize the projection--based quantum measurement--driven $k$--SAT algorithm of Benjamin, Zhao, and Fitzsimons (BZF, \cite{benjamin2017measurement}) to arbitrary strength quantum measurements, including the limit of continuous monitoring.
In doing so, we clarify that this algorithm is a particular case of the measurement--driven quantum control strategy elsewhere referred to as ``Zeno dragging''. 
We argue that the algorithm is most efficient with finite time and measurement resources in the continuum limit, where measurements have an infinitesimal strength and duration. 
Moreover, for solvable $k$-SAT problems, the dynamics generated by the algorithm converge deterministically towards target dynamics in the long--time (Zeno) limit, implying that the algorithm can successfully operate autonomously via Lindblad dissipation, without detection.
We subsequently study both the conditional and unconditional dynamics of the algorithm implemented via generalized measurements, quantifying the advantages of detection for heralding errors.
These strategies are investigated first in a computationally--trivial $2$-qubit $2$-SAT problem to build intuition, and then we consider the scaling of the algorithm on $3$-SAT problems encoded with $4 - 10$ qubits.
The average number of shots needed to obtain a solution scales with qubit number as $\lambda^n$. For vanishing dragging time (with final readout only), we find $\lambda = 2$ (corresponding to a brute--force search over possible solutions). 
However, the deterministic (autonomous) property of the algorithm in the adiabatic (Zeno) limit implies that we can drive $\lambda$ arbitrarily close to $1$, at the cost of a growing pre-factor. 
We numerically investigate the tradeoffs in these scalings with respect to algorithmic runtime and assess their implications for using this analog measurement--driven approach to quantum computing in practice.  
\end{abstract}

\maketitle

%%%%%%%%%%%%%%%%%%%%%%%%%%%%%%%%%%%%%%%%%%%%%%%%%%%%%%%%%%%%%%
%%%%%%%%%%%%%%%%%%%%%%%%%%%%%%%%%%%%%%%%%%%%%%%%%%%%%%%%%%%%%%
\section{Introduction}

Since the remarks by Feynman \cite{Feynman_1982}, and algorithms by Shor \cite{Shor} and Grover \cite{Grover}, quantum computation \cite{Benioff_1980, BookNielsen} has drawn increasing research interest, spurring the rapid development of quantum technologies across a variety of experimental platforms. 
Even independent of any particular hardware implementation, a variety of approaches have been developed, including circuit--based quantum computers \cite{BookNielsen}, adiabatic-- \cite{farhi2000quantum, Das_2008} or  annealing--based \cite{morita2008mathematical, hauke2020perspectives} quantum computing, and measurement--based quantum computing \cite{Briegel_2009}, in addition to many flavors of quantum simulators. 
Below we will explore an instance of a slightly different approach: In contrast with ``measurement--based'' computation, where a highly entangled state is prepared and then measured, we will here study an example of what may instead be called ``measurement--driven'' \cite{Childs_2002, benjamin2017measurement, Verstraete_2009, zhao2019measurement, berwald2023grover, berwald2023zenoeffect, ding2023singleancilla} computation.
Measurement--driven computation here refers to an approach in which a variety of measurements or dissipators are used to create the quantum dynamics that solve a computational problem. 
Related approaches to computation involving or simulating open quantum systems have also been explored \cite{Hu_2020, Schlimgen_2022, chen2023quantum, *chen2023efficient, Ding_2024}.

Our point of departure in this manuscript is the quantum measurement--driven approach to solving $k$--SAT problems proposed by Benjamin, Zhao, and Fitzsimons \cite{benjamin2017measurement} (henceforth BZF). 
$k$--SAT problems involve $n$ Boolean variables and require satisfaction of $m$ clauses each containing $k$ Boolean variables.
The complexity of $k$-SAT is well understood, and Boolean satisfiability problems were among the first to be classified as NP-complete \cite{Cook_1971, Karp_1972, Levin_1973}. 
For example, using results from statistical physics, it can be shown that as the clause density $\alpha = m/n$ increases, the random SAT instances undergo a satisfiable--unsatisfiable phase transition in the asymptotic limit of large bit number ($n \rightarrow \infty$) \cite{doi:10.1126/science.1073287}. 
Many rigorous and heuristic classical algorithms have been proposed to solve $k$-SAT, particularly $3$-SAT. 
While not the best classical algorithm, the celebrated Sch\"{o}ning's algorithm \cite{schoning1999probabilistic}, which repeatedly checks and randomly corrects the clauses, is probably the most well-known. 
This has a provable runtime upper bound that scales as $1.334^n$ for $3$-SAT \cite{schoning1999probabilistic}. Our study in this work is a quantum generalization of Sch\"{o}ning's algorithm.

BZF proposed a method by which a $k$--SAT problem may be solved through repeated cycles of $k$--qubit projective measurements, with a measurement representing each clause of the logical proposition. 
Through gradual adjustment of the measurement axes defining {\tt true} or {\tt false} on a given qubit, the clause measurements are able to negotiate a solution. 
This may be regarded as a form of analog quantum computation (most similar to adiabatic computation), where open system dynamics (specifically, the invasiveness of measurement) is used to create the solution dynamics, instead of the closed system (unitary) dynamics that are more often considered for such purposes.
Our aim in this manuscript is to generalize the BZF scheme to the cases of sequential general--strength measurements, and weak measurements or dissipators. 
We highlight two main motivations for such a generalization: 
\begin{enumerate}
    \item Projective measurements imply idealized and discrete quantum operations that exist at the limit of infinite resource consumption \cite{Guryanova2020idealprojective}, in contrast with laboratory situations which typically involve continuous dissipation of a system via imperfectly--monitored channels, 
    \item We will be able to formally connect the BZF $k$--SAT algorithm to ``Zeno Dragging'' \cite{Aharonov_1980, ZenoDragging, lewalle2023optimal}, which is a method for measurement--driven quantum control based on the quantum Zeno effect.
\end{enumerate}
As we formally develop these ideas, we will be able to quantify some of the algorithm's convergence properties as a function of the quantum measurement resources it requires. 

Our interest in generalizing the types of measurements used in the BZF algorithm partly rests on the enormous progress made in continuous quantum monitoring over the past few decades. 
Generalized quantum measurements are well understood theoretically \cite{Kraus_1983} in the language of Positive Operator--Valued Measures (POVMs). 
Generalized measurements include not only projectors, but also minimally invasive (and minimally informative) weak quantum measurements. 
Continuous quantum monitoring arises in the limit of continuous infinitesimal--strength quantum measurements \cite{BookCarmichael, BookWiseman, BookBarchielli, BookJacobs, BookJordan}, and has been extensively explored in experiments, primarily on superconducting qubit platforms \cite{Gambetta2008, Murch2013, LeighShay2020, Blais_CQED}. 
The ``quantum trajectories'' conditioned on sequences of measurement readouts are accessible in real-time in such experiments, and the ensemble average of such trajectories reduces to Lindbladian dissipation of the monitored channel(s). 
Continuous monitoring enables such features as simultaneous monitoring of non-commuting qubit observables \cite{ShayLeigh2016, Chantasri_2017, Lewalle_Caustic-thy, Ficheux2018, Lewalle_chaos, Atalaya2018, Atalaya2018-2, Atalaya2019}, and measurement--dependent feedback control \cite{BookWiseman, BookJacobs, Jacobs_Shabani_2008, Gough_2012, ZhangFeedback2017, Minev2019} that enables such diverse capabilities as e.g.,~measurement--driven entanglement generation \cite{martin2015deterministic, martin2017optimal, zhang2020locally, Lewalle:21, martin2019single, Lewalle_2020_cycle}, quantum state or subspace stabilization \cite{Facchi_2002, Mirrahimi_2007, Ticozzi_2008, amini2011stability, Ticozzi_2013, Benoist_2017, Cardona_2018, liang2019exponential, Cardona_2020, amini2023exponential, liang2023exploring}, dissipative control via continuously--modified measurements (i.e.~``Zeno Dragging'') \cite{Aharonov_1980, ZenoDragging, Guillaud_2019, lewalle2023optimal}, or real--time error detection and correction \cite{Ahn_2002, Ahn_2003, Ahn_2004, Sarovar_2004, vanhandel2005optimal, Oreshkov_2007, Mascarenhas_2010, Atalaya_CQEC, Mohseninia2020alwaysquantumerror, Atalaya_2021_CQEC, Convy2022logarithmicbayesian, Livingston_2022, Convy_2022}. 

The quantum Zeno effect \cite{misra1977zeno} describes the inhibition of quantum dynamics that occur on a timescale which is slow compared to measurement or dissipation. 
This too has been examined for weak measurements and/or continuous dissipation \cite{Presilla_1996, Facchi_2002, Sarandy2005adiabaticapprox, facchi2008quantum, Venuti2016adiabaticinopen, Burgarth2020quantumzenodynamics, Kumar2020Zeno, Snizhko2020Zeno, Burgarth2022oneboundtorulethem}, and has moreover found a number of applications as a form of dissipation engineering \cite{Albert_2016, Harrington_2022}, for quantum control \cite{blumenthal2021, ZenoGateTheory, gautier2023designing}, and for error correction \cite{gautier2023designing, Paz-Silva_2012, Dominy_2013, Cohen_2014, Mirrahimi_2014, Lihm_2018, Guillaud_2019, lebreuilly2021autonomous, Gertler_2021, shtanko2023bounds, Mirrahimi_2007, Ticozzi_2008, amini2011stability, Ticozzi_2013, Benoist_2017, Cardona_2018, liang2019exponential, Cardona_2020, amini2023exponential, liang2023exploring}. 
The extensive literature concerning quantum measurement contributes to the understanding of the generalized BZF algorithm that we develop below because 
i) the algorithm of interest requires monitoring non-commuting observables corresponding to different clauses of a $k$--SAT problem, and 
ii) the observables are changed over time as in the Zeno dragging approach to control, such that a solution state or subspace is dissipatively stabilized by the set of clause measurements. In particular, whenever a solution exists, this follows a pure--state kernel of the Liouvillian in the adiabatic regime, as in Refs.~\cite{Sarandy2005adiabaticapprox, Venuti2016adiabaticinopen}.

The plan and aims of this paper are as follows: In Sec.~\ref{SEC:k-SAT} we review the construction of $k$--SAT problems, and the construction of quantum measurements used in the BZF projective algorithm. 
In Sec.~\ref{SEC:General-BZF}, we describe how the projectors in the BZF algorithm may be generalized to finite strength and/or continuous monitoring. 
Specifically, we write down Kraus operators and state update rules for generalized clause measurements in Sec.~\ref{sec-kraus-up}, and then write the corresponding time--continuous version of the dynamics in Sec.~\ref{sec-continuous-dynamics}. 
In Sec.~\ref{sec-Convergence} we then formalize the parallel between these continuous dynamics and Zeno dragging, and describe the convergence properties of the time--continuous algorithm in the Zeno limit (analogous to the adiabatic limit). 
These convergence properties imply that the algorithm is capable of functioning autonomously. 
We are then in a position to begin putting forward several generalized BZF algorithms: In Sec.~\ref{sec_avg_algo} we describe a dissipative (autonomous) algorithm, and in Sec.~\ref{sec_filter_algo} we describe a heralded algorithm that explicitly uses the clause measurement records. 
Finally, in Sec.~\ref{sec_ro_tts}, we describe the final (local) qubit readout, used to turn a state in the qubit register into a candidate solution bitstring to a $k$-SAT proposition, and describe how we quantify the time-to-solution (TTS).
The methods of Sec.~\ref{SEC:General-BZF} are applied first for the simple case of a two--qubit 2-SAT problem in Sec.~\ref{SEC:2Q-2SAT}, with the aim of building intuition, and then applied to larger--scale 3-SAT problems in Sec.~\ref{SEC:BZF-kSAT-examples}. 
As we investigate larger 3-SAT problems, we comment increasingly on the TTS and its scaling with regard to qubit number. 
Concluding remarks are offered in Sec.~\ref{SEC:conclude}. 

%%%%%%%%%%%%%%%%%%%%%%%%%%%%%%%%%%%%%%%%%%%%%%%%%%%%%%%%%%%%%%
%%%%%%%%%%%%%%%%%%%%%%%%%%%%%%%%%%%%%%%%%%%%%%%%%%%%%%%%%%%%%%
\section{Towards Formulating the BZF Scheme via Continuous Measurement \label{SEC:k-SAT}}

%%%%%%%%%%%%%%%%%%%%%%%%%%%%%%%%%%%%%%%%%%%%%%%%%%%%%%%%%%%%%%
\subsection{SAT Problems} 

In this subsection, we briefly introduce the Boolean satisfiability (SAT) problem. A SAT problem involves $n$ Boolean variables $\{b_j \}_{ j= 1}^{n}$, where each variable $b_j$ takes values from either {\tt true} (denoted by 0 or +) or {\tt false} (denoted by 1 or -). 
A literal $x_j$ can take values from $\{b_j, \bar{b}_j \}$, where $\bar{b}_j$ is the negation of $b_j$. In $k$-SAT, a clause $C_i$ contains $k$ literals connected by logical OR $\vee$. A $k$-SAT instance is then defined by a Boolean formula $F$ in the conjunctive normal form (CNF), which involves $m$ clauses connected by logical AND $\wedge$
\begin{equation}
    F = C_1 \wedge C_2 \wedge \cdots \wedge C_m
\end{equation}
where $C_i$ can be, for example, $C_1 = b_2 \vee b_3 \vee \bar{b}_5$ for $k = 3$. 
A clause is satisfied when at least one of the literals is {\tt true}, and we say a given SAT instance is satisfiable iff there exists an assignment of the $n$ Boolean variables $b_{soln} \in \{0, 1\}^n$ such that all $m$ clauses are satisfied simultaneously. 
A SAT problem is then a decision problem, where the goal is to decide whether a given SAT instance is satisfiable or not. 

Note that, while 2-SAT can be solved by classical algorithms such as \cite{krom1967decision} in polynomial time, $k$-SAT for $k \geq 3$ is NP-complete. 
However, the classical complexity arguments are based on the worst case analysis. In practice, the difficulties of solving random SAT  are not uniformly distributed for the clause density $\alpha$, which is defined as $\alpha = m/n$. 
It has been shown that, when $\alpha$ is increasing from $0$, there is a phase transition exhibiting easy-hard-easy behavior for SAT \cite{gent1994easy}. 
In this work, we will focus both on the aggregate algorithm effectiveness for random instances with $\alpha$ being close to the critical value (away from which the $k$--SAT problem can be solved easily), and on case studies of the measurement--driven quantum dynamics for some specific $k$--SAT problems. 

\subsection{The BZF Algorithm}

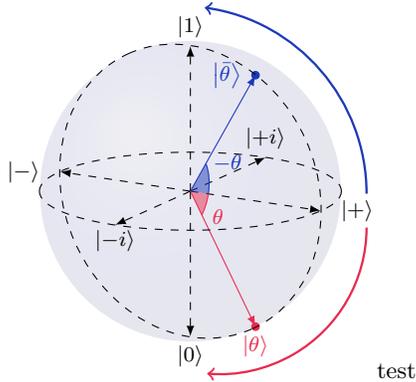
\begin{figure}
\def\rotationSphere{-30}
\def\radiusSphere{2cm}
\def\psiLat{60}
\def\psiLon{0}
\begin{blochsphere}[radius=\radiusSphere, opacity=0.05, rotation=\rotationSphere, color = blue]
  \drawLatitudeCircle[style={dashed}]{0}
  \drawLongitudeCircle[style={dashed}]{0}
  % Define the different points on the bloch sphere
  \labelLatLon{ket0}{90}{0};
  \labelLatLon{ket1}{-90}{0};
  \labelLatLon{ketminus}{0}{180};
  \labelLatLon{ketplus}{00}{0};
  \labelLatLon{ketpluspi2}{0}{-90};  % Longitude seems to be defined in the "wrong" direction, hence the minus
  \labelLatLon{ketplus3pi2}{0}{-270};
  \labelLatLon{psi}{\psiLat}{-\psiLon};
  \labelLatLon{psi_bar}{-\psiLat}{-\psiLon};
  % Draw and label the axis
  \draw[-latex, style = dashed] (0,0) -- (ket0) node[above,inner sep=.5mm] at (ket0) {\footnotesize $\ket{1}$};
  \draw[-latex, style = dashed] (0,0) -- (ket1) node[below,inner sep=.5mm] at (ket1) {\footnotesize $\ket{0}$};
  \draw[-latex, style = dashed] (0,0) -- (ketplus) node[right,inner sep=2mm] at (ketplus) {\footnotesize$\ket{+}$};
  \draw[-latex, style = dashed] (0,0) -- (ketminus) node[left,inner sep=2mm] at (ketminus) {\footnotesize$\ket{-}$};
  \draw[-latex, style = dashed] (0,0) -- (ketpluspi2) node[above,inner sep=.5mm] at (ketpluspi2) {\footnotesize $\ket{+i}$};
  \draw[-latex, style = dashed] (0,0) -- (ketplus3pi2) node[below,inner sep=.01mm] at (ketplus3pi2) {\footnotesize $\ket{-i}$};
  % Draw |psi>
  \draw[-latex, color = persianblue] (0,0) -- (psi) node[left, inner sep = 1.2mm, color = persianblue]{\footnotesize $\ket{\bar{\theta}}$};
  % Draw |psi>
  \draw[-latex, color = amaranth] (0,0) -- (psi_bar) node[below, inner sep = 0.5mm, color = amaranth]{\footnotesize $\ket{{\theta}}$};
  \draw[draw = persianblue, fill = persianblue] (psi) circle (0.05cm);
  \draw[draw = amaranth, fill = amaranth] (psi_bar) circle (0.05cm);
  
  % Draw the angles
  \coordinate (origin) at (0,0);
  { \setLongitudinalDrawingPlane{\psiLon}
    % Draw the angle
    \pic[current plane, draw,fill=persianblue,fill opacity=.5, text opacity=1,"\footnotesize $-\theta$", angle eccentricity=2, color = persianblue]{angle=ketplus--origin--psi};
  }
  { \setLongitudinalDrawingPlane{\psiLon}
    % Draw the angle
    \pic[current plane, draw,fill=amaranth,fill opacity=.5, text opacity=1,"\footnotesize $\theta$", angle eccentricity=1.8, color = amaranth]{angle=psi_bar--origin--ketplus};
  }
\end{blochsphere} 
\begin{tikzpicture}[overlay]
\node[] at (0,0) {test};
\draw[line width = 0.03cm, ->, persianblue] (-0.4,2.35) arc (0:82:2.5cm);
\draw[line width = 0.03cm, ->, amaranth] (-0.4,1.9) arc (0:-95:1.95cm);
\end{tikzpicture}
\caption{
A visual representation of the two qubit states Eq.~\eqref{qubit_encode}
%-- \eqref{eq:R_y}
in the $xz$ Bloch plane contributing to the reduced density matrix of a single qubit following rotation from the initial state $\ket{+}$ according to  $\hat{R}_Y(\theta)$, Eq. \eqref{eq:R_y}.
The {\tt true} and {\tt false} states are initialized in parallel along $\ket{+}$ for $\theta = 0$, and are then gradually rotated so as to become orthogonal for $\theta = \pi/2$. 
}\label{fig-Rotation}
\end{figure}

\begin{table}
\begin{tabular}{llcc}
index & name & min & max \\\hline\hline
$k$ & clause size & & \\ 
$j$ & bit / qubit & $1$ & $n$ \\ 
$i$ & clause index & $1$ & $m$\\ 
$q$ & literal & $1$ & $k$ \\
$c$ & cycle no. & $1$ & $\ell$
\end{tabular}
\caption{
We list the various index variables used throughout the manuscript. 
For example, one may have a $k$--SAT problem defined by $m$ clauses of length $k$, operating on a register of $n$ qubits. The $q^\mathrm{th}$ literal in the $i^\mathrm{th}$ clause will be notated $x_{i_q}$. The cycle number $c$ is only relevant for time--discrete realizations, and is replaced by the time $t$ in the continuum case.   
}\label{tab-indices}
\end{table}

We now describe the way we encode the classical Boolean variables into qubits, following the scheme due to Benjamin, Zhao, and Fitzsimons (BZF) \cite{benjamin2017measurement}. 
We will then briefly review the BZF algorithm for 3-SAT based on projective measurements before proceeding to Sec.~\ref{SEC:General-BZF}  where we present the extended algorithm based on generalized measurement schemes that constitute the focus of this work.

For a classical Boolean variable $b_j$, the two possible values ({\tt true} and {\tt false}) are represented by two $\theta$-dependent pure states of a qubit $j$ as per
\begin{equation}
    \begin{aligned}
\text { {\tt true} } \rightarrow|\theta\rangle_{(j)} & =\hat{R}_Y(+\theta)|+\rangle_{(j)}, \\
\text { {\tt false} } \rightarrow|\bar{\theta}\rangle_{(j)} & =\hat{R}_Y(-\theta)|+\rangle_{(j)},
\end{aligned}\label{qubit_encode}
\end{equation}
where $\theta$ is a control parameter taking its value from $[0, \pi/2 ]$. Here $|+\rangle$ is the equal superposition state $|+\rangle = \frac{1}{\sqrt{2}} (|0\rangle + |1\rangle)$ that lies on the equator in the $xz$-plane of the Bloch sphere. The $\theta$-dependence is obtained by rotating $|+\rangle$ around the $y$-axis of the Bloch sphere by an angle $\pm \theta$, with the sign corresponding to {\tt true} or {\tt false}. 
The rotation operator is given by 
\begin{equation} \label{eq:R_y}
    \hat{R}_Y(\theta)=\left(\begin{array}{rr}
\cos \theta/2 & -\sin \theta/2 \\
\sin \theta/2 & \cos \theta/2
\end{array}\right).
\end{equation}
See Fig. \ref{fig-Rotation} for an illustration.
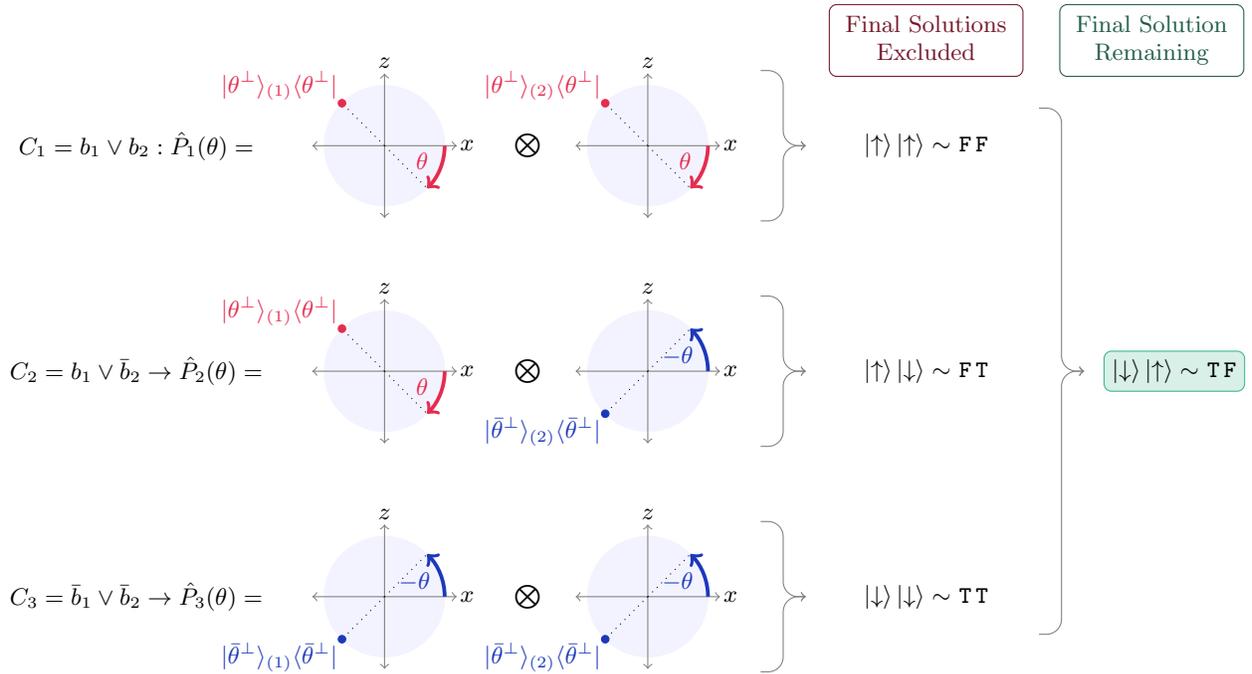
\begin{figure*}
\begin{tikzpicture}

% bloch sphere, axes, labels
\draw[fill, color = blue, opacity=.05] (0.3,0) circle (0.8cm);
\draw[black!50!white,<->] (-1.2*0.8+0.3,0) -- (1.2*0.8+0.3,0);
\draw[black!50!white,<->] (0+0.3,-1.2*0.8) -- (0+0.3,1.2*0.8);
\node[] at (0.3,1.1) {$z$};
\node[] at (1.4,0) {$x$};
% state
\draw[dotted] (0.3,0) -- (0.8/1.414+0.3,-0.8/1.414);
\draw[line width = 0.05cm,->,amaranth] (1.1,0) arc (0:-45:0.8cm);
\node[] at (0.8,-0.2) {\color{amaranth} $\theta$};
\draw[dotted] (0.3,0) -- (-0.8/1.414 +0.3, 0.8/1.414);
\draw[draw = amaranth, fill = amaranth] (-0.8/1.414 +0.3, 0.8/1.414) circle (0.05cm);
\node[] at (-1.1,0.8) {\color{amaranth} $|\theta^{\perp}\rangle_{(1)} \langle \theta^{\perp}|$};

\node[] at (-3,0) {$C_1 = b_1 \vee b_2 : \hat{P}_1(\theta) = $};
\node[] at (2.2,0) {$\bigotimes$};

% bloch sphere, axes, labels
\draw[fill, color = blue, opacity=.05] (3.8,0) circle (0.8cm);
\draw[black!50!white,<->] (3.8 - 1.2*0.8,0) -- (3.8 + 1.2*0.8,0);
\draw[black!50!white,<->] (3.8,-1.2*0.8) -- (3.8,1.2*0.8);
\node[] at (3.8,1.1) {$z$};
\node[] at (4.9,0) {$x$};
% state
\draw[dotted] (3.8,0) -- (0.8/1.414+3.8,-0.8/1.414);
\draw[line width = 0.05cm,->,amaranth] (4.6,0) arc (0:-45:0.8cm);
\node[] at (4.3,-0.2) {\color{amaranth} $\theta$};
\draw[dotted] (3.8,0) -- (-0.8/1.414+3.8, 0.8/1.414);
\draw[draw = amaranth, fill = amaranth] (-0.8/1.414 + 3.8, 0.8/1.414) circle (0.05cm);
\node[] at (2.4,0.8) {\color{amaranth} $|\theta^{\perp}\rangle_{(2)} \langle \theta^{\perp}|$};

% bloch sphere, axes, labels
\draw[fill, color = blue, opacity=.05] (0.3,0-3) circle (0.8cm);
\draw[black!50!white,<->] (-1.2*0.8+0.3,0-3) -- (1.2*0.8+0.3,0-3);
\draw[black!50!white,<->] (0+0.3,-1.2*0.8-3) -- (0+0.3,1.2*0.8-3);
\node[] at (0.3,1.1-3) {$z$};
\node[] at (1.4,0-3) {$x$};
% state
\draw[dotted] (0.3,0-3) -- (0.8/1.414+0.3,-0.8/1.414-3);
\draw[line width = 0.05cm,->,amaranth] (1.1,0-3) arc (0:-45:0.8cm);
\node[] at (0.8,-0.2-3) {\color{amaranth} $\theta$};
\draw[dotted] (0.3,0-3) -- (-0.8/1.414 +0.3, 0.8/1.414-3);
\draw[draw = amaranth, fill = amaranth] (-0.8/1.414 +0.3, 0.8/1.414-3) circle (0.05cm);
\node[] at (-1.1,0.8-3) {\color{amaranth} $|\theta^{\perp}\rangle_{(1)} \langle \theta^{\perp}|$};

\node[] at (-3,0-3) {$C_2 = b_1 \vee \bar{b}_2 \rightarrow \hat{P}_2(\theta) = $};
\node[] at (2.2,0-3) {$\bigotimes$};

% bloch sphere, axes, labels
\draw[fill, color = blue, opacity=.05] (3.8,0-3) circle (0.8cm);
\draw[black!50!white,<->] (3.8 - 1.2*0.8,0-3) -- (3.8 + 1.2*0.8,0-3);
\draw[black!50!white,<->] (3.8,-1.2*0.8-3) -- (3.8,1.2*0.8-3);
\node[] at (3.8,1.1-3) {$z$};
\node[] at (4.9,0-3) {$x$};
% state
\draw[dotted] (3.8,0-3) -- (0.8/1.414+3.8,0.8/1.414-3);
\draw[line width = 0.05cm,->,persianblue] (4.6,0-3) arc (0:45:0.8cm);
\node[] at (4.2,0.2-3) {\color{persianblue} $-\theta$};
\draw[dotted] (3.8,0-3) -- (-0.8/1.414+3.8, -0.8/1.414-3);
\draw[draw = persianblue, fill = persianblue] (-0.8/1.414 + 3.8, -0.8/1.414-3) circle (0.05cm);
\node[] at (2.4,-0.8-3) {\color{persianblue} $|\bar{\theta}^{\perp}\rangle_{(2)} \langle \bar{\theta}^{\perp}|$};

% bloch sphere, axes, labels
\draw[fill, color = blue, opacity=.05] (0.3,0-6) circle (0.8cm);
\draw[black!50!white,<->] (-1.2*0.8+0.3,0-6) -- (1.2*0.8+0.3,0-6);
\draw[black!50!white,<->] (0+0.3,-1.2*0.8-6) -- (0+0.3,1.2*0.8-6);
\node[] at (0.3,1.1-6) {$z$};
\node[] at (1.4,0-6) {$x$};
% state
\draw[dotted] (0.3,0-6) -- (0.8/1.414+0.3,0.8/1.414-6);
\draw[line width = 0.05cm,->,persianblue] (1.1,0-6) arc (0:45:0.8cm);
\node[] at (0.7,0.2-6) {\color{persianblue} $-\theta$};
\draw[dotted] (0.3,0-6) -- (-0.8/1.414 +0.3, -0.8/1.414-6);
\draw[draw = persianblue, fill = persianblue] (-0.8/1.414 +0.3, -0.8/1.414-6) circle (0.05cm);
\node[] at (-1.1,-0.8-6) {\color{persianblue} $|\bar{\theta}^{\perp}\rangle_{(1)} \langle \bar{\theta}^{\perp}|$};

\node[] at (-3,0-6) {$C_3 = \bar{b}_1 \vee \bar{b}_2 \rightarrow \hat{P}_3(\theta) = $};
\node[] at (2.2,0-6) {$\bigotimes$};

% bloch sphere, axes, labels
\draw[fill, color = blue, opacity=.05] (3.8,0-6) circle (0.8cm);
\draw[black!50!white,<->] (3.8 - 1.2*0.8,0-6) -- (3.8 + 1.2*0.8,0-6);
\draw[black!50!white,<->] (3.8,-1.2*0.8-6) -- (3.8,1.2*0.8-6);
\node[] at (3.8,1.1-6) {$z$};
\node[] at (4.9,0-6) {$x$};
% state
\draw[dotted] (3.8,0-6) -- (0.8/1.414+3.8,0.8/1.414-6);
\draw[line width = 0.05cm,->,persianblue] (4.6,0-6) arc (0:45:0.8cm);
\node[] at (4.2,0.2-6) {\color{persianblue} $-\theta$};
\draw[dotted] (3.8,0-6) -- (-0.8/1.414+3.8, -0.8/1.414-6);
\draw[draw = persianblue, fill = persianblue] (-0.8/1.414 + 3.8, -0.8/1.414-6) circle (0.05cm);
\node[] at (2.4,-0.8-6) {\color{persianblue} $|\bar{\theta}^{\perp}\rangle_{(2)} \langle \bar{\theta}^{\perp}|$};

\draw[->,rounded corners = 0.2 cm,black!50!white] (5.3,1) -- (5.6,1) -- (5.6,0) -- (5.9,0);
\draw[->,rounded corners = 0.2 cm,black!50!white] (5.3,-1) -- (5.6,-1) -- (5.6,0) -- (5.9,0);
\draw[->,rounded corners = 0.2 cm,black!50!white] (5.3,-2) -- (5.6,-2) -- (5.6,-3) -- (5.9,-3);
\draw[->,rounded corners = 0.2 cm,black!50!white] (5.3,-4) -- (5.6,-4) -- (5.6,-3) -- (5.9,-3);
\draw[->,rounded corners = 0.2 cm,black!50!white] (5.3,-5) -- (5.6,-5) -- (5.6,-6) -- (5.9,-6);
\draw[->,rounded corners = 0.2 cm,black!50!white] (5.3,-7) -- (5.6,-7) -- (5.6,-6) -- (5.9,-6);
\node[draw = black!50!amaranth,rounded corners = 0.1cm] at (7.5,1.4) {\color{black!50!amaranth} $\begin{array}{c} \text{Final Solutions}\\ \text{Excluded}\end{array}$};
\node[] at (7.5,0) {$\ket{\uparrow}\ket{\uparrow} \sim$ {\tt F\,F}};
\node[] at (7.5,-3) {$\ket{\uparrow}\ket{\downarrow} \sim$ {\tt F\,T}};
\node[] at (7.5,-6) {$\ket{\downarrow}\ket{\downarrow} \sim$ {\tt T\,T}};

\draw[->,rounded corners = 0.2 cm,black!50!white] (9.0,0.5) -- (9.3,0.5) -- (9.3,-3) -- (9.6,-3);
\draw[->,rounded corners = 0.2 cm,black!50!white] (9.0,-6.5) -- (9.3,-6.5) -- (9.3,-3) -- (9.6,-3);
\node[draw = black!50!mint,rounded corners = 0.1cm] at (10.5,1.4) {\color{black!50!mint} $\begin{array}{c} \text{Final Solution}\\ \text{Remaining}\end{array}$};
\node[draw = mint, fill = mint!20!white, rounded corners = 0.1cm] at (10.8,-3) {$\ket{\downarrow}\ket{\uparrow} \sim$ {\tt T\,F}};

\end{tikzpicture}
\caption{
Visual representation of Eq.~\eqref{eq-projector} for the 2-qubit $2$-SAT problem, with each qubit located in the $xz$-plane of its Bloch sphere.
The three projectors corresponding to the three clauses in the 2-qubit $2$-SAT defined by Eq.~\eqref{2q2SAT_clause} are shown as tensor products of $|\theta^{\perp}\rangle = \hat{R}(\pi+\theta)|+\rangle$ (indicated as a red dot) and $|\bar{\theta}^{\perp}\rangle=\hat{R}(\pi-\theta)|+\rangle$ (indicated as a blue dot). 
Upon measuring the observable corresponding to each $\hat{P}_j(\theta)$, an outcome $+1$ heralds success (and corresponds to the $\pm\theta$ states), while an outcome $-1$ heralds failure (and corresponds to the $\pm\theta^\perp$ states). 
Thus, each measurement serves to \emph{rule out} a possible final solution, shown to the right of drawings of each projector. The unique solution that remains is the solution to this 2-SAT problem.
}\label{fig-projectors}
\end{figure*}

A $\theta$-dependent projection operator $\hat{P}_i(\theta)$ is assigned to each clause $C_i$. Specifically, for $k$-SAT, the projector is given by 
\begin{equation}
    \hat{P}_i(\theta) = \bigotimes_{q=1}^{k} |l_{i_q} \theta^{\perp} \rangle_{(i_q)} \langle l_{i_q}\theta^{\perp} |, \label{eq-projector}
\end{equation}
where the integer $l_{i_q}$ is $+1$ if the $q^{\mathrm{th}}$ literal in the $i^{\mathrm{th}}$ clause $x_{i_q}$ is equal to the Boolean variable $b_{i_q}$ itself, and $l_{i_q}$ is $-1$ if $x_{i_q}$ is equal to the negated corresponding Boolean variable $\bar{b}_{i_q}$. The states $|\theta^{\perp} \rangle$ and $|-\theta^{\perp} \rangle$ are defined by 
\begin{equation}
    \begin{aligned}
    &|\theta^{\perp} \rangle = \hat{R}_Y(\pi +\theta)|+\rangle, \\ 
    &|\bar{\theta}^{\perp} \rangle = |-\theta^{\perp} \rangle = \hat{R}_Y(\pi -\theta)|+\rangle,
    \end{aligned}
\end{equation}
which are orthogonal to $|\theta\rangle$ and $|\bar{\theta}\rangle$ defined in Eq.~\eqref{qubit_encode} respectively, i.e.,~$\langle \theta ^{\perp} | \theta \rangle =0= \langle \bar{\theta} ^{\perp} | \bar{\theta} \rangle $. As an example, the projector of the clause $C_1 = b_2 \vee b_3 \vee \bar{b}_5$ is given by
\begin{equation}
    \hat{P}_1(\theta) = |\theta^{\perp} \rangle_{(2)} \langle \theta^{\perp} | \otimes |\theta^{\perp} \rangle_{(3)} \langle \theta^{\perp} | \otimes |(-\theta)^{\perp} \rangle_{(5)} \langle (-\theta)^{\perp} | .
\end{equation}
Essentially, the projector $\hat{P}_i(\theta)$ checks whether the clause $C_i$ is violated: $\hat{P}_i(\theta)$ defines a measurement of a Hermitian observable 
\begin{equation}\label{eq_hermitian_X}
    \hat{X}_i(\theta) = \hat{\openone} - 2\hat{P}_i(\theta),
\end{equation} 
where $\hat{\openone}$ is the identity operator and $\hat{X}_i(\theta)$ has two eigenvalues $\pm 1$. If the measurement gives $+1$ (Success), then $C_i$ is not violated, while if the measurement gives $-1$ (Fail), then $C_i$ is violated and the $k$-qubit subsystem of the quantum state is projected into the only possible state, $ \bigotimes_{q=1}^{k} |l_{i_q}  \theta^{\perp} \rangle _{(i_q)}\langle l_{i_q}  \theta^{\perp} |$, which corresponds to the only assignment of the relevant $k$ Boolean variables violating the clause $C_i$. 

The BZF algorithm \cite{benjamin2017measurement} can be summarized as follows. The input quantum state is the equal superposition of all computational basis states $|+\rangle ^{\otimes n}$. In the $c^{\mathrm{th}}$ cycle of clause measurements, one sequentially checks all the clauses $\{C_i\}_{i=1}^m$ via projectors $\{\hat{P}_i(\theta_c)\}_{i=1}^m$ in a pre-determined order. 
At the end of each cycle of clause measurements, $\theta_c$ is updated according to some schedule moving from $\theta = 0$ to $\theta = \pi/2$ over the course of the algorithm. 
The result is that after many clause measurement cycles, with the {\tt true/false} measurement angles getting further apart, the qubits' states ``fan out'', and eventually arrive at $z = \pm 1$ in the computational basis when $\theta$ reaches $\pi/2$.
The algorithm terminates at this point, and then all of the qubits are individually measured in the computational basis to give the assignment of Boolean variables as the output. 
During the entire process, if any clause measurement has failed, then one restarts the algorithm from the very beginning. 

Intuitively speaking, when one fixes $\theta = 0$, the quantum state cannot distinguish {\tt true} and {\tt false} of the Boolean variable, i.e., $|\theta\rangle = |\bar{\theta}\rangle$ and $|\theta^{\perp}\rangle = |\bar{\theta}^{\perp}\rangle$, but the algorithm will never fail. 
On the other hand, when $\theta = \pi/2$, the quantum states corresponding to {\tt true} and {\tt false} are orthogonal $\langle \bar{\theta}|\theta\rangle = \langle \bar{\theta}^{\perp}|\theta^{\perp}\rangle = 0$ and thus can be perfectly distinguished. 
However, implementing the algorithm at fixed $\theta = \pi/2$ is equivalent to a randomized classical brute force search. 
This is the advantage of the BZF algorithm, where $\theta$ is varied from $\theta_i$ to $\pi/2$ so that the success probability throughout the algorithm is higher than classical brute force searching, while the value $\theta = \pi/2$ at the end of the algorithm ensures the complete information of the solution assignment can be obtained. 
In particular, BZF showed that for $\theta \in (0, \pi/2)$, the number of quantum states that can satisfy all clause checks is equal to the number of solution assignments of the Boolean variables in the SAT problem. 
Furthermore, if there is a solution assignment $(s_1, \dots, s_n)$ with $s_j \in \{\pm 1\}$ satisfying the CNF Boolean formula, then there is a corresponding quantum solution given by \cite{benjamin2017measurement}
\begin{equation} \label{general-solution-state}
    |\phi_{soln}(\theta)\rangle  = \bigotimes_{j = 1}^n | s_j \cdot \theta_j\rangle.
\end{equation}
We note that this solution state is both pure and separable, and necessarily also a +1 simultaneous eigenstate of all of the clause observables $\hat{X}_j(\theta)$ defined in Eq.~\eqref{eq_hermitian_X}.
BZF also showed that the fidelity of the instantaneous quantum state to the solution is monotonically increasing as one continues to implement clause check cycles that herald success.
BZF's numerical results indicate that the running time of their projection--based quantum algorithm scales like $(1.19)^n$, outperforming the classical Sch\"{o}ning algorithm which is well known for its provable upper bound of $(1.334)^n$ \cite{schoning1999probabilistic}.

One can now begin to appreciate how this measurement--driven algorithm is operating in the spirit of adiabatic--like Zeno dynamics. BZF have essentially proposed a discrete and projective algorithm in which success amounts to following a particular eigenstate common to many observables. 
This is a discrete version of what we have elsewhere called Zeno Dragging (see \cite{lewalle2023optimal} and references therein).\footnote{Zeno dragging involves following an eigenstate of a monitored or dissipated observable(s) as that observable varies in time. This allows for high--probability / high--fidelity control as long as the observable is moved slowly compared to the strength with which it is measured \cite{Venuti2016adiabaticinopen, zhao2019measurement}. We comment further on this in later sections.} 
As we generalize the measurement strength below, we will be moving the discrete and projective scheme of BZF towards a different scheme of generalized measurement, including the limit of weak continuous measurement, in which the measurement or dissipation is composed of a sequence of infinitesimal--strength open--system processes occurring on infinitesimal timesteps.
With finite measurement strength, perfect projectors exist only in the limit of operations that take an infinitely long time to complete; such ideal operations do not really exist in any laboratory. 
Explicitly considering tradeoffs between measurement quality and the time expended to carry them out will later prove to be important in quantifying the performance and speed of the measurement--driven algorithm.

%%%%%%%%%%%%%%%%%%%%%%%%%%%%%%%%%%%%%%%%%%%%%%%%%%%%%%%%%%%%%%
\section{Generalizing the Measurement Strength \label{SEC:General-BZF}}

In this section, we introduce a measurement model that will allow us to scale between the limits of projective measurement and weak continuous measurement, in the context of the BZF algorithm for solving $k$-SAT. 
This will allow us to generalize the BZF algorithm beyond the projective limit. 
We will ultimately define two main variants of the generalized algorithm, involving i) the average (dissipative) dynamics, or ii) the true measurement dynamics in which errors can be heralded. These two approaches are eventually presented in four algorithmic subroutines below.

\subsection{Conditional and Un-conditional Dynamics Under Generalized Measurement and Dissipation \label{sec-kraus-up}}

In general, a measurement process can be described by a set of Kraus operators $\{ \hat{M}_r\}$ satisfying $\sum_r \hat{M}_r^{\dagger} \hat{M}_r = \hat{\openone}$, where $r$ is the label for the measurement record.  Given the prior-measurement state described by a density matrix $\rho_0$, the probability of obtaining measurement result $r$ is given by the Born's rule $\mathbb{P}(r) = \mathrm{Tr}(\hat{M}_r \rho_0 \hat{M}_r^{\dagger})$, and the post-measurement state $\rho_r$ conditioned on the readout $r$ is 
\begin{equation}
    \rho_r = \frac{\hat{M}_r \rho_0 \hat{M}_r^{\dagger}}{\mathrm{Tr}(\hat{M}_r \rho_0 \hat{M}_r^{\dagger})}.\label{generalized_meas}
\end{equation}
We will consider here Kraus operators of the form
\begin{equation} \label{Kraus-Algo}
    \begin{aligned}
    \hat{M}_r(\theta)=&\left(\frac{\Delta t}{2 \pi}\right)^{\frac{1}{4}}\bigg\lbrace \exp\left[-\frac{\Delta t}{4}\left(r+\frac{1}{\sqrt{\tau}}\right)^2\right] \frac{\hat{\openone}- \hat{X}(\theta)}{2} \\
    &+\exp\left[-\frac{\Delta t}{4}\left(r-\frac{1}{\sqrt{\tau}}\right)^2\right]\frac{\hat{\openone}+ \hat{X}(\theta)}{2}\bigg\rbrace,
    \end{aligned}
\end{equation}
where $\hat{X}(\theta)$ is defined in Eq.~\eqref{eq_hermitian_X}.
Here, $\tau$ is the ``characteristic measurement time'' representing the measurement strength and $\Delta t$ is the duration of the measurement process. Short $\tau$ denotes the fast ``collapse'' or strong measurement, while large $\tau$ denotes the slow ``collapse'' or weak measurement. 
The particular form of these Kraus operators assumes a measurement apparatus generating a continuous--valued readout $r$, and is based on e.g.,~the Kraus operators derived in quantum optical contexts \cite{Korotkov1, *Korotkov2, *Korotkov3, *Korotkov4, Steinmetz_2022}. 
In particular, in the limit $\Delta t/\tau \gg 1$ the above--defined generalized measurement is effectively projective, while $\Delta t/\tau \ll 1$ is the limit of weak measurement. 
The latter leads to diffusive quantum trajectories when infinitesimal--strength measurements are continuously made over time (i.e.,~in the limit of continuous monitoring) \cite{BookBarchielli, BookWiseman, BookJacobs, BookJordan}. 
When $\Delta t/\tau$ is between the two limits, the Kraus operators describe a generalized discrete measurement with finite strength. 

When one considers the measurement process without measurement records, the dynamics are described by the average over all the possible trajectories, weighted by their probabilities. 
In this case, given the density operator $\rho(t)$ at $t$, the average post-measurement density operator $\bar{\rho}(t+\Delta t)$ is given by
\begin{equation}
\begin{aligned}
    \bar{\rho}(t+\Delta t) &= \int_{-\infty}^{+\infty} dr \hat{M}_r \rho(t) \hat{M}_r^{\dagger} \\
    &= \frac{1+\beta}{2}\rho(t) + \frac{1-\beta}{2}\hat{X}(\theta)\rho(t) \hat{X}(\theta)  ,
\end{aligned}\label{avg_dynamics}
\end{equation}
where $\beta = e^{-\Delta t/2\tau}$. 

\subsection{Weak Continuous Limit \label{sec-continuous-dynamics}}

In this work, we are particularly interested in the limit when $\Delta t/\tau \ll 1$, which is the limit of weak continuous measurement. 
In this case, the measurement strength in infinitesimal time interval $\Delta t$ approaches $0$. By expanding Eq. \eqref{avg_dynamics} to first order in $\Delta t$, we can derive the Lindblad master equation (LME) for the average dynamics
\begin{equation}
    \dot{\bar{\rho}} =  \frac{1}{4\tau}\mathcal{L}[ \hat{X}(\theta)]\bar{\rho} , \label{Lindblad_eq}
\end{equation}
where $\mathcal{L}[ \hat{X}]$ is the Lindbladian generator
$\mathcal{L}[ \hat{X}]\rho =   \hat{X} \rho  \hat{X}^{\dagger} - \frac{1}{2}( \hat{X}^{\dagger} \hat{X}\rho + \rho  \hat{X}^{\dagger} \hat{X})$. 
This is an expected and general property of Markovian quantum trajectories \cite{BookBarchielli, BookWiseman, BookJacobs, BookJordan, Lindblad, GKS}. 
Note that we will be able to assume $ \hat{X}(\theta)^{\dagger} =  \hat{X}(\theta)$ and $ \hat{X}(\theta)^2 =  \hat{\openone}$ throughout this work. 
The second property is guaranteed for $\hat{X}$ of the form $\hat{\openone} - 2\,\hat{P}$.

The individual trajectories conditioned on the measurement record, which can be expressed $r \,dt = \langle  \hat{X} \rangle dt/\sqrt{\tau} + dW$ are also of interest. 
From Eq.~\eqref{generalized_meas} we can derive an It\^{o} stochastic master equation (SME) describing such dynamics under the weak continuous limit \cite{BookBarchielli, BookWiseman, BookJacobs}
\begin{equation}
    d \rho = \frac{1}{4\tau}\mathcal{L}[ \hat{X}(\theta)]\rho \, dt + \frac{1}{2\sqrt{\tau}}\mathcal{H}[ \hat{X}(\theta)]\rho \,dW, \label{quantum_filter_eq}
\end{equation}
where $\mathcal{H}[ \hat{X}] \rho = \rho \hat{X} +  \hat{X} \rho - 2\langle  \hat{X} \rangle \rho$ is the measurement backaction with $\langle  \hat{X} \rangle = \mathrm{Tr}( \hat{X}\rho)$ the expectation value of $ \hat{X}$. Here $dW$ is the Wiener increment satisfying $dW^2 = dt$ by It\^{o}'s lemma. 
Notice that the dynamics satisfying Eq.~\eqref{Lindblad_eq} are recovered by averaging over all possible trajectories in Eq.~\eqref{quantum_filter_eq}, consistent with the fact that $dW$ has zero mean.

A striking difference between the algorithm dynamics for SAT under continuous measurement and projective measurement is the effect of measurement ordering. 
To see this, we first notice that the observables corresponding to different clauses do not necessarily commute. This happens for $\theta \in (0, \pi/2)$, due to the common Boolean variable involved in the two clauses being of complementary form. 
For example, the observables for $C_1 = x_1 \vee x_2$ and $C_2 = x_1 \vee \bar{x}_2$ do not commute. 
Therefore, in the dynamics under projective measurement, the order of measurements will be important. 
However, for weak continuous measurement, non-commutativity plays no role in the master equations Eq.~\eqref{Lindblad_eq} and Eq.~\eqref{quantum_filter_eq}, which are first order in $\Delta t$. 
Non-commutativity only appears in terms that are of order higher than $O( \Delta t)$. 
In other words, one can simultaneously measure all the clause observables under weak continuous measurement, and any effects due to clause ordering must vanish in the limit $\Delta t/ \tau \rightarrow 0$. 
The average dynamics under such $m$ simultaneous clause measurements is described by the master equation
\begin{equation}
    \dot \rho =  \sum_{i=1}^{m}\frac{1}{4\tau}\mathcal{L}[ \hat{X}_i(\theta)]\rho, \label{eq_Lindblad_multi},
\end{equation}
and the individual trajectory conditioned on  $\{ dW_i \}_{i=1}^{m}$ is described by the SME
\begin{equation}
    d \rho = \sum_{i=1}^{m}\frac{1}{4\tau}\mathcal{L}[ \hat{X}_i(\theta)]\rho \, dt + \sum_{i=1}^{m}\frac{1}{\sqrt{4\tau}}\mathcal{H}[ \hat{X}_i(\theta)]\rho \,dW_i,  \label{eq_SME_multi}
\end{equation}
where each observable $\hat{X}_i(\theta)$ corresponds to a clause $C_i$.
Each of the readouts on which these dynamics are conditioned is a sum of signal and noise, namely,
\be \label{eq_SME_ro}
r_i\,dt = \frac{1}{\sqrt{\tau}}\,\mathrm{Tr}\left(\rho(t)\,\hat{X}_i(\theta) \right)\,dt + dW_i(t),
\ee
where the first term is the expected signal content of the measurement outcome, and the Wiener process $dW_i$ is pure noise. 

\subsection{Convergence in the Zeno Limit \label{sec-Convergence}}

We now elaborate on the convergence properties of Zeno dragging necessary for an understanding of our algorithms. 
Let us consider the case where there is a unique solution to our $k$--SAT problem, i.e.,~we assume there exists a unique solution of the form of Eq.~\eqref{general-solution-state}. 
We may then define a frame change by the rotation 
\be \label{Qframe}
\hat{Q} = \bigotimes_{j = 1}^n \hat{R}_Y^{(j)}(\mp(\tfrac{\pi}{2}-\theta)),
\ee
where $\pm$ rotations are assigned to each qubit according to the solution bitstring $\mathbf{s}$, such that the ideal solution dynamics become static in the $\hat{Q}$--frame. 
In this frame all of the $\hat{X}_i(\theta)$ are partially diagonalized, in the sense that the row/column of each $\hat{\mathcal{X}}_i = \hat{Q}^\dag\,\hat{X}_i(\theta)\,\hat{Q}$ which corresponds to the solution state at $\theta = \pi/2$ (or $t = T$) in the computational basis will now be occupied only by its $\theta$--independent diagonal element. 
The schedule $\theta(t)$ is here assumed to vary in a continuous and differentiable way.

Let $\varrho = \hat{Q}^\dag\,\rho\,\hat{Q}$, such that the It\^{o} $\hat{Q}$--frame dynamics read
\begin{subequations}\be 
d\varrho =
i[\hat{H}_Q,\varrho]\,dt + \frac{1}{4\tau} \sum_i^m \mathcal{L}[\hat{\mathcal{X}}_i]\varrho\,dt + \frac{1}{2\sqrt{\tau}}\sum_i^m \mathcal{H}[\hat{\mathcal{X}}_i]\varrho\,dW_i 
\ee
with
\be 
\hat{H}_Q = \tfrac{1}{2}\,\dot{\theta}\sum_j^n s_j\,\hat{\sigma}_y^{(j)}. 
\ee\end{subequations}
The additional Hamiltonian term $\hat{H}_Q$ encodes diabatic motion due to movement of the $\hat{Q}$--frame as the observables are rotated, with $s_j \in \{ \pm 1\}^{\otimes n}$ representing the solution bitstring corresponding to Eq. \eqref{Qframe}. 
The solution state Eq. \eqref{general-solution-state} is a simultaneous $+1$ eigenstate of the clause observables $\hat{X}_i$, which we now notate as $\tilde{\rho} = \ket{\phi_{soln}(\theta)}\bra{\phi_{soln}(\theta)}$. 
In the new frame, this eigenstate is $\theta$--independent and therefore time--independent, i.e.,~we have $\hat{\mathcal{X}}_i\,\tilde{\varrho}\,\hat{\mathcal{X}}_i = \tilde{\varrho}$ for all $i$. 

We can now consider the algorithm dynamics in the $\hat{Q}$--frame, in the Zeno limit $T/\tau \rightarrow \infty$ (which is an adiabatic limit).  
We initialize our system in $\tilde{\varrho}$, corresponding to $\rho = (\ket{+}\bra{+})^{\otimes n}$ at $\theta = 0$. 
Notice that because $\tilde{\varrho}$ is an eigenstate of all the $\hat{\mathcal{X}}_i$, we have $\mathcal{L}[\hat{\mathcal{X}}_i]\tilde{\varrho} = 0$ and $\mathcal{H}[\hat{\mathcal{X}}_i]\tilde{\varrho} = 0$ for all $i$. 
Then in the limit $T/\tau \rightarrow \infty$, we have $\dot{\theta} \rightarrow 0$, and $\tilde{\varrho}$ is therefore a fixed point of both the conditional and unconditional dynamics. 
This means that in the Zeno limit, our algorithm converges \emph{in probability} to the desired solution dynamics via perfect Zeno pinning in the $\hat{Q}$--frame, thereby achieving deterministic (diffusion--free) evolution.  
In other words, when the $\hat{Q}$--frame can be constructed, namely, when a unique solution exists, one may think of the rotation of the clause observables as generating a diabatic perturbation about the solution Eq.~\eqref{general-solution-state} when $T/\tau \gg 1$. 
We point out that these limiting--case solution dynamics are both pure and separable in the case of a unique solution.
Furthermore, \emph{the arguments above imply that the scaling of the Lindbladian algorithm with qubit number goes to $1^n$ in the Zeno limit, since any solution that exists is found deterministically, independent of the number of qubits}. 
See Appendix \ref{sec-app-converge} and Refs.~\cite{lewalle2023optimal,Mirrahimi_2007, Ticozzi_2008, amini2011stability, Ticozzi_2013, Benoist_2017, Cardona_2018, liang2019exponential, Cardona_2020, amini2023exponential, liang2023exploring} for further comments and context. 
We will revisit questions related to algorithmic scaling in later sections. 

We now proceed by clarifying the connection between such continuous BZF algorithms and the measurement--driven approach to quantum control known as ``Zeno Dragging''.  
In Ref.~\cite{lewalle2023optimal} we defined ``Zeno Dragging'' in terms of a quantity 
\be \label{little-g} \begin{split}
\mathsf{g}(\rho,\theta) & = \sum_i \mathsf{g}_i(\rho,\theta) = \frac{1}{2\tau}\sum_i \left(\ensavg{\hat{X}_i^2(\theta)} - \ensavg{\hat{X}_i(\theta)}^2 \right) \\ & = \tfrac{1}{2\tau}\sum_i \mathrm{var}\left(\ensavg{\hat{X}_i(\theta)}\right), 
\end{split} \ee 
claiming that ``Zeno dragging is a viable approach to driving a quantum system from some initial $\ket{\psi_i}$ to a final $\ket{\psi_f}$, if and only if there exists a parameter(s) $\theta$ controlling the choice of measurement such that i) a continuous sweep in $\theta$ is possible, and ii) that this generates a continuous deformation of a local minimum of $\mathsf{g}(\rho,\theta)$ which traces a path from $\ket{\psi_i}$ to $\ket{\psi_f}$''. It is easy to verify that each $\mathsf{g}_j$ vanishes at an eigenstate of $\hat{X}_j(\theta)$, and hence that $\mathsf{g}(\rho,\theta)$ vanishes at a common eigenstate of all the $\hat{X}_j(\theta)$, if such an eigenstate exists. 
One can readily see the connection between this definition with definitions employing the Liouvillian kernel \cite{Sarandy2005adiabaticapprox, Venuti2016adiabaticinopen}). 
We conclude that the time--continuum version of the finite--time generalized--measurement BZF algorithm is a Zeno Dragging operation by the definition of Ref.~\cite{lewalle2023optimal}, where Zeno dragging here means that we continuously and quasi-adiabatically deform the initial state $\ket{+}^{\otimes n}$ to the distinguishable solution state $\ket{\phi_{soln}(\theta = \tfrac{\pi}{2})}$ by rotating $\theta$ from $0$ to $\pi/2$. 

Finally, we note that in the case with multiple solutions, each solution will still take the form of Eq.~\eqref{general-solution-state}, and together they will span the solution subspace.
The convergnce and stability arguments given above will apply to the entire solution subspace in so far as they apply for each of the individual solutions Eq.~\eqref{general-solution-state} spanning that space.
Even in the multi--solution case, the solution space should here be defined as the root of $\mathsf{g}$, i.e.,~the common eigenspace of all the clause observables. From a control perspective, $\mathsf{g}$ is an objective function, and minimization of $\mathsf{g}$ implies staying as close to our target solution eigenspace as possible \cite{lewalle2023optimal}. See Appendix \ref{sec-app-converge} for more extended remarks.

To summarize: the adiabatic theorem for Lindbladian dynamics, which is equivalent to Zeno dragging in the time continuum limit,  ensures that if one starts in the kernel of $\mathcal{L}(\theta)$, and if the total evolution time $T$ is long enough compared to the minimum Liouvillian gap of $\mathcal{L}(\theta)$, then the quantum system will stay near the instantaneous pure--state kernel of $\mathcal{L}(\theta)$ \cite{Sarandy2005adiabaticapprox, Venuti2016adiabaticinopen}. 
In our case, it is easy to check that the kernel of $\mathcal{L}(\theta)$ is spanned by the solution state(s) $|\phi_{soln}(\theta)\rangle$ \cite{benjamin2017measurement}, since a state being in the kernel is equivalent to its passing all clause checks with certainty.
It is natural from a control perspective to imagine Zeno dragging in terms of a single measurement that isolates some target state or subspace (see \cite{lewalle2023optimal} and references therein). 
Here however, we have a sort of mirror image of that scenario: instead of a single measurement that positively identifies some target dynamics, the continuous BZF algorithm contains a collection of measurements that each \emph{rule out} a possible solution, with the desired solution emerging as the remaining option from the collective dynamics. We might term this ``Zeno exclusion control''.  
We have essentially shown that a collective ``ruling out'' of all states that fail clause checks leads to dynamics with the same autonomous / stabilizing properties on the remaining solution subspace as controlled Zeno dragging. 
While having many measurements appears experimentally cumbersome, this strategy makes sense from an algorithmic perspective, because it implements Zeno dragging \emph{without our knowing the target dynamics} a priori (and here knowing target dynamics $\ket{\phi_{soln}(\theta)}$ would amount to already knowing a solution to the $k$-SAT problem). 

%%%%%%%%%%%%%%%%%%%%%%%%%%%%%%%%%%%%%%%%%%%%
%%%%%%%%%%%%%%%%%%%%%%%%%%%%%%%%%%%%%%%%%%%%
\subsection{Un-conditional BZF Algorithm With Finite Measurement Strength \label{sec_avg_algo}}

The autonomous stability we have just described motivates us to further investigate a BZF-type algorithm based on Lindbladian dissipation $\mathcal{L}(\theta) = \sum_{i=1}^{m}\frac{1}{4\tau}\mathcal{L}[ \hat{X}_i(\theta)]$ (from Eq.~\eqref{eq_Lindblad_multi}) alone. 
Recall that Eq.~\eqref{avg_dynamics} gives the average dynamics of our quantum system under a single clause check measurement with finite measurement strength. 
We use this to propose an algorithm based on the average dynamics for general $\Delta t/\tau$ (where Eq. \eqref{Lindblad_eq} is recovered in the $\Delta t \rightarrow dt$ limit of Eq. \eqref{avg_dynamics}).
Algorithmically, one can perform such clause check measurements for all \bw{$m$} clauses in a predetermined order, which consists of a clause check cycle. 
At the end of each clause check cycle, one then updates the control parameter $\theta$, proceeding monotonically from $\theta = 0$ to $\pi/2$. Finally, one reads out all the qubits, in the computational basis. This is summarized in Algorithm \ref{alg1} (with the readout procedure deferred to Algorithm \ref{alg4}). 

\begin{algorithm}[H]
\caption{$k$-SAT by average dynamics}
\label{alg1}
\begin{algorithmic}[1]
    \State \textbf{input}: $\tau$,  $T_f$, $\Delta t$,  and a schedule function $\theta(t/T_f)$
    \State initialize $t \gets 0$ and the state to $\rho(0) = |+\rangle \langle + |^{\otimes n}$
    \While{$t \leq T_f$}
    \State $t \gets t + \Delta t$
    \State $\theta \gets \theta(t/T_f)$
    \State sequentially dissipate clause checks defined by Eq.~\eqref{avg_dynamics} for all clauses
    \EndWhile
    \State \Return $\rho(T_f)$
\end{algorithmic}
\end{algorithm}

One can immediately check that the solution state defined in Eq.~\eqref{general-solution-state} is a fixed point of Eq.~\eqref{avg_dynamics} applied over all clause dissipators.
However, we also point out that the algorithm using average dynamics nevertheless allows  the possibility of reaching the final solution state at $\theta = \pi/2$ via some diabatic path, where the measurement (if recorded) has failed and thus the state deviates from the solution state at some intermediate time.
A Lindbladian BZF algorithm is a special case of Algorithm \ref{alg1}, operating in the time--continuum limit. 
It relies on the fact that convergence in mean and probability in the Zeno limit, as described in Sec.~\ref{sec-Convergence}, effectively indicates that our algorithm can function autonomously (without recording or using the measurement outcomes in any way). 

%%%%%%%%%%%%%%%%%%%%%%%%%%%%%%%%%%%%%%%%%%%%%%%
%%%%%%%%%%%%%%%%%%%%%%%%%%%%%%%%%%%%%%%%%%%%%%%
\subsection{Heralding BZF Algorithm With Finite Measurement Strength Via Filtering \label{sec_filter_algo}}

In the preceding section, we considered a dissipation--only version of our algorithm.  
We next turn our attention to a version where clause failure information is actually collected and used in individual experimental runs.
In other words, we now consider a generalized--measurement BZF algorithm with detection rather than merely dissipation, with the aim of ascertaining how much more effective the algorithm can be in the case that the observer has the option to abort failed experimental runs in real time, and thereby gains a success herald via clause detection before making a final readout in the computational basis. 
Formally, this amounts to saying that we will have a detector granting us access to the pure state conditional dynamics Eq.~\eqref{generalized_meas} instead of just the average dynamics Eq.~\eqref{avg_dynamics}. 
It is possible that propagation of the conditional dynamics may be computationally very expensive in contexts where solving a $k$-SAT problem is of interest, even though it is always possible in principle. 
This is no barrier to the scheme presented here however, since it does not require computation of the conditional $\rho(t)$, but requires instead only access to the clause readouts $r_i$ (followed by local $z$ readouts $r_j$ after $\theta$ reaches $\pi/2$; see Sec.~\ref{sec_ro_tts} for details). 
Note that in the event that the conditional dynamics can actually be tracked through the entire evolution with high efficiency, the terminal local $z$ readout may no longer be necessary.

We see two primary benefits of an algorithm employing heralded success of clause readout $r_i$.
On the one hand, we might quickly terminate trajectories that have already collapsed into the subspace associated with failure of clauses, which means they are no longer able to adiabatically follow the instantaneous solution state. 
Another reason to consider real--time detection is the possibility of feedback. With feedback, one could in principle not only herald errors in the course of running the algorithm, but also then intervene to correct them immediately instead of starting the run over again. 
In this work we focus on detection only, and leave investigation of feedback correction to future work (see further comments in Sec.~\ref{SEC:conclude}).

Unlike projective measurements where one can easily diagnose the collapse into some subspace of the measured observable using the measurement result, in weak measurement the dynamics are diffusive, which complicates fast diagnosis of errors directly from the noisy measurement record. In order to overcome this, a filter is needed.
We first use the time continuous situation to develop some intuition for the filter we are going to use.
Such filtering is used in continuous quantum error correction to detect and correct errors in real-time \cite{Ahn_2002, Ahn_2003, Ahn_2004, Sarovar_2004, vanhandel2005optimal, Oreshkov_2007, Mascarenhas_2010, Atalaya_CQEC, Mohseninia2020alwaysquantumerror, Atalaya_2021_CQEC, Convy2022logarithmicbayesian, Livingston_2022, Convy_2022} (CQEC). 
In the language of error correction, an ideal Zeno dragging procedure conditioned on $r_j = +1/\sqrt{\tau}~\forall~j$ defines the ``codespace'' that we attempt to follow. 
A failed clause measurement will return readouts with a mean signal centered around $-1/\sqrt{\tau}$ instead of $+1/\sqrt{\tau}$, corresponding to an ``error subspace'' in error correction language. The main difficulty in realizing an effective CQEC implementation is in managing the tradeoff between rapid error detection and statistical confidence in the detection and characterization of the error. 
We consider filter functions on the readout of the form 
\be \label{log-bayes-factor-windowed}
\mathcal{B}_j(t) = \frac{1}{2\,\mathcal{N}_\mathcal{W}\,\sqrt{\tau}} \int_0^t dt'\,r_j(t')\,\mathcal{W}(t,t'),
\ee
where $\mathcal{W}(t,t')$ is a window function to be chosen below, and $\mathcal{N}_\mathcal{W} = \int_0^t \mathcal{W}(t,t')\,dt'$ is a normalization factor. 
For $\mathcal{W}/\mathcal{N}_\mathcal{W} = 1$, and in the continuum limit where $dt$ is infinitesimal, $\mathcal{B}_j$ can be interpreted as a time--continuum approximation of the log-likelihood ratio for a sequence of clause measurement heralding success $\ensavg{r_j} = +1/\sqrt{\tau}$ versus failure $\ensavg{r_j} = -1/\sqrt{\tau}$, over the entire measurement record. 
Thus, \eqref{log-bayes-factor-windowed} should be similarly understood as being like a log-likelihood, where the role of the window function $\mathcal{W}(t,t')$ is to weight that likelihood towards the ``recent history'' of the measurement record in an appropriate way.

For the measurement signal $r_i(t)$ for each clause $i$,  we obtained the filtered signal $\bar{r}_i(t)$ obtained via an exponential filter with a finite integration window
\begin{equation}\label{eq:filter}
    \bar{r}_i(t) = \frac{1}{N_{be}}\int_{t-T_{be}}^t dt'\,\frac{e^{-\frac{t-t^{\prime}}{T_{be}}}}{T_{be}}\,r_i(t^{\prime}),
\end{equation}
where $N_{be} = 1 - e^{-1}$ is the normalization constant, and $T_{be}$ is the response time. 
This is essentially an exponential filter inside a single threshold boxcar filter \cite{Mohseninia2020alwaysquantumerror, Atalaya_CQEC}. 

Recall the general form of the clause readouts Eq. \eqref{eq_SME_ro} in the time--continuum limit.
If the signal has reached a steady value $s_0$ before $t_0$, i.e., $\langle \bar{r}_i(t_0) \rangle = \langle r_i(t_0) \rangle = s_0$, and the ``signal part'' of the raw signal changes from $s_0$ to $\langle r_i(t>t_0)\rangle = s_1$ at $t_0$, then the expectation value of the filtered signal will approach to the new steady value $s_1$ exponentially as
\begin{equation}
    \langle\bar{r}_i(t)\rangle = \frac{1}{N_{be}} \left[s_1 \left(1- e^{-\frac{t-t_0}{T_{be}}} \right) + s_0\left(e^{-\frac{t-t_0}{T_{be}}} - e^{-1} \right)\right],
\end{equation}
for $t_0 \leq t \leq t_0 + T_{be}$, with $\langle \cdot \rangle$ understood as an ensemble average. One can check that $\langle \bar{r}_i(t+T_{be}) \rangle = s_1$, which will remain for $t > t_0+ T_{be}$. In the steady state, the variance of the filtered signal is
\begin{equation}
    \mathrm{Var}[\bar{r}_i(t)] = \frac{e+1}{2(e-1)} \frac{1}{T_{be}}\approx \frac{1}{T_{be}},
\end{equation}
which is consistent with the intuition that in the filtered signal, a larger $T_{be}$ value results in smaller fluctuations but a longer response time, while a smaller $T_{be}$ value results in a quicker response but bigger fluctuations.

When implementing the error detection for a discrete-time algorithm, one then discretizes Eq. \eqref{eq:filter}
 to get an update equation for the filtered signal $\bar{r}_i(t)$ as
\begin{equation}\label{eq:filter_discrete}\begin{split}
\bar{r}_i(t + \Delta t) =\,& \bar{r}_i(t)\left(1-\frac{\Delta t}{T_{be}}\right) \\&+ \left[ e \cdot r_i(t)  - r_i(t-T_{be})\right]\frac{\Delta t}{(e-1)T_{be}}.
\end{split}\end{equation}

The error-detection strategy using the filtered signal depends on a threshold value $r_{\mathrm{thr}}$. During the evolution of the system under Eq. \eqref{generalized_meas} for all clauses, if any of the filtered signal corresponding to the $i^{\mathrm{th}}$ clause is below the threshold, i.e., $\bar{r}_i(t) < r_{th}$, at time $t$, then the algorithm is terminated at time $t$ and diagnosed with ``FAILED''. This subroutine is summarized in Algorithm \ref{alg2}.

\begin{algorithm}[H]
	\caption{$k$-SAT by single trial heralded dynamics}
	\label{alg2}
    \begin{algorithmic}[1]
        \State \textbf{input}: $\tau$,  $T_f$, $\Delta t$,  $T_{be}$, $r_{th}$, and a schedule  $\theta(t/T_f)$
        
        \State initialize $t \gets 0$ and the state to $\rho(0) = |+\rangle \langle + |^{\otimes n}$
        \While{$t \leq T_f$}
        \State $t \gets t + \Delta t$
        \State $\theta \gets \theta(t/T_f) $
        \longstate{sequentially measure all clauses as defined by Eq. \eqref{generalized_meas} and get measurement records $\{ r_i(t)\}_{i=1}^m$.}
        \State update the filtered signals $\{\bar{r}_i(t+\Delta t)\}_{i=1}^m$ via Eq. \eqref{eq:filter_discrete}.
        \If{any $\bar{r}_i(t) < r_{th}$}
        \State \textbf{terminate} and  \Return ``FAILED'', $t$
        \EndIf
        \EndWhile
        \State\Return $\rho(T_f)$
\end{algorithmic}
\end{algorithm}
Notice that in the projective limit where $\Delta t /\tau \gg 1$, one can choose $T_{be} = \Delta t$, so that the projection into the failed subspace can be detected immediately. In the continuum limit where $\Delta t /\tau \ll 1$, one can typically choose $T_{be} \sim \tau$, so that the collapse caused by the measurement is sufficiently far along to be confidently detectable through the readout noise.

As discussed above, a heralded dynamics benefits from earlier detection of the failure. This advantage allows us to define a heralded algorithm that restarts when the failure is detected early, and thus is more likely to follow the correct trajectory given a fixed amount of temporal computational resource (total algorithm running time), even without feedback. This heralded algorithm is summarized in  Algorithm \ref{alg3}. 
\begin{algorithm}[H]
	\caption{$k$-SAT by heralded dynamics}
	\label{alg3}
    \begin{algorithmic}[1]
        \State \textbf{input}: $\tau$,  $T_f$, $\Delta t$, $T_{be}$,  $T_{min}$, $r_{th}$, and a schedule  $\theta(t/T_f)$
        
        \State initialize $t_{rest} \gets T_f$ and the state to $\rho(0) = |+\rangle \langle + |^{\otimes n}$ 
        \While{$t_{rest} \geq T_{min}$}
        \State run Algorithm \ref{alg2} with input $\tau$, $t_{rest}$, $\Delta t$, $T_{be}$, $r_{th}$,  $\theta(t)$
        \If{Obtained ``Failed'' and $t$}
        \State $t_{rest} \gets t_{rest} - t$
        \Else
        \State \Return $\rho(t_{rest})$
        \EndIf
        \EndWhile
        \State run Algorithm \ref{alg2} without failure detection and with schedule $\theta(t/t_{rest})$ from $t=0$ until time $t = t_{rest}$
        \State\Return $\rho(t_{rest})$     
\end{algorithmic}
\end{algorithm}
Notice that in Algorithm \ref{alg3}, we do not terminate the dynamics when the remaining time $t_{rest}$ is smaller than a minimum value $T_{min}$. This is because when the total dragging time is too small (comparable to $T_{be}$ and $\tau$), the quantum Zeno effect is not strong and the failure detection based on the filter is also not reliable anymore. However, in this situation we can still use conditional dynamics, i.e., Eq \eqref{generalized_meas} rather than the averaged dynamics of Eq. \eqref{avg_dynamics}. 

We reiterate that although the clause measurement outcomes are recorded in
the implementation of Algorithm \ref{alg3}, we do not assume that these measurement records are used directly to estimate the solution bitstring or conditional state, even if the solution state might be successfully prepared at the end of the algorithm. We only use these measurement signals to herald the success of trajectories. 
Therefore, for both Algorithm \ref{alg1} and Algorithm \ref{alg3}, we would need to perform local Pauli-$z$ measurements to read out the solution bitstring at the final time $T_f$. We shall discuss the final readout measurements in the next section.

We would like to emphasize that although Algorithms \ref{alg1} -- \ref{alg3} are defined with finite $\Delta t$, one can also obtain continuous--time versions of each algorithm by going into the limit of $\Delta t \rightarrow 0$. In this case, the dynamics are described by Eq. \eqref{eq_Lindblad_multi} and \eqref{eq_SME_multi}, and the filter is given by Eq. \eqref{eq:filter}.

\subsection{Readout Scheme and Time-to-Solution \label{sec_ro_tts}}

Before we move on to illustrate the algorithms, we shall first introduce some performance metrics that we use to benchmark the algorithms in Secs.~\ref{SEC:2Q-2SAT} and \ref{SEC:BZF-kSAT-examples}.

For $n$-qubit systems, if we make the weak measurements in the computational basis, the Kraus operator for obtaining a readout signal tuple $\mathbf{r} = (r_1, \cdots , r_n) \in \mathbb{R}^{n}$ is
\begin{equation}\label{eq_readout_Kraus} \begin{split}
    M_{\mathbf{r}} &= \bigotimes_{j=1}^n \left(\frac{\Delta t_m}{2\pi} \right)^\frac{1}{4} \bigg\lbrace e^{-\frac{\Delta t_m}{4}\left(r_j-\frac{1}{\sqrt{\tau}} \right)^2}\frac{\hat{\openone}+\hat{\sigma}_z^{(j)}}{2}
    \\ &\quad\quad\quad\quad\quad\quad\quad\quad + e^{-\frac{\Delta t_m}{4}\left(r_j+\frac{1}{\sqrt{\tau}} \right)^2}\frac{\hat{\openone}-\hat{\sigma}_z^{(j)}}{2}\bigg\rbrace
    \\ &= \left( \frac{\Delta t_m}{2\pi}\right)^{\frac{n}{4}}\sum_{\mathbf{b} \in \{-1, +1 \}^n}e^{\frac{-\Delta t}{4}\sum_{i=1}^{n}(r_j - \frac{b_j}{\sqrt{\tau}})^2}\hat{P}_{\mathbf{b}},
\end{split} \end{equation}
where $\Delta t_m$ is the duration of the readout. 
Here $\mathbf{b} = (b_1, \cdots , b_n) \in \{-1, +1 \}^{n}$ represents a possible solution bitstring, and 
\be 
\hat{P}_{\mathbf{b}} = \bigotimes_{j=1}^n \left[\tfrac{1}{2}\left(\hat{\openone} + b_j\,\hat{\sigma}_z^{(j)} \right) \right]  
\ee
is the corresponding projector in the computational basis ($z$ basis). 
Suppose that upon performing local readouts of the state $\rho = \rho(T_f)$ returned by one of Algorithms \ref{alg1}--\ref{alg3} to obtain $\mathbf{r}$, we then generate the our candidate solution bitstring via $\tilde{\mathbf{s}} = \mathrm{sign}(\mathbf{r})$. 
Let $\mathbf{s} = (s_1, \cdots , s_n) \in \{-1, +1 \}^{n}$ be the actual solution bitstring of the $k$-SAT problem. Then the probability of getting $\tilde{\mathbf{s}} = \mathbf{s}$, i.e.,~the probability that our algorithm and readout return a correct solution, is 
\begin{subequations} \label{PrS_general}\begin{equation}
\begin{aligned}
    \mathbb{P}_{\mathbf{s}} &= \int_{0}^{\text{sign}(\mathbf{s})\infty} d \mathbf{r} \,\, \,\text{Tr}\left(M_{\mathbf{r}} \rho M_{\mathbf{r}}^{\dagger} \right)\\
    &= \sum_{\mathbf{b}}\mathrm{Tr}(\rho \hat{P}_{\mathbf{b}})\left [\prod_{j=1}^n \int_0^{\mathrm{sign}(s_j)\infty} dr_j \mathcal{G}_j \right ],
\end{aligned}
\end{equation}
where we have abbreviated 
\be 
\mathcal{G}_j = \sqrt{\frac{\Delta t_m}{2\pi}} e^{\frac{-\Delta t_m}{2}(r_j - \frac{b_j}{\sqrt{\tau}})^2}.
\ee
Simplifying this yields
\begin{equation} \begin{split}
    \mathbb{P}_{\mathbf{s}} &= \frac{1}{2^n}\sum_{\mathbf{b}}\left\{\mathrm{Tr}(\rho \hat{P}_{\mathbf{b}})\left[1+\mathrm{erf}\left(\sqrt{\frac{\Delta t_m}{2\tau}}\right) \right]^{n - h_{\mathbf{s},\mathbf{b}}} \right. \\
    &\left. \quad\quad\quad\quad\quad\quad \times\left[1-\mathrm{erf}\left(\sqrt{\frac{\Delta t_m}{2\tau}} \right) \right]^{ h_{\mathbf{s},\mathbf{b}}} \right\},
\end{split} \end{equation} \end{subequations}
where $h_{\mathbf{s},\mathbf{b}}$ is the Hamming distance between $\mathbf{s}$ and each possible bitstring $\mathbf{b}$. The above derivation gives a procedure of reading out a bitstring via generalized measurement from the final state of either quantum Algorithm \ref{alg1} or Algorithm \ref{alg3}. 
The overall general algorithm using any one of these two generalized measurement algorithms together with the final state readout is summarized in Algorithm \ref{alg4} below.

\begin{algorithm}[H]
	\caption{$k$-SAT with generalized qubit readout}
	\label{alg4}
    \begin{algorithmic}[1]
        \State \textbf{input}: a CNF Boolean formula $F$, $\tau$,  $T_f$, $\Delta t$, $\Delta t_m$,  a schedule  $\theta(t/T_f)$: if Algorithm \ref{alg3} is used, values of $T_{be}$,  $T_{min}$, $r_{th}$ are also input.
        \State run either Algorithm \ref{alg1} or Algorithm \ref{alg3}, and obtain a final state $\rho(T_f)$
        \State read out local $\hat{\sigma}_z$ of $\rho(T_f)$ with Kraus operators given in Eq. \eqref{eq_readout_Kraus}, obtaining readouts $\mathbf{r} = \{ r _i\}_{i = 1}^n$
        \State define a candidate bitstring $\mathbf{b} = \text{sign}(\mathbf{r})$
        \If{$\mathbf{b}$ is classically verified to be a solution bitstring} \State \Return verified solution $\mathbf{b}$
        \Else 
        \State repeat $2 - 4$
        \EndIf
\end{algorithmic}
\end{algorithm}

Recall from the arguments of Sec.~\ref{sec-Convergence} that in the limit of long dragging times $T_f$, $\rho$ is expected to concentrate itself in the solution subspace, i.e.,~$\mathrm{Tr}(\rho\,\hat{P}_\mathbf{b})$ will only contribute to bitstrings that are actually solutions. 
Moreover, as the readout time also becomes long ($\Delta t_m \gg \tau$, such that the local readout is effectively projective), $\mathrm{erf}(\sqrt{\Delta t_m/2\tau})$ also tends to one, so that the leading $2^{-n}$ is canceled out, and we expect to deterministically recover a correct solution bitstring.

We analyze the tradeoffs between the success probability $\mathbb{P}_{\mathbf{s}}$ in an individual run of Algorithm \ref{alg4} and algorithm runtime by first defining a time-to-solution (TTS) as 
\be \label{basic-tts}
\mathrm{TTS}^{(m)} = N\cdot (T_f + \Delta t_m),
\ee
where an algorithm realized within a dragging time $T_f$ is followed by qubit readout of duration $\Delta t_m$ and is repeated $N$ times. 
The superscript $(m)$ denotes that the final measurement time $\Delta t_m$ is included. 
Suppose we allow ourselves to perform these $N$ shots with the assumption that classically verifying whether each output bitstring $\mathbf{b}$ is actually a solution to our $k$-SAT problem is easy. 
This is true since $k$-SAT is in $\bold{\text{NP}}$.
How much time, and how many shots $N$, should be required to guarantee that the correct solution were to appear at least once with probability greater than some confidence cutoff $\mathbb{P}_{\star}$?
We can say that the probability that $M$ shots out of $N$ are successful are governed by binomial statistics, i.e.,
\begin{subequations}\be 
\mathbb{P}^{(M/N)}_\mathbf{s} = \left( \begin{array}{c} N \\ M \end{array} \right) \mathbb{P}_{\mathbf{s}}^M\,(1-\mathbb{P}_{\mathbf{s}})^{N-M}. 
\ee
Then the probability of \emph{at least one successful run} is 
\be 
1 - \mathbb{P}_{\mathbf{s}}^{(0/N)} = 1-(1-\mathbb{P}_{\mathbf{s}})^N,
\ee
where $\mathbb{P}_{\mathbf{s}}^{(0/N)}$ is the probability that all $N$ runs fail. 
Consequently our condition of interest is
\be 
1- (1-\mathbb{P}_{\mathbf{s}})^N \geq \mathbb{P}_{\star},
\ee \end{subequations}
such that the expected number of runs needed to achieve a solution with probability at least $\mathbb{P}_{\star}$ is
\be \label{Nstar}
N_\star = \mathrm{ceil}\left[\frac{\log(1-\mathbb{P}_{\star})}{\log(1-\mathbb{P}_{\mathbf{s}})} \right].
\ee 
Here, $\mathbb{P}_{\star}$ should be understood as a desired level of solution confidence specified by the experimenter. 
The expression $N_\star$ may also be understood as a means of assessing whether a given set of parameters $T/\tau$ and $\Delta t_m/\tau$ are sufficient to achieve any useful advantage between this measurement--driven algorithm and some other option. 
For example, a choice of parameters requiring $N_\star > 2^n$ to achieve the desired confidence level is clearly useless, because it would be faster to simply guess solution bitstrings at random and check whether they satisfy all $m$ clauses. 

Given the above, we may now define a modified TTS that replaces $N$ by $N_\star$,
\be \label{TTS-star} 
\mathrm{TTS}_\star^{(m)} = N_\star\cdot(T_f + \Delta t_m).
\ee
which is now consistent with the confidence bound $\mathbb{P}_{\star}$, i.e., it measures the time to solution for achieving a solution with probability at least $\mathbb{P}_{\star}$.
The readout time $\Delta t_m$ is often neglected \cite{ronnow2014defining}, in order to emphasize the \emph{scaling} of the TTS with the \emph{pure dragging} time $T_f$ alone, which is consistent with common analyses of asymptotic scaling of quantum algorithms. 

%%%%%%%%%%%%%%%%%%%%%%%%%%%%%%%%%%%%%%%%%%%%%%%%%%%%%%%%%%%%%%
%%%%%%%%%%%%%%%%%%%%%%%%%%%%%%%%%%%%%%%%%%%%%%%%%%%%%%%%%%%%%%
\section{Two--Qubit 2-SAT: From Discrete to Continuous Measurement \label{SEC:2Q-2SAT}}

In order to gain some intuition about the dynamics under our measurement-driven algorithm, it is insightful to analyze the minimal version of the SAT problem. 
2-SAT belongs to complexity class $\mathrm{P}$ and thus might be viewed as less interesting compared with other intractable SAT problems. The particular situation in which there are only two Boolean variables involved in the 2-SAT CNF formula is nevertheless of interest, here because the application of our generalized measurement BZF algorithms 1-3 illustrates the continuous form of BZF dynamics, while remaining relatively simple and accessible to detailed analysis. 
We will term this problem the 2-qubit 2-SAT.

%%%%%%%%%%%%%%%%%%%%%%%%%%%%%%%%%%%%%%%%%%%%%%%%%%%%%%%%%%%%%%
\subsection{The Simplest 2-SAT Problem \label{sec-2SAT}}

2-qubit 2-SAT is the minimal version of the SAT problem that captures most of the significant features of SAT while involving a minimum number of qubits and clause measurements. 
In this section, we will analyze the algorithmic dynamics of the two-qubit system under generalized measurements. 
In particular, we will show that for a given total algorithm running time, weak continuous measurements lead to the highest success probability for finite $T/\tau$. This motivates a detailed study of those continuous dynamics. We first examine the unconditioned average dynamics in Algorithm \ref{alg1} with LME \eqref{eq_Lindblad_multi}, 
and then consider the heralded dynamics of Algorithm \ref{alg3} with the SME of Eq. \eqref{eq_SME_multi}.

More specifically, we consider a 2-qubit 2-SAT problem with a single satisfying solution. Without loss of generality such a problem can be defined by the following CNF formula
\begin{equation}
    F = C_1 \wedge C_2 \wedge C_3, 
\end{equation}
where 
\begin{equation} 
    \begin{aligned}
        C_1 &= b_1 \vee b_2 \\
        C_2 &= b_1 \vee \bar{b}_2\\
        C_3 &= \bar{b}_1 \vee \bar{b}_2.
    \end{aligned}\label{2q2SAT_clause}
\end{equation}
It can be easily checked that the only solution to this simple 2-qubit 2-SAT problem is $(b_1, b_2) = (0, 1)$. Classically, one can imagine solving this by the following procedure.  In the space $\{ 0, 1\}^2$, each clause excludes an assignment that will violate it. For example, $C_1= b_1 \vee b_2$ will exclude  $(b_1, b_2) = (1, 1)$ as a solution assignment. After the last clause check, the legal assignments that survive this procedure correspond to the solution. The BZF algorithm is similar, in that each clause measurement checks whether the quantum state is projected into the subspace it excludes as a failure subspace and the surviving quantum state conditioned on successful clause checks is the solution subspace. 
The three observables corresponding to the three clause checks in Eq.~\eqref{2q2SAT_clause} are given by
\begin{subequations}\begin{equation} \label{eq-2SAT-observables}
    \begin{aligned}
         \hat{X}_1(\theta) &= \frac{1}{2} \left\{\hat{\openone} +  \hat{\sigma}^{(1)}_{+\theta} +  \hat{\sigma}^{(2)}_{+\theta} +  \hat{\sigma}^{(1)}_{+\theta}  \,\hat{\sigma}^{(2)}_{+\theta} \right\} \\
         \hat{X}_2(\theta) &= \frac{1}{2} \left\{\hat{\openone} +  \hat{\sigma}^{(1)}_{+\theta} +  \hat{\sigma}^{(2)}_{-\theta} +  \hat{\sigma}^{(1)}_{+\theta}\,  \hat{\sigma}^{(2)}_{-\theta} \right\} \\
         \hat{X}_3(\theta) &= \frac{1}{2} \left\{\hat{\openone} +  \hat{\sigma}^{(1)}_{-\theta} +  \hat{\sigma}^{(2)}_{-\theta} +  \hat{\sigma}^{(1)}_{-\theta}\,  \hat{\sigma}^{(2)}_{-\theta} \right\},
    \end{aligned}
\end{equation}
with
\begin{equation}
    \hat{\sigma}^{(i)}_{\pm \theta} = \mathrm{cos}(\theta)\hat{\sigma}_x^{(i)} \mp \mathrm{sin}(\theta)\hat{\sigma}_z^{(i)}
\end{equation} \end{subequations}
where $\hat{\sigma}_x^{(i)}$ and $\hat{\sigma}_z^{(i)}$ are Pauli-$x$ and Pauli-$z$ operators that only acts on the $i^\mathrm{th}$ qubit. $\hat{\sigma}_x^{(i)}(\theta)$ is then the Pauli-$x$ operator rotated clockwise in the $x$-$z$ plane by $\theta$ as indicated in Fig. \ref{fig-Rotation}.

The quantum system is then driven by the generalized measurements of these three observables, and we adopt a simple linear schedule for the control parameter $\theta$: for the $c^{\mathrm{th}}$ cycle of clause measurements, $\theta_c = \frac{c}{N}\cdot \frac{\pi}{2}$ where $N$ is the total number of cycles of clause measurements. The instantaneous solution state $|\phi_{soln}(\theta)\rangle$, which is the common ``$\mathrm{+1}$"-eigenstate of $ \hat{X}_1(\theta),  \hat{X}_2(\theta),  \hat{X}_3(\theta)$ is given by
\begin{equation} \label{2sat-solstate}
\begin{aligned}
    |\phi_{soln}(\theta)\rangle &= \left[\mathrm{cos}\left(\frac{\theta}{2} + \frac{\pi}{4}\right)|0\rangle + \mathrm{sin}\left(\frac{\theta}{2} + \frac{\pi}{4}\right)|1\rangle  \right]\\
    & \otimes \left[\mathrm{cos}\left(-\frac{\theta}{2} + \frac{\pi}{4}\right)|0\rangle + \mathrm{sin}\left(-\frac{\theta}{2} + \frac{\pi}{4}\right)|1\rangle  \right].
\end{aligned}
\end{equation}
The probability of successfully implementing the algorithm and thus also the running time then depend primarily on the ability to follow this instantaneous solution state in an adiabatic (Zeno) sense.

%%%%%%%%%%%%%%%%%%%%%%%%%%%%%%%%%%%%%%%%%%%%%%%%%%%%%%%%%%%%%%
\subsection{Zeno Dragging for Discrete and Continuous Generalized Measurements \label{sec:2Q-2SAT_ZenoDrag}}

\textcite{Burgarth2020quantumzenodynamics} showed that the weak continuous limit is favorable for generating Zeno dynamics in the case of single measurement channel. We now show that a similar result can be obtained for the dynamics generated by multiple measurement channels. In particular, we will demonstrate that dynamics driven by weak continuous measurements approaches the target dynamics $|\phi_{soln}(\theta)\rangle$ better than discrete stronger measurements executed using the same total measurement strength and duration. 

\begin{figure}
    \centering
    \includegraphics[width = \columnwidth,trim = {10 20 20 10},clip]{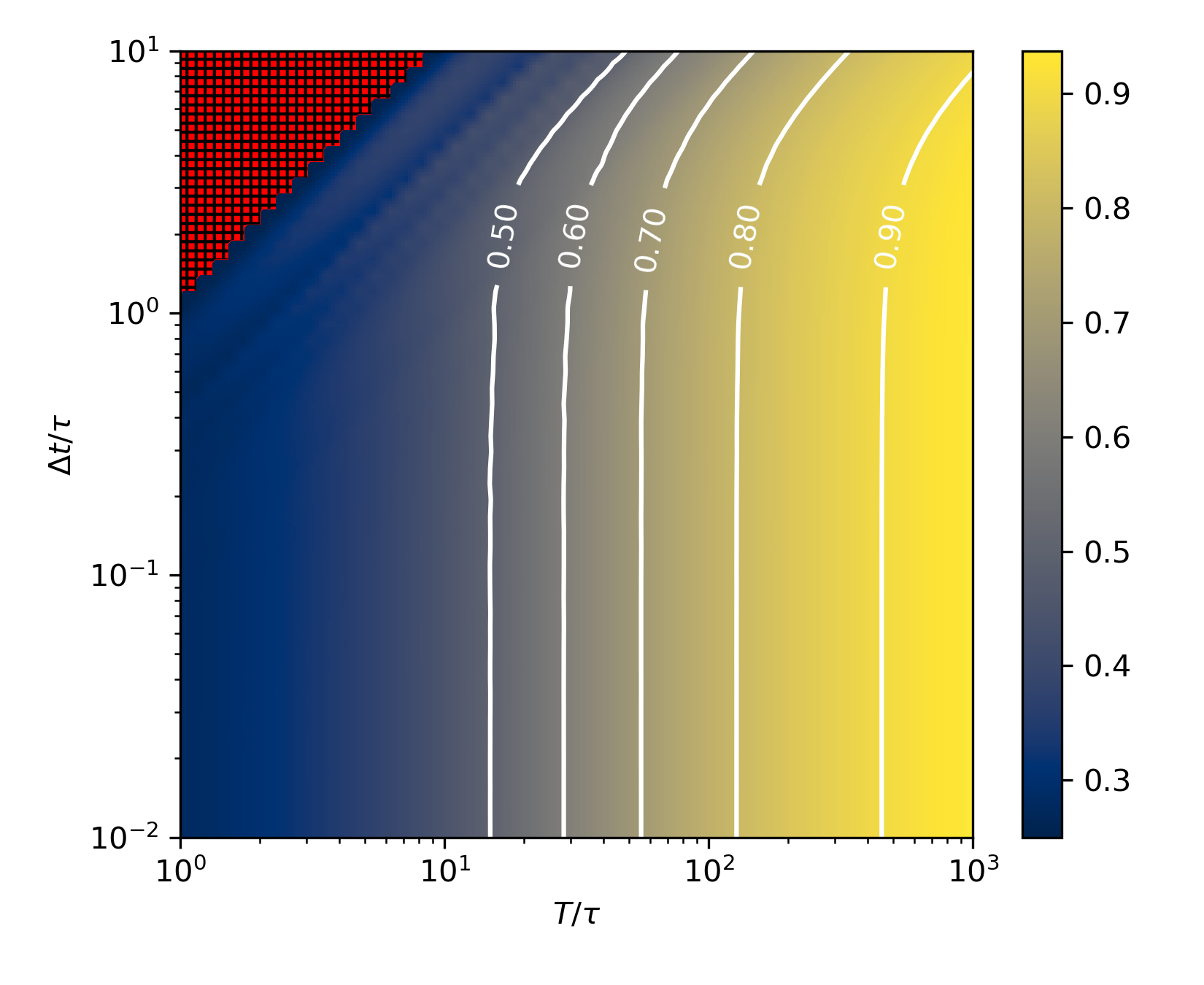}
    \includegraphics[width = \columnwidth,trim = {10 20 50 32},clip]{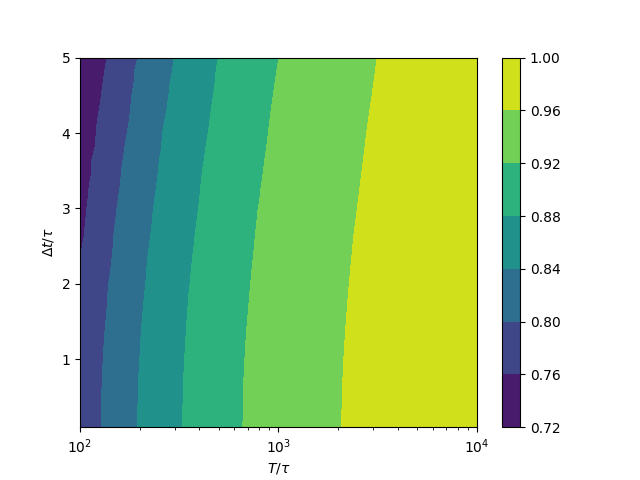} \\
    \begin{tikzpicture}[overlay]
    \node[draw = white, fill = white] at (-0.1,0.0) {\small $T_f/\tau$};
    \node[draw = white, fill = white] at (-0.1,6.9) {\small $T_f/\tau$};
    \end{tikzpicture} 
    \caption{ 
    Fidelity of the final state $\mathscr{F} = \langle 10 | \rho_f |10\rangle $ for the 2-SAT problem on 2 qubits Eq.~\eqref{2q2SAT_clause}, shown as a function of single clause measurement time $\Delta t/\tau$ and total dragging time $T_f/\tau$,
    where we set the time scale for the generalized measurement $\tau = 1$  to be the reference time scale.
    The two panels differ only in the scale and range of their axes.
    These results are obtained from simulating the average dynamics driven by measuring the three clause observables $ \hat{X}_1(\theta)$, $ \hat{X}_2(\theta)$ and $ \hat{X}_3(\theta)$ in the 2-qubit 2-SAT problem Eq.~\eqref{2q2SAT_clause} according to Eq.~\eqref{avg_dynamics}. 
    For a given dragging time $T_f = \ell \cdot \Delta t$, where $T_f$ is subdivided into $\ell$ measurements each of duration $\Delta t$, the optimal value of $\Delta t/\tau$ is found at $\Delta t/\tau \rightarrow 0$, which implies that $\ell \rightarrow \infty$.
    The upper panel also shows small values of $T_f \sim \Delta t$: in this region we set $\Delta t$ to the nearest $\ell$ value that evenly subdivides $T$. 
    The small fidelity oscillations seen in this region are due to the discrete nature of this operation. 
    For $\Delta t > T_f$ (red hatched region, which is forbidden), we perform only the clause measurement at $\theta = \pi/2$, yielding the lowest possible value of fidelity $\mathscr{F} = \tfrac{1}{4}$ for a two-qubit problem. 
    This forbidden region is a general feature of such an analysis, and informs the behavior of the surrounding contours. 
    The maximization of the fidelity for a given $T_f/\tau$ via taking $\Delta t\rightarrow 0$ illustrates the advantages of operating in the limit of continuous weak measurements that is located towards the bottom edge of each plot.
    } \label{fig:avg_weak_meas_countor}
\end{figure}
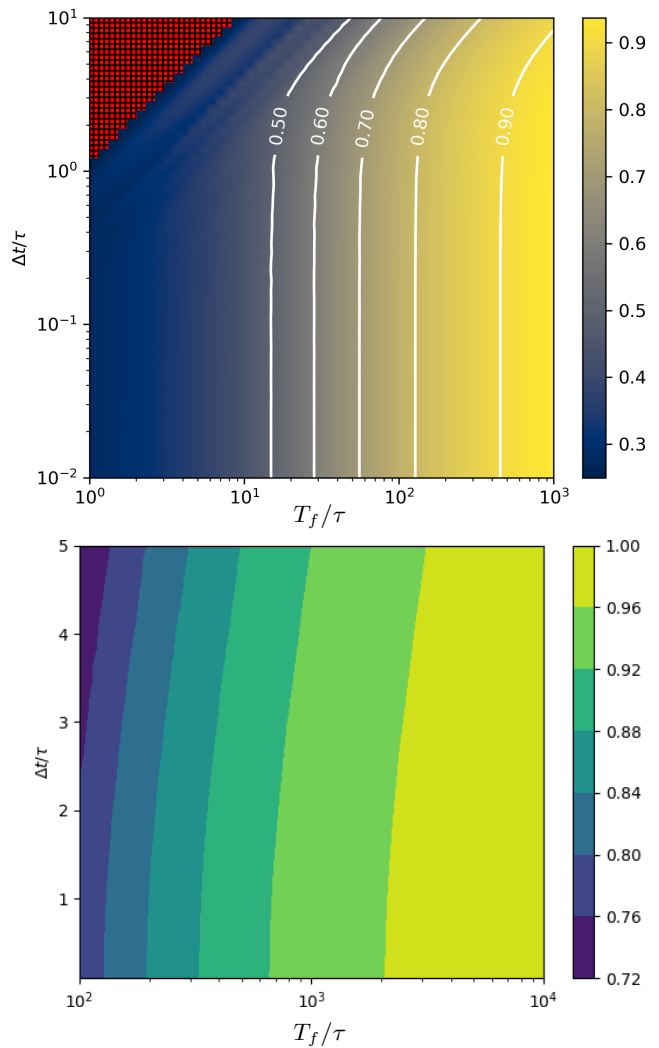

It will be adequate to consider the average dynamics given by Eq.~\eqref{avg_dynamics} to make this point. 
Let the final state following the average dynamics driven by the generalized measurements be $\rho_f$. We then calculate the fidelity $f$ between $\rho_f$ and $|\phi_{soln}(\theta = \frac{\pi}{2})\rangle = |10\rangle$, i.e., $f = \langle 10 | \rho_f |10\rangle$. 
More specifically, we calculate the fidelity $f(\Delta t/\tau, T_f/\tau)$ as a function of both $\Delta t/\tau$, which controls the effective individual measurement strength, with $\tau$ the measurement time, and $T_f/\tau$, where $T_f = l\cdot \Delta t$ is the total duration of Zeno dragging.
Fig.~\ref{fig:avg_weak_meas_countor} illustrates the relevant simulation of the average dynamics Eq.~\eqref{avg_dynamics} for the 2-SAT problem on 2 qubits. 
This computation is consistent with the intuition that for a given measurement strength, increasing the total duration $T_f$ (thus dragging more slowly) brings us towards the adiabatic / Zeno regime. 
However, for a given fixed and finite value of $T_f$, the fidelity finds its maximum value at $\Delta t/\tau = 0$. This means that rather than multiple pulsed and discrete strong measurements, Zeno dragging in our two--qubit 2-SAT problem is most effective in the continuum limit, for any given value of $T_f/\tau$.

%%%%%%%%%%%%%%%%%%%%%%%%%%%%%%%%%%%%%%%%%%%%%%%%%%%%%%%%%%%%%%
%%%%%%%%%%%%%%%%%%%%%%%%%%%%%%%%%%%%%%%%%%%%%%%%%%%%%%%%%%%%%%
\subsection{Illustration of Adiabatic Convergence via Lindblad Dynamics\label{sec-Convergence-2Q2SAT}}

\begin{figure*}
\centering
~\\
\hspace{-0.75cm}\includegraphics[height = .25\textheight, width = .225\textheight, trim = {51 0 41 45}, clip]{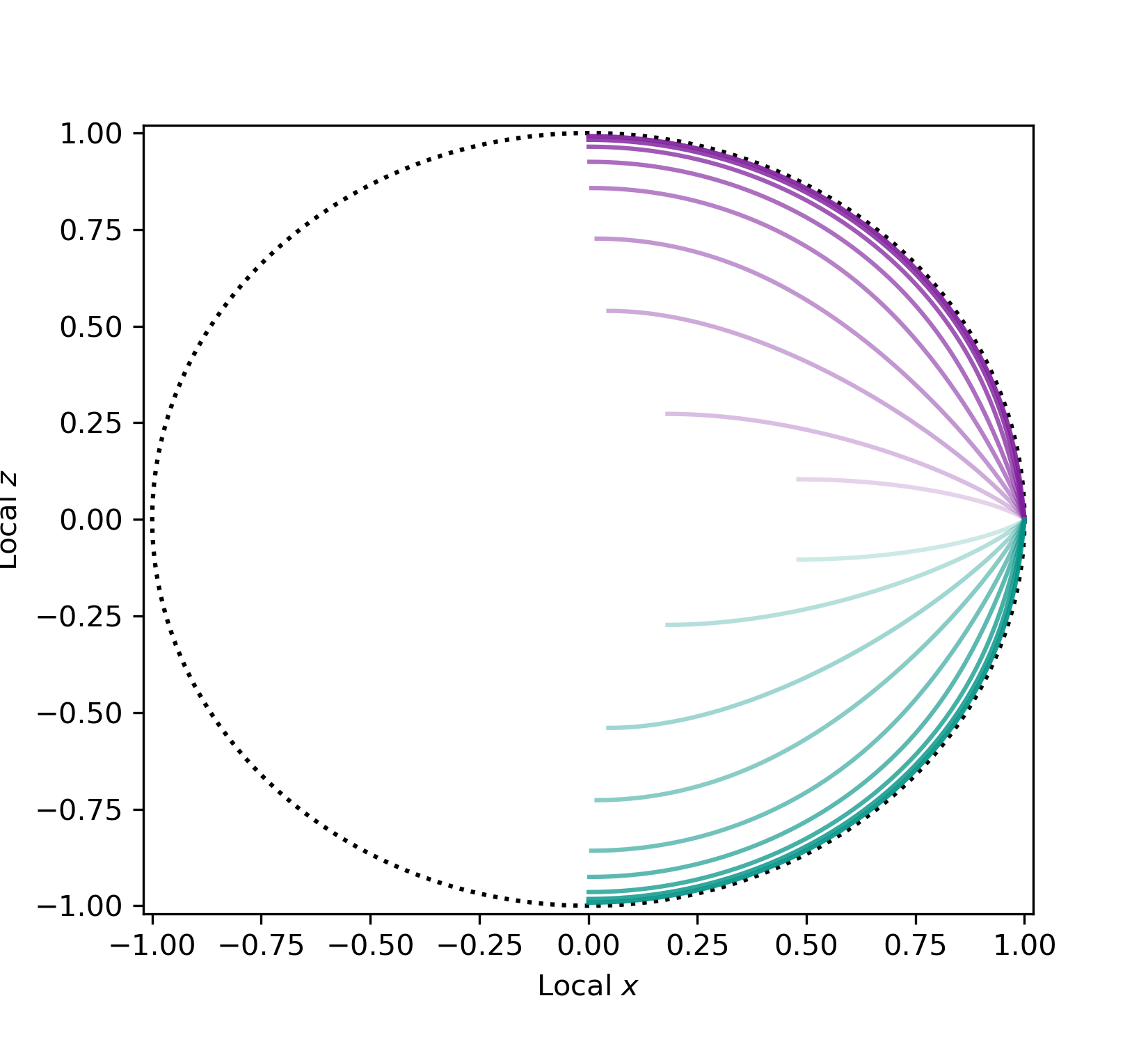}
\includegraphics[width = .49\textwidth]{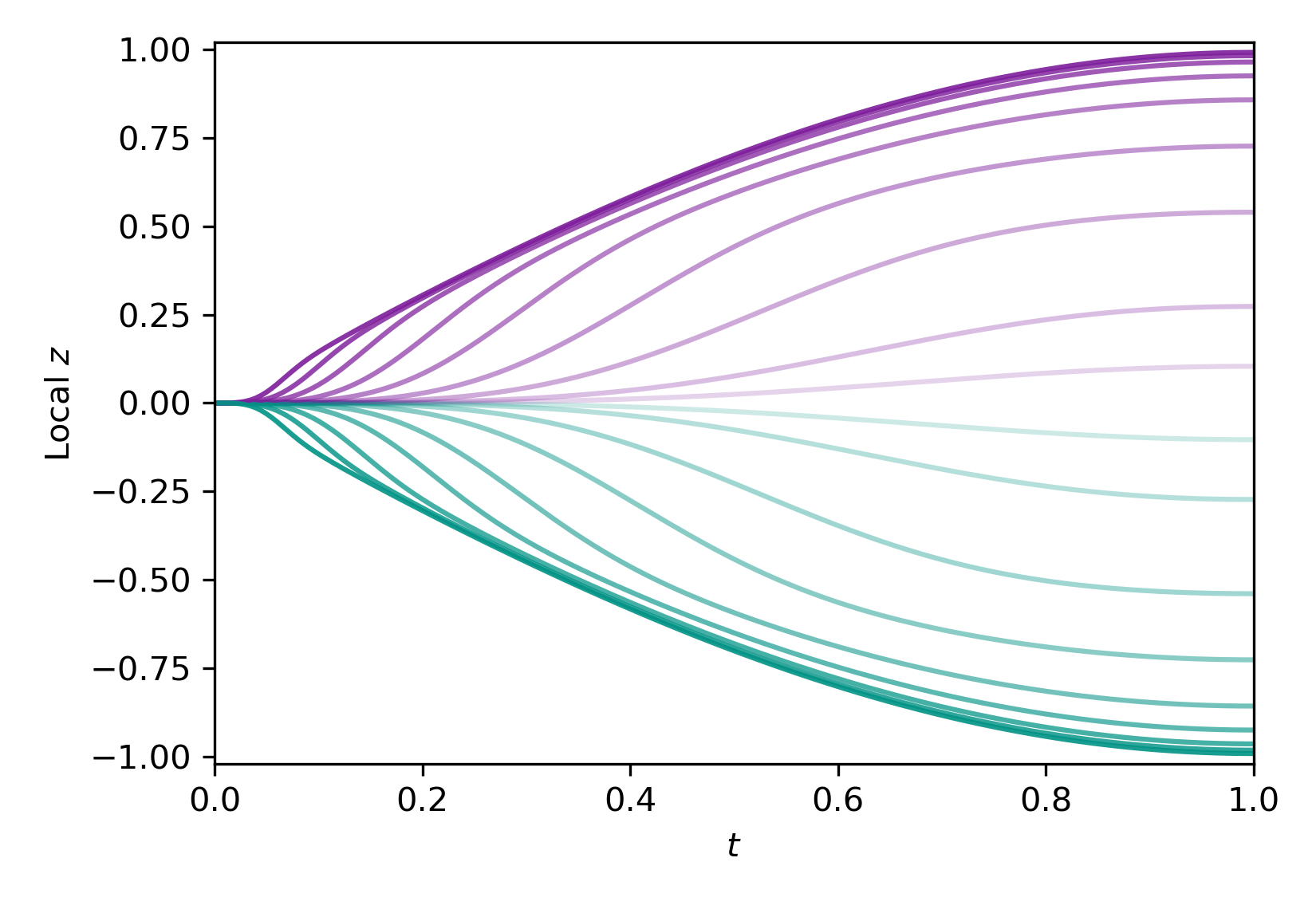} \hspace{1cm}
\\
\includegraphics[width = .48\textwidth]{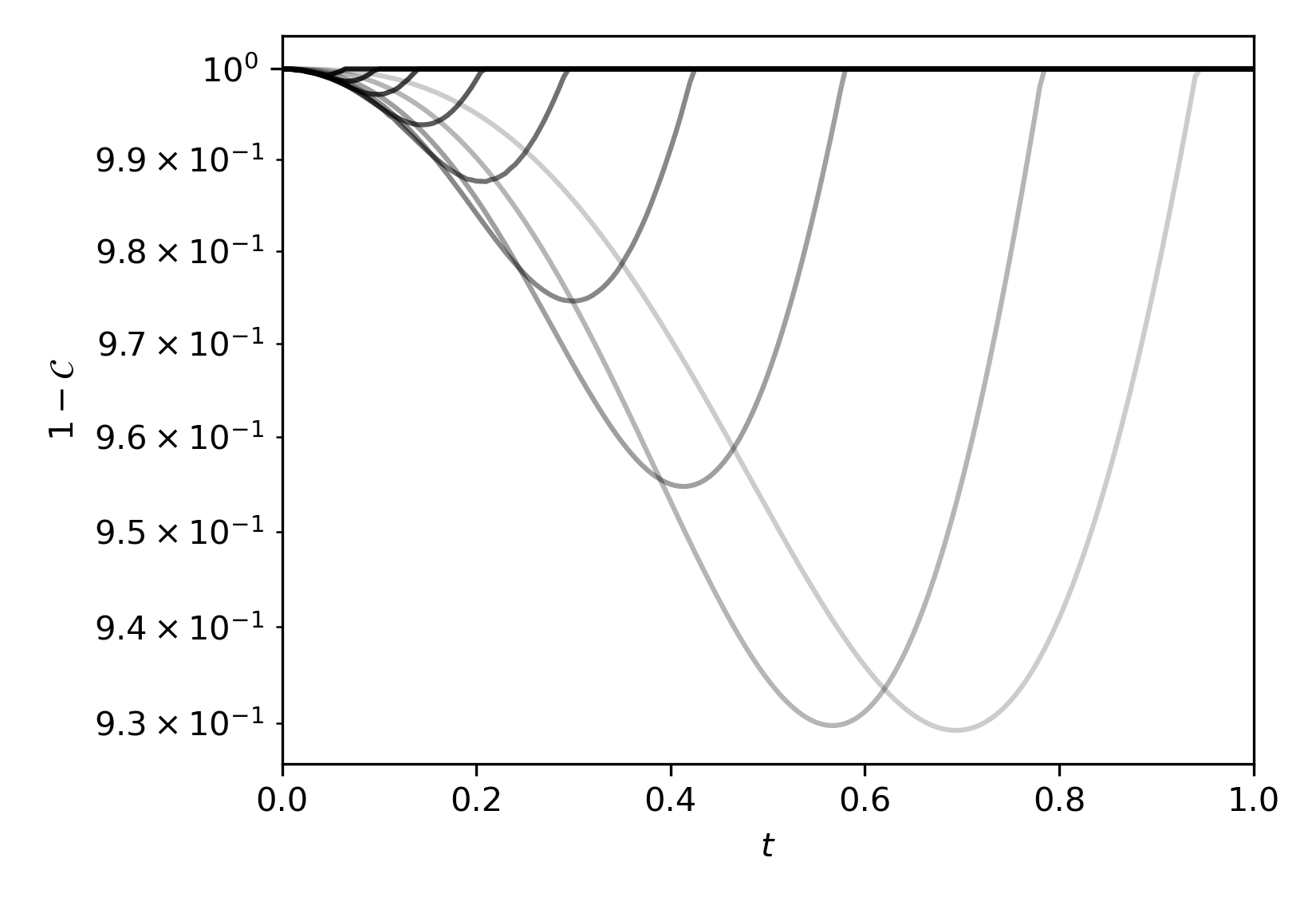} 
\includegraphics[width = .48\textwidth]{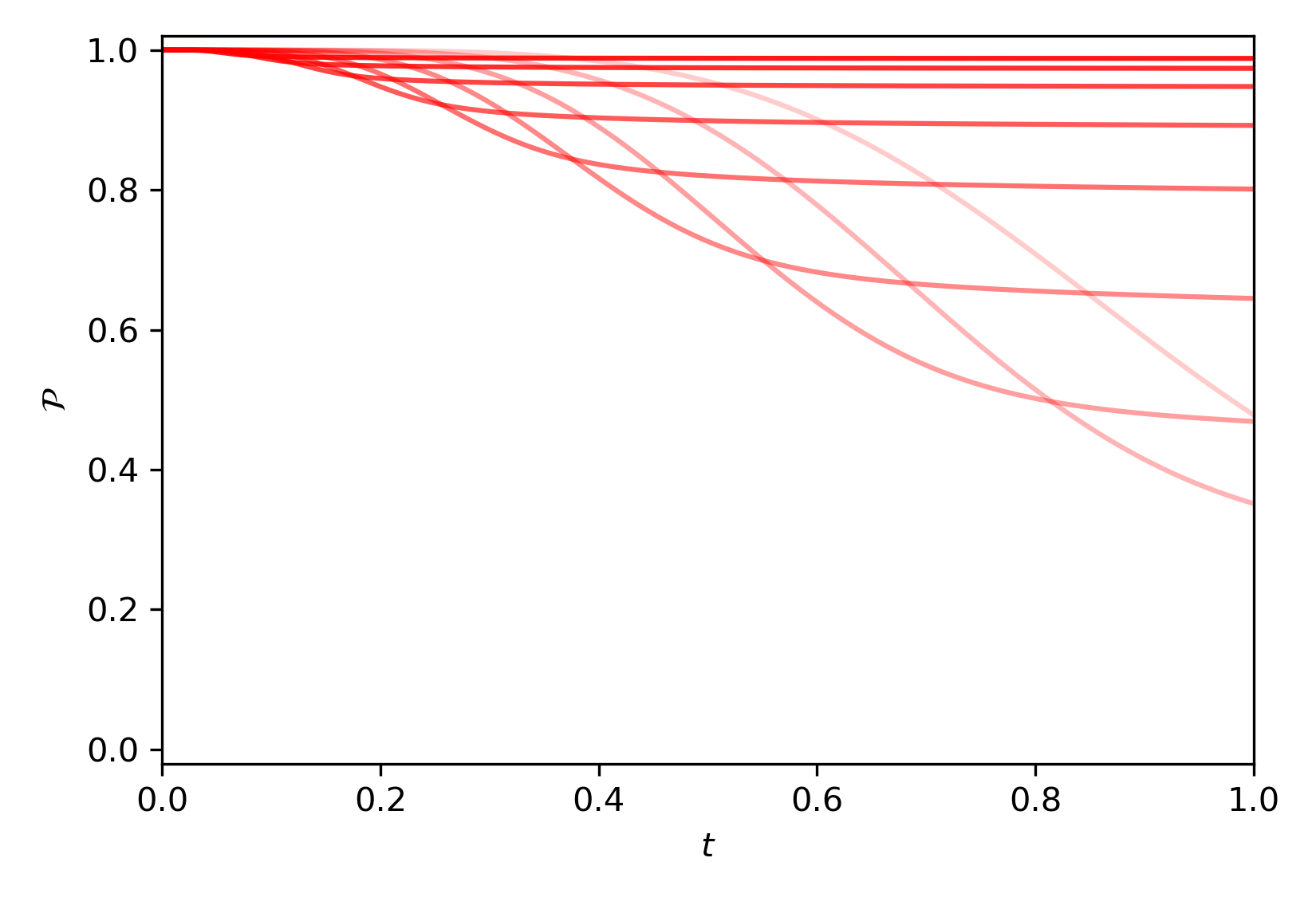} \\
\begin{tikzpicture}[overlay, yshift = -21pt]
\draw[fill = white, draw = white] (-8,7.3) rectangle (-1,7.95);
\node[] at (-7,12.6) {(a)};
\node[] at (-0.2,12.6) {(b)};
\node[] at (-6.5,2.3) {(c)};
\node[] at (1.5,2.3) {(d)};
%%%%%%%%%%%%%%%%%
\node[] at (-4.8,13.2) {$\ket{1}$};
\node[] at (-4.8,7.7) {$\ket{0}$};
\node[] at (-2,12) {\color{electricviolet} $\rho_2$};
\node[] at (-2,9) {\color{mint} $\rho_1$};
%%%%%%%%%%%%%%%%%
\node[] at (1.5,12) {\color{electricviolet} $z_2$};
\node[] at (1.5,9) {\color{mint} $z_1$};
\node[] at (7.5,13.0) {\color{patriarch!75!black} \footnotesize $\Gamma\,T_f = 1\times 10^4$};
\node[] at (7.5,12.7) {\color{patriarch!90!black} \footnotesize $\Gamma\,T_f = 3\times 10^3$};
\node[] at (7.5,12.4) {\color{patriarch!90!white} \footnotesize $\Gamma\,T_f = 1\times 10^3$};
\node[] at (7.5,12.1) {\color{patriarch!75!white} \footnotesize $\Gamma\,T_f = 3\times 10^2$};
\node[] at (7.5,11.8) {\color{patriarch!60!white} \footnotesize $\Gamma\,T_f = 1\times 10^2$};
\node[] at (7.5,11.5) {\color{patriarch!45!white} \footnotesize $\Gamma\,T_f = 3\times 10^1$};
\node[] at (7.5,11.2) {\color{patriarch!37!white} \footnotesize $\Gamma\,T_f = 1\times 10^1$};
\node[] at (7.5,10.9) {\color{patriarch!30!white} \footnotesize $\Gamma\,T_f = 3\times 10^0$};
\node[] at (7.5,10.6) {\color{patriarch!25!white} \footnotesize $\Gamma\,T_f = 1\times 10^0$};
\node[] at (7.5,7.9) {\color{mint!75!black} \footnotesize $\Gamma\,T_f = 1\times 10^4$};
\node[] at (7.5,8.2) {\color{mint!90!black} \footnotesize $\Gamma\,T_f = 3\times 10^3$};
\node[] at (7.5,8.5) {\color{mint!90!white} \footnotesize $\Gamma\,T_f = 1\times 10^3$};
\node[] at (7.5,8.8) {\color{mint!75!white} \footnotesize $\Gamma\,T_f = 3\times 10^2$};
\node[] at (7.5,9.1) {\color{mint!60!white} \footnotesize $\Gamma\,T_f = 1\times 10^2$};
\node[] at (7.5,9.4) {\color{mint!45!white} \footnotesize $\Gamma\,T_f = 3\times 10^1$};
\node[] at (7.5,9.7) {\color{mint!37!white} \footnotesize $\Gamma\,T_f = 1\times 10^1$};
\node[] at (7.5,10.0) {\color{mint!30!white} \footnotesize $\Gamma\,T_f = 3\times 10^0$};
\node[] at (7.5,10.3) {\color{mint!25!white} \footnotesize $\Gamma\,T_f = 1\times 10^0$};
\end{tikzpicture} \vspace{-10pt}
\caption{
Properties of the Lindblad dynamics for the $2$--SAT problem of Eq.~\eqref{2q2SAT_clause} illustrated in  Fig.~\ref{fig-projectors}, on scanning through a range of values of measurement strength $\Gamma = 1/(4\tau)$. The same measurement strengths and linear schedule $\theta = \pi\,t/2\,T_f$ are used through all four panels. The time axis is displayed here in units of the total evolution time $T_f$.
Panel (a) illustrates the reduced density matrix evolution for qubit 1 (teal) and qubit 2 (purple), shown for both in the $xz$ Bloch plane; results are shown for a variety of measurement strengths spanning from $\Gamma\,T_f = 1$ (pale) to $\Gamma\,T_f = 1\times 10^4$ (dark). 
Panel (b) shows the corresponding time traces of the local $z$ coordinates for the two qubits.
Panel (c) shows the behavior of the separability ($1 - \mathcal{C}$, where $\mathcal{C}$ is the concurrence~\cite{Wooters_1998}) throughout the Lindblad evolution. 
Panel (d) plots the purity of the two--qubit state as it undergoes Lindblad evolution. 
These results show that in the adiabatic regime $\Gamma\,T_f\gg 1$, we converge towards a pure--state and completely separable evolution (see Eq. \eqref{2sat-solstate}) that deterministically generates the solution of this very simple $2$--SAT problem, i.e., $z_1 = 1$ and $z_2 = -1$, corresponding to $(b_1, b_2) = (0, 1)$ as expected from Eq.~\eqref{2q2SAT_clause}. 
We showed in Sec.~\ref{sec-Convergence} that this is in fact a general feature of $k$--SAT problems with a unique solution in this measurement--driven algorithm. 
Even far from the adiabatic regime ($\Gamma\,T \sim 1$), we still see that some information about this solution manifests itself on average, since the reduced density matrix traces (local $z$ coordinates in panel (b)) still drift discernibly apart and move towards their respective solution states. 
}\label{fig:gam-scan-2SAT-lind}
\end{figure*}
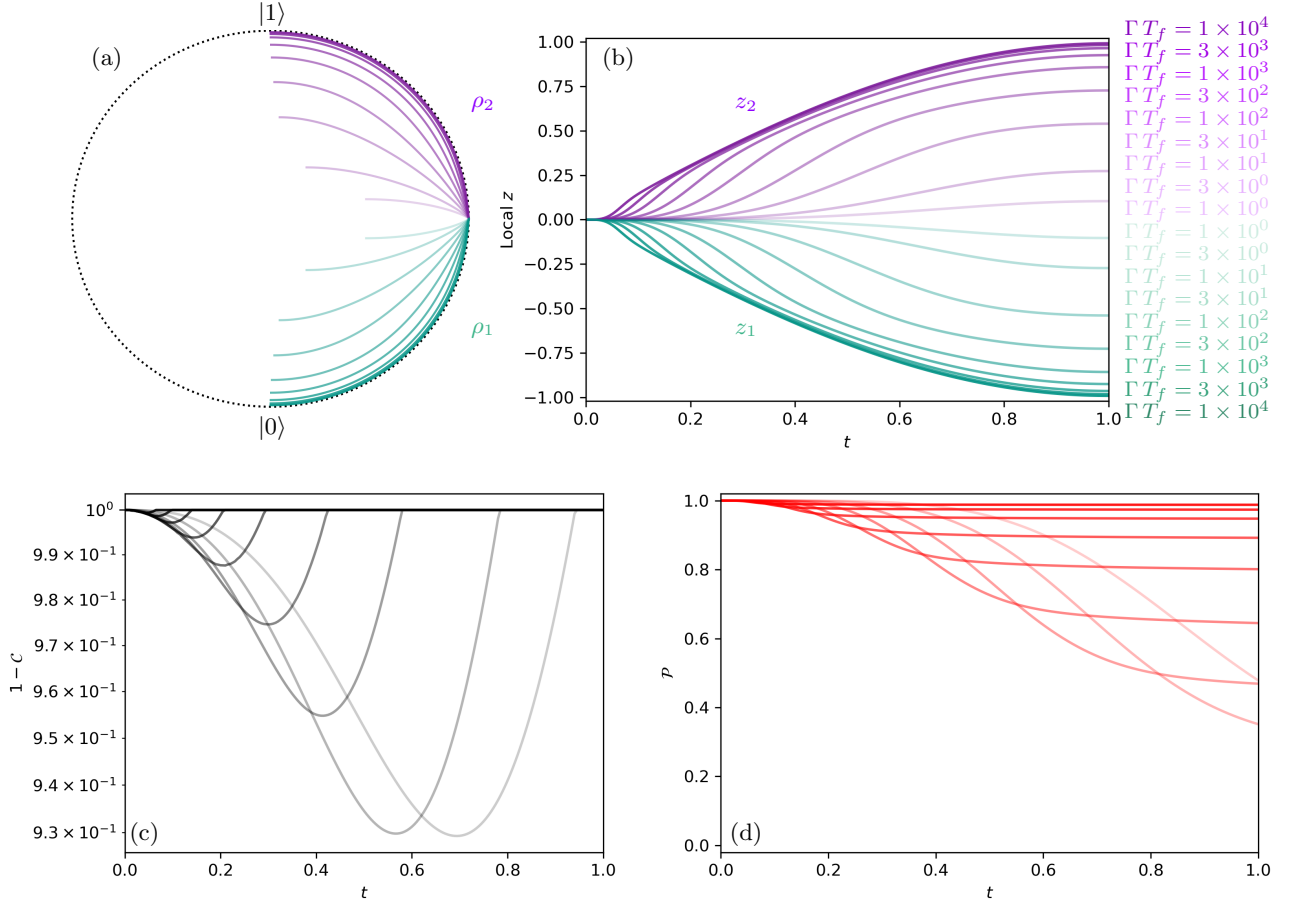

We demonstrate here the convergence of the LME Eq.~\eqref{eq_Lindblad_multi} to the adiabatic limit in the long time using the 2-qubit 2-SAT problem. Firstly, it %would be 
is illustrative to show the exact diagonal form of the solution subspace in the Zeno frame defined by Eq.~\eqref{Qframe}. For the 2-qubit 2-SAT example in Fig.~\ref{fig-projectors}, we have $\hat{Q} = \hat{R}_Y(\theta - \pi/2)\otimes \hat{R}_Y(\pi/2-\theta)$, and the observables of Eq. \eqref{eq-2SAT-observables}  are then equal to
\begin{subequations}\label{eq_2q1sat_obs_Q} \be 
\hat{\mathcal{X}}_1 =  \left(
\begin{array}{cccc}
\cos (2 \theta ) & \sin (2 \theta ) & 0 & 0 \\
\sin (2 \theta ) & -\cos (2 \theta ) & 0 & 0 \\
 0 & 0 & {\color{mint}1} & 0 \\
 0 & 0 & 0 & 1 \\
\end{array}
\right), \ee \be 
\hat{\mathcal{X}}_2 =   \left(
\begin{array}{cccc}
 1 & 0 & 0 & 0 \\
 0 & -1 & 0 & 0 \\
 0 & 0 & {\color{mint}1} & 0 \\
 0 & 0 & 0 & 1 \\
\end{array}
\right), \ee \be 
\hat{\mathcal{X}}_3 = \left(
\begin{array}{cccc}
 1 & 0 & 0 & 0 \\
 0 & - \cos (2 \theta ) & 0 &  \sin (2 \theta ) \\
 0 & 0 & {\color{mint}1} & 0 \\
 0 &  \sin (2 \theta ) & 0 & 1 \cos (2 \theta ) \\
\end{array}
\right) 
\ee\end{subequations}
in the $\hat{Q}$--frame, represented in the $\lbrace \ket{11}, \ket{10}, \ket{01}, \ket{00}\rbrace^\top$ basis. 
The solution state, which is now $\theta$--independent, is marked in green in Eq.~\eqref{eq_2q1sat_obs_Q}. 
Clearly, the solution subspace is isolated in a block diagonal form and becomes $\theta$-independent, as discussed in previous sections.

We may now consider the dynamics of the algorithm on average, i.e,~as modelled by Eq. \eqref{Lindblad_eq}. 
These dynamics represent the average performance, but can also be interpreted as the dynamics arising in the event that we dissipate information to the environment without actually detecting it \cite{BookWiseman, BookJacobs, BookJordan, BookBarchielli}. 
This reflects the perspective that the average / Lindbladian dynamics are equivalently the conditional dynamics in the limit of vanishing measurement efficiency. 
In looking at the Lindblad dynamics, we are then looking at an ``un-heralded'' version of the continuous--time BZF algorithm, which is the time-continuous limit $\Delta t /\tau \rightarrow 0$ of Algorithm \ref{alg1}.  
Note that from the discussion of convergence in Sec.~\ref{sec-Convergence}, we can expect to deterministically achieve perfect solution dynamics in the limit of long dragging time, even in the case
without detection. 

Some key features of these Lindblad algorithm dynamics are illustrated in Fig.~\ref{fig:gam-scan-2SAT-lind}.
Here we integrate the Lindblad dynamics for a variety of $\Gamma\,T_f$ values (where $\Gamma = 1/4\tau$ is the measurement rate), scanning from the quasi-adiabatic / Zeno regime $\Gamma\,T_f \gg 1$, to the diabatic regime $\Gamma\,T_f \sim 1$. 
We see that in the adiabatic limit $\Gamma\,T_f \rightarrow \infty$, the dynamics converge to the pure and separable solution Eq. \eqref{2sat-solstate}, as expected. 
For smaller values of $\Gamma\,T_f$, the fidelity is clearly reduced. We show that the reduced density matrices become less and less pure, with reduced contrast along the local $z$--axes along which we wish to read out a final 2-SAT solution. 
This loss of purity in the reduced density matrices is due \emph{both} to overall purity loss of the two--qubit state due to dissipation, \emph{and} due to the formation of spurious measurement--induced entanglement between the qubits when the Lindblad algorithm is constrained to finish within a finite time.

In Fig.~\ref{fig:gam-scan-2SAT-lind}(a,b) (and also in Fig.~\ref{fig:2Q-2SAT-TTS}(a) below) one may immediately observe that as the dragging time $T_f$ increases, the system enters a time regime where lengthening the algorithm evolution time leads to rapid improvement in the contrast between the output states (around $T_f \lesssim 20$). 
However, the benefit of increasing $T_f$ is comparatively small beyond this point ($T_f \gtrsim 20$). 
Therefore, to obtain the solution state with a desired high probability, we expect that there should exist an optimal runtime value $T_f^{opt}$, such that repeating this Lindbladian dragging process some number of times leads to extraction of the solution at the desired confidence level. 

%%%%%%%%%%%%%%%%%%%%%%%%%%%%%%%%%%%%%%%%%%%%%%%%%%%%%%%%%%%%%%

\subsection{Illustration of Readout Scheme and TTS}

In the Lindbladian setting the observer gains \emph{no} information during the algorithm evolution time itself, due to performing dissipation without detection. 
All solution information must then be obtained from local measurements to read out the qubit states at the end of the algorithm, as described in Algorithm \ref{alg4} and Sec.~\ref{sec_ro_tts}. 

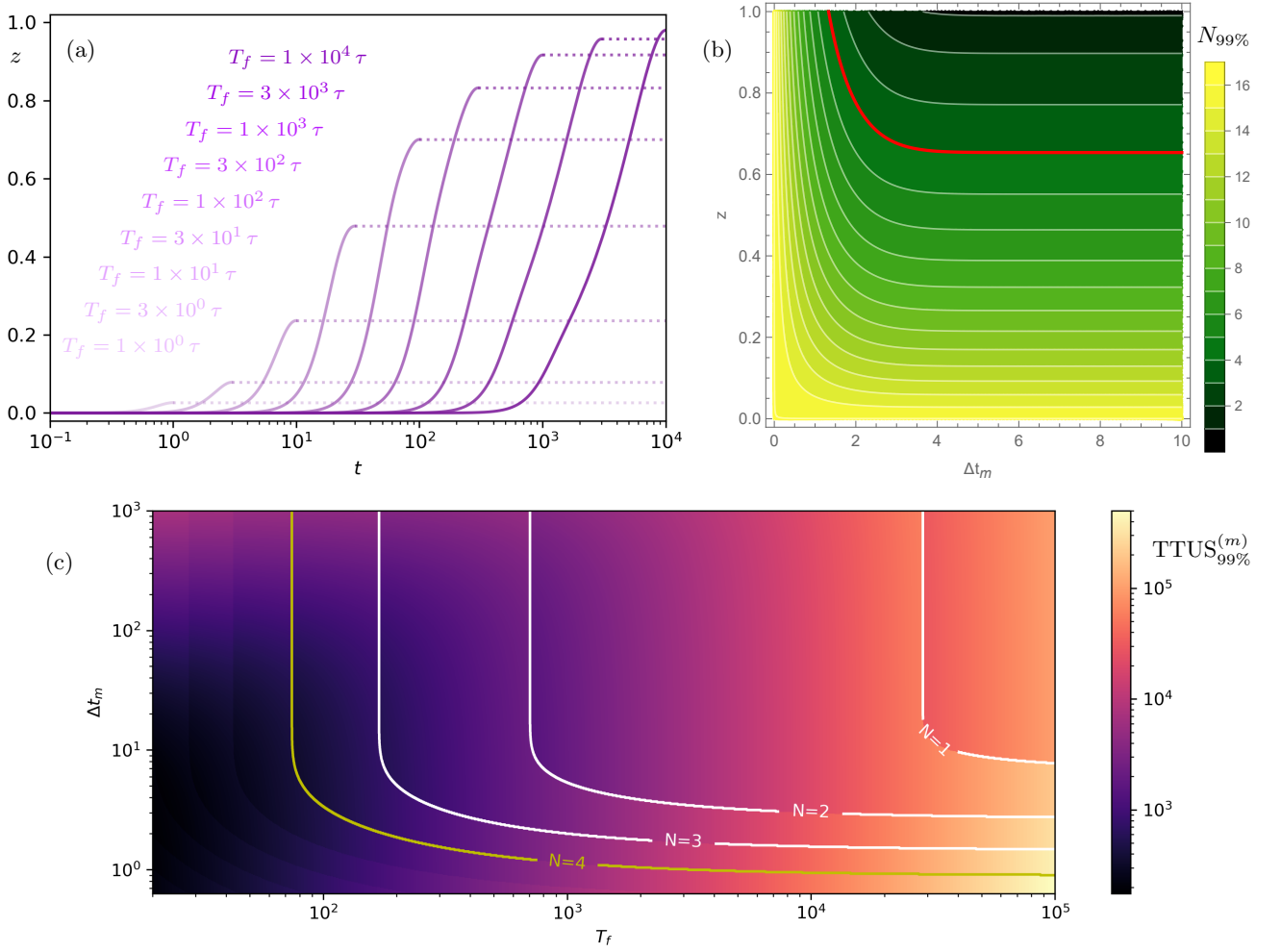
\begin{figure*}
\includegraphics[width = .54\textwidth, trim = {25 16 0 0},clip]{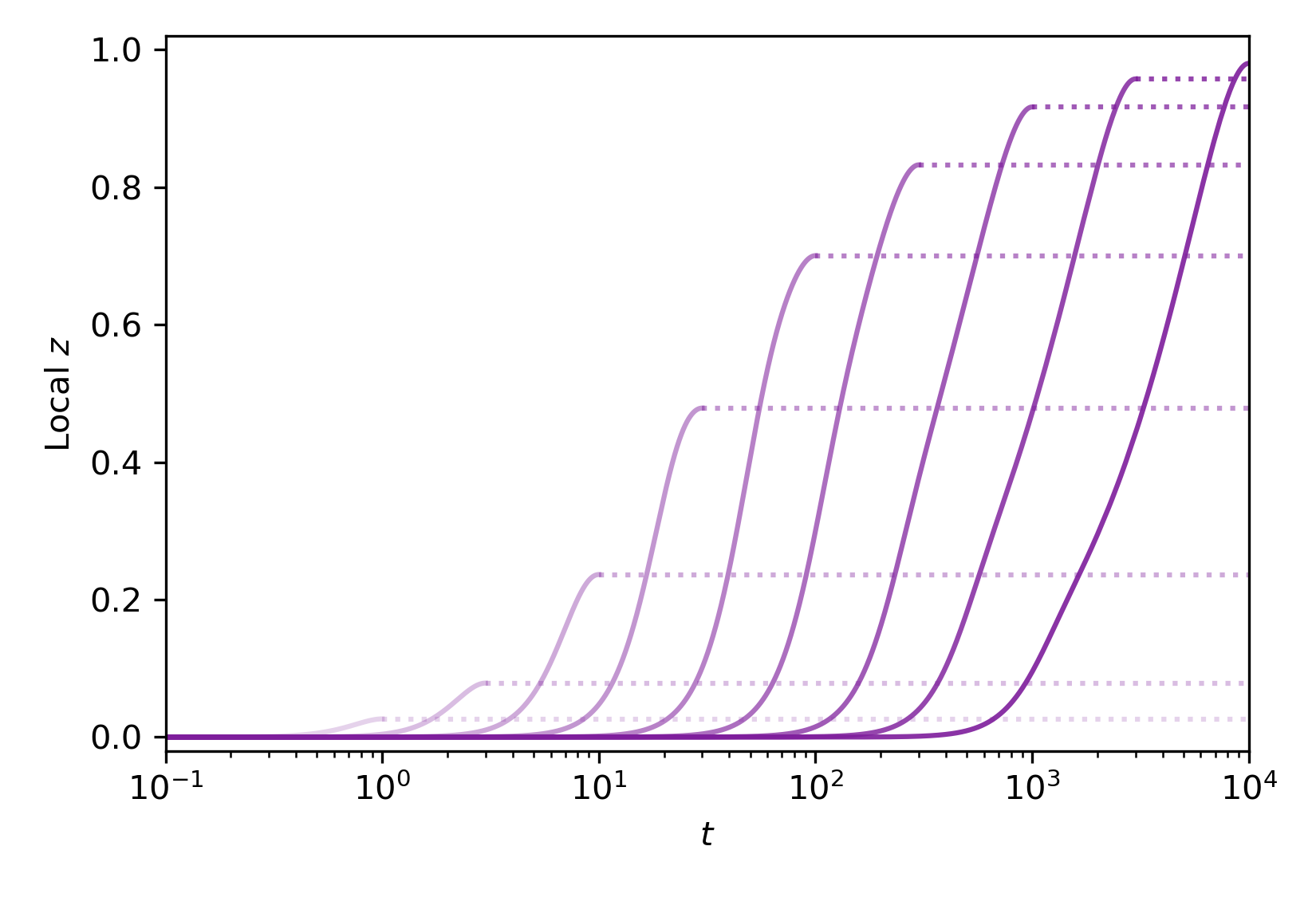}
\includegraphics[width = .37\textwidth]{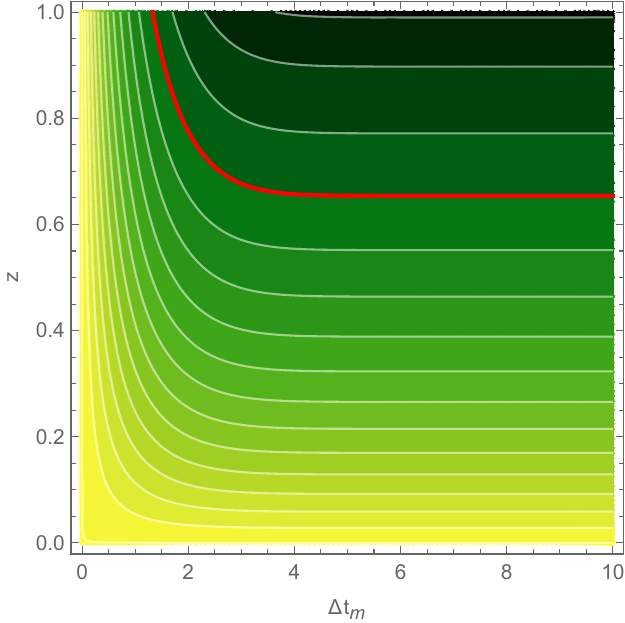} \includegraphics[height = .25\textheight]{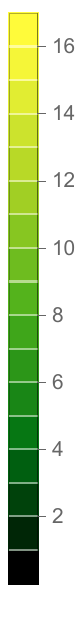} \\
\includegraphics[width = .85\textwidth, trim = {10 0 70 0}, clip]{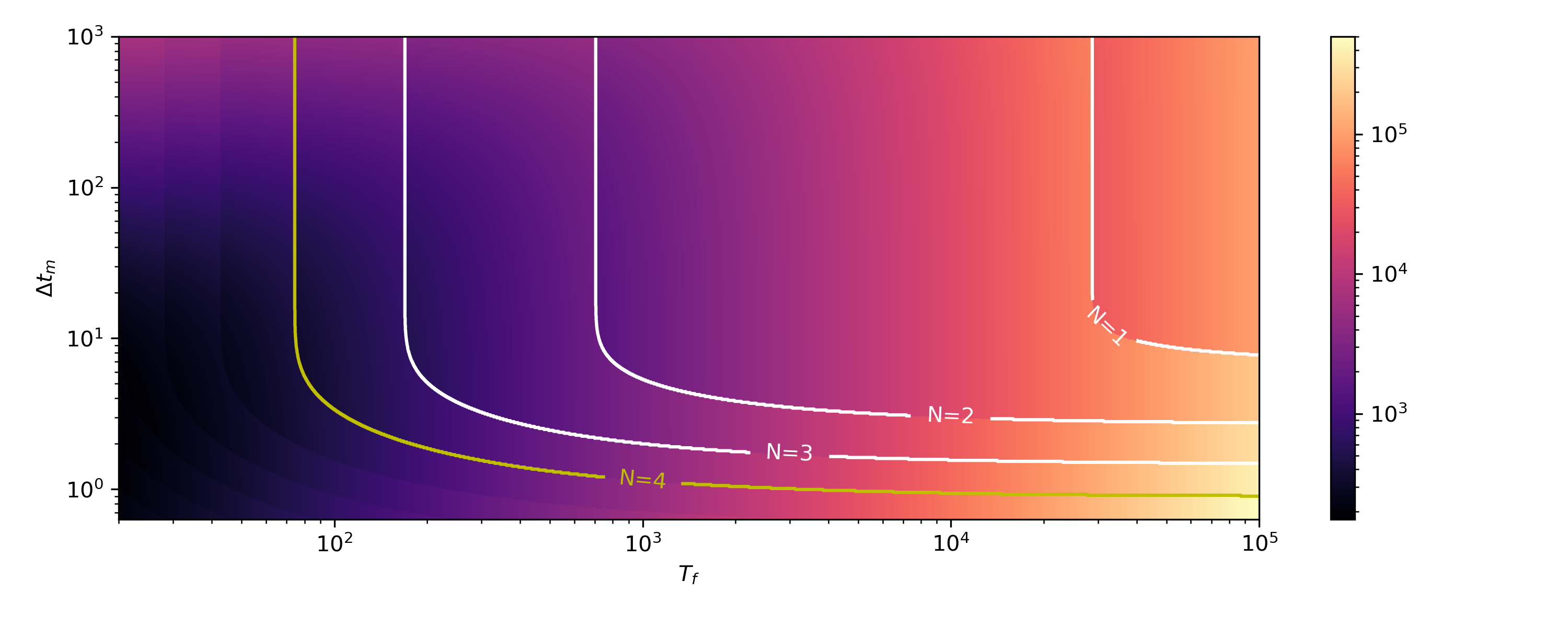} \\
\begin{tikzpicture}[overlay, yshift = -0.3cm]
\node at (-7.5,13.5) {(a)};
\node at (1.3,13.5) {(b)};
\node at (-7.8,6.4) {(c)};
\node at (8,6.6) {$\mathrm{TTUS}^{(m)}_{99\%}$};
\node at (8.3,13.7) {$N_{99\%}$};
\node at (-8.4,13.4) {$z$};
%%%%%%%%%%%%%%%%%%%%%%%%
\node[] at (-4.5,13.4) {\color{patriarch!75!black} \footnotesize $T_f = 1\times 10^4\,\tau$};
\node[] at (-4.8,12.9) {\color{patriarch!90!black} \footnotesize $T_f = 3\times 10^3\,\tau$};
\node[] at (-5.1,12.4) {\color{patriarch!90!white} \footnotesize $T_f = 1\times 10^3\,\tau$};
\node[] at (-5.4,11.9) {\color{patriarch!75!white} \footnotesize $T_f = 3\times 10^2\,\tau$};
\node[] at (-5.7,11.4) {\color{patriarch!60!white} \footnotesize $T_f = 1\times 10^2\,\tau$};
\node[] at (-6.0,10.9) {\color{patriarch!45!white} \footnotesize $T_f = 3\times 10^1\,\tau$};
\node[] at (-6.3,10.4) {\color{patriarch!37!white} \footnotesize $T_f = 1\times 10^1\,\tau$};
\node[] at (-6.5,9.9) {\color{patriarch!30!white} \footnotesize $T_f = 3\times 10^0\,\tau$};
\node[] at (-6.8,9.4) {\color{patriarch!25!white} \footnotesize $T_f = 1\times 10^0\,\tau$};
%%%%%%%%%%%%%%%%%%%%%%
\end{tikzpicture} \vspace{-20pt}
\caption{
Illustration of the expected time to solution for the two-qubit 2-SAT problem of Eq.~\eqref{2q2SAT_clause} and Fig.~\ref{fig-projectors}, as obtained by the Lindblad solver (Algorithm \ref{alg4} with Algorithm \ref{alg1}) using a linear schedule $\theta = \pi\,t/2\,T_f$. 
The times $T_f$, $\Delta t_m$, and $\mathrm{TTUS}_{99\%}$ (Eq. \eqref{TTUS-star-special}) are all given in units of the characteristic measurement time $\tau$, which is assumed to characterize the strength of both the clause measurements and subsequent local readout measurements. 
Panel (a) shows the time evolution of the local $z$ coordinate (see Fig.~\ref{fig:gam-scan-2SAT-lind}(b)), and illustrates the average $z$ resolution we can expect to obtain on any one of the two qubits, and illustrates how this depends on the Lindblad dragging times $T_f$ expended on the analog measurement--driven computation. 
Panel (b) analyzes the readout time $\Delta t_m$ required to achieve a solution in $N$ shots, given the local bias $|z|$ on the qubits. We plot Eq. \eqref{N-star_unique} for $n=2$ and required confidence level $\mathbb{P}_{\star} = 0.99$, which we denote $N_{99\%}$. 
Since we consider a two-qubit problem ($n = 2$), it is effectively meaningless to consider performing $N > 2^n = 4$ shots; the contour separating the useful (lower) region from this region is highlighted in red. 
Panel (c) aggregates the results of panels (a) and (b). Here we plot Eq. \eqref{TTUS-star-special} (a special case of Eq. \eqref{TTS-star}) for $\mathbb{P}_{\star} = 0.99$, notated as $\mathrm{TTS}^{(m)}_{99\%}$.
Some selected contours of $N_{99\%}$ Eq.~\eqref{N-star_unique} are overlaid, highlighting the discrete nature of the ceiling function introduced to enforce the confidence threshold.
For this very small toy problem, we observe that there is not an optimal $\mathrm{TTS}$ visible.
Because $2^n$ is small in this instance, the stipulation that $N$ be at most $2^n$ is quite restrictive (it is, after all, very easy to solve this toy problem just by checking all candidate solutions). 
We will see in later sections that the importance of these small-size effects falls away as we consider larger problems and TTS scaling. 
}\label{fig:2Q-2SAT-TTS}
\end{figure*}

Let us consider the statistics of the measurement outcomes when such a measurement is applied to the reduced density matrix $\rho_j$ for the $j^\mathrm{th}$ qubit in our register. 
This means that we apply Eq. \eqref{eq_readout_Kraus} under the simplifying assumption that there is negligible correlation between any of the qubits (Fig.~\ref{fig:gam-scan-2SAT-lind} indicates that for relatively slow dragging times, this is a reasonable assumption to make).
We have the probability density
\be \label{eq-prz-sep} \begin{split}
\wp(r_j|\rho_j) =& \left( \frac{1+z_j}{2} \right) \left(\frac{\Delta t_m}{2 \pi}\right)^{\frac{1}{2}}e^{-\frac{\Delta t_m}{4}\left(r_j-\frac{1}{\sqrt{\tau}}\right)^2} \\ \quad& + \left( \frac{1-z_j}{2}\right)\left(\frac{\Delta t_m}{2 \pi}\right)^{\frac{1}{2}}e^{-\frac{\Delta t_m}{4}\left(r_j+\frac{1}{\sqrt{\tau}}\right)^2}
\\ =& \mathcal{P}(0)\wp(r|0) + \mathcal{P}(1)\wp(r|1)
\end{split} \ee
for each qubit, where $\wp(r|0)$ and $\wp(r|1)$ are Gaussians with positive and negative mean signals respectively, and variances $1/\Delta t_m$ (such models are commonly used for dispersive or longitudinal readout of individual qubits \cite{Korotkov1, *Korotkov2, *Korotkov3, *Korotkov4, Steinmetz_2022}). 
As in Sec.~\ref{sec_ro_tts}, suppose that upon reading out $\mathbf{r}_z$ (a vector of all the $r_j$), we obtain the solution bitstring from that run of the experiment via $\tilde{\mathbf{s}} = \mathrm{sign}(\mathbf{r}_z)$. 
The probability that each $\tilde{s}_j$ correctly reflects the underlying biases $z_j$, such that $\tilde{\mathbf{s}}$ matches the correct solution $\mathbf{s}$, is given by
\be\label{Pbs}\begin{split} 
P(\mathbf{b} = \mathbf{s}|\lbrace z_j \rbrace) &= \left(\prod_j^n \left|\int_0^{\mathrm{sign}(b_j)\infty} dr_j\wp(r_j|\rho_j)  \right|\right) 
\\ &= \prod_j^n \frac{1+|z_j|\,\mathrm{erf}\left( \sqrt{\Delta t_m/2\tau} \right)}{2}
\\ &= \left(\frac{1+|z_T|\,\mathrm{erf}\left( \sqrt{\Delta t_m/2\tau} \right)}{2}\right)^n,
\end{split}\ee
where in the last line the biases $|z_j| = |z_T|$ are assumed the same up to a sign. 
We note that this expression is a special case of Eq.~\eqref{PrS_general}, under the simplifying assumptions that the state of the different qubits is approximately separable, and that the local biases are uniform and correctly reflect the solution state.
The assumptions of separability and correct solution bias are generically valid for sufficiently long $T_f$, and when applied to problems with a unique solution.
Notice that in the limit of deterministic dragging $T/\tau \rightarrow \infty$ and strong readout $\Delta t_m/\tau \rightarrow \infty$, this expression scales as $1^n$.
In the opposing limit of $T \rightarrow 0$ (so that $|z_T| \rightarrow 0$), the scaling instead goes as $2^n$.  
This corresponds to the readout essentially choosing one of the $2^n$ candidate solution bitstrings at random, in analogy with the worst, i.e., brute-force classical approach to $k$-SAT. 

Let us adapt the number of runs $N_\star$ required to achieve a solution probability at least $\mathbb{P}_{\star}$, Eq.~\eqref{Nstar}, to this special case of 2-qubit 2-SAT.
For $k$-SAT problems with unique solution and uniform local bias, we may write
\be \label{N-star_unique}
\check{N}_\star = \mathrm{ceil}\left[ \frac{\log(1-\mathbb{P}_{\star})}{\log\left(1 -  \tfrac{1}{2^n}\left(1+|z_T|\,\mathrm{erf}\left( \sqrt{\Delta t_m/2\tau} \right)\right)^n \right)} \right],
\ee
using Eq.~\eqref{Pbs}. 
A visualization of this expression for $n =2$ 2-SAT appears in Fig.~\ref{fig:2Q-2SAT-TTS}(b).  
We remark that if a bound on the bias $|z_T|$ could be systematically derived in the case of a unique solution, then the behavior of this dissipation--driven algorithm could in turn be systematically bounded using the expressions above.  

Given the above, we may now define the time to a \emph{unique} solution,
\be \label{TTUS-star-special} \begin{split} 
\mathrm{TTUS}_\star^{(m)} &= \check{N}_\star\cdot(T_f + \Delta t_m).
\end{split} \ee
This is a special case of Eq.~\eqref{TTS-star}, using the further assumptions about the local qubit bias implicit in Eq.~\eqref{N-star_unique}. 
These assumptions are compatible with dynamics like those of Fig.~\ref{fig:gam-scan-2SAT-lind}, and this expression is consequently used in the analysis of Fig.~\ref{fig:2Q-2SAT-TTS} for that same example.
Panels (b) and (c) of Fig.~\ref{fig:2Q-2SAT-TTS} are instructive with regards to the process of estimating a time to solution, illustrating how one might derive regions of $T_f$ and $\Delta t_m$ required to solve a $k$-SAT problem with a continuous measurement--driven approach, within a certain number of shots and/or with a specified confidence threshold.

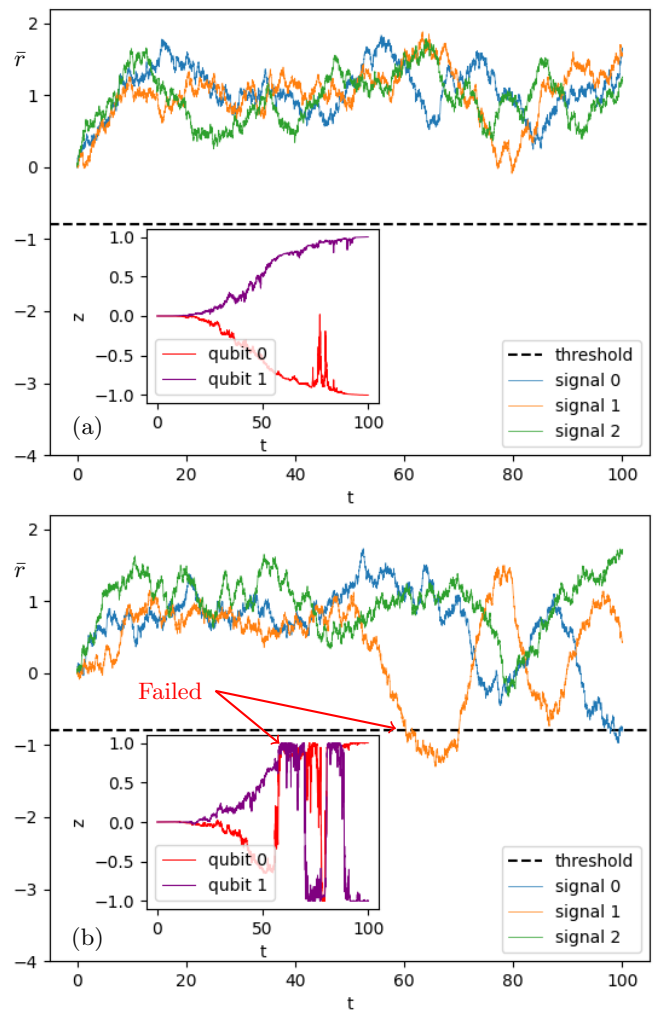
\begin{figure}
    \centering
    ~\includegraphics[width = 0.48\textwidth, trim = {35 5 40 40}, clip]{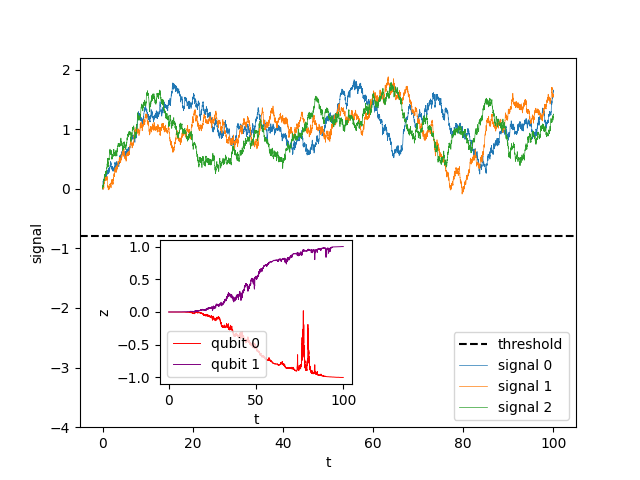}
    \\ 
    ~\includegraphics[width = 0.48\textwidth, trim = {35 5 40 40}, clip]{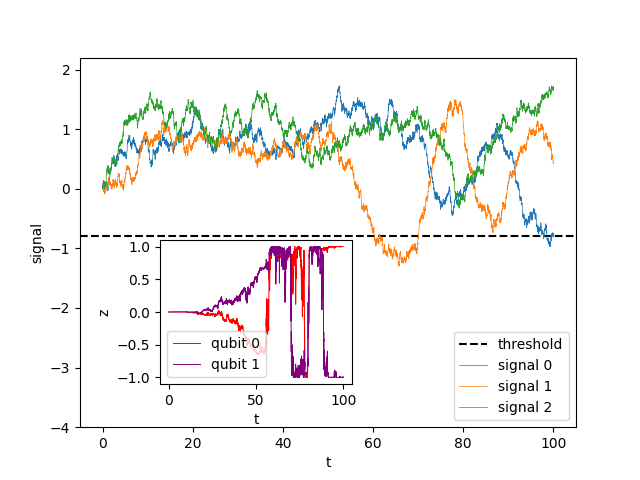}\\
    \begin{tikzpicture}[overlay,xshift = -4.9cm,yshift = 0.1cm]
    \node[] at (1.7,8.1) {(a)};
    \node[] at (1.7,1.3) {(b)};
    \node[color = red] at (2.8, 4.6)  {Failed};
    \draw[->, red, thick] (3.4,4.6) -- (5.8,4.1) ;
    \draw[->, red, thick] (3.4,4.6) -- (4.25,3.9) ;
    \node[] at (0.8,13) {$\bar{r}$};
    \node[] at (0.8,6.2) {$\bar{r}$};
    \end{tikzpicture} \vspace{-12pt}
    \caption{
    Qubit dynamics and filtered clause readout signals under the heralded Algorithm \ref{alg2}. 
    The SAT problem here is the same as the one defined in Eq.~\eqref{2q2SAT_clause}. 
    We set the measurement collapse time $\tau =1$ and the total dragging time $T_f = 100$. The filter response time is set to be $T_{be} = \mathrm{max} \{2\tau, 0.1T_f\}$, and the threshold is $r_{th} = -2.5/\sqrt{T_{be}}$. Panel (a) shows the evolution of a successful run, where the filtered signals $\bar{r}$ Eq. \eqref{eq:filter} never reach the threshold value and the reduced conditional qubit dynamics (inset) diffuse relatively cleanly towards the correct solution state. Panel (b) shows the evolution of a failed run. In this example, qubit 0 is collapsed into an incorrect subspace near $t/\tau \approx 50$ (see inset) and the filtered signal 1 reaches the threshold near $t/\tau \approx 60$, heralding violation of a particular clause, and therefore a failure of the algorithm. The failure is detected when the threshold is crossed: the algorithm is terminated at this point. We also show the evolution for some time past this point, up to $t = T_f$, for illustrative purposes.}
    \label{fig:conditional_2q2SAT}
\end{figure} 
%%%%%%%%%%%%%%%%%%%%%%%%%%%%%%%%%%%%%%%%%%%%%%%%%%%%%%%%%
%%%%%%%%%%%%%%%%%%%%%%%%%%%%%%%%%%%%%%%%%%%%%%%%%%%%%%%%%
\subsection{Remarks: Multi-Solution and Unsatisfiable Problems}

In the main text here, we have emphasized the sample two--qubit 2-SAT problem of Fig.~\ref{fig-projectors}, which has a unique solution. 
However, variants on this problem containing either more than one solution, or no solutions at all, are also illuminating to consider. 
We describe such alternative two--qubit 2-SAT problems in detail in Appendix \ref{app-multi-unsat}, restricting ourselves here to a brief summary. 

When multiple solutions satisfy our two--qubit 2-SAT problem, or when no solution can satisfy our two--qubit 2-SAT problem, we are no longer guaranteed any bias in the local $z$ coordinates on average at the end of a Lindbladian dragging interval -- even in the Zeno limit. 
However, the output of the two problems, i.e., when multiple solutions exist or when no solution exists, still differ in meaningful ways. 
In the case of multiple solutions, we will generate relatively high purity entangled states within the solution subspace, i.e.,~superpositions of the possible classical solutions, one of which is highly likely to be correctly resolved by reading out the qubits in the desirable regime $T_f \gg \tau$ and $\Delta t_m \gg \tau$. 
This means that although the individual readout measurements of $\hat{\sigma}_z^{(j)}$ will appear random run-to-run, information about the solution space will still appear in the structure of the \emph{correlations} between those $z_j$ outcomes on a run-by-run basis. 
On the other hand, when the SAT problem is unsatisfiable, we obtain the maximally--mixed state on average, so that the local qubit readouts at the end will be random and not exhibit any mutual qubit correlations. 

%%%%%%%%%%%%%%%%%%%%%%%%%%%%%%%%%%%%%%%%%%%%%%%%%%%%%%%%%%%%%%
%%%%%%%%%%%%%%%%%%%%%%%%%%%%%%%%%%%%%%%%%%%%%%%%%%%%%%%%%%%%%%

\subsection{Illustration of  Heralded Algorithm \label{sec_2Q-herald}}

Finally, we demonstrate the heralded algorithm (Algorithm 3) using the 2-qubit 2-SAT defined by Eq. \eqref{2q2SAT_clause}. We have set the threshold to be $r_{th} = -2.5/\sqrt{T_{be}}$ so that some level of fluctuation due to noise is allowed by the filtered signal. 
This corresponds to correctly identifying the failure with a confidence probability of $99.65\%$ in the steady state. 
Fig.~\ref{fig:conditional_2q2SAT} shows the evolution of the qubit dynamics $z_i(t)$ and the filtered signals $\tilde{r}_i(t)$ under the heralded algorithmic dynamics.  
A successful run of the algorithm will drag the qubits to their corresponding solution states, analogously to the fixed point algorithm driven by the Lindbladian. The difference is that with heralded algorithm, we can now detect any failure shortly after it occurred instead of at the end of the algorithm. This can be seen in Fig. \ref{fig:conditional_2q2SAT}, where the failed trajectory is detected within a period $\sim T_{be}$ after the failure occurred.

%%%%%%%%%%%%%%%%%%%%%%%%%%%%%%%%%%%%%%%%%%%%%%%%%%%%%%%%%%%%%%%%%
%%%%%%%%%%%%%%%%%%%%%%%%%%%%%%%%%%%%%%%%%%%%%%%%%%%%%%%%%%%%%%%%%
%%%%%%%%%%%%%%%%%%%%%%%%%%%%%%%%%%%%%%%%%%%%%%%%%%%%%%%%%%%%%%%%%

\section{Scaling the Problem Up \label{SEC:BZF-kSAT-examples}}

We have gained some intuition about the workings of the continuous BZF algorithm for $k$-SAT by investigating the simplest version of it, namely for the $k=2$ with 2 qubits. 
We now extend our analysis towards cases of interest, to study the algorithm's performance on $k$-SAT problems with both $k=2$ and $k=3$, on 4-10 qubits. In this section, we will benchmark the Zeno dragging algorithms at various values of clause density $\alpha$, as well as for different parameters $\Delta t$ and $T_f$. In order to study the dependence only on the dragging time $T_f$, we will then assume projective measurement at the final readout phase, and thus we will make calls to Algorithm \ref{alg1} and Algorithm \ref{alg3} instead of to Algorithm \ref{alg4}.

%%%%%%%%%%%%%%%%%%%%%%%%%%%%%%%%%%%%%%%%%%%%%%%%%%%%%%%%%%%%%%
%%%%%%%%%%%%%%%%%%%%%%%%%%%%%%%%%%%%%%%%%%%%%%%%%%%%%%%%%%%%%%

\subsection{Quantum Computational Phase Transition}\label{sec_qcpt}
It has been rigorously proven that in the large $n$ limit, the probability of satisfying a random SAT problem exhibits a phase transition from satisfiable to unsatisfiable (SAT-UNSAT) when the number of clauses per qubit number, i.e., $\alpha = m/n$, exceeds a critical value $\alpha_c$ \cite{doi:10.1126/science.1073287}. 
For 2-SAT, the critical clause density is analytically determined to be $\alpha_c = 1$ \cite{GOERDT1996469}. For 3-SAT, the critical value is empirically evaluated to be $\alpha_c \approx 4.26$ \cite{10.1145/2594413.2594424, CRAWFORD199631}.  The computational cost for solving these random SAT problems correspondingly exhibits an easy-hard-easy pattern, with the computational cost transition occurring at the critical value of $\alpha_c$ \cite{CRAWFORD199631, 10.5555/1867135.1867206}. Quantum algorithms that optimize solutions for $k$-SAT, such as the quantum approximate optimization algorithm (QAOA), show a similar computational phase transition near $\alpha_c$, even for systems as small as 6 qubits \cite{PhysRevLett.124.090504, zhang2022quantum}. 

Here, we show evidence of an analogous quantum computational phase transition for both random 2-SAT and 3-SAT under our measurement-driven quantum algorithms. Specifically, we numerically estimate the probability of successfully determining the satisfiability of a random instance as a function of the clause density $\alpha$, under different values of dragging time $T_f$. The empirical estimation of this success probability $P_{\mathrm{succ}}(\alpha, T_f)$ is given by
\begin{equation}
    P_{\mathrm{succ}}(\alpha, n, T_f) = \frac{N_{\mathrm{succ}}(\alpha, n, T_f)}{N_{\mathrm{prob}}(\alpha, n)},\label{eq_PT_probability}
\end{equation}
where $N_{\mathrm{prob}}(\alpha, n)$ is the number of random instances generated \bw{for} clause density $\alpha$ and number of variables $n$, and $N_{\mathrm{succ}}(\alpha, n, T_f)$ is the corresponding number of instances whose satisfiability are correctly determined by the measurement-driven quantum algorithm with total dragging time $T_f$.

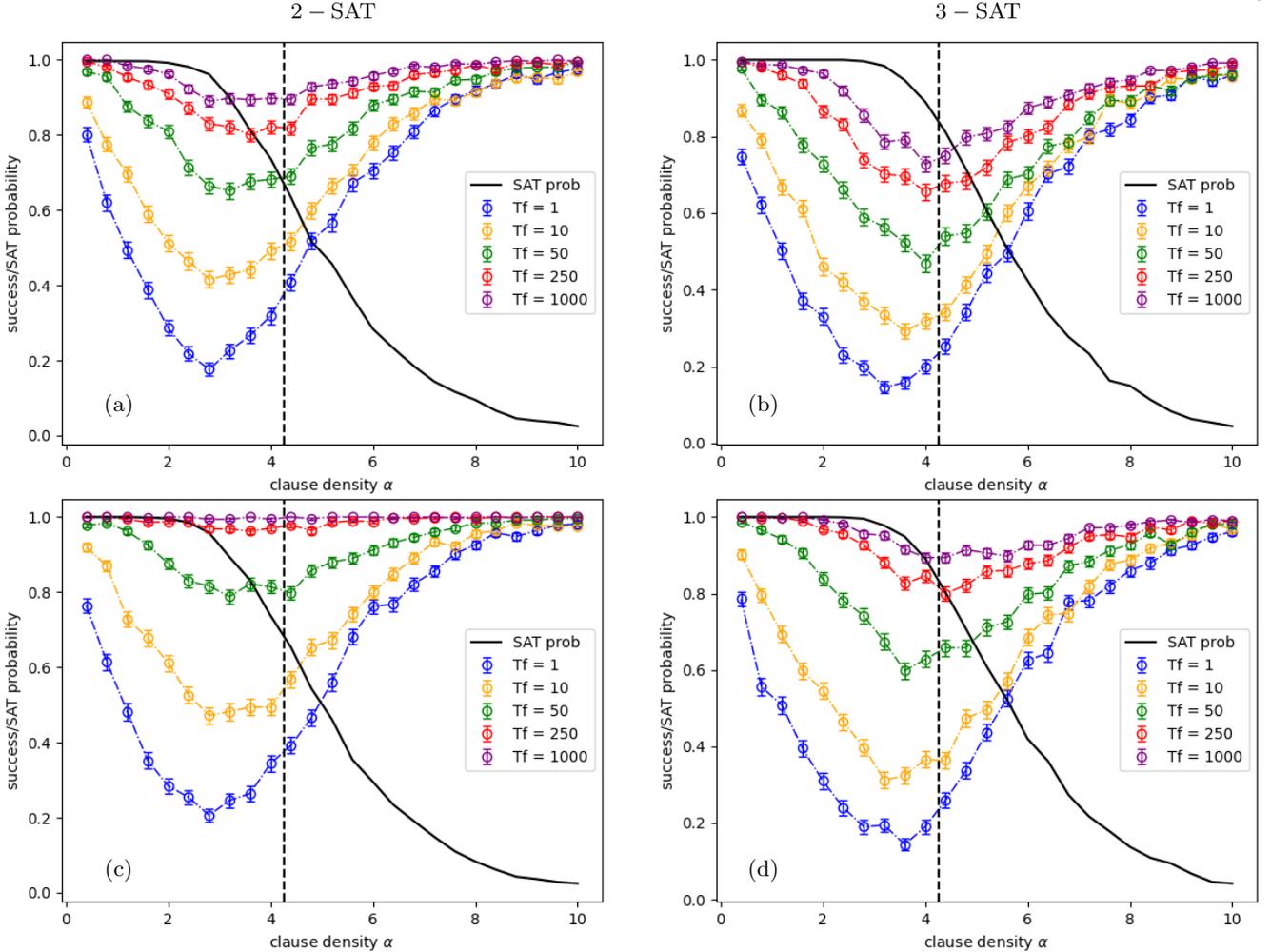
\begin{figure*}
    \centering
    \includegraphics[width = 0.49\textwidth, trim = {5 5 40 40}, clip]{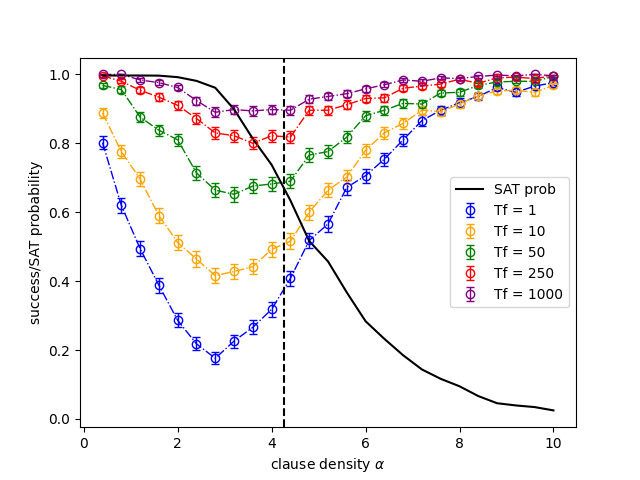}
    \hfill
    \includegraphics[width = 0.49\textwidth, trim = {5 5 40 40}, clip]{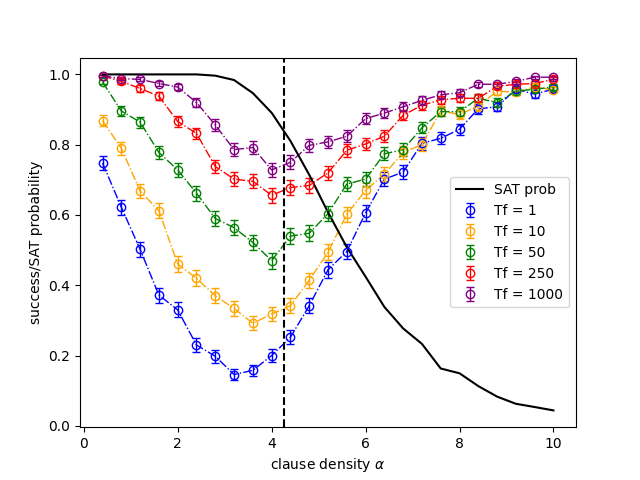}\\
    \includegraphics[width = 0.49\textwidth, trim = {5 5 40 40}, clip]{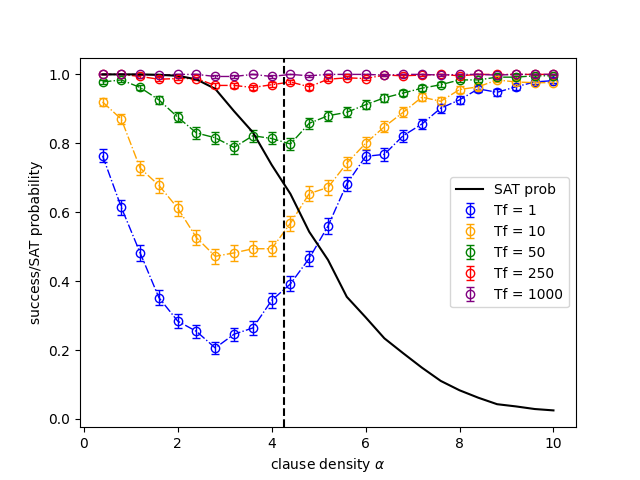}
    \hfill
    \includegraphics[width = 0.49\textwidth, trim = {5 5 40 40}, clip]{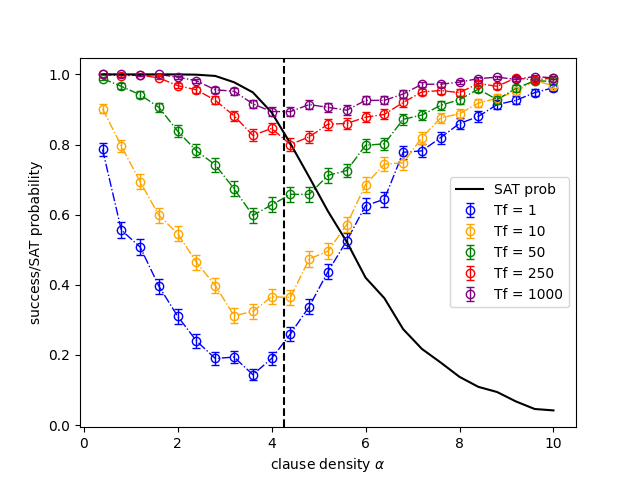}\\
    \begin{tikzpicture}[overlay]
    \node[] at (-7,8) {(a)};
    \node[] at (2,8) {(b)};
    \node[] at (-7,1.5) {(c)};
    \node[] at (2,1.5) {(d)};
    \node[] at (-4,13.5) {\color{black} $2-\mathrm{SAT}$};
    \node[] at (5,13.5) {\color{black} $3-\mathrm{SAT}$};
    \end{tikzpicture} \vspace{-12pt}
    \caption{We demonstrate the quantum computational phase transition by evaluating the probability of successfully determining the satisfiability $P_{\mathrm{succ}}(\alpha, n, T_f)$ as a function of clause density $\alpha$, i.e., Eq. \eqref{eq_PT_probability}, for $n = 5$ qubits, using a range of values of the total dragging time $T_f$. Panels (a) and (b) are for 2-SAT and 3-SAT with the Lindblad Algorithm \ref{alg1} ($\Delta t \rightarrow dt$) respectively, while panels (c) and (d) are with the heralded Algorithm \ref{alg3}. The same measurement strength $\Gamma = 1/(4\tau)$ is used for all panels, together with a linear schedule $\theta(t) = \pi t/2T_f$ for $0\leq t \leq T_f$. We have set the measurement time $\tau = 1$ throughout all the simulations. In the heralded algorithm, we have additionally set $T_{be} = \mathrm{max}\{2\tau, 0.1T_f \}$, $T_{min} = 5\tau$, and $r_{th} = -2.5/\sqrt{T_{be}}$. The black dotted vertical lines are the locations of critical clause densities in the limit of large $n$, i.e. $\alpha_c = 1$ for 2-SAT and $\alpha_c \approx 4.26$ for 3-SAT. The black curves are the probabilities of satisfiability as a function of $\alpha$ for the randomly generated SAT instances. The colored curves are $P_{\mathrm{succ}}(\alpha, n, T_f)$ under various values of $T_f$. Each datum on the curve is obtained using a sample size $N_{\mathrm{prob}}(\alpha, n) = 500$. The non-zero width of the critical region and the deviation of the location of the lowest value of $P_{\mathrm{succ}}$ from $\alpha_c$ result from the finite system size $n$.
    } \label{fig:phase_transition}
\end{figure*}

Panels (a) and (b) of Fig,~\ref{fig:phase_transition} show results for calculations with $n=5$ qubits using the Lindblad Algorithm \ref{alg1}.
We can see that even with such a relatively small system, for both 2-SAT and 3-SAT problems the SAT probability undergoes a clear SAT-UNSAT transition crossing at a distinct value $\alpha_c$, while $P_{\mathrm{succ}}(\alpha, n, T_f)$ shows an easy-hard-easy transition, with the hardest part located near $\alpha_c$. We also see that when more quantum computational resources are provided, specifically, for larger $T_f$, the algorithm can obtain higher values of $P_{\mathrm{succ}}(\alpha, n, T_f)$, indicating better performance.

As discussed in Sec.~\ref{sec_filter_algo}, the heralded dynamics benefit from real-time detection of any failures. This advantage allows us to define the heralded algorithm (Algorithm \ref{alg3}) that restarts when the failure is detected early, and thus is more likely to follow the correct trajectory given a fixed amount of computation time $T_f$ even without feedback.
Notice that in Algorithm \ref{alg3}, we do not terminate the dynamics when the time left $t_{rest}$ is smaller then a minimum value $T_{min}$. This is because when the total dragging time is too small (comparable to $T_{be}$ and $\tau$), the quantum Zeno effect is not strong and the failure detection based on the filter is also not reliable anymore. However, we can still use conditional dynamics, i.e., SME Eq \eqref{eq_SME_multi} rather than the averaged dynamics LME Eq. \eqref{eq_Lindblad_multi}, in this situation.

We perform the same type of calculation for the success probability Eq. \eqref{eq_PT_probability} as a function of the clause density $\alpha$ under the heralded algorithm \ref{alg3}. The results are shown in panels (c) and (d) of Fig. \ref{fig:phase_transition}. We can see that the heralded algorithm shows the same type of computational phase transition as the Lindblad algorithm in panels (a) and (b). However, the performance of the heralded algorithm is better than that of the Lindblad algorithm, especially near the critical clause density $\alpha_c$, and also when $T_f$ is large. We can understand the latter as a consequence of the combination of earlier detection of failure and the possibility of running multiple trials. 

This analysis has shown that for the continuous measurement--driven quantum algorithms, it would be most difficult to successfully solve SAT problems having the critical clause density $\alpha_c$, similar to known results for QAOA and for classical algorithms. Therefore, in order to establish the scaling of the algorithm with respect to the system size $n$, we shall focus on the hardest SAT instances for the quantum algorithm,  i.e., instances with $\alpha = \alpha_c$ from now on.

%%%%%%%%%%%%%%%%%%%%%%%%%%%%%%%%%%%%%%%%%%%%%%%%%%%%%%%%%%%%%%
%%%%%%%%%%%%%%%%%%%%%%%%%%%%%%%%%%%%%%%%%%%%%%%%%%%%%%%%%%%%%%
\subsection{Scaling with Qubit Number for 3-SAT\label{SEC:scaling}}
In this section, we study the scaling of the TTS with the qubit number $n$, which is central to quantifying the algorithm performance in terms of computational complexity. Specifically, we do not expect the TTS to scale polynomially with the qubit number $n$, which would imply $\textbf{NP} \subseteq \textbf{BQP}$. Since there are no strong computational complexity arguments implying this, we will then instead assume that TTS scales exponentially with the qubit number $n$ throughout the rest of this work, i.e., TTS $\sim \lambda^n$. We shall focus on studying the base number $\lambda$ for the algorithms with different parameter settings.

In Sec.~\ref{sec_2Q-herald} we demonstrated the behavior of the TTS as a function of the sum of the dragging time $T_f$ and the final readout time $\Delta t_m$. However, in the context of algorithmic scaling, the TTS is generally characterized without taking the final readout time into account, which corresponds to using the dragging time $T_f$ alone.
To enable comparison with the literature, we therefore define the algorithmic time to solution
as the following \cite{ronnow2014defining}
\begin{equation}\label{eq_TTS_99}
    \text{TTS}_{99\%} = T_f \frac{\text{log}(1-0.99)}{\text{log}(1 - \mathbb{P}_{\mathbf{s}})} = T_f\,N_{99\%},
\end{equation}
where $\mathbb{P}_{\mathbf{s}}$ is the solution state probability at the end of one run of the algorithm. 

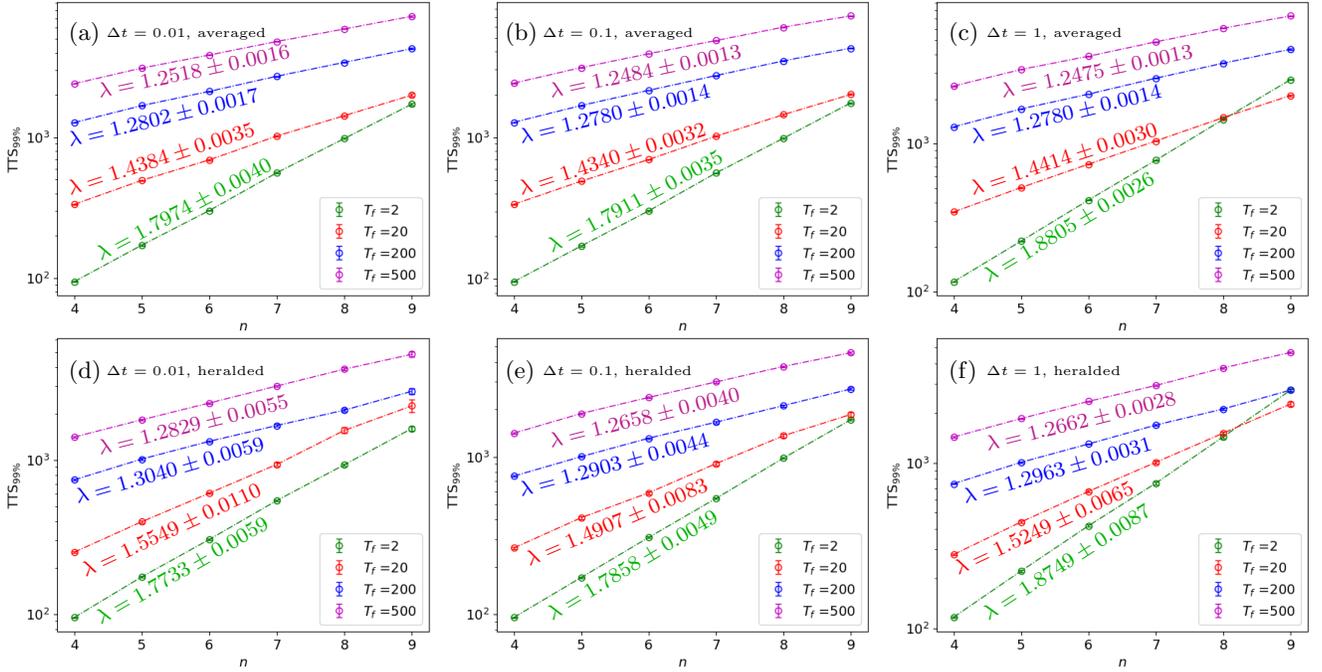
\begin{figure*}
    \centering
    \includegraphics[width = 0.32\textwidth, trim = {5 5 40 40}, clip]{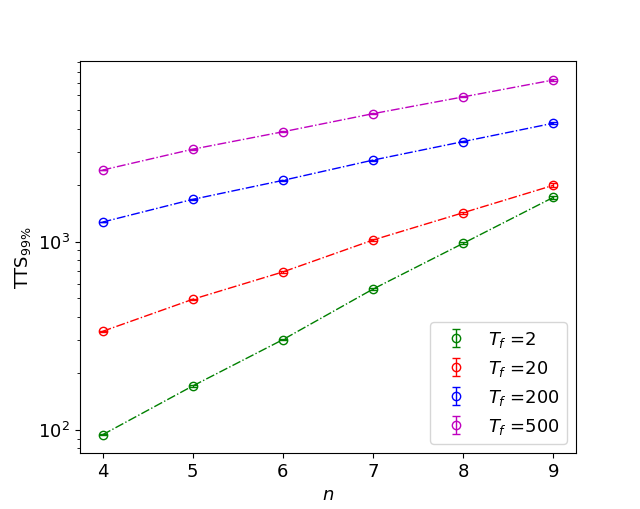}
    \includegraphics[width = 0.32\textwidth, trim = {5 5 40 40}, clip]{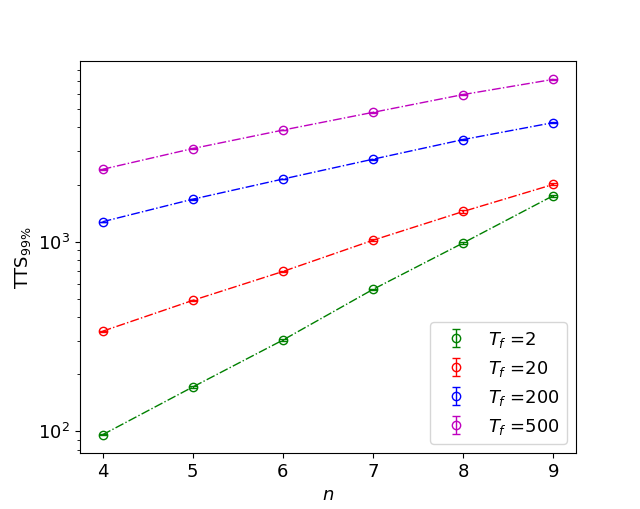}
    \includegraphics[width = 0.32\textwidth, trim = {5 5 40 40}, clip]{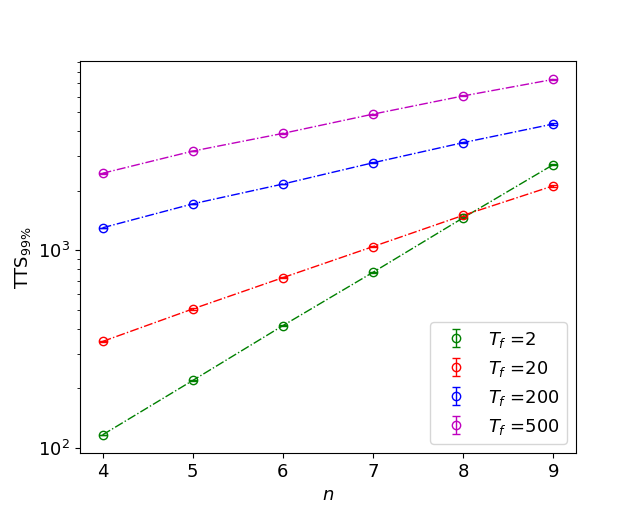} \\
    \includegraphics[width = 0.32\textwidth, trim = {5 5 40 40}, clip]{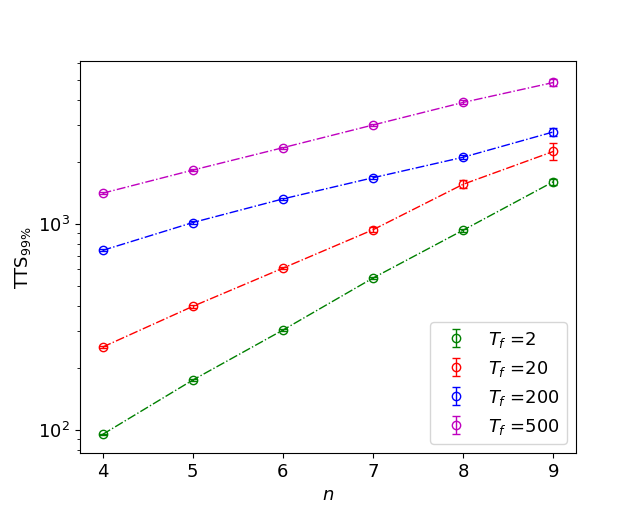}
    \includegraphics[width = 0.32\textwidth, trim = {5 5 40 40}, clip]{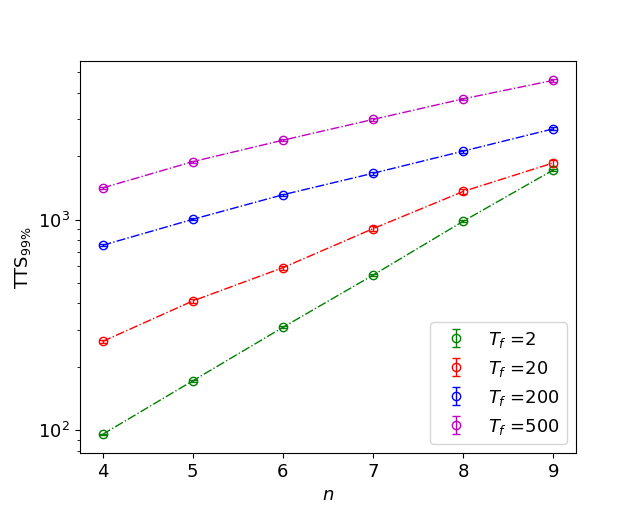}
    \includegraphics[width = 0.32\textwidth, trim = {5 5 40 40}, clip]{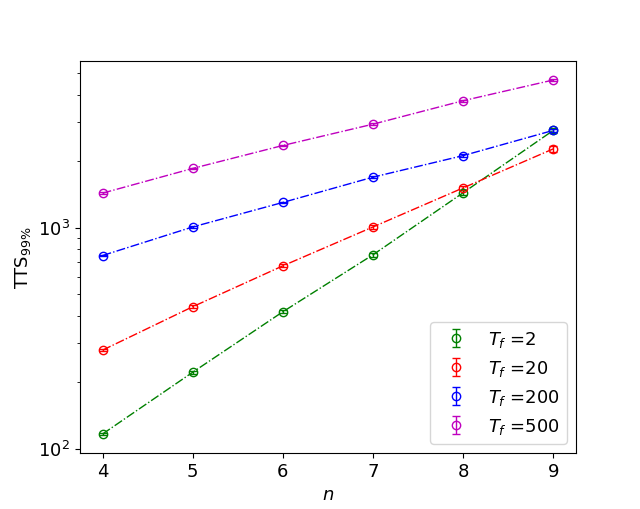}
    \\
    \begin{tikzpicture}[overlay]
    \node[anchor = west] at (-7.9,8.8) {(a)};
    \node[anchor = west] at (-2.1,8.8) {(b)};
    \node[anchor = west] at (3.8,8.8) {(c)};
    \node[anchor = west] at (-7.9,4.3) {(d)};
    \node[anchor = west] at (-2.1,4.3) {(e)};
    \node[anchor = west] at (3.8,4.3) {(f)};
    %%%%%%%%%%%%%%%%%%%%%%
    \node[anchor = west] at (-7.4,8.8) {\tiny $\Delta t= 0.01$, averaged};
    \node[anchor = west] at (-1.6,8.8) {\tiny $\Delta t= 0.1$, averaged};
    \node[anchor = west] at (4.3,8.8) {\tiny $\Delta t= 1$, averaged};
    \node[anchor = west] at (-7.4,4.3) {\tiny $\Delta t= 0.01$, heralded};
    \node[anchor = west] at (-1.6,4.3) {\tiny $\Delta t= 0.1$, heralded};
    \node[anchor = west] at (4.3,4.3) {\tiny $\Delta t= 1$, heralded};
    %%%%%%%%%%%%%%%%%%%%%
    \node[anchor = west, rotate = 28] at (-7.5,5.8) {\footnotesize\color{green!70!black} $\lambda = 1.7974 \pm 0.0040$};
    \node[anchor = west, rotate = 18] at (-7.9,6.7) {\footnotesize\color{red} $\lambda = 1.4384 \pm 0.0035$};
    \node[anchor = west, rotate = 13.5] at (-7.9,7.35) {\footnotesize\color{blue} $\lambda = 1.2802 \pm 0.0017$};
    \node[anchor = west, rotate = 12] at (-7.5,8) {\footnotesize\color{magenta!40!electricviolet} $\lambda = 1.2518 \pm 0.0016$};
    %%%%%%%%%%%%%%%%%%%%%
    \node[anchor = west, rotate = 28] at (-7.5+6,5.9) {\footnotesize\color{green!70!black} $\lambda = 1.7911 \pm 0.0035$};
    \node[anchor = west, rotate = 18] at (-7.9+6,6.7) {\footnotesize\color{red} $\lambda = 1.4340 \pm 0.0032$};
    \node[anchor = west, rotate = 13.5] at (-7.9+6,7.4) {\footnotesize\color{blue} $\lambda = 1.2780 \pm 0.0014$};
    \node[anchor = west, rotate = 12] at (-7.5+6,8) {\footnotesize\color{magenta!40!electricviolet} $\lambda = 1.2484 \pm 0.0013$};
    %%%%%%%%%%%%%%%%%%%%%
    \node[anchor = west, rotate = 31] at (-7.7+12,5.5) {\footnotesize\color{green!70!black} $\lambda = 1.8805 \pm 0.0026$};
    \node[anchor = west, rotate = 18] at (-7.9+12,6.7) {\footnotesize\color{red} $\lambda = 1.4414 \pm 0.0030$};
    \node[anchor = west, rotate = 13.5] at (-7.9+12,7.4) {\footnotesize\color{blue} $\lambda = 1.2780 \pm 0.0014$};
    \node[anchor = west, rotate = 12] at (-7.5+12,8) {\footnotesize\color{magenta!40!electricviolet} $\lambda = 1.2475 \pm 0.0013$};
    %%%%%%%%%%%%%%%%%%%%%
    \node[anchor = west, rotate = 30] at (-7.5,0.95) {\footnotesize\color{green!70!black} $\lambda = 1.7733 \pm 0.0059$};
    \node[anchor = west, rotate = 25] at (-7.7,1.6) {\footnotesize\color{red} $\lambda = 1.5549 \pm 0.0110$};
    \node[anchor = west, rotate = 16] at (-7.8,2.6) {\footnotesize\color{blue} $\lambda = 1.3040 \pm 0.0059$};
    \node[anchor = west, rotate = 14] at (-7.5,8-4.75) {\footnotesize\color{magenta!40!electricviolet} $\lambda = 1.2829 \pm 0.0055$};
    %%%%%%%%%%%%%%%%%%%%%
    \node[anchor = west, rotate = 31] at (-7.5+6,5.9-4.9) {\footnotesize\color{green!70!black} $\lambda = 1.7858 \pm 0.0049$};
    \node[anchor = west, rotate = 23] at (-7.8+6,1.75) {\footnotesize\color{red} $\lambda = 1.4907 \pm 0.0083$};
    \node[anchor = west, rotate = 15] at (-7.9+6,7.4-4.7) {\footnotesize\color{blue} $\lambda = 1.2903 \pm 0.0044$};
    \node[anchor = west, rotate = 14] at (-7.5+6,8-4.65) {\footnotesize\color{magenta!40!electricviolet} $\lambda = 1.2658 \pm 0.0040$};
    %%%%%%%%%%%%%%%%%%%%%
    \node[anchor = west, rotate = 33] at (-7.7+12,5.5-4.5) {\footnotesize\color{green!70!black} $\lambda = 1.8749 \pm 0.0087$};
    \node[anchor = west, rotate = 25] at (-8.1+12,1.625) {\footnotesize\color{red} $\lambda = 1.5249 \pm 0.0065$};
    \node[anchor = west, rotate = 16] at (-8+12,7.4-4.8) {\footnotesize\color{blue} $\lambda = 1.2963 \pm 0.0031$};
    \node[anchor = west, rotate = 14] at (-7.7+12,3.3) {\footnotesize\color{magenta!40!electricviolet} $\lambda = 1.2662 \pm 0.0028$};
    \end{tikzpicture} \vspace{-12pt}
    \caption{The 3-SAT time to solution $\text{TTS}_{99\%}$ defined by Eq.~\eqref{eq_TTS_99} is plotted as a function of the number of qubits, for different values of the total dragging time $T_f$ and for different time duration $\Delta t$ of a single clause measurement under both Algorithm \ref{alg1} with average dynamics (upper panels (a), (b), (c)) and Algorithm \ref{alg3} with heralded dynamics (lower panels (d), (e), (f)). 
    The 3-SAT instances are randomly generated at $\alpha \approx \alpha_c$ with a unique solution
    We have set $\tau = 1$ and used a linear schedule $\theta(t) = \pi t/2T_f$ for all simulations. 
    We use $\Delta t = 0.01$ (panels (a), (d)), $\Delta t = 0.1$ (panels (b), (e)), and $\Delta t = 1$ (panels (c), (f)), all in units of $\tau$.
    For the heralded algorithm, we have additionally set $T_{be} = \mathrm{max}\{2\tau, 0.1T_f \}$, $T_{min} = 5\tau$, and $r_{th} = -2.5/\sqrt{T_{be}}$. 
    The data for the heralded algorithm are averaged over more than 10000 trajectories while the data for averaged algorithm are averaged over more than 150 trajectories. We see that $\text{TTS}_{99\%}$ generally increases exponentially as a function of the number of qubits. Notice the appearance of line-crossings, which is a signature of non-monotonic dependence of $\text{TTS}_{99\%}$ on $T_f$ that indicates the possible existence of optimal values of $\text{TTS}_{99\%}$. 
    This figure continues with larger values of $\Delta t$ in Fig.~\ref{fig_TTS_vs_nq-2}. 
   }\label{fig_TTS_vs_nq}
\end{figure*}

\begin{figure*}
    \centering
    \includegraphics[width = 0.32\textwidth, trim = {5 5 40 34}, clip]{TTS_vs_nq_average_dt_1.png} 
    \includegraphics[width = 0.32\textwidth, trim = {5 5 40 34}, clip]{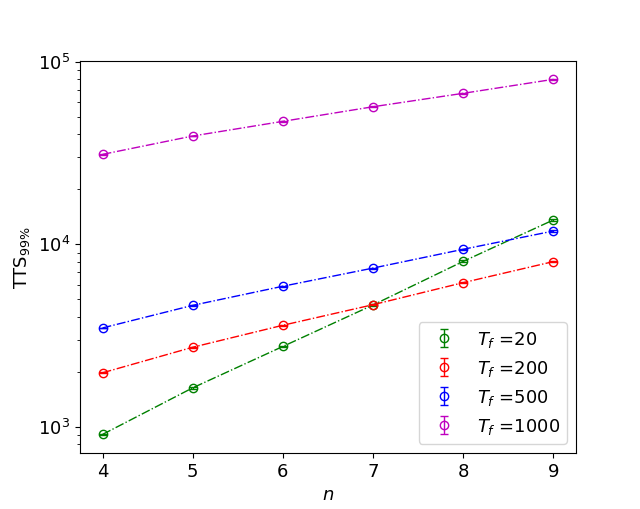}
    \includegraphics[width = 0.32\textwidth, trim = {5 5 40 34}, clip]{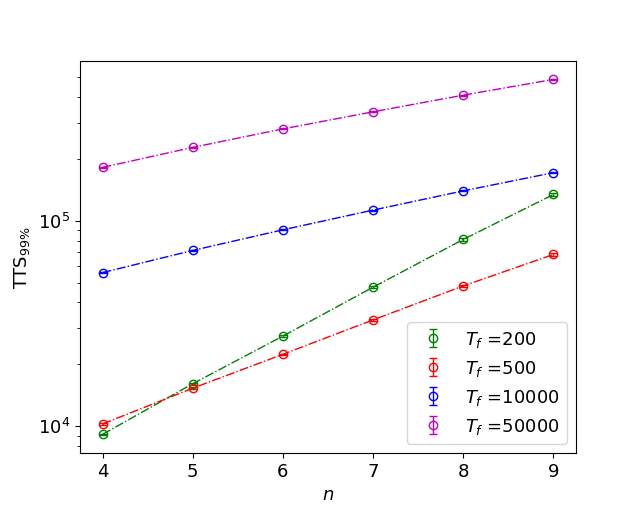} \\
    \includegraphics[width = 0.32\textwidth, trim = {5 5 40 34}, clip]{TTS_vs_nq_heralded_dt_1.png} 
    \includegraphics[width = 0.32\textwidth, trim = {5 5 40 34}, clip]{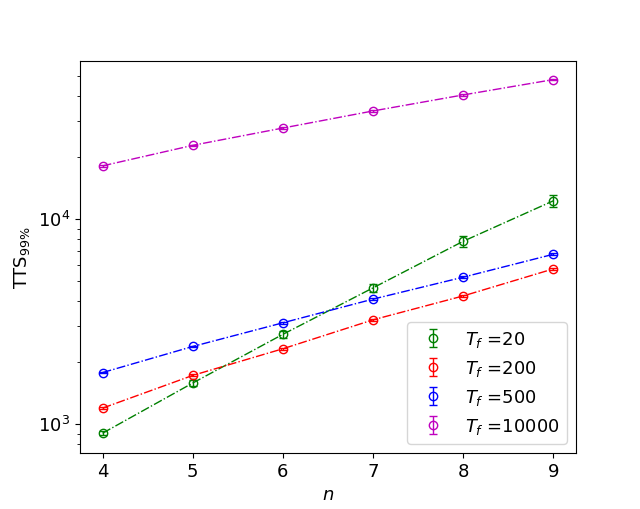} 
    \includegraphics[width = 0.32\textwidth, trim = {5 5 40 34}, clip]{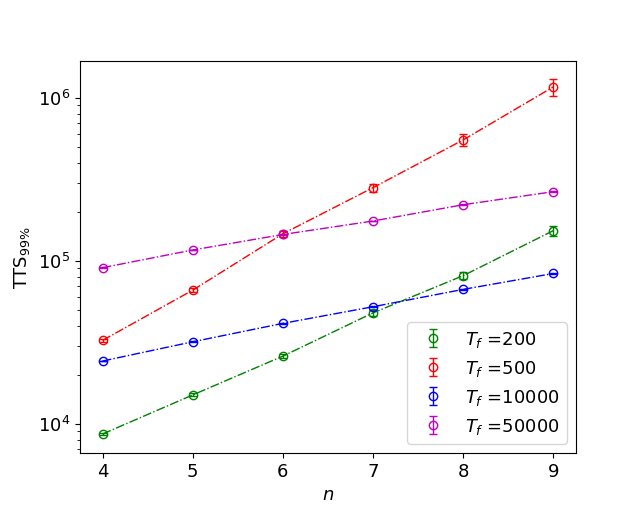} 
    \\
    \begin{tikzpicture}[overlay]
    \node[anchor = west] at (-7.9,9) {(a)};
    \node[anchor = west] at (-2.1,9) {(b)};
    \node[anchor = west] at (3.8,9) {(c)};
    \node[anchor = west] at (-7.9,4.5) {(d)};
    \node[anchor = west] at (-2.1,4.5) {(e)};
    \node[anchor = west] at (3.8,4.5) {(f)};
    %%%%%%%%%%%%%%%%%%%%%%
    \node[anchor = west] at (-7.4,8.9) {\tiny $\Delta t= 1$, averaged};
    \node[anchor = west] at (-1.6,8.9) {\tiny $\Delta t= 10$, averaged};
    \node[anchor = west] at (4.3,8.9) {\tiny $\Delta t= 100$, averaged};
    \node[anchor = west] at (-7.4,4.4) {\tiny $\Delta t= 1$, heralded};
    \node[anchor = west] at (-1.6,4.4) {\tiny $\Delta t= 10$, heralded};
    \node[anchor = west] at (4.3,4.4) {\tiny $\Delta t= 100$, heralded};
    %%%%%%%%%%%%%%%%%%%%%
    \node[anchor = west, rotate = 33] at (-7.7+0.3,5.5-4.5) {\footnotesize\color{green!70!black} $\lambda = 1.8749 \pm 0.0087$};
    \node[anchor = west, rotate = 25] at (-8.1+0.3,1.625) {\footnotesize\color{red} $\lambda = 1.5249 \pm 0.0065$};
    \node[anchor = west, rotate = 16] at (-8+0.3,7.4-4.8) {\footnotesize\color{blue} $\lambda = 1.2963 \pm 0.0031$};
    \node[anchor = west, rotate = 14] at (-7.7+0.3,3.3) {\footnotesize\color{magenta!40!electricviolet} $\lambda = 1.2662 \pm 0.0028$};
    %%%%%%%%%%%%%%%%%%%%%
    \node[anchor = west, rotate = 31] at (-7.7+0.3,5.6) {\footnotesize\color{green!70!black} $\lambda = 1.8805 \pm 0.0026$};
    \node[anchor = west, rotate = 18] at (-7.9+0.3,6.8) {\footnotesize\color{red} $\lambda = 1.4414 \pm 0.0030$};
    \node[anchor = west, rotate = 13.5] at (-7.9+0.3,7.5) {\footnotesize\color{blue} $\lambda = 1.2780 \pm 0.0014$};
    \node[anchor = west, rotate = 12] at (-7.5+0.3,8.1) {\footnotesize\color{magenta!40!electricviolet} $\lambda = 1.2475 \pm 0.0013$};
    %%%%%%%%%%%%%%%%%%%%%
    \node[anchor = west, rotate = 25] at (-1.7,5.48) {\footnotesize\color{green!70!black} $\lambda = 1.7151 \pm 0.0047$};
    \node[anchor = west, rotate = 14] at (-7.9+6,6.4) {\footnotesize\color{red} $\lambda = 1.3224 \pm 0.0020$};
    \node[anchor = west, rotate = 13] at (-7.9+6,6.8) {\footnotesize\color{blue} $\lambda = 1.2761 \pm 0.0017$};
    \node[anchor = west, rotate = 10] at (-7.5+6,8.2) {\footnotesize\color{magenta!40!electricviolet} $\lambda = 1.2086 \pm 0.0013$};
    %%%%%%%%%%%%%%%%%%%%%
    \node[anchor = west, rotate = 27] at (0.3,2.5) {\footnotesize\color{green!70!black} $\lambda = 1.7042 \pm 0.0167$};
    \node[anchor = west, rotate = 18] at (-1.7,0.95) {\footnotesize\color{red} $\lambda = 1.3669 \pm 0.0040$};
    \node[anchor = west, rotate = 14] at (-8+6,2.3) {\footnotesize\color{blue} $\lambda = 1.3035 \pm 0.0031$};
    \node[anchor = west, rotate = 12] at (-7.0+6,3.6) {\footnotesize\color{magenta!40!electricviolet} $\lambda = 1.2109 \pm 0.0024$};
    %%%%%%%%%%%%%%%%%%%%%
    \node[anchor = west, rotate = 28.5] at (-7.7+12.3,6.2) {\footnotesize\color{green!70!black} $\lambda = 1.7163 \pm 0.0047$};
    \node[anchor = west, rotate = 20] at (-7.9+12,5.5) {\footnotesize\color{red} $\lambda = 1.4651 \pm 0.0035$};
    \node[anchor = west, rotate = 13] at (-7.9+12,7.5) {\footnotesize\color{blue} $\lambda = 1.2533 \pm 0.0015$};
    \node[anchor = west, rotate = 11.5] at (-7.5+12,8.1) {\footnotesize\color{magenta!40!electricviolet} $\lambda = 1.2198 \pm 0.0014$};
    %%%%%%%%%%%%%%%%%%%%%
    \node[anchor = west, rotate = 23.5] at (4.1,0.9) {\footnotesize\color{green!70!black} $\lambda = 1.7598 \pm 0.0147$};
    \node[anchor = west, rotate = 29] at (-8.1+13.8,3.2) {\footnotesize\color{red} $\lambda = 2.0564 \pm 0.0250$};
    \node[anchor = west, rotate = 11] at (-8+12.4,7.4-5.35) {\footnotesize\color{blue} $\lambda = 1.2813 \pm 0.0022$};
    \node[anchor = west, rotate = 10] at (-7.7+13.6,2.8) {\footnotesize\color{magenta!40!electricviolet} $\lambda = 1.2382 \pm 0.0020$};
    \end{tikzpicture} \vspace{-12pt}
    \caption{
    We continue Fig.~\ref{fig_TTS_vs_nq} for larger values of $\Delta t$. We again have Algorithm \ref{alg1} with average dynamics (panels (a), (b), (c)), and Algorithm \ref{alg3} with heralded dynamics (panels (d), (e), (f)). 
    Here $\Delta t = 1$ (panels (a), (d)), $\Delta t = 10$ (panels (b), (e)), and $\Delta t = 100$ (panels (c), (f)), in units of $\tau$. All other considerations and comments match those of Fig.~\ref{fig_TTS_vs_nq}. 
    }\label{fig_TTS_vs_nq-2}
\end{figure*}

We have shown that for infinitesimal $dt$, there exists a limit of deterministic and pure--state dynamics following the solution subspace; this implies that there exists a sufficiently long $T_f$ for any satisfiable problem such that $N_{99\%}$ becomes 1. 
Our challenge now however is to navigate the tradeoff between $T_f$ and $N$ that determines the $\text{TTS}_{99\%}$.
In order to do this, we consider 3-SAT problem instances that are chosen to have critical clause density $\alpha_c \approx 4.26$ and a unique satisfying assignment. Those problems are among the hardest to solve and thus serve as good testing problems for benchmarking our generalized measurement algorithms.

\subsubsection{Scaling of $\text{TTS}_{99\%}$ as a Function of $T_f$ and $\Delta t$}
In this subsection, we now study how the base number $\lambda$ of the scaling with $n$ depends on the total pure dragging time $T_f$ and on the individual clause measurement duration $\Delta t$ for Algorithm \ref{alg1} with average dynamics as well as  for Algorithm \ref{alg3} with heralded dynamics. 

\begin{figure}
    \includegraphics[width = \columnwidth, trim = {5 5 40 40}, clip]{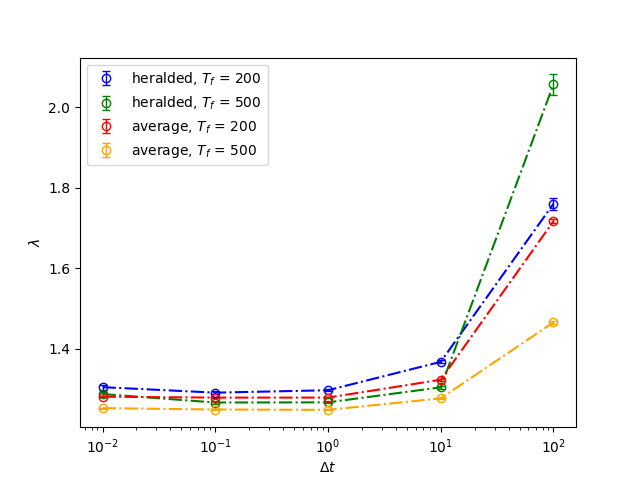}
    \caption{
     The fitted values of the base scaling parameter $\lambda$ (not $\lambda_{opt}$) shown in Figs.~\ref{fig_TTS_vs_nq} and \ref{fig_TTS_vs_nq-2} are aggregated and plotted here as a function of the duration $\Delta t$ of clause measurement, for different dragging times $T_f$ under both Algorithm \ref{alg1} with average dynamics (red and orange points) and Algorithm \ref{alg3} with heralded dynamics (blue and green points). 
    } \label{fig_lambda_vs_dt}
\end{figure}

We simulated both algorithms using the same sets of parameters as in Sec.~\ref{sec_qcpt}, for up to $n=9$ qubits. The results are shown in Figs.~\ref{fig_TTS_vs_nq} and~\ref{fig_TTS_vs_nq-2}, from which it is evident that
 as the number of qubits increases, the $\text{TTS}_{99\%}$ increases exponentially. Furthermore, for different single-shot total pure dragging times $T_f$, the $\text{TTS}_{99\%}$ exhibits different scaling base numbers $\lambda$. 
Typically, one can reach a better (smaller) value of
$\lambda$ by implementing the algorithms with a larger value of $T_f$: 
this is observed over a range of $\Delta t$ values, for both algorithms. This makes sense because for a fixed $T_f$, the TTS scaling parameter $\lambda$ derives only from the number of runs $N_{99\%} =\text{log}(1-0.99)/\text{log}(1 - \mathbb{P}_\mathbf{s})$ required to achieve 99$\%$  confidence threshold as per Eq.~\eqref{eq_TTS_99} and one would therefore expect better performance (smaller $N$) with more computational resources (larger $T_f$). 

Another interesting behavior is the dependence of $\lambda$ on $\Delta t$. We recall that $\Delta t /\tau \ll 1$ corresponds to weak measurement and $\Delta t/\tau \gg 1$ corresponds to strong measurement. 
Fig.~\ref{fig_lambda_vs_dt} shows the scaling base $\lambda$ as a function of $\Delta t$ for different values of dragging time $T_f$. 
We observe that for fixed $T_f$, both average and heralded algorithms 
%would 
display better scaling $\lambda$ in the weak continuum limit ($\Delta t \rightarrow 0$ than in the strong measurement limit $\Delta t \rightarrow \infty$. This indicates the advantage of weak continuous measurement for the $\text{TTS}_{99\%}$ scaling, in addition to the advantage demonstrated in Fig. \ref{fig:avg_weak_meas_countor} in terms of final state fidelity with respect to the solution state.

It is worth noting here that the $\text{TTS}_{99\%}$ scaling $\lambda$ 
studied in this subsection should not be interpreted as a performance metric of the algorithm in the asymptotic sense. For the limited size of systems that we are able to simulate, the absolute value of the scaling base number $\lambda$ can be easily manipulated. To see this, consider either Algorithm \ref{alg1} or Algorithm \ref{alg3} for some set of  finite system sizes $n_{list} = \{n_1, \dots, n_d \}$. 
One can choose $T_f$ finite but large enough such that the dynamics for all  $n_i$ in $n_{list}$ is in the quasi-adiabatic limit, where the final solution state probability is close to $1$. 
In this case, one only needs one attempt of the algorithm in order to find the solution bitstring, and thus the benchmarked scaling parameter $\lambda$ is masked by the finite-size effect and appears to be $1$, which is not reasonable. 
Nevertheless, $\lambda$ can serve as a good metric for demonstrating performance improvement when using different parameters in the algorithm or when comparing different variants of the algorithm, e.g., the heralded and unheralded (average) generalized measurement algorithms (Algorithm~\ref{alg3} and Algorithm~\ref{alg1}, respectively).

\subsubsection{Optimal $\text{TTS}_{99\%}$ Scaling}

 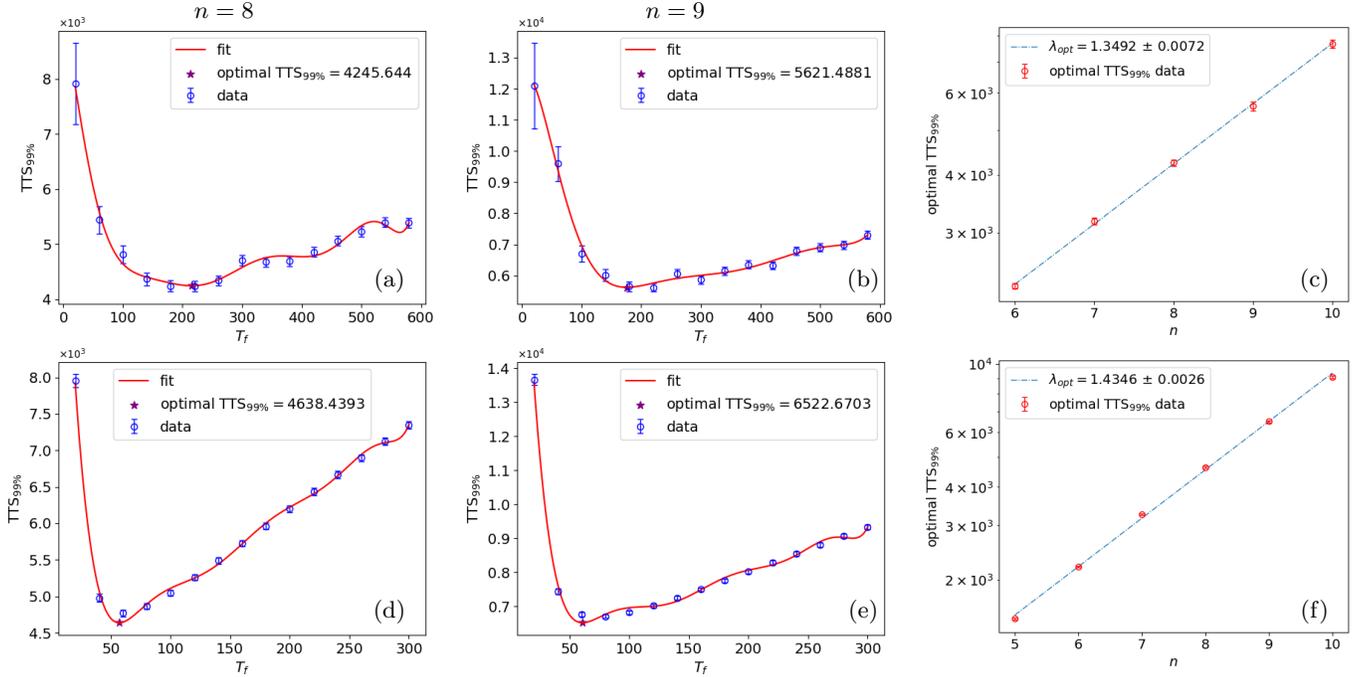
\begin{figure*}
     \centering
    \includegraphics[width = 0.32\textwidth, trim = {5 0 35 25}, clip]{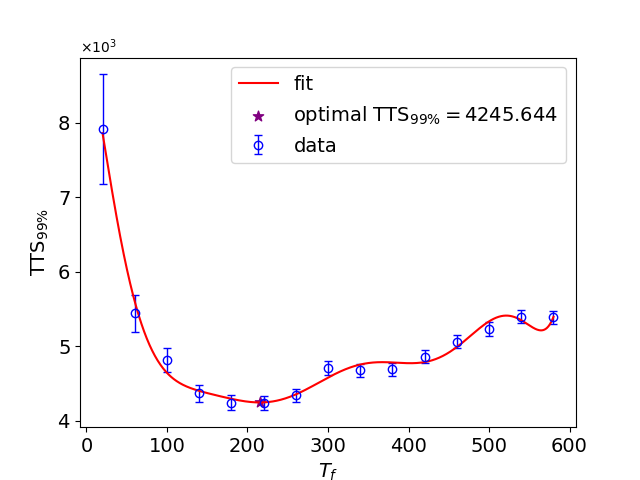}
    \hfill
    \includegraphics[width = 0.32\textwidth, trim = {5 0 35 25}, clip]{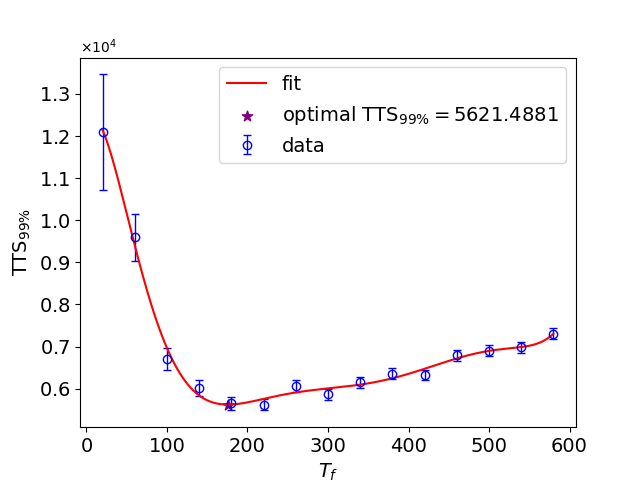}
    \hfill
    \includegraphics[width = 0.32\textwidth, trim = {5 5 5 5}, clip]{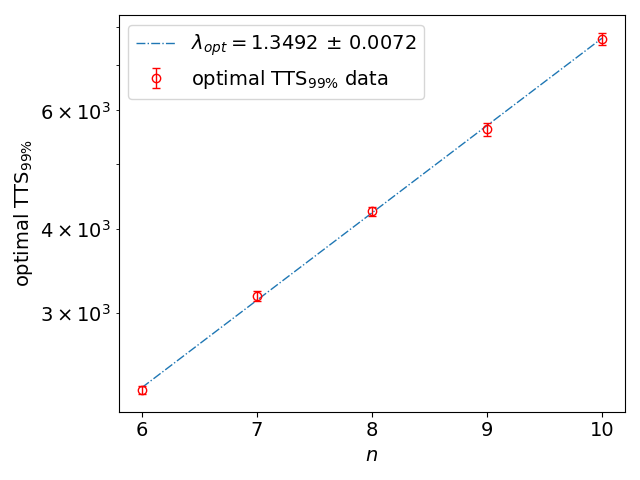}
    \\
    \includegraphics[width = 0.32\textwidth, trim = {5 0 35 25}, clip]{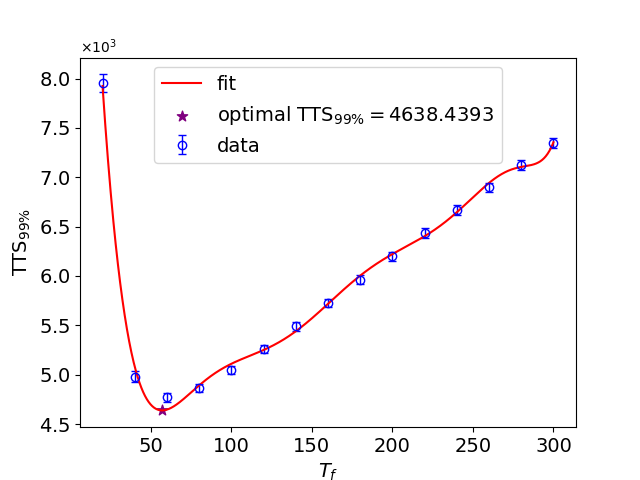}
    \hfill
    \includegraphics[width = 0.32\textwidth, trim = {5 0 35 25}, clip]{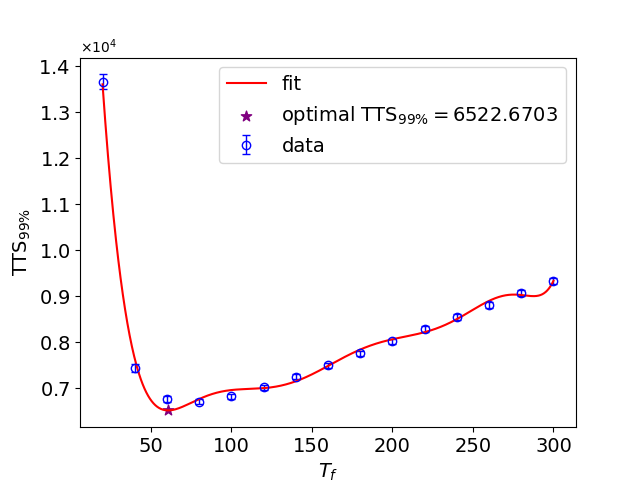}
    \hfill
    \includegraphics[width = 0.32\textwidth, trim = {5 5 5 5}, clip]{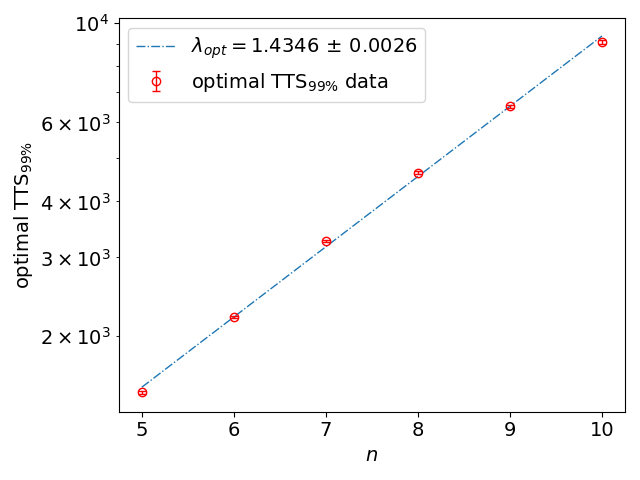}
    \\
    \begin{tikzpicture}[overlay]
    \node[] at (-3.8,5.6) {(a)};
    \node[] at (2.5,5.6) {(b)};
    \node[] at (8.5,5.6) {(c)};
    \node[] at (-3.8,1.2) {(d)};
    \node[] at (2.5,1.2) {(e)};
    \node[] at (8.5,1.2) {(f)};
    \node[] at (-6,9.2) {\color{black} $n = 8$};
    \node[] at (0,9.2) {\color{black} $n = 9$};
    \end{tikzpicture} \vspace{-12pt}
    \caption{We plot $\text{TTS}_{99\%}$ as a function of the total dragging time $T_f$ under both Algorithm \ref{alg3} with heralded dynamics (panels (a) - (c)), and Algorithm \ref{alg1} with average dynamics (panels (d) - (f)). 
    Simulation parameters are the same as Fig. \ref{fig_TTS_vs_nq} with $\Delta t = 10$ and a linear schedule is also used. 
    We observe clear optima for $\text{TTS}_{99\%}$ in panels (a), (b), (d), and (e). We locate these optimal $\text{TTS}_{99\%}$ values by fitting the data with a polynomial function. In panel (c) and panel (f), we plot the obtained optimal $\text{TTS}_{99\%}$ as a function of qubit number $n$, and fit the data with an exponential function to evaluate the scaling base number $\lambda_{opt}$. These figures with $\Delta t = 10$ serve as a demonstration of how $\lambda_{opt}$ is extracted. Similar calculations were  made for $\Delta t \in \{0.01, 0.1, 1, 10, 100\}$ to allow construction of Fig. \ref{fig_optimal_lambda_TTS_vs_dt} below.
    }\label{fig_TTS_vs_T_f_heralded_dt_10}
 \end{figure*}

\begin{figure*}
    \centering
    \includegraphics[width = 0.49\textwidth, trim = {5 5 40 40}, clip]{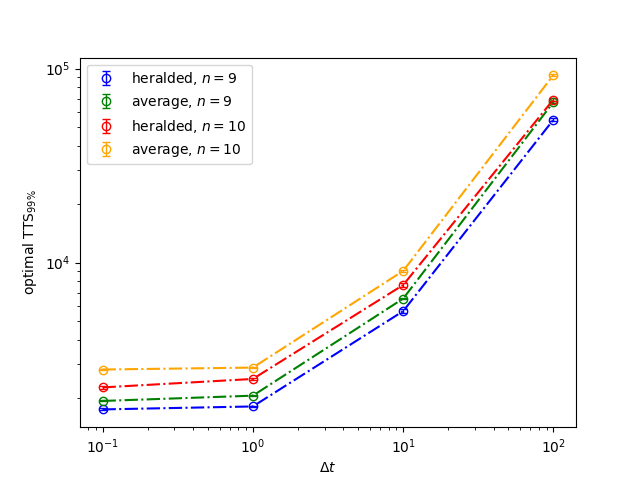}
    \hfill
    \includegraphics[width = 0.49\textwidth, trim = {5 5 40 40}, clip]{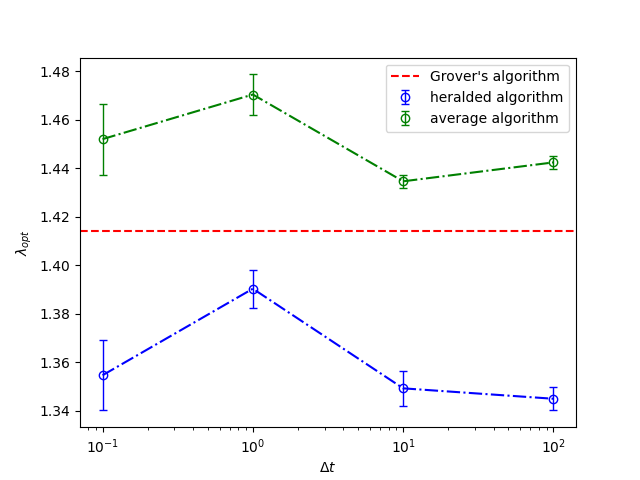}
    \\
    \begin{tikzpicture}[overlay]
    \node[] at (-4,1.5) {(a)};
    \node[] at (5.1,1.5) {(b)};
    \end{tikzpicture} \vspace{-12pt}
    \caption{Dependence of the performance of the generalized measurement-based algorithms on $\Delta t$ for both the average algorithm, Algorithm \ref{alg1} and the heralded algorithm, Algorithm \ref{alg3}. Panel (a) displays the optimal $\text{TTS}_{99\%}$ as a function of $\Delta t$ for different system sizes. We see that the optimal $\text{TTS}_{99\%}$ increases as $\Delta t$ increases, and the heralded algorithm has a consistently smaller optimal $\text{TTS}_{99\%}$ than the corresponding average algorithms. Panel (b) shows $\lambda_{opt}$ as a function of $\Delta t$ for both algorithms. The scaling $\lambda_{opt}$ is seen to be generically better in the strong measurement region than in the weak measurement region. It also evident that $\lambda_{opt}$ for the heralded algorithm is uniformly better than that for the average algorithm, consistent with the behavior in panel (a). 
    }\label{fig_optimal_lambda_TTS_vs_dt}
\end{figure*}
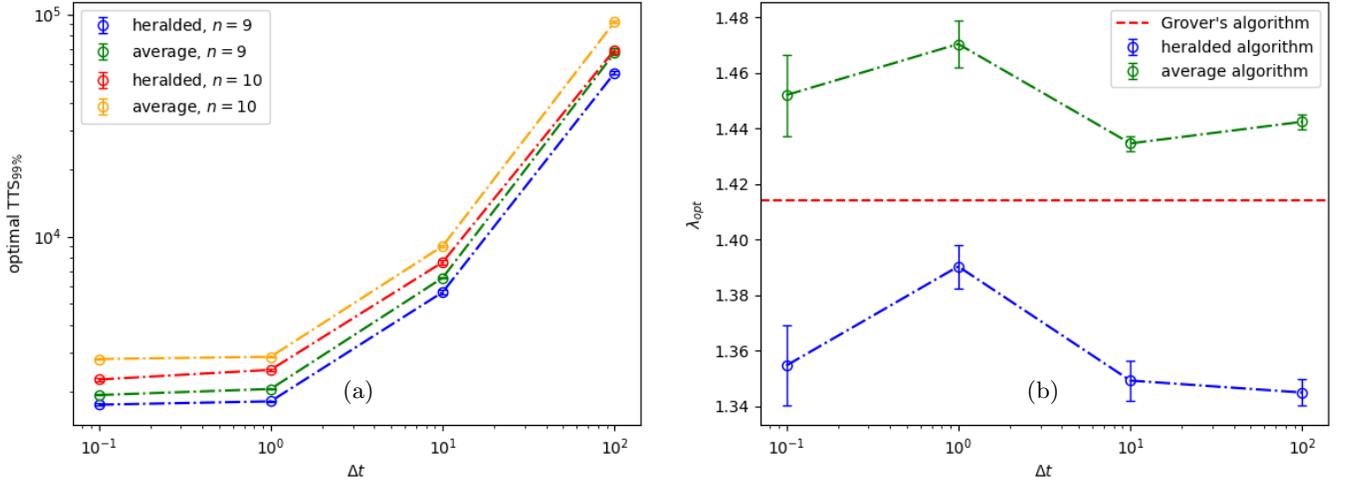
 
As shown in \cite{ronnow2014defining, albash2018demonstration} for quantum annealing, to benchmark the scaling of such algorithms in a manner independent of the value of the single-shot running time, one needs to identify an optimal $\text{TTS}_{99\%}$ for each system size $n$ and then obtain the scaling parameter $\lambda_{opt}$, such that $\mathrm{TTS}_{opt} \sim \lambda_{opt}^n$.
The existence of an optimal $\text{TTS}_{99\%}$ value is intuitive from the following arguments. 
When $T_f$ is small, one typically needs a diverging (or $2^n$) number of repeats of the algorithm in order to find the solution bitstring, therefore in this case the $\text{TTS}_{99\%}$ is also diverging or as large as $\sim 2^n$. 
In contrast, when $T_f$ is large and divergent, while one only needs a single attempt of the algorithm, the $\text{TTS}_{99\%}$ is then also divergent.
Therefore, one would expect a sweet spot where $T_f$ is finite and the $\text{TTS}_{99\%}$ attains an optimal value by balancing the trade-off of using a large $T_f$ with a single attempt versus using a small $T_f$ with many attempts.

We demonstrate here that similar to quantum annealing, both the unheralded and heralded Zeno-dragging algorithms also exhibit an optimal $\text{TTS}_{99\%}$ that depends on the system size $n$ and on $\Delta t$. 
To understand how the performance of the algorithms scales with the system size in a well-defined sense, we study here the scaling of this optimal $\text{TTS}_{99\%}$ with $n$.
 
In Fig.~\ref{fig_TTS_vs_nq} it is evident that when the number of qubits is larger than some value, there can be some crossings of the $\text{TTS}_{99\%}$ lines, e.g., for both algorithms with $\Delta t = 10$. 
This implies that when increasing $T_f$, the $\text{TTS}_{99\%}$ exhibits a 
nonlinear 
behavior and does not necessarily increase or decrease with $T_f$, which is an indication of the possible existence of an optimal $\text{TTS}_{99\%}$ value. We investigate this nonlinearity by plotting the $\text{TTS}_{99\%}$ as a function of $T_f$ for system size up to $n = 10$ in Fig.~\ref{fig_TTS_vs_T_f_heralded_dt_10}. 
This shows the $\text{TTS}_{99\%}$ versus $T_f$ for both unheralded and heralded algorithms with $\Delta t = 10$, for system sizes $n = 8$ and $n = 9$.
For these sizes the $\text{TTS}_{99\%}$ clearly exhibits an optimum value. 
By fitting the optimal values obtained from different system sizes $n \in [5, 10]$ with an exponential function (panels (c) and (f) in Fig.~\ref{fig_TTS_vs_T_f_heralded_dt_10}), we determined the optimal $\text{TTS}_{99\%}$ to scale as $\sim (\lambda_{opt})^n$, with 
$\lambda_{opt} \approx 1.4346$ for Algorithm \ref{alg1} (average dynamics) and
$\lambda_{opt} \approx 1.3492$ for Algorithm \ref{alg3} (heralded dynamics).

We then repeat the procedure for a range of measurement times $\Delta t \in \{0.01, 0.1, 1, 10, 100 \}$ for both algorithms, and obtain the values of $\lambda_{opt}$ for each value of $\Delta t$. 
The results are shown in Fig.~\ref{fig_optimal_lambda_TTS_vs_dt}, where we see that for $n=9, 10$ qubits, the $\text{TTS}_{99\%}$ from the heralded algorithm is smaller (better) than for the average algorithm (left panel a), and more importantly, the scaling base parameter $\lambda_{opt}$ is significantly smaller for the heralded algorithm, over a wide range of $\Delta t$ values (right panel b).
We notice here that the scaling $\lambda_{opt}$ achieved by the heralded algorithm is systematically smaller than that for Grover's algorithm \cite{Grover} which has a scaling base number $\lambda  = \sqrt{2} \approx 1.414$. This is a result of the proper usage of structure in our algorithm (recall that Grover is for unstructured search). 
Overall, it is evident that the heralded algorithm using signal filtering shows better $\text{TTS}_{99\%}$ behavior than the unconditional algorithm for all values of $\Delta t$. 
This indicates the advantages of implementing the earlier error detection of the heralded algorithm in the quasi-adiabatic regime, rather than relying solely on the autonomous features on which the unconditional algorithm is based, deep in the adiabatic regime.

%%%%%%%%%%%%%%%%%%%%%%%%%%%%%%%%%%%%%%%%%%%%%%%%%%%%%%%%%%%%%%
%%%%%%%%%%%%%%%%%%%%%%%%%%%%%%%%%%%%%%%%%%%%%%%%%%%%%%%%%%%%%%
\section{Discussion and Outlook \label{SEC:conclude}}

In this paper we have extended a quantum algorithm first proposed by Benjamin, Zhao, and Fitzsimons \cite{benjamin2017measurement} (BZF), 
that uses projective measurements for solving Boolean satisfiability problems, to use instead generalized measurements for measurements of the Boolean clauses and for readout. 
First, we proposed a dissipation-only variant of the generalized algorithm, which is equivalent to discarding all the measurement results along the dynamics, or equivalently, using average dynamics.
Second, we developed a filtering protocol for detecting clause failure based on the noisy signals of the generalized measurements.
We discussed the statistics of solution readout following both versions of the generalized algorithm, again using generalized measurements, and showed how this readout time can be incorporated into the estimation of the expected time to solution (TTS).
We then established the convergence of the algorithm in the limit of long dragging time and time--continuous measurement, noting that the time--continuous algorithm is an instance of ``Zeno dragging'' \cite{lewalle2023optimal}, where autonomous and deterministic convergence of the dynamics to a solution subspace is guaranteed in the Zeno limit.
Finally, we illustrated the algorithmic dynamics, including the convergence in the Zeno limit and the filtering protocol.  
We have done this in detail for an illustrative and relatively simple 2-qubit 2-SAT problem, and then performed extensive numerical simulations on larger 2-SAT and 3-SAT problems involving 4--10 qubits.

With these larger simulations we numerically benchmarked the performance of both the heralded and unheralded algorithms as a function of the total pure dragging time $T_f$, the clause density $\alpha$, as well as the duration of the clause measurements, $\Delta t$. 
We found that both algorithms exhibit a computational phase transition, as many other classical and quantum algorithms do, where the $k$-SAT problem is only hard for the algorithms to solve in the vicinity of the region of a critical value of $\alpha$. 
We found strong evidence that working in the weak continuous limit is advantageous over operating in the strong measurement limit, at least for problems on moderate numbers of qubits. 
This result can be extracted from the observations that:
i) for fixed $T_f$, the scaling base number $\lambda$ decreases as $\Delta t$ decreases,
and ii) the optimal TTS decreases as $\Delta t$ decreases. 
Finally, we also demonstrated that the algorithm performance can be systematically improved using clause detection and filtering in order to herald errors and restarting runs.
This is shown in the improvement of the success probability for identifying a satisfying solution near the critical $\alpha$ that is enabled by heralded dynamics, as well as the smaller value of the optimal TTS scaling base parameter $\lambda_{opt}$ in the heralded dynamics relative to the average dynamics.

This generalized measurement--driven approach shows some unique advantages and potential.
For example, even though there are still many strategies by which one might optimize the algorithm, the optimal TTS scalings that we evaluate here are better than those achieved by Grover's algorithm \cite{Grover},  characterized by $\lambda \approx 1.414$, and are comparable to those of Sch\"{o}ning's celebrated results \cite{schoning1999probabilistic}, characterized by $\lambda \approx 1.33$, respectively. 
BZF \cite{benjamin2017measurement} have shown that their projective algorithm admits a solution state of the form of Eq. \eqref{general-solution-state} for every solution bitstring to a $k$-SAT problem. 
Adapting this to the case of time--continuous monitoring, we have shown in Sec.~\ref{sec-Convergence} and Appendix \ref{sec-app-converge} that in the limit of slow Zeno dragging, we can expect to deterministically converge to such a solution, assuming we apply the dynamics to a satisfiable $k$-SAT problem, i.e., when a solution does exist.
This implies that for large $T_f/\tau$, we can push $\lambda$ arbitrarily close to $1$ (albeit with a diverging pre-factor in the TTS). 
This is a direct consequence of the fact that operations that use the Zeno effect for stabilization generally do so in the mean, i.e.,~they stabilize autonomously \cite{Facchi_2002, facchi2008quantum, Mirrahimi_2007, Ticozzi_2008, amini2011stability, Ticozzi_2013, Benoist_2017, Cardona_2018, liang2019exponential, Cardona_2020, amini2023exponential, liang2023exploring, lewalle2023optimal, Burgarth2020quantumzenodynamics, Burgarth2022oneboundtorulethem}.
The use of feedback in addition to this autonomous stabilization offers the possibility of not only detecting errors in real time, as we have discussed in Secs.~\ref{sec_2Q-herald} and \ref{SEC:BZF-kSAT-examples}, but also of correcting them immediately.
Such feedback would be a natural subject for theoretical work following up on the present investigation, and would be closely related to continuous quantum error correction \cite{Lihm_2018, lebreuilly2021autonomous, Mirrahimi_2007, Ticozzi_2008, amini2011stability, Ticozzi_2013, Benoist_2017, Cardona_2018, liang2019exponential, Cardona_2020, amini2023exponential, liang2023exploring, lewalle2023optimal, Ahn_2002, Oreshkov_2007, Mascarenhas_2010, Atalaya_CQEC, Mohseninia2020alwaysquantumerror, Atalaya_2021_CQEC, Convy2022logarithmicbayesian, Livingston_2022, Convy_2022, Ahn_2003, Ahn_2004, Sarovar_2004, vanhandel2005optimal}. 

There are still many modifications that could be made to optimize the performance of the algorithm beyond what we have developed above. 
For instance, based on the equivalence of Zeno--dragging control and Zeno--exclusion control, one can exploit the newly developed CDJ--P method \cite{lewalle2023optimal}, which is a open-loop control technique to optimize the schedule function $\theta(t)$. 
Another direction for future work could be developing better filters for error detection, such as a Bayesian filter or machine learning filter, for use in the feedback strategies mentioned in the preceding paragraph.

Implementation of this algorithm on real hardware will require the engineering of several complicated measurements (or dissipators), which do not commute and will have to be implemented simultaneously. 
We suggest that progress on this front could realistically aim to follow a similar roadmap to the one we have followed in this theoretical paper, namely, to start with implementation of a two--qubit 2-SAT problem. Successful realizing of such a 2-SAT problem would be a strong step towards understanding and solving some of the engineering challenges involved and establishing proof-of-principle, at which point one could reasonably then consider attempting to scale the relevant experimental methods towards problems of a computationally more interesting size. 
Several experimental results that suggest ways forward here already exist. In particular, measuring or dissipating dynamic observables has been realized in a few contexts \cite{ZenoDragging, Touzard_2018, Leigh_Phase}, as has measurement of non-commuting observables \cite{ShayLeigh2016, Livingston_2022}. 
Ref.~\cite{Livingston_2022}, focused on continuous quantum error correction, is especially relevant to our current context since the real-time error correction is implemented in that work on three qubits by simultaneously monitoring the parity of overlapping pairs of qubits, thereby connecting with several conceptual aspects of the generalized BZF algorithm presented in this work. 
In addition, the Zeno effect has been used for control, demonstrating that it is possible to engineer measurements that divide a larger system into specific subspaces \cite{blumenthal2021}, much like the clause measurements used here.

Taken together, we believe that our results indicate that the generalized BZF $k$-SAT algorithm shows promise for realization of a measurement-driven $k$-SAT solver, suggesting new avenues along which measurement--driven quantum computation might be further developed.

\subsection*{Acknowledgements}

This material is based upon work supported by the U.S. Department of Energy, Office of Science, National Quantum Information Science Research Centers, Quantum Systems Accelerator. 
PL is grateful to the UMass Lowell department of Physics \& Applied Physics for their hospitality during part of this manuscript's preparation. 
This document was written without the use of AI. Simulations and calculations were performed with the help of Python, C++, and Mathematica.

%%%%%%%%%%%%%%%%%%%%%%%%%%%%%%%%%%%%%%%%%%%%%%%%%%%%%%%%%%%%%%%%%
%%%%%%%%%%%%%%%%%%%%%%%%%%%%%%%%%%%%%%%%%%%%%%%%%%%%%%%%%%%%%%%%%
%%%%%%%%%%%%%%%%%%%%%%%%%%%%%%%%%%%%%%%%%%%%%%%%%%%%%%%%%%%%%%%%%
\appendix

%%%%%%%%%%%%%%%%%%%%%%%%%%%%%%%%%%%%%%%%%%%%%%%%%%%%%%%%%%%%%%%%%
%%%%%%%%%%%%%%%%%%%%%%%%%%%%%%%%%%%%%%%%%%%%%%%%%%%%%%%%%%%%%%%%%
%%%%%%%%%%%%%%%%%%%%%%%%%%%%%%%%%%%%%%%%%%%%%%%%%%%%%%%%%%%%%%%%%
\section{Convergence to Unique Solution States in the Zeno Limit \label{sec-app-converge}}

This appendix elaborates on Sec.~\ref{sec-Convergence}. 
We here use the CDJ stochastic action \cite{Chantasri2013, Chantasri2015} to illustrate that the continuous version of the BZF $k$--SAT algorithm described in the main text converges in probability to the desired solution dynamics. 
For simplicity, we focus on the frustration--free case, emphasizing first the situation in which there is a unique solution. 
This work draws extensively on the theoretical formalism we recently developed for measurement--driven quantum control \cite{lewalle2023optimal}; other conceptually relevant works include \cite{Mirrahimi_2007, Ticozzi_2008, amini2011stability, Ticozzi_2013, Benoist_2017, Cardona_2018, liang2019exponential, Cardona_2020, amini2023exponential, liang2023exploring}. 

Consider the pure--state evolution under perfect clause monitoring, as per \begin{widetext}
\begin{subequations} \label{strato-sme} \be 
\dot{\rho} = \sum_i^m \left\lbrace \widehat{\mathsf{A}}(\rho,\hat{X}_i) + \widehat{\mathsf{b}}(\rho,\hat{X}_i) \circ \frac{dW_i}{dt} \right\rbrace,\quad\text{with}
\ee \be \begin{split}
\widehat{\mathsf{A}}(\rho,\hat{L}_i) &= \mathsf{s}_i\,\widehat{\mathsf{b}}(\rho,\hat{L}_i) + \frac{1}{4\tau}\left(2\,\rho\,\tr{\hat{L}_i\,\rho\,\hat{L}_i} - \hat{L}_i^2\,\rho - \rho\,\hat{L}_i^2\right), \\
\widehat{\mathsf{b}}(\rho,\hat{L}_i) &= \frac{1}{2\sqrt{\tau}}\left(\hat{L}_i\,\rho + \rho\,\hat{L}_i\right) - \rho\,\mathsf{s}_i, \quad\&\quad \mathsf{s}_i = \frac{1}{\sqrt{\tau}}\,\tr{\rho\,\hat{L}_i}.
\end{split} \ee \end{subequations}
Here $\widehat{\mathsf{A}}_i$ is the \emph{Stratonovich} drift associated with each Hermitian monitored observable $\hat{L}_i$, and $\widehat{\mathsf{b}}_i$ is the corresponding diffusion, with $\mathsf{s}_i$ the expected signal from the clause measurement, i.e.~$r_i\,dt = \mathsf{s}_i\,dt + dW_i$. 
Recall also the definition Eq.~\eqref{little-g}
\be 
\mathsf{g}_i(\rho) = \frac{1}{2\tau}\left(\ensavg{\hat{L}_i^2} - \ensavg{\hat{L}_i}^2 \right)  = \frac{1}{2}\mathrm{var}\left(\mathsf{s}_i\right).  
\ee 
We note the following properties of these functions, that hold for time--independent $\hat{L}_i$ (or hold instantaneously for time--dependent $\hat{L}_i$): 
\begin{itemize}
\item If $\tilde{\rho}_i$ is the $+1$ eigenstate of $\hat{L}_i$, then $\widehat{\mathsf{A}}_i(\tilde{\rho}_i) = 0$, $\widehat{\mathsf{b}}_i(\tilde{\rho}_i) = 0$, and $\mathsf{g}_i(\tilde{\rho}_i) = 0$. 
\item If $\tilde{\rho}$ is a simultaneous $+1$ eigenstate of \emph{all} the $\hat{L}_i$, then all of the $\widehat{\mathsf{A}}_i$, $\widehat{\mathsf{b}}_i$, and $\mathsf{g}_i$ vanish, such that $\tilde{\rho}$ is a fixed point of the measurement--conditioned dynamics \eqref{strato-sme}. 
\end{itemize}
These properties hold at any given instant for time--dependent $\hat{L}_i$, or in general for time--independent $\hat{L}_i$. 

Now let us introduce the CDJ path integral to this analysis (for the derivation of the equations below, see \cite{Chantasri2013, lewalle2023optimal}). 
We have some coordinates $\mathbf{q}$ parameterizing the density matrix $\rho$ \cite{Bertlmann2008}, and co-states $\boldsymbol{\Lambda}$ conjugate to the $\mathbf{q}$. 
After marginalizing away (or equivalently optimizing) the noises $dW_i$, the trajectory probability may be expressed as
\be 
\wp(\mathbf{q}(t),t\in[0,T]|\rho_0) = \int \mathcal{D}[\boldsymbol{\Lambda}] \exp\left[ -\int\limits_0^T \left( \boldsymbol{\Lambda}\cdot\dot{\mathbf{q}} - \mathcal{H}_\mathrm{CDJ}^\star\right)\right] \quad\text{for}\quad \mathcal{H}_\mathrm{CDJ}^\star = \tfrac{1}{2}\,\boldsymbol{\Lambda}^\top\,\boldsymbol{\mathsf{B}}\,\boldsymbol{\Lambda} + \boldsymbol{\Lambda}^\top\,\boldsymbol{\mathsf{A}} - \mathsf{g},
\ee 
where $\boldsymbol{\mathsf{B}} = \boldsymbol{\mathsf{b}}_i^\top\,\boldsymbol{\mathsf{b}}_i$, $\boldsymbol{\mathsf{A}} = \sum_i \boldsymbol{\mathsf{A}}_i$, and $\mathsf{g} = \sum_i \mathsf{g}_i$, are all expressed in the $\mathbf{q}$ coordinates. 
\begin{itemize}
\item At $\tilde{\rho}\rightarrow \tilde{\mathbf{q}}$, a simultaneous $+1$ eigenstate of all the $\hat{L}_i$, we have $\dot{\mathbf{q}} = 0$ and $\mathcal{H}_\mathrm{CDJ}^\star = 0$.
Then for time--independent $\hat{L}_i$ and initial state $\rho_0 = \tilde{\rho}$, we have $\wp(\tilde{\mathbf{q}}) = 1$ for all time. 
\end{itemize}
Note that all of the $\mathsf{g}_i(\rho)$ are non-negative (they are variances) that effectively quantify the distance between $\rho$ and the nearest $\hat{L}_i$ eigenstate. 
These can be interpreted as an information gain rate from each measurement \cite{Cylke_2022inprep,Karmakar_2022,Philippe_Thesis,BookBarchielli} and their collective effect is to statistically punish paths which go far from the eigenstate $\tilde{\rho}$. 
The presence of this term is thus important in understanding how the Zeno limit converges \emph{in probability} towards following $\tilde{\rho}$, i.e.,~the Zeno pinning becomes deterministic. 
See \cite{lewalle2023optimal,Mirrahimi_2007, Ticozzi_2008, amini2011stability, Ticozzi_2013, Benoist_2017, Cardona_2018, liang2019exponential, Cardona_2020, amini2023exponential, liang2023exploring} for further related comments. 

We now recapitulate how the static measurements described above apply to our $k$-SAT scenario involving moving observables. 
We consider here for simplicity $k$-SAT problems with a unique solution, i.e.,~we assume that the ideal dynamics can be expressed by a unique solution of the form of Eq.~\eqref{general-solution-state}. 
Then we can define a frame change by the rotation Eq.~\eqref{Qframe}
\be \label{Q-definition-app}
\hat{Q} = \bigotimes_{j = 1}^n \hat{R}_Y^{(j)}(\mp(\tfrac{\pi}{2}-\theta))
\ee
such that the ideal solution dynamics are effectively rendered static in this frame.
Note that the assignment of $+$ or $-$ to the rotation on any given qubit requires knowing a satisfying solution bitstring $\mathbf{s}$.
The Stratonovich dynamics in this frame can be expressed via $\varrho = \hat{Q}^\dag\,\rho\,\hat{Q}$, and read
\be 
\dot{\varrho} = i\,[\hat{H}_Q,\varrho] + \hat{Q}^\dag\,\dot{\rho}\,\hat{Q} 
= i[\hat{H}_Q,\varrho] + \sum_i^m \left\lbrace \widehat{\mathsf{A}}(\varrho,\hat{\mathcal{X}}_i) + \widehat{\mathsf{b}}(\varrho,\hat{\mathcal{X}}_i) \circ \frac{dW_i}{dt} \right\rbrace
\quad\text{with}\quad \hat{H}_Q = \tfrac{1}{2}\,\dot{\theta}\sum_j^n s_j\,\hat{\sigma}_y^{(j)},
\ee \end{widetext}
where $\mathbf{s} \in \{-1,+1\}^n$ is the solution bitstring corresponding to the $\hat{Q}$--frame.
We can now easily adapt the arguments from the previous paragraphs.
Specifically, it is evident that if we initialize our system in $\tilde{\varrho}$, corresponding to $\rho = (\ket{+}\bra{+})^{\otimes n}$, then the only contribution to the dynamics is $\dot{\varrho} = i[\hat{H}_Q,\varrho]$. 
For slow rotation $\dot{\theta}\,\tau \ll 1$, this is a small perturbation about the fixed point of the adiabatic dynamics as discussed in the main text, and by the arguments above, the fixed point is thus preserved perfectly in the Zeno limit $\dot{\theta}\rightarrow 0$. 

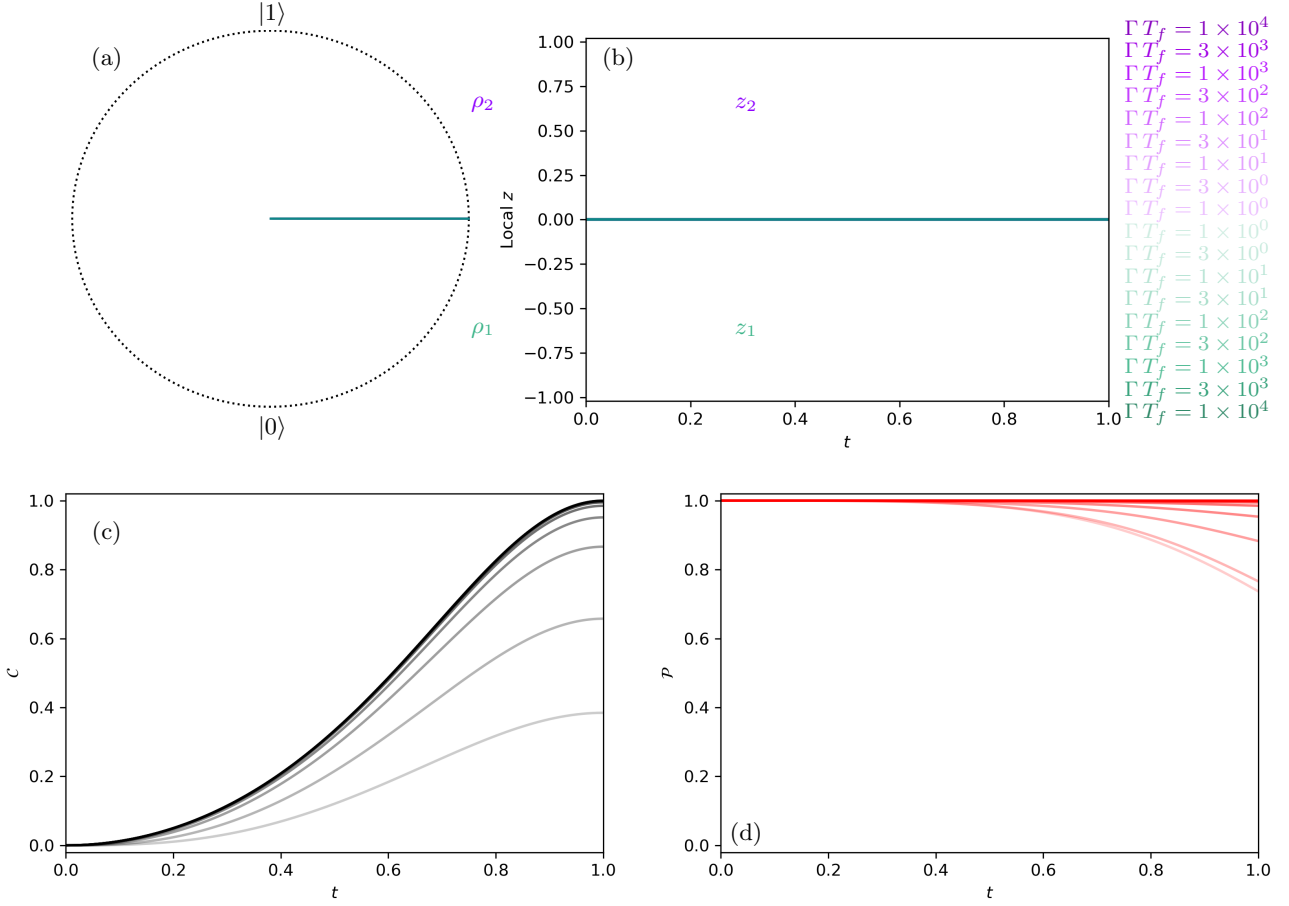
\begin{figure*}
\centering
~\\
\hspace{-0.75cm}\includegraphics[height = .25\textheight, width = .225\textheight, trim = {51 0 41 45}, clip]{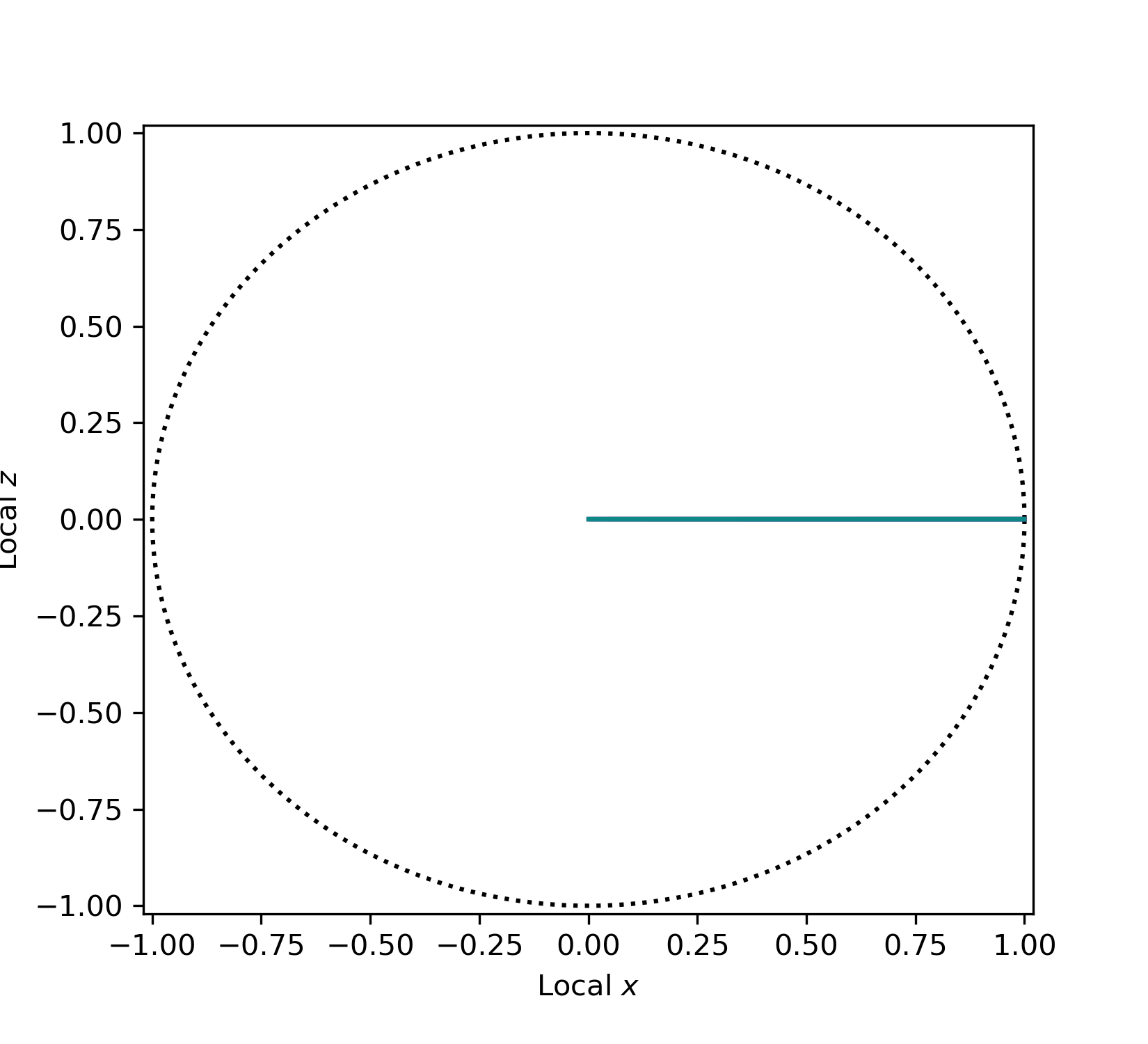}
\includegraphics[width = .49\textwidth]{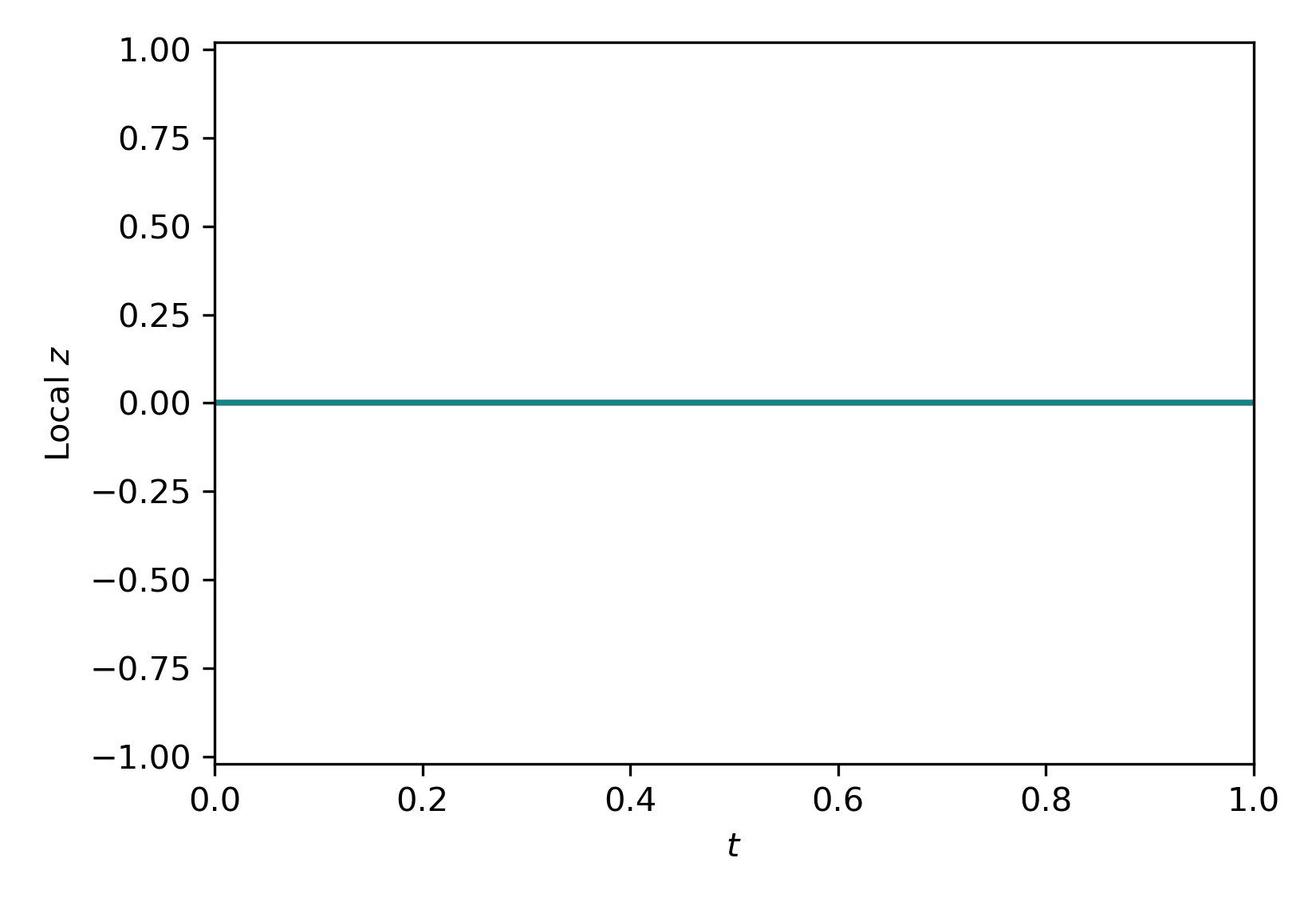} \hspace{1cm}
\\
\includegraphics[width = .48\textwidth]{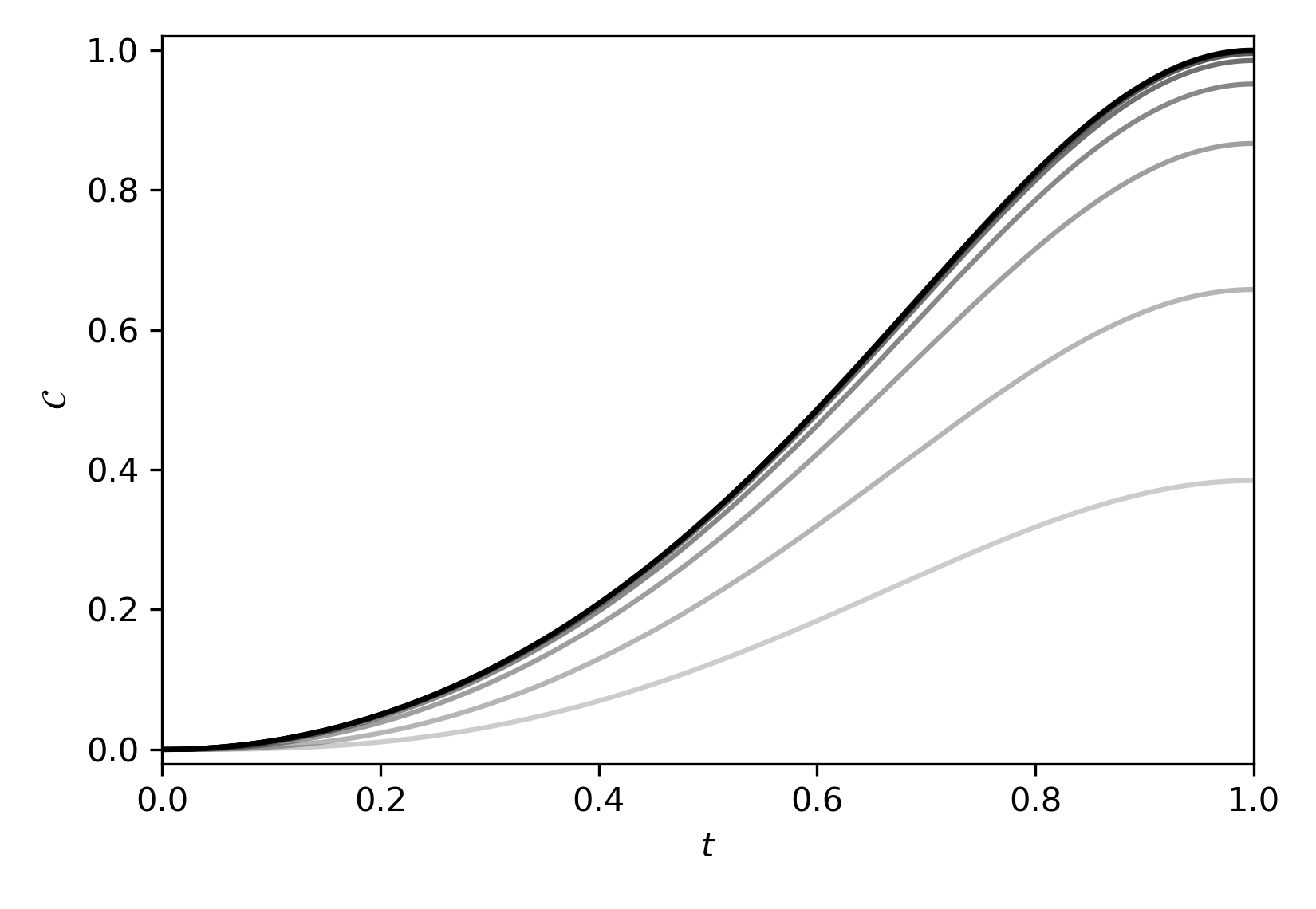} 
\includegraphics[width = .48\textwidth]{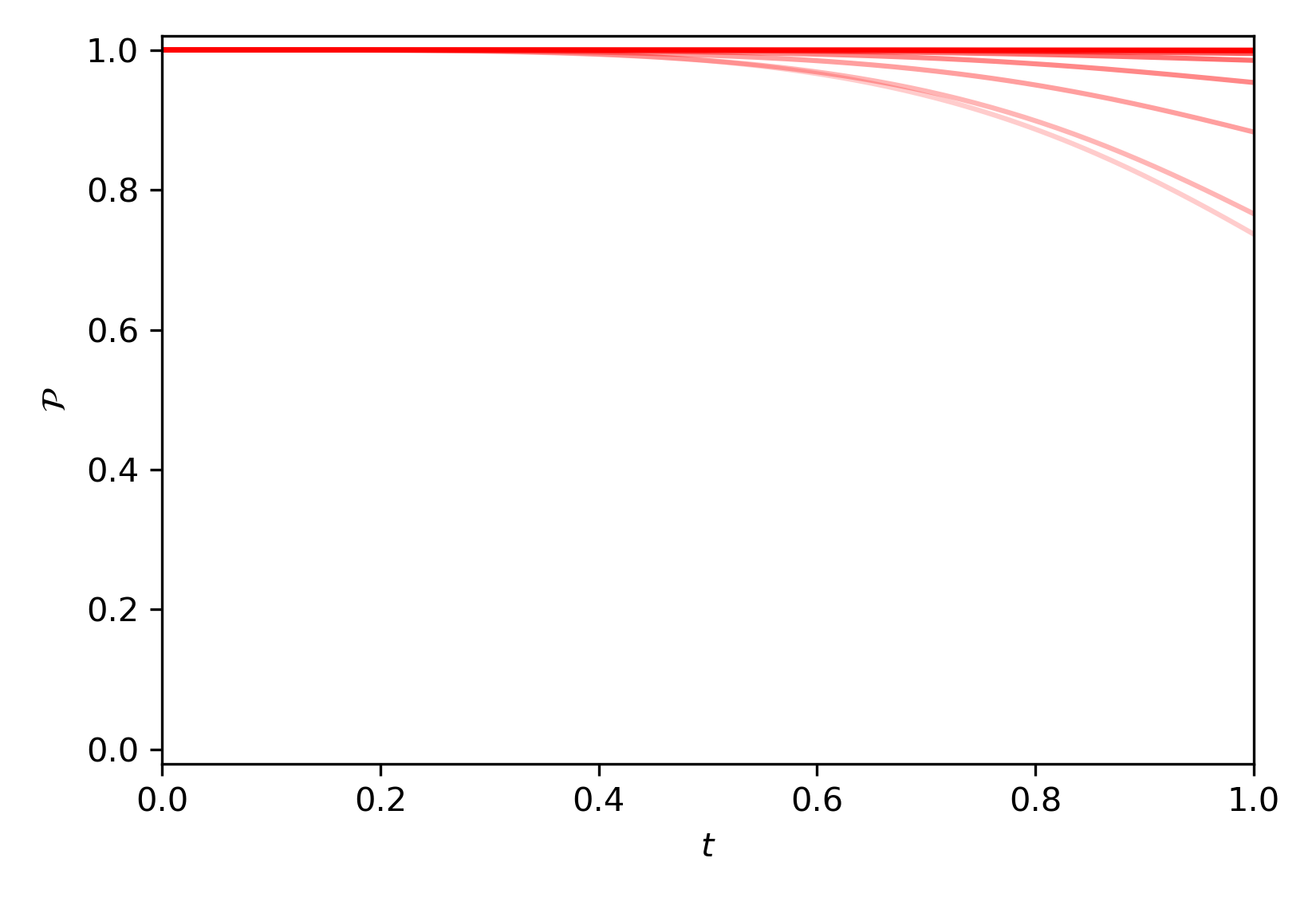} \\
\begin{tikzpicture}[overlay, yshift = -21pt]
\draw[fill = white, draw = white] (-8,7.3) rectangle (-1,7.95);
\node[] at (-7,12.6) {(a)};
\node[] at (-0.2,12.6) {(b)};
\node[] at (-7,6.3) {(c)};
\node[] at (1.5,2.3) {(d)};
%%%%%%%%%%%%%%%%%
\node[] at (-4.8,13.2) {$\ket{1}$};
\node[] at (-4.8,7.7) {$\ket{0}$};
\node[] at (-2,12) {\color{electricviolet} $\rho_2$};
\node[] at (-2,9) {\color{mint} $\rho_1$};
%%%%%%%%%%%%%%%%%
\node[] at (1.5,12) {\color{electricviolet} $z_2$};
\node[] at (1.5,9) {\color{mint} $z_1$};
\node[] at (7.5,13.0) {\color{patriarch!75!black} \footnotesize $\Gamma\,T_f = 1\times 10^4$};
\node[] at (7.5,12.7) {\color{patriarch!90!black} \footnotesize $\Gamma\,T_f = 3\times 10^3$};
\node[] at (7.5,12.4) {\color{patriarch!90!white} \footnotesize $\Gamma\,T_f = 1\times 10^3$};
\node[] at (7.5,12.1) {\color{patriarch!75!white} \footnotesize $\Gamma\,T_f = 3\times 10^2$};
\node[] at (7.5,11.8) {\color{patriarch!60!white} \footnotesize $\Gamma\,T_f = 1\times 10^2$};
\node[] at (7.5,11.5) {\color{patriarch!45!white} \footnotesize $\Gamma\,T_f = 3\times 10^1$};
\node[] at (7.5,11.2) {\color{patriarch!37!white} \footnotesize $\Gamma\,T_f = 1\times 10^1$};
\node[] at (7.5,10.9) {\color{patriarch!30!white} \footnotesize $\Gamma\,T_f = 3\times 10^0$};
\node[] at (7.5,10.6) {\color{patriarch!25!white} \footnotesize $\Gamma\,T_f = 1\times 10^0$};
\node[] at (7.5,7.9) {\color{mint!75!black} \footnotesize $\Gamma\,T_f = 1\times 10^4$};
\node[] at (7.5,8.2) {\color{mint!90!black} \footnotesize $\Gamma\,T_f = 3\times 10^3$};
\node[] at (7.5,8.5) {\color{mint!90!white} \footnotesize $\Gamma\,T_f = 1\times 10^3$};
\node[] at (7.5,8.8) {\color{mint!75!white} \footnotesize $\Gamma\,T_f = 3\times 10^2$};
\node[] at (7.5,9.1) {\color{mint!60!white} \footnotesize $\Gamma\,T_f = 1\times 10^2$};
\node[] at (7.5,9.4) {\color{mint!45!white} \footnotesize $\Gamma\,T_f = 3\times 10^1$};
\node[] at (7.5,9.7) {\color{mint!37!white} \footnotesize $\Gamma\,T_f = 1\times 10^1$};
\node[] at (7.5,10.0) {\color{mint!30!white} \footnotesize $\Gamma\,T_f = 3\times 10^0$};
\node[] at (7.5,10.3) {\color{mint!25!white} \footnotesize $\Gamma\,T_f = 1\times 10^0$};
\end{tikzpicture} \vspace{-20pt}
\caption{
We re-create Fig.~\ref{fig:gam-scan-2SAT-lind} (Lindblad dynamics, Algorithm \ref{alg1}) for the 2-SAT problem Eq. \eqref{2-SAT_2soln}, which admits two solutions instead of one. 
We now see no bias in the local $z$ components of the density matrix, because the solution must terminate in the subspace spanned by $\lbrace \ket{eg},\ket{ge} \rbrace$ in the Zeno limit $\Gamma\,T \rightarrow \infty$; see panels (a) and (b). However, panel (d) shows that we still retain ideal state purity in the Zeno limit. Furthermore, the state is no longer separable, but tends towards the Bell states $\ket{\Psi^\pm} \propto \ket{eg} \pm \ket{ge}$ in the Zeno limit, as illustrated by the concurrence in panel (c). 
While in general, statistical analysis of the final readout would be made more complicated by the lack of \emph{average} bias in panels (a) and (b), the creation of a Bell state in this case implies that the $z_j$ will take on a bias in individual runs of the experiment and will retain correlations on a run-by-run basis. 
}\label{fig:gam-scan-2SAT-multi}
\end{figure*}

\begin{figure*}
\centering
\includegraphics[width = .48\textwidth]{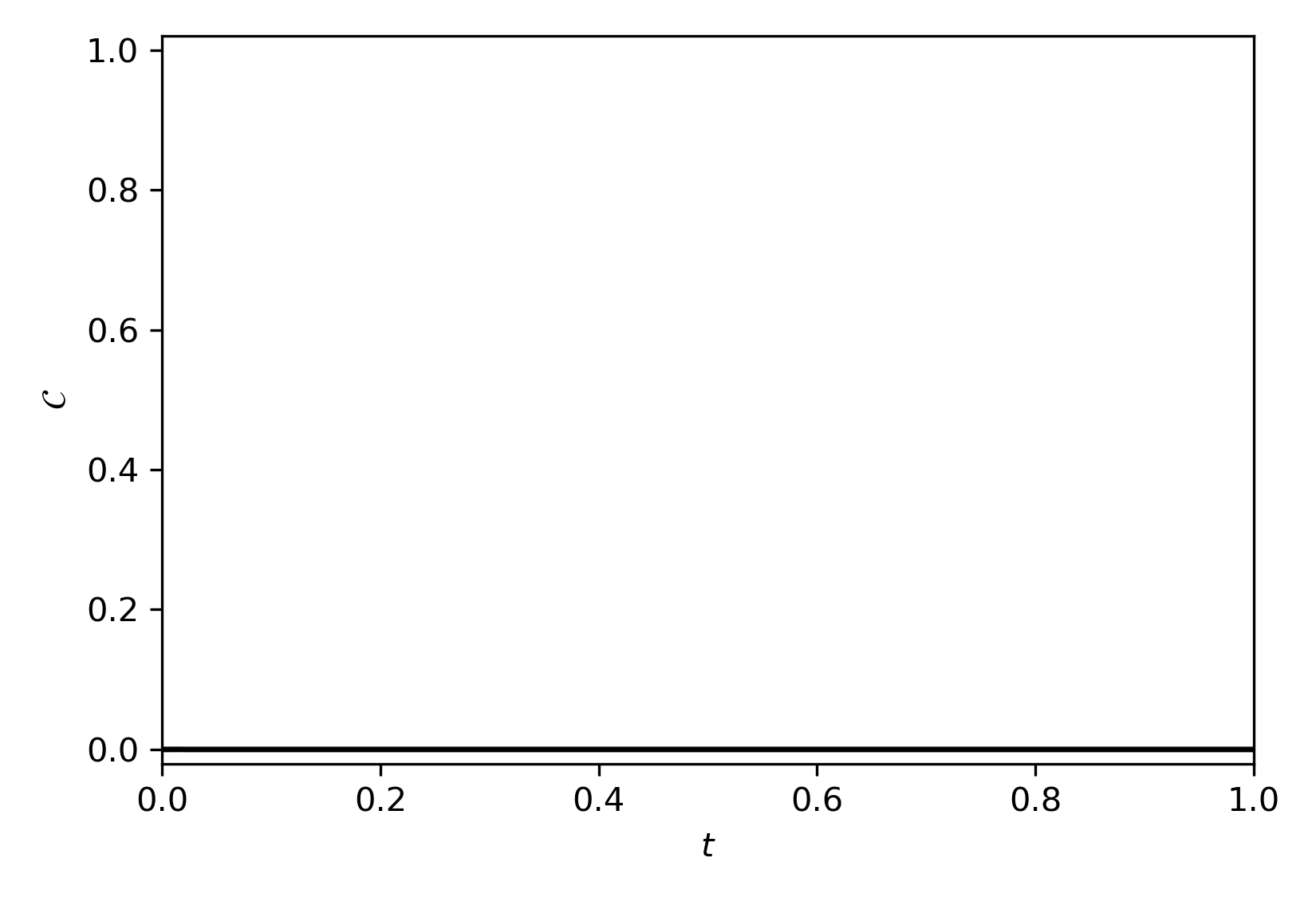} 
\includegraphics[width = .48\textwidth]{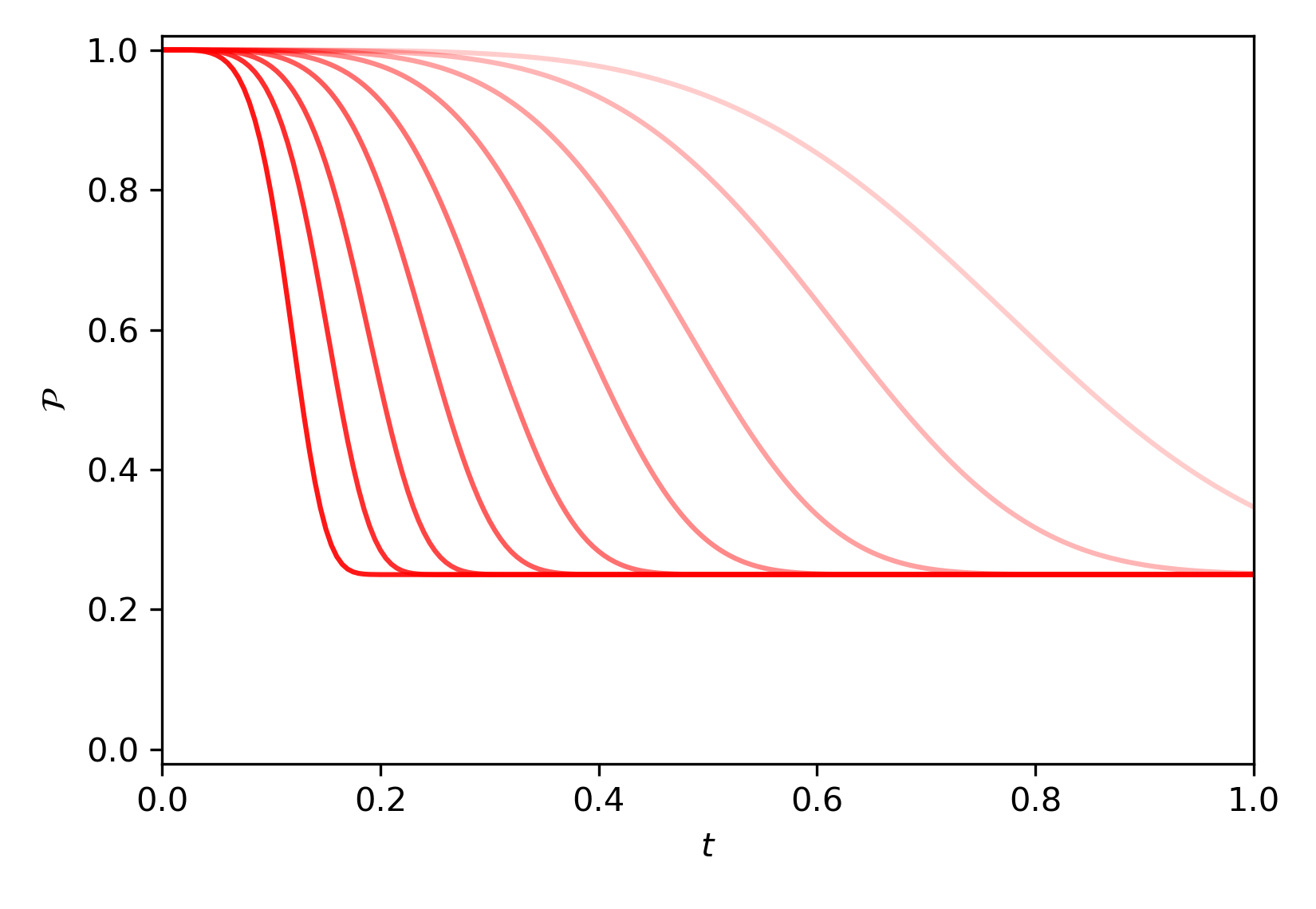} \\
\begin{tikzpicture}[overlay, yshift = -21pt]
\node[] at (-7,6.3) {(c)};
\node[] at (1.5,2.3) {(d)};
\end{tikzpicture} \vspace{-20pt}
\caption{
We re-create Fig.~\ref{fig:gam-scan-2SAT-lind} (Lindblad dynamics, Algorithm \ref{alg1}) for the 2-SAT problem Eq. \eqref{2-SAT_unsat}, which has no satisfying solution.
There is no local bias in the qubits, so that panels (a) and (b) are identical to the corresponding panels in Fig.~\ref{fig:gam-scan-2SAT-multi}; they are consequently not shown again.
However, the final state is markedly different. In contrast with the multi--solution case of Fig.~\ref{fig:gam-scan-2SAT-multi}, here we have no entanglement between our qubits (panel (c)), and significant loss of overall state purity (panel (d)). In other words, for this problem our generalizd-measurement-based algorithm generates the maximally mixed state for sufficiently long evolution times, and the final readouts of both qubits will be individually random and completely uncorrelated. 
}\label{fig:gam-scan-2SAT-unsat}
\end{figure*}
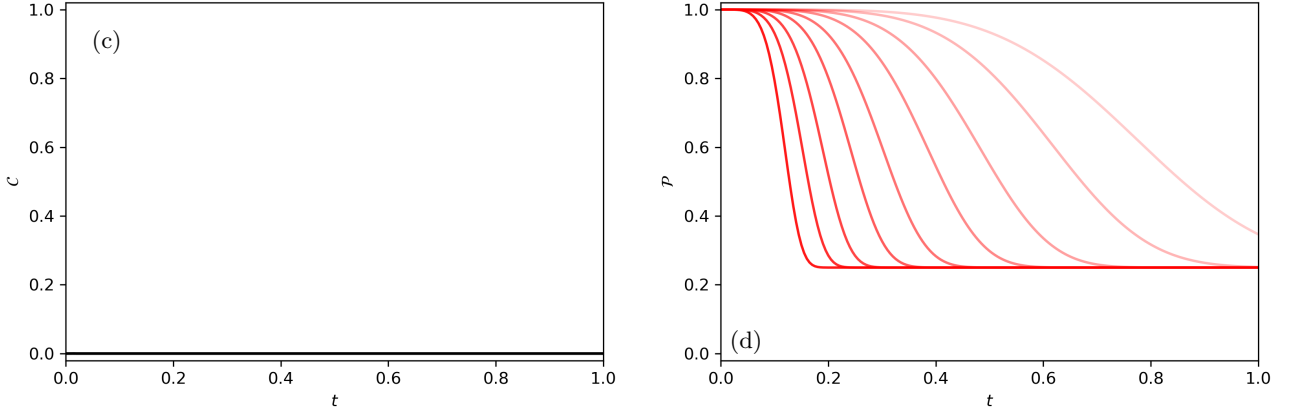

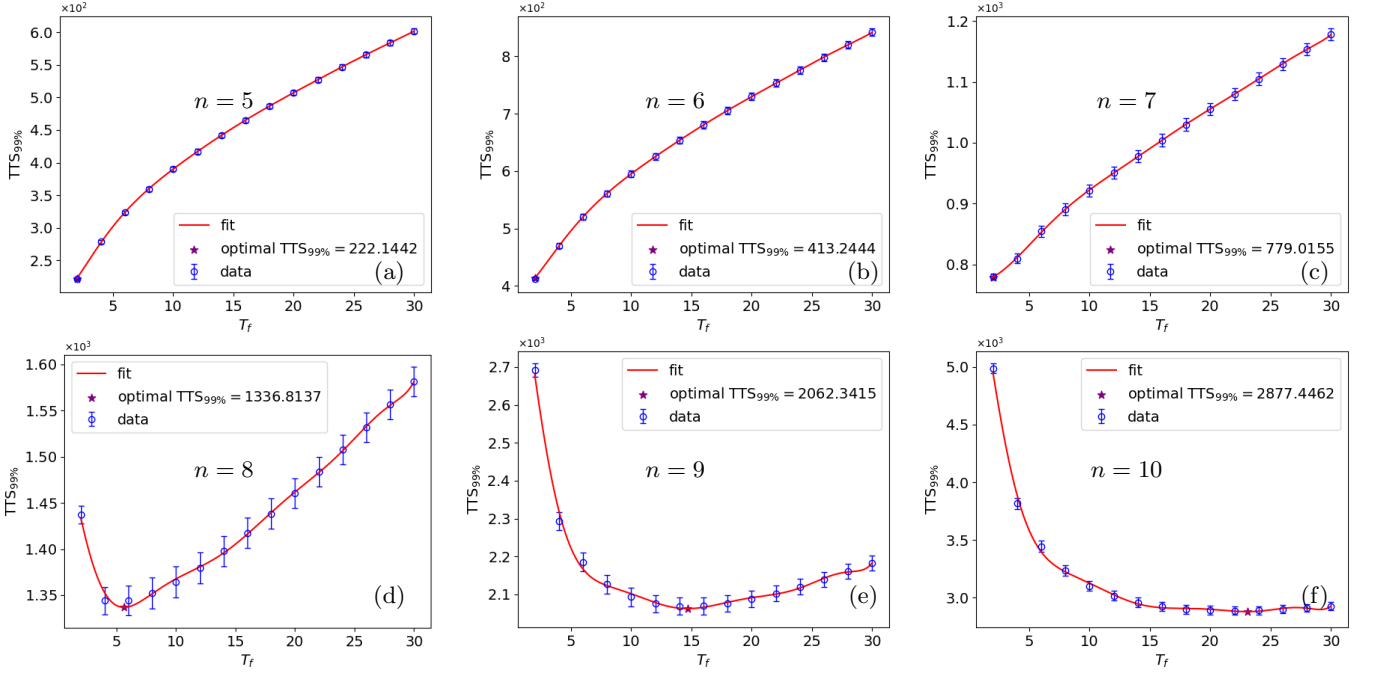
\begin{figure*}
     \centering
    \includegraphics[width = 0.32\textwidth, trim = {5 0 40 25}, clip]{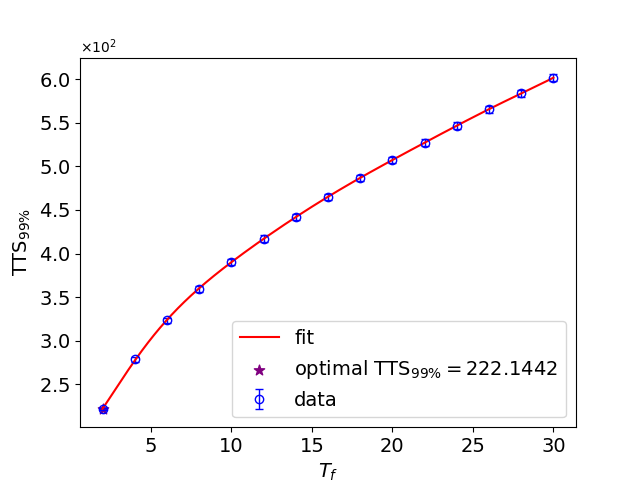}
    \hfill
    \includegraphics[width = 0.32\textwidth, trim = {5 0 40 25}, clip]{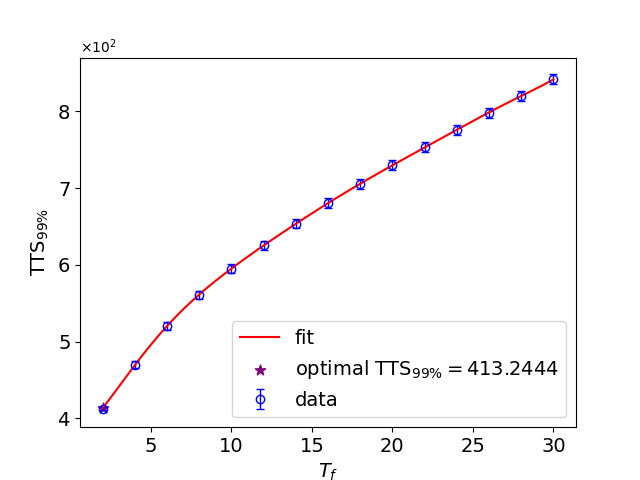}
    \hfill
    \includegraphics[width = 0.32\textwidth, trim = {5 0 40 25}, clip]{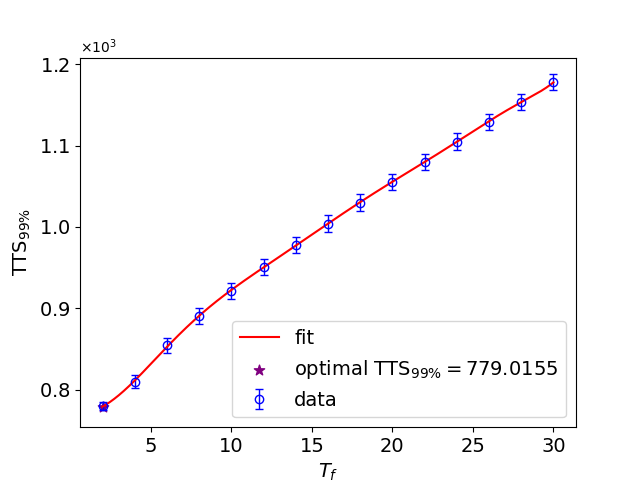}
    \\
    \includegraphics[width = 0.32\textwidth, trim = {0 0 40 25}, clip]{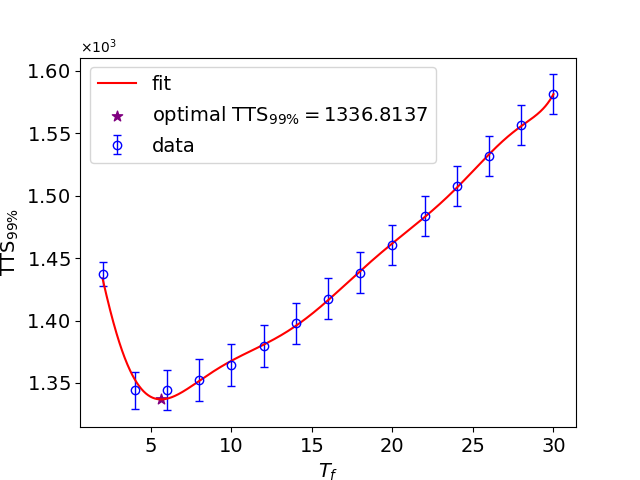}
    \hfill
    \includegraphics[width = 0.32\textwidth, trim = {5 0 40 25}, clip]{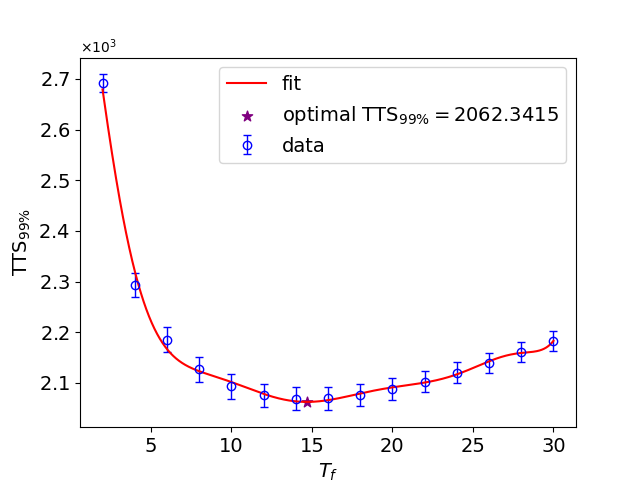}
    \hfill
    \includegraphics[width = 0.32\textwidth, trim = {5 0 40 25}, clip]{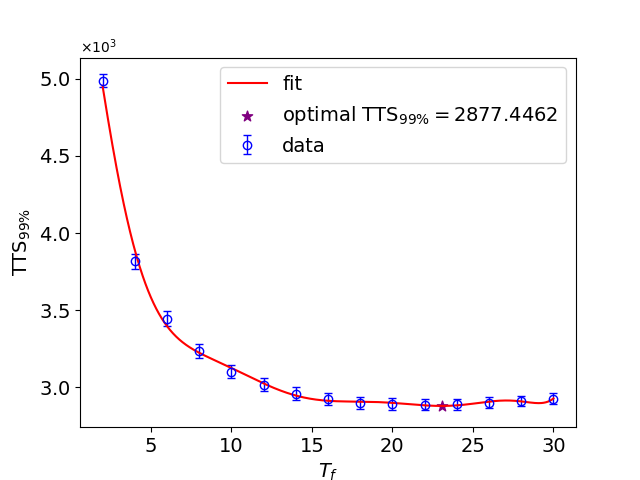}
    \\
    \begin{tikzpicture}[overlay]
    \node[] at (-3.8,5.6) {(a)};
    \node[] at (2.5,5.6) {(b)};
    \node[] at (8.5,5.6) {(c)};
    \node[] at (-3.8,1.3) {(d)};
    \node[] at (2.5,1.3) {(e)};
    \node[] at (8.5,1.3) {(f)};
    \node[] at (-6,7.9) {\color{black} $n = 5$};
    \node[] at (0,7.9) {\color{black} $n = 6$};
    \node[] at (6,7.9) {\color{black} $n = 7$};
    \node[] at (-6,3) {\color{black} $n = 8$};
    \node[] at (0,3) {\color{black} $n = 9$};
    \node[] at (6,3) {\color{black} $n = 10$};
    \end{tikzpicture} \vspace{-12pt}
    \caption{We plot the 3-SAT $\text{TTS}_{99\%}$ for $n=5 - 10$ qubits as a function of the total dragging time $T_f$ under Algorithm \ref{alg1} with average dynamics. The simulation parameters are the same as Fig. \ref{fig_TTS_vs_nq} with $\Delta t = 1$ and a linear schedule is also used. 
   % Panels (a) - (f) are for $n = 5$ to $n = 10$. 
    Panels (a) - (c) for $n=5-7$ show no optimal values of $\text{TTS}_{99\%}$, but we observe clear optimal values for $\text{TTS}_{99\%}$ for qubit numbers $n=8-10$ in panels (d) - (e). We locate these optimal $\text{TTS}_{99\%}$ values by fitting the data in each case with a polynomial function and differentiating this to find the minimum.
    }\label{fig_TTS_vs_T_f_average_dt_1}
 \end{figure*}

What if we have more than one satisfying solution?
Let us consider the generic case of two possible solutions.
Suppose that we have $\tilde{\varrho}_a = \hat{Q}_a^\dag\,\tilde{\rho}_a\,\hat{Q}_a$ and $\tilde{\varrho}_b= \hat{Q}_b^\dag\,\tilde{\rho}_b\,\hat{Q}_b$,
where $\hat{Q}_a$ and $\hat{Q}_b$ are constructed like  Eq. \eqref{Q-definition-app} from solution states $\rho_a = \ket{\phi_a}\bra{\phi_a}$, $\rho_b = \ket{\phi_b}\bra{\phi_b}$ of the form of Eq. \eqref{general-solution-state}, as above. 
These solution states can be assumed to satisfy two important properties:
\begin{subequations} \be \label{multi-eigenstate}
\hat{\mathcal{X}}_i\ket{\tilde{\phi}_a} = \ket{\tilde{\phi}_a} \quad\&\quad
\hat{\mathcal{X}}_i\ket{\tilde{\phi}_b} = \ket{\tilde{\phi}_b} \quad\forall~i,
\ee \be \label{ortho-eigenstates}
\text{and}\quad \tr{\tilde{\varrho}_a\,\tilde{\varrho}_b} = 0.
\ee \end{subequations}
Eq.~\eqref{multi-eigenstate} states that both solutions are eigenstates of the clause observables \emph{with the same eigenvalue}. 
Futhermore, by virtue of our two solutions being distinct, we can assume that the states are orthogonal at the terminal dragging angle $\ip{\phi_a(\theta = \pi/2)}{\phi_b(\theta = \pi/2)} = 0$, such that Eq. \eqref{ortho-eigenstates} is true.
Under these assumptions, it is simple to check that the conditions $\widehat{\mathsf{A}} = 0$, $\widehat{\mathsf{b}} = 0$, and $\mathsf{g} = 0$ apply not only at $\tilde{\varrho}_a = \ket{\tilde{\phi}_a}\bra{\tilde{\phi}_a}$ and $\tilde{\varrho}_b = \ket{\tilde{\phi}_b}\bra{\tilde{\phi}_b}$, but also at any pure state $\check{\varrho}$ in the $\mathrm{span}(\ket{\phi_a},\ket{\phi_b})$. 
In other words, for multiple solution states Eq. \eqref{general-solution-state} (that are $+1$ eigenstates and mutually orthonormal at $\theta = \pi/2$), our dynamics stabilize the entire solution subspace (the subspace spanned by the solution states), in much the same way as was just described for the single--solution case. 

As a final remark, the $\hat{Q}$--frame above is the analog of the ``Zeno frame'' described in \cite{lewalle2023optimal}.
We have shown in that previous work that the Zeno frame has considerable use for quantum control, for instance by allowing one to accelerate the deterministic dynamics via a ``shortcut to Zeno'' (STZ). 
While we have shown above that the underpinnings of STZ can be used fruitfully here to reason about the dynamical behavior of the $k$-SAT system relative to its solution states, it turns out that such control is not actually useful for solving a $k$--SAT problem faster.
This is because, as noted in Sec.~\ref{sec-Convergence}, construction of the $\hat{Q}$--frame / Zeno frame requires pre-existing knowledge of the target subspace dynamics, which in the case of our continuous BZF algorithm implies that one already knows the solution to the computational problem!
Therefore, while STZ may here be useful for proofs and/or for rapid solution \emph{verification} in its existing, form it cannot help us to accelerate solution finding.
Quantum control for algorithm acceleration will have to go beyond the approach of Ref. \cite{lewalle2023optimal}, e.g., by restricting the controls to those that do not depend \emph{a priori} on knowledge of the target solution dynamics. 
We speculate that if such controls exist, they are likely to take the form of measurement--based feedback depending on the clause readouts $r_i$. 

%%%%%%%%%%%%%%%%%%%%%%%%%%%%%%%%%%%%%%%%%%%%%%%%%%%%%%%%%%%%%%%%%
%%%%%%%%%%%%%%%%%%%%%%%%%%%%%%%%%%%%%%%%%%%%%%%%%%%%%%%%%%%%%%%%%
%%%%%%%%%%%%%%%%%%%%%%%%%%%%%%%%%%%%%%%%%%%%%%%%%%%%%%%%%%%%%%%%%
\section{Two--Qubit Case Study: Multi-Solution and Unsatisfiable Problems \label{app-multi-unsat}}

%%%%%%%%%%%%%%%%%%%%%%%%%%%%%%%%%%%%%%%%%%%%%%%%%%%%%%%%%%%%%%%%%
%%%%%%%%%%%%%%%%%%%%%%%%%%%%%%%%%%%%%%%%%%%%%%%%%%%%%%%%%%%%%%%%%
\subsection{Two--Solution two--qubit 2-SAT}

Suppose we delete $C_2$ from the problem of Eq. \eqref{2q2SAT_clause} (see also Fig.~\ref{fig-projectors}). 
Then both {\tt FT} and {\tt TF} are possible solutions to the revised problem
\be \label{2-SAT_2soln}
C_1 = b_1 \vee b_2, \quad C_3 = \bar{b}_1 \vee \bar{b}_2 \quad\rightarrow\quad C_1 \wedge C_3. 
\ee
In the quantum formulation of our algorithm, this implies that we rule out the components of the two--qubit basis with support on $\lbrace \ket{11},\ket{00} \rbrace$ by the end of the algorithm (these correspond to {\tt TT} and {\tt FF}, which are ruled out by clauses $C_3$ and $C_1$, respectively), but leave open possible solutions in the subspace spanned by $\lbrace \ket{01},\ket{10} \rbrace$. 
We show in Fig.~\ref{fig:gam-scan-2SAT-multi} that we obtain a Bell state $\ket{\Psi^\pm} \propto \ket{01}\pm\ket{10}$ in the long--dragging limit under this scheme. 
This implies that while the individual qubits have no bias \emph{on average}, the values of $z_j$ in individual runs do still have bias, and the entangled state guarantees that they will be measured to be anti-correlated in individual runs of the experiment, with statistics similar to those of the individual runs. It is in these correlations that the information about the solution subspace is encoded.
In particular, the readout scheme of Eq. \eqref{PrS_general} will still function perfectly well in this situation. Even though for $T_f \gg \tau$ and $\Delta t_m \gg \tau$ it is completely random whether our readouts will suggest a bitstring {\tt TF} or {\tt FT} in any given run, they will almost certainly return one of those two possible solutions upon readout, due to $\rho$ converging towards the subspace spanned precisely by these solution states.

While not the emphasis of this paper, we remark that the above analysis presents, to our knowledge, a novel way of creating measurement--driven entanglement, quite distinct from other approaches using continuous measurements, see e.g., \cite{martin2015deterministic, martin2017optimal, martin2019single, zhang2020locally, Lewalle:21, Lewalle_2020_cycle, ZenoGateTheory, blumenthal2021, Doucet_2020, doucet2023scalable, Greenfield_2024} and references therein. 
The most--closely related of these approaches is arguably \cite{doucet2023scalable}, or the single--measurement version we have recently proposed in \cite{lewalle2023optimal}. 
In contrast with the two--qubit example of Ref.~\cite{lewalle2023optimal} where we generate a Bell state by engineering Zeno confinement to an entanglement--producing path, we have here imposed a kind of ``Zeno exclusion'' control, where two measurements eliminate a solution subspace, and entanglement is generated in the remaining subspace. 

%%%%%%%%%%%%%%%%%%%%%%%%%%%%%%%%%%%%%%%%%%%%%%%%%%%%%%%%%%%%%%%%%
%%%%%%%%%%%%%%%%%%%%%%%%%%%%%%%%%%%%%%%%%%%%%%%%%%%%%%%%%%%%%%%%%
\subsection{Unsatisfiable two--qubit 2-SAT}

The multi-solution case is best understood in contrast to the behavior of the algorithm when attempting to solve an unsatisfiable problem.
Suppose we take the problem Eq. \eqref{2q2SAT_clause} and/or Fig.~\ref{fig-projectors}, and add a clause $C_4$ that rules out the unique solution that we had in Sec.~\ref{SEC:2Q-2SAT}, i.e.,
\be \label{2-SAT_unsat}
C_1 = b_1 \vee b_2, \quad C_2 = b_1 \vee \bar{b}_2, \quad C_3 = \bar{b}_1 \vee \bar{b}_2, \quad C_4 = \bar{b}_1 \vee b_2 
\ee
together constitute $C_1 \wedge C_2 \wedge C_3 \wedge C_4$, with no possible solution. 
The behavior of the Lindbladian algorithm in this situation is illustrated in Fig.~\ref{fig:gam-scan-2SAT-unsat}. While the qubit biases are eliminated as in Fig.~\ref{fig:gam-scan-2SAT-multi}, we accomplish this here by going to the maximally mixed state in the Zeno limit, instead of converging to a pure Bell state in the solution basis. 
Essentially, the symmetry of this example guarantees that the steady state of the Liouvillian is the maximally--mixed state; in the absence of any pure--state solution subspace to follow, there is no fixed point left for the dynamics to settle to except for this mixed steady state. 

\section{More numerics on the optimal $\text{TTS}_{99\%}$}
Here we present additional numerical results contained in a complete series of plots for $\text{TTS}_{99\%}$ as a function of $T_f$ for system sizes $n=5$ to $n=10$ qubits. 
Fig. \ref{fig_TTS_vs_T_f_average_dt_1} presents results for Algorithm \ref{alg1} with $\Delta t = 1$, which we take as representative examples.  
We can see that for the smallest system sizes $n \leq n_c$ (here $n_c = 7$), the $\text{TTS}_{99\%}$ increases monotonically as a function of $T_f$. 
However, for $n > n_c$, there appears for each size to be an optimum  value of $T_f$ where $\text{TTS}_{99\%}$ finds a minimum value. 
We may understand this behavior with the following analysis. For small system sizes, because $\mathbb{P}_s$ is finite even for $T_f = 0$, the dependence of the final solution state probability $\mathbb{P}_s$ on $T_f$ is already in a ``saturating phase" where increasing $T_f$ does not bring any additional benefit, so that the $\text{TTS}_{99\%}$ value will only increase with $T_f$.  
On the other hand, for large system sizes, the initial value of $\mathbb{P}_s$ for $T_f = 0$ is very small, and increasing $T_f$ can initially improve  $\mathbb{P}_s$ significantly, up to an
%$T_f$ smaller than its 
optimal value, after which the $\text{TTS}_{99\%}$ also reaches a ``saturating phase". 
There is therefore a sweet spot in $T_f$ such that an optimal (minimal) value of $\text{TTS}_{99\%}$ is achieved.
We note that the appearance of the optimal $\text{TTS}_{99\%}$ varies with the measurement time $\Delta t$, and we further observe a trend that for smaller values of $\Delta t$, the optimal $\text{TTS}_{99\%}$ value starts to emerge at larger values of $n$. 
This is another indication of different algorithmic behaviors for different measurement strength.

%%%%%%%%%%%%%%%%%%%%%%%%%%%%%%%%%%%%%%%%%%%%%%%%%%%%%%%%%%%%%%%%%
%%%%%%%%%%%%%%%%%%%%%%%%%%%%%%%%%%%%%%%%%%%%%%%%%%%%%%%%%%%%%%%%%
%%%%%%%%%%%%%%%%%%%%%%%%%%%%%%%%%%%%%%%%%%%%%%%%%%%%%%%%%%%%%%%%%
%%%%%%%%%%%%%%%%%%%%%%%%%%%%%%%%%%%%%%%%%%%%%%%%%%%%%%%%%%%%%%%%%
%%%%%%%%%%%%%%%%%%%%%%%%%%%%%%%%%%%%%%%%%%%%%%%%%%%%%%%%%%%%%%%%%
%%%%%%%%%%%%%%%%%%%%%%%%%%%%%%%%%%%%%%%%%%%%%%%%%%%%%%%%%%%%%%%%%
\bibliography{refs}

\begin{thebibliography}{135}%
\makeatletter
\providecommand \@ifxundefined [1]{%
 \@ifx{#1\undefined}
}%
\providecommand \@ifnum [1]{%
 \ifnum #1\expandafter \@firstoftwo
 \else \expandafter \@secondoftwo
 \fi
}%
\providecommand \@ifx [1]{%
 \ifx #1\expandafter \@firstoftwo
 \else \expandafter \@secondoftwo
 \fi
}%
\providecommand \natexlab [1]{#1}%
\providecommand \enquote  [1]{``#1''}%
\providecommand \bibnamefont  [1]{#1}%
\providecommand \bibfnamefont [1]{#1}%
\providecommand \citenamefont [1]{#1}%
\providecommand \href@noop [0]{\@secondoftwo}%
\providecommand \href [0]{\begingroup \@sanitize@url \@href}%
\providecommand \@href[1]{\@@startlink{#1}\@@href}%
\providecommand \@@href[1]{\endgroup#1\@@endlink}%
\providecommand \@sanitize@url [0]{\catcode `\\12\catcode `\$12\catcode `\&12\catcode `\#12\catcode `\^12\catcode `\_12\catcode `\%12\relax}%
\providecommand \@@startlink[1]{}%
\providecommand \@@endlink[0]{}%
\providecommand \url  [0]{\begingroup\@sanitize@url \@url }%
\providecommand \@url [1]{\endgroup\@href {#1}{\urlprefix }}%
\providecommand \urlprefix  [0]{URL }%
\providecommand \Eprint [0]{\href }%
\providecommand \doibase [0]{https://doi.org/}%
\providecommand \selectlanguage [0]{\@gobble}%
\providecommand \bibinfo  [0]{\@secondoftwo}%
\providecommand \bibfield  [0]{\@secondoftwo}%
\providecommand \translation [1]{[#1]}%
\providecommand \BibitemOpen [0]{}%
\providecommand \bibitemStop [0]{}%
\providecommand \bibitemNoStop [0]{.\EOS\space}%
\providecommand \EOS [0]{\spacefactor3000\relax}%
\providecommand \BibitemShut  [1]{\csname bibitem#1\endcsname}%
\let\auto@bib@innerbib\@empty
%</preamble>
\bibitem [{\citenamefont {Benjamin}\ \emph {et~al.}(2017)\citenamefont {Benjamin}, \citenamefont {Zhao},\ and\ \citenamefont {Fitzsimons}}]{benjamin2017measurement}%
  \BibitemOpen
  \bibfield  {author} {\bibinfo {author} {\bibfnamefont {S.~C.}\ \bibnamefont {Benjamin}}, \bibinfo {author} {\bibfnamefont {L.}~\bibnamefont {Zhao}},\ and\ \bibinfo {author} {\bibfnamefont {J.~F.}\ \bibnamefont {Fitzsimons}},\ }\href@noop {} {\bibinfo {title} {Measurement-driven quantum computing: Performance of a 3-{SAT} solver}} (\bibinfo {year} {2017}),\ \Eprint {https://arxiv.org/abs/1711.02687} {arXiv:1711.02687 [quant-ph]} \BibitemShut {NoStop}%
\bibitem [{\citenamefont {Feynman}(1982)}]{Feynman_1982}%
  \BibitemOpen
  \bibfield  {author} {\bibinfo {author} {\bibfnamefont {R.}~\bibnamefont {Feynman}},\ }\bibfield  {title} {\bibinfo {title} {Simulating physics with computers},\ }\href {https://doi.org/10.1007/BF02650179} {\bibfield  {journal} {\bibinfo  {journal} {Int J Theor Phys}\ }\textbf {\bibinfo {volume} {21}},\ \bibinfo {pages} {467–488} (\bibinfo {year} {1982})}\BibitemShut {NoStop}%
\bibitem [{\citenamefont {Shor}(1994)}]{Shor}%
  \BibitemOpen
  \bibfield  {author} {\bibinfo {author} {\bibfnamefont {P.}~\bibnamefont {Shor}},\ }\bibfield  {title} {\bibinfo {title} {Algorithms for quantum computation: discrete logarithms and factoring},\ }in\ \href {https://doi.org/10.1109/SFCS.1994.365700} {\emph {\bibinfo {booktitle} {Proceedings 35th Annual Symposium on Foundations of Computer Science}}}\ (\bibinfo {year} {1994})\ pp.\ \bibinfo {pages} {124--134}\BibitemShut {NoStop}%
\bibitem [{\citenamefont {Grover}(1996)}]{Grover}%
  \BibitemOpen
  \bibfield  {author} {\bibinfo {author} {\bibfnamefont {L.~K.}\ \bibnamefont {Grover}},\ }\bibfield  {title} {\bibinfo {title} {A fast quantum mechanical algorithm for database search},\ }in\ \href {https://doi.org/10.1145/237814.237866} {\emph {\bibinfo {booktitle} {Proceedings of the Twenty-Eighth Annual ACM Symposium on Theory of Computing}}},\ \bibinfo {series and number} {STOC '96}\ (\bibinfo  {publisher} {Association for Computing Machinery},\ \bibinfo {address} {New York, NY, USA},\ \bibinfo {year} {1996})\ p.\ \bibinfo {pages} {212–219}\BibitemShut {NoStop}%
\bibitem [{\citenamefont {Benioff}(1980)}]{Benioff_1980}%
  \BibitemOpen
  \bibfield  {author} {\bibinfo {author} {\bibfnamefont {P.}~\bibnamefont {Benioff}},\ }\bibfield  {title} {\bibinfo {title} {The computer as a physical system: A microscopic quantum mechanical hamiltonian model of computers as represented by turing machines},\ }\href {https://doi.org/10.1007/BF01011339} {\bibfield  {journal} {\bibinfo  {journal} {J Stat Phys}\ }\textbf {\bibinfo {volume} {22}},\ \bibinfo {pages} {563–591} (\bibinfo {year} {1980})}\BibitemShut {NoStop}%
\bibitem [{\citenamefont {Nielsen}\ and\ \citenamefont {Chuang}(2000)}]{BookNielsen}%
  \BibitemOpen
  \bibfield  {author} {\bibinfo {author} {\bibfnamefont {M.~A.}\ \bibnamefont {Nielsen}}\ and\ \bibinfo {author} {\bibfnamefont {I.~L.}\ \bibnamefont {Chuang}},\ }\href {https://doi.org/10.1017/CBO9780511976667} {\emph {\bibinfo {title} {Quantum Computation and Quantum Information}}}\ (\bibinfo  {publisher} {Cambridge University Press},\ \bibinfo {year} {2000})\BibitemShut {NoStop}%
\bibitem [{\citenamefont {Farhi}\ \emph {et~al.}(2000)\citenamefont {Farhi}, \citenamefont {Goldstone}, \citenamefont {Gutmann},\ and\ \citenamefont {Sipser}}]{farhi2000quantum}%
  \BibitemOpen
  \bibfield  {author} {\bibinfo {author} {\bibfnamefont {E.}~\bibnamefont {Farhi}}, \bibinfo {author} {\bibfnamefont {J.}~\bibnamefont {Goldstone}}, \bibinfo {author} {\bibfnamefont {S.}~\bibnamefont {Gutmann}},\ and\ \bibinfo {author} {\bibfnamefont {M.}~\bibnamefont {Sipser}},\ }\href@noop {} {\bibinfo {title} {Quantum computation by adiabatic evolution}} (\bibinfo {year} {2000}),\ \Eprint {https://arxiv.org/abs/quant-ph/0001106} {arXiv:quant-ph/0001106 [quant-ph]} \BibitemShut {NoStop}%
\bibitem [{\citenamefont {Das}\ and\ \citenamefont {Chakrabarti}(2008)}]{Das_2008}%
  \BibitemOpen
  \bibfield  {author} {\bibinfo {author} {\bibfnamefont {A.}~\bibnamefont {Das}}\ and\ \bibinfo {author} {\bibfnamefont {B.~K.}\ \bibnamefont {Chakrabarti}},\ }\bibfield  {title} {\bibinfo {title} {Colloquium: Quantum annealing and analog quantum computation},\ }\href {https://doi.org/10.1103/RevModPhys.80.1061} {\bibfield  {journal} {\bibinfo  {journal} {Rev. Mod. Phys.}\ }\textbf {\bibinfo {volume} {80}},\ \bibinfo {pages} {1061} (\bibinfo {year} {2008})}\BibitemShut {NoStop}%
\bibitem [{\citenamefont {Morita}\ and\ \citenamefont {Nishimori}(2008)}]{morita2008mathematical}%
  \BibitemOpen
  \bibfield  {author} {\bibinfo {author} {\bibfnamefont {S.}~\bibnamefont {Morita}}\ and\ \bibinfo {author} {\bibfnamefont {H.}~\bibnamefont {Nishimori}},\ }\bibfield  {title} {\bibinfo {title} {Mathematical foundation of quantum annealing},\ }\bibfield  {journal} {\bibinfo  {journal} {Journal of Mathematical Physics}\ }\textbf {\bibinfo {volume} {49}},\ \href {https://doi.org/10.1063/1.2995837} {10.1063/1.2995837} (\bibinfo {year} {2008})\BibitemShut {NoStop}%
\bibitem [{\citenamefont {Hauke}\ \emph {et~al.}(2020)\citenamefont {Hauke}, \citenamefont {Katzgraber}, \citenamefont {Lechner}, \citenamefont {Nishimori},\ and\ \citenamefont {Oliver}}]{hauke2020perspectives}%
  \BibitemOpen
  \bibfield  {author} {\bibinfo {author} {\bibfnamefont {P.}~\bibnamefont {Hauke}}, \bibinfo {author} {\bibfnamefont {H.~G.}\ \bibnamefont {Katzgraber}}, \bibinfo {author} {\bibfnamefont {W.}~\bibnamefont {Lechner}}, \bibinfo {author} {\bibfnamefont {H.}~\bibnamefont {Nishimori}},\ and\ \bibinfo {author} {\bibfnamefont {W.~D.}\ \bibnamefont {Oliver}},\ }\bibfield  {title} {\bibinfo {title} {Perspectives of quantum annealing: Methods and implementations},\ }\href {https://doi.org/10.1088/1361-6633/ab85b8} {\bibfield  {journal} {\bibinfo  {journal} {Reports on Progress in Physics}\ }\textbf {\bibinfo {volume} {83}},\ \bibinfo {pages} {054401} (\bibinfo {year} {2020})}\BibitemShut {NoStop}%
\bibitem [{\citenamefont {Briegel}\ \emph {et~al.}(2009)\citenamefont {Briegel}, \citenamefont {Browne}, \citenamefont {D\"{u}r}, \citenamefont {Raussendorf},\ and\ \citenamefont {den Nest}}]{Briegel_2009}%
  \BibitemOpen
  \bibfield  {author} {\bibinfo {author} {\bibfnamefont {H.~J.}\ \bibnamefont {Briegel}}, \bibinfo {author} {\bibfnamefont {D.~E.}\ \bibnamefont {Browne}}, \bibinfo {author} {\bibfnamefont {W.}~\bibnamefont {D\"{u}r}}, \bibinfo {author} {\bibfnamefont {R.}~\bibnamefont {Raussendorf}},\ and\ \bibinfo {author} {\bibfnamefont {M.~V.}\ \bibnamefont {den Nest}},\ }\bibfield  {title} {\bibinfo {title} {Measurement-based quantum computation},\ }\href {https://doi.org/10.1038/nphys1157} {\bibfield  {journal} {\bibinfo  {journal} {Nature Phys}\ }\textbf {\bibinfo {volume} {5}},\ \bibinfo {pages} {19–26} (\bibinfo {year} {2009})}\BibitemShut {NoStop}%
\bibitem [{\citenamefont {Childs}\ \emph {et~al.}(2002)\citenamefont {Childs}, \citenamefont {Deotto}, \citenamefont {Farhi}, \citenamefont {Goldstone}, \citenamefont {Gutmann},\ and\ \citenamefont {Landahl}}]{Childs_2002}%
  \BibitemOpen
  \bibfield  {author} {\bibinfo {author} {\bibfnamefont {A.~M.}\ \bibnamefont {Childs}}, \bibinfo {author} {\bibfnamefont {E.}~\bibnamefont {Deotto}}, \bibinfo {author} {\bibfnamefont {E.}~\bibnamefont {Farhi}}, \bibinfo {author} {\bibfnamefont {J.}~\bibnamefont {Goldstone}}, \bibinfo {author} {\bibfnamefont {S.}~\bibnamefont {Gutmann}},\ and\ \bibinfo {author} {\bibfnamefont {A.~J.}\ \bibnamefont {Landahl}},\ }\bibfield  {title} {\bibinfo {title} {Quantum search by measurement},\ }\href {https://doi.org/10.1103/PhysRevA.66.032314} {\bibfield  {journal} {\bibinfo  {journal} {Phys. Rev. A}\ }\textbf {\bibinfo {volume} {66}},\ \bibinfo {pages} {032314} (\bibinfo {year} {2002})}\BibitemShut {NoStop}%
\bibitem [{\citenamefont {Verstraete}\ \emph {et~al.}(2009)\citenamefont {Verstraete}, \citenamefont {Wolf},\ and\ \citenamefont {Ignacio~Cirac}}]{Verstraete_2009}%
  \BibitemOpen
  \bibfield  {author} {\bibinfo {author} {\bibfnamefont {F.}~\bibnamefont {Verstraete}}, \bibinfo {author} {\bibfnamefont {M.}~\bibnamefont {Wolf}},\ and\ \bibinfo {author} {\bibfnamefont {J.}~\bibnamefont {Ignacio~Cirac}},\ }\bibfield  {title} {\bibinfo {title} {Quantum computation and quantum-state engineering driven by dissipation},\ }\href {https://doi.org/10.1038/nphys1342} {\bibfield  {journal} {\bibinfo  {journal} {Nature Phys.}\ }\textbf {\bibinfo {volume} {5}},\ \bibinfo {pages} {633–636} (\bibinfo {year} {2009})}\BibitemShut {NoStop}%
\bibitem [{\citenamefont {Zhao}\ \emph {et~al.}(2019)\citenamefont {Zhao}, \citenamefont {P\'erez-Delgado}, \citenamefont {Benjamin},\ and\ \citenamefont {Fitzsimons}}]{zhao2019measurement}%
  \BibitemOpen
  \bibfield  {author} {\bibinfo {author} {\bibfnamefont {L.}~\bibnamefont {Zhao}}, \bibinfo {author} {\bibfnamefont {C.~A.}\ \bibnamefont {P\'erez-Delgado}}, \bibinfo {author} {\bibfnamefont {S.~C.}\ \bibnamefont {Benjamin}},\ and\ \bibinfo {author} {\bibfnamefont {J.~F.}\ \bibnamefont {Fitzsimons}},\ }\bibfield  {title} {\bibinfo {title} {Measurement-driven analog of adiabatic quantum computation for frustration-free hamiltonians},\ }\href {https://doi.org/10.1103/PhysRevA.100.032331} {\bibfield  {journal} {\bibinfo  {journal} {Phys. Rev. A}\ }\textbf {\bibinfo {volume} {100}},\ \bibinfo {pages} {032331} (\bibinfo {year} {2019})}\BibitemShut {NoStop}%
\bibitem [{\citenamefont {Berwald}\ \emph {et~al.}(2023{\natexlab{a}})\citenamefont {Berwald}, \citenamefont {Chancellor},\ and\ \citenamefont {Dridi}}]{berwald2023grover}%
  \BibitemOpen
  \bibfield  {author} {\bibinfo {author} {\bibfnamefont {J.}~\bibnamefont {Berwald}}, \bibinfo {author} {\bibfnamefont {N.}~\bibnamefont {Chancellor}},\ and\ \bibinfo {author} {\bibfnamefont {R.}~\bibnamefont {Dridi}},\ }\href@noop {} {\bibinfo {title} {Grover speedup from many forms of the zeno effect}} (\bibinfo {year} {2023}{\natexlab{a}}),\ \Eprint {https://arxiv.org/abs/2305.11146} {arXiv:2305.11146 [quant-ph]} \BibitemShut {NoStop}%
\bibitem [{\citenamefont {Berwald}\ \emph {et~al.}(2023{\natexlab{b}})\citenamefont {Berwald}, \citenamefont {Chancellor},\ and\ \citenamefont {Dridi}}]{berwald2023zenoeffect}%
  \BibitemOpen
  \bibfield  {author} {\bibinfo {author} {\bibfnamefont {J.}~\bibnamefont {Berwald}}, \bibinfo {author} {\bibfnamefont {N.}~\bibnamefont {Chancellor}},\ and\ \bibinfo {author} {\bibfnamefont {R.}~\bibnamefont {Dridi}},\ }\href@noop {} {\bibinfo {title} {Zeno-effect computation: Opportunities and challenges}} (\bibinfo {year} {2023}{\natexlab{b}}),\ \Eprint {https://arxiv.org/abs/2311.08432} {arXiv:2311.08432 [quant-ph]} \BibitemShut {NoStop}%
\bibitem [{\citenamefont {Ding}\ \emph {et~al.}(2023)\citenamefont {Ding}, \citenamefont {Chen},\ and\ \citenamefont {Lin}}]{ding2023singleancilla}%
  \BibitemOpen
  \bibfield  {author} {\bibinfo {author} {\bibfnamefont {Z.}~\bibnamefont {Ding}}, \bibinfo {author} {\bibfnamefont {C.-F.}\ \bibnamefont {Chen}},\ and\ \bibinfo {author} {\bibfnamefont {L.}~\bibnamefont {Lin}},\ }\href@noop {} {\bibinfo {title} {Single-ancilla ground state preparation via lindbladians}} (\bibinfo {year} {2023}),\ \Eprint {https://arxiv.org/abs/2308.15676} {arXiv:2308.15676} \BibitemShut {NoStop}%
\bibitem [{\citenamefont {Hu}\ \emph {et~al.}(2020)\citenamefont {Hu}, \citenamefont {Xia},\ and\ \citenamefont {Kais}}]{Hu_2020}%
  \BibitemOpen
  \bibfield  {author} {\bibinfo {author} {\bibfnamefont {Z.}~\bibnamefont {Hu}}, \bibinfo {author} {\bibfnamefont {R.}~\bibnamefont {Xia}},\ and\ \bibinfo {author} {\bibfnamefont {S.}~\bibnamefont {Kais}},\ }\bibfield  {title} {\bibinfo {title} {A quantum algorithm for evolving open quantum dynamics on quantum computing devices},\ }\href {https://doi.org/10.1038/s41598-020-60321-x} {\bibfield  {journal} {\bibinfo  {journal} {Sci. Rep.}\ }\textbf {\bibinfo {volume} {10}},\ \bibinfo {pages} {3301} (\bibinfo {year} {2020})}\BibitemShut {NoStop}%
\bibitem [{\citenamefont {Schlimgen}\ \emph {et~al.}(2022)\citenamefont {Schlimgen}, \citenamefont {Head-Marsden}, \citenamefont {Sager-Smith}, \citenamefont {Narang},\ and\ \citenamefont {Mazziotti}}]{Schlimgen_2022}%
  \BibitemOpen
  \bibfield  {author} {\bibinfo {author} {\bibfnamefont {A.~W.}\ \bibnamefont {Schlimgen}}, \bibinfo {author} {\bibfnamefont {K.}~\bibnamefont {Head-Marsden}}, \bibinfo {author} {\bibfnamefont {L.~M.}\ \bibnamefont {Sager-Smith}}, \bibinfo {author} {\bibfnamefont {P.}~\bibnamefont {Narang}},\ and\ \bibinfo {author} {\bibfnamefont {D.~A.}\ \bibnamefont {Mazziotti}},\ }\bibfield  {title} {\bibinfo {title} {Quantum state preparation and nonunitary evolution with diagonal operators},\ }\href {https://doi.org/10.1103/PhysRevA.106.022414} {\bibfield  {journal} {\bibinfo  {journal} {Phys. Rev. A}\ }\textbf {\bibinfo {volume} {106}},\ \bibinfo {pages} {022414} (\bibinfo {year} {2022})}\BibitemShut {NoStop}%
\bibitem [{\citenamefont {Chen}\ \emph {et~al.}(2023{\natexlab{a}})\citenamefont {Chen}, \citenamefont {Kastoryano}, \citenamefont {Brandão},\ and\ \citenamefont {Gilyén}}]{chen2023quantum}%
  \BibitemOpen
  \bibfield  {author} {\bibinfo {author} {\bibfnamefont {C.-F.}\ \bibnamefont {Chen}}, \bibinfo {author} {\bibfnamefont {M.~J.}\ \bibnamefont {Kastoryano}}, \bibinfo {author} {\bibfnamefont {F.~G. S.~L.}\ \bibnamefont {Brandão}},\ and\ \bibinfo {author} {\bibfnamefont {A.}~\bibnamefont {Gilyén}},\ }\href@noop {} {\bibinfo {title} {Quantum thermal state preparation}} (\bibinfo {year} {2023}{\natexlab{a}}),\ \Eprint {https://arxiv.org/abs/2303.18224} {arXiv:2303.18224} \BibitemShut {NoStop}%
\bibitem [{\citenamefont {Chen}\ \emph {et~al.}(2023{\natexlab{b}})\citenamefont {Chen}, \citenamefont {Kastoryano},\ and\ \citenamefont {Gilyén}}]{chen2023efficient}%
  \BibitemOpen
  \bibfield  {author} {\bibinfo {author} {\bibfnamefont {C.-F.}\ \bibnamefont {Chen}}, \bibinfo {author} {\bibfnamefont {M.~J.}\ \bibnamefont {Kastoryano}},\ and\ \bibinfo {author} {\bibfnamefont {A.}~\bibnamefont {Gilyén}},\ }\href@noop {} {\bibinfo {title} {An efficient and exact noncommutative quantum gibbs sampler}} (\bibinfo {year} {2023}{\natexlab{b}}),\ \Eprint {https://arxiv.org/abs/2311.09207} {arXiv:2311.09207} \BibitemShut {NoStop}%
\bibitem [{\citenamefont {Ding}\ \emph {et~al.}(2024)\citenamefont {Ding}, \citenamefont {Li},\ and\ \citenamefont {Lin}}]{Ding_2024}%
  \BibitemOpen
  \bibfield  {author} {\bibinfo {author} {\bibfnamefont {Z.}~\bibnamefont {Ding}}, \bibinfo {author} {\bibfnamefont {X.}~\bibnamefont {Li}},\ and\ \bibinfo {author} {\bibfnamefont {L.}~\bibnamefont {Lin}},\ }\bibfield  {title} {\bibinfo {title} {Simulating open quantum systems using hamiltonian simulations},\ }\href {https://doi.org/10.1103/PRXQuantum.5.020332} {\bibfield  {journal} {\bibinfo  {journal} {PRX Quantum}\ }\textbf {\bibinfo {volume} {5}},\ \bibinfo {pages} {020332} (\bibinfo {year} {2024})}\BibitemShut {NoStop}%
\bibitem [{\citenamefont {Cook}(1971)}]{Cook_1971}%
  \BibitemOpen
  \bibfield  {author} {\bibinfo {author} {\bibfnamefont {S.~A.}\ \bibnamefont {Cook}},\ }\bibfield  {title} {\bibinfo {title} {The complexity of theorem-proving procedures},\ }in\ \href {https://doi.org/10.1145/800157.805047} {\emph {\bibinfo {booktitle} {Proceedings of the Third Annual ACM Symposium on Theory of Computing}}},\ \bibinfo {series and number} {STOC '71}\ (\bibinfo  {publisher} {Association for Computing Machinery},\ \bibinfo {address} {New York, NY, USA},\ \bibinfo {year} {1971})\ p.\ \bibinfo {pages} {151–158}\BibitemShut {NoStop}%
\bibitem [{\citenamefont {Karp}(1972)}]{Karp_1972}%
  \BibitemOpen
  \bibfield  {author} {\bibinfo {author} {\bibfnamefont {R.~M.}\ \bibnamefont {Karp}},\ }\bibinfo {title} {Reducibility among combinatorial problems},\ in\ \href {https://doi.org/10.1007/978-1-4684-2001-2_9} {\emph {\bibinfo {booktitle} {Complexity of Computer Computations}}},\ \bibinfo {editor} {edited by\ \bibinfo {editor} {\bibfnamefont {R.~E.}\ \bibnamefont {Miller}}, \bibinfo {editor} {\bibfnamefont {J.~W.}\ \bibnamefont {Thatcher}},\ and\ \bibinfo {editor} {\bibfnamefont {J.~D.}\ \bibnamefont {Bohlinger}}}\ (\bibinfo  {publisher} {Springer US},\ \bibinfo {address} {Boston, MA},\ \bibinfo {year} {1972})\ pp.\ \bibinfo {pages} {85--103}\BibitemShut {NoStop}%
\bibitem [{\citenamefont {Levin}(1973)}]{Levin_1973}%
  \BibitemOpen
  \bibfield  {author} {\bibinfo {author} {\bibfnamefont {L.~A.}\ \bibnamefont {Levin}},\ }\bibfield  {title} {\bibinfo {title} {Universal sequential search problems},\ }\href {http://mathscinet.ams.org/mathscinet-getitem?mr=340042} {\bibfield  {journal} {\bibinfo  {journal} {Problems Inform. Transmission}\ }\textbf {\bibinfo {volume} {9}},\ \bibinfo {pages} {265} (\bibinfo {year} {1973})}\BibitemShut {NoStop}%
\bibitem [{\citenamefont {Mézard}\ \emph {et~al.}(2002)\citenamefont {Mézard}, \citenamefont {Parisi},\ and\ \citenamefont {Zecchina}}]{doi:10.1126/science.1073287}%
  \BibitemOpen
  \bibfield  {author} {\bibinfo {author} {\bibfnamefont {M.}~\bibnamefont {Mézard}}, \bibinfo {author} {\bibfnamefont {G.}~\bibnamefont {Parisi}},\ and\ \bibinfo {author} {\bibfnamefont {R.}~\bibnamefont {Zecchina}},\ }\bibfield  {title} {\bibinfo {title} {Analytic and algorithmic solution of random satisfiability problems},\ }\href {https://doi.org/10.1126/science.1073287} {\bibfield  {journal} {\bibinfo  {journal} {Science}\ }\textbf {\bibinfo {volume} {297}},\ \bibinfo {pages} {812} (\bibinfo {year} {2002})}\BibitemShut {NoStop}%
\bibitem [{\citenamefont {Schoning}(1999)}]{schoning1999probabilistic}%
  \BibitemOpen
  \bibfield  {author} {\bibinfo {author} {\bibfnamefont {T.}~\bibnamefont {Schoning}},\ }\bibfield  {title} {\bibinfo {title} {A probabilistic algorithm for k-sat and constraint satisfaction problems},\ }in\ \href@noop {} {\emph {\bibinfo {booktitle} {40th Annual Symposium on Foundations of Computer Science (Cat. No. 99CB37039)}}}\ (\bibinfo {organization} {IEEE},\ \bibinfo {year} {1999})\ pp.\ \bibinfo {pages} {410--414}\BibitemShut {NoStop}%
\bibitem [{\citenamefont {Guryanova}\ \emph {et~al.}(2020)\citenamefont {Guryanova}, \citenamefont {Friis},\ and\ \citenamefont {Huber}}]{Guryanova2020idealprojective}%
  \BibitemOpen
  \bibfield  {author} {\bibinfo {author} {\bibfnamefont {Y.}~\bibnamefont {Guryanova}}, \bibinfo {author} {\bibfnamefont {N.}~\bibnamefont {Friis}},\ and\ \bibinfo {author} {\bibfnamefont {M.}~\bibnamefont {Huber}},\ }\bibfield  {title} {\bibinfo {title} {Ideal {P}rojective {M}easurements {H}ave {I}nfinite {R}esource {C}osts},\ }\href {https://doi.org/10.22331/q-2020-01-13-222} {\bibfield  {journal} {\bibinfo  {journal} {{Quantum}}\ }\textbf {\bibinfo {volume} {4}},\ \bibinfo {pages} {222} (\bibinfo {year} {2020})}\BibitemShut {NoStop}%
\bibitem [{\citenamefont {Aharonov}\ and\ \citenamefont {Vardi}(1980)}]{Aharonov_1980}%
  \BibitemOpen
  \bibfield  {author} {\bibinfo {author} {\bibfnamefont {Y.}~\bibnamefont {Aharonov}}\ and\ \bibinfo {author} {\bibfnamefont {M.}~\bibnamefont {Vardi}},\ }\bibfield  {title} {\bibinfo {title} {Meaning of an individual "feynman path"},\ }\href {https://doi.org/10.1103/PhysRevD.21.2235} {\bibfield  {journal} {\bibinfo  {journal} {Phys. Rev. D}\ }\textbf {\bibinfo {volume} {21}},\ \bibinfo {pages} {2235} (\bibinfo {year} {1980})}\BibitemShut {NoStop}%
\bibitem [{\citenamefont {Hacohen-Gourgy}\ \emph {et~al.}(2018)\citenamefont {Hacohen-Gourgy}, \citenamefont {Garc\'{\i}a-Pintos}, \citenamefont {Martin}, \citenamefont {Dressel},\ and\ \citenamefont {Siddiqi}}]{ZenoDragging}%
  \BibitemOpen
  \bibfield  {author} {\bibinfo {author} {\bibfnamefont {S.}~\bibnamefont {Hacohen-Gourgy}}, \bibinfo {author} {\bibfnamefont {L.~P.}\ \bibnamefont {Garc\'{\i}a-Pintos}}, \bibinfo {author} {\bibfnamefont {L.~S.}\ \bibnamefont {Martin}}, \bibinfo {author} {\bibfnamefont {J.}~\bibnamefont {Dressel}},\ and\ \bibinfo {author} {\bibfnamefont {I.}~\bibnamefont {Siddiqi}},\ }\bibfield  {title} {\bibinfo {title} {Incoherent qubit control using the quantum zeno effect},\ }\href {https://doi.org/10.1103/PhysRevLett.120.020505} {\bibfield  {journal} {\bibinfo  {journal} {Phys. Rev. Lett.}\ }\textbf {\bibinfo {volume} {120}},\ \bibinfo {pages} {020505} (\bibinfo {year} {2018})}\BibitemShut {NoStop}%
\bibitem [{\citenamefont {Lewalle}\ \emph {et~al.}(2023{\natexlab{a}})\citenamefont {Lewalle}, \citenamefont {Zhang},\ and\ \citenamefont {Whaley}}]{lewalle2023optimal}%
  \BibitemOpen
  \bibfield  {author} {\bibinfo {author} {\bibfnamefont {P.}~\bibnamefont {Lewalle}}, \bibinfo {author} {\bibfnamefont {Y.}~\bibnamefont {Zhang}},\ and\ \bibinfo {author} {\bibfnamefont {K.~B.}\ \bibnamefont {Whaley}},\ }\href@noop {} {\bibinfo {title} {Optimal zeno dragging for quantum control: A shortcut to zeno with action-based scheduling optimization}} (\bibinfo {year} {2023}{\natexlab{a}}),\ \Eprint {https://arxiv.org/abs/2311.01631} {arXiv:2311.01631 [quant-ph]} \BibitemShut {NoStop}%
\bibitem [{\citenamefont {Kraus}\ \emph {et~al.}(1983)\citenamefont {Kraus}, \citenamefont {B\"{o}hm}, \citenamefont {Dollard},\ and\ \citenamefont {Wootters}}]{Kraus_1983}%
  \BibitemOpen
  \bibfield  {author} {\bibinfo {author} {\bibfnamefont {K.}~\bibnamefont {Kraus}}, \bibinfo {author} {\bibfnamefont {A.}~\bibnamefont {B\"{o}hm}}, \bibinfo {author} {\bibfnamefont {J.~D.}\ \bibnamefont {Dollard}},\ and\ \bibinfo {author} {\bibfnamefont {W.~H.}\ \bibnamefont {Wootters}},\ }\href {https://doi.org/10.1007/3-540-12732-1} {\emph {\bibinfo {title} {States, Effects, and Operations Fundamental Notions of Quantum Theory}}},\ Lecture Notes in Physics, 190\ (\bibinfo  {publisher} {Springer},\ \bibinfo {address} {Berlin, Heidelberg},\ \bibinfo {year} {1983})\BibitemShut {NoStop}%
\bibitem [{\citenamefont {Carmichael}(1993)}]{BookCarmichael}%
  \BibitemOpen
  \bibfield  {author} {\bibinfo {author} {\bibfnamefont {H.~J.}\ \bibnamefont {Carmichael}},\ }\href {https://doi.org/10.1007/978-3-540-47620-7} {\emph {\bibinfo {title} {An Open Systems Approach to Quantum Optics}}}\ (\bibinfo  {publisher} {Springer},\ \bibinfo {address} {Berlin},\ \bibinfo {year} {1993})\BibitemShut {NoStop}%
\bibitem [{\citenamefont {Wiseman}\ and\ \citenamefont {Milburn}(2009)}]{BookWiseman}%
  \BibitemOpen
  \bibfield  {author} {\bibinfo {author} {\bibfnamefont {H.~M.}\ \bibnamefont {Wiseman}}\ and\ \bibinfo {author} {\bibfnamefont {G.~J.}\ \bibnamefont {Milburn}},\ }\href {https://doi.org/10.1017/CBO9780511813948} {\emph {\bibinfo {title} {Quantum Measurement and Control}}}\ (\bibinfo  {publisher} {Cambridge University Press},\ \bibinfo {year} {2009})\BibitemShut {NoStop}%
\bibitem [{\citenamefont {Barchielli}\ and\ \citenamefont {Gregoratti}(2009)}]{BookBarchielli}%
  \BibitemOpen
  \bibfield  {author} {\bibinfo {author} {\bibfnamefont {A.}~\bibnamefont {Barchielli}}\ and\ \bibinfo {author} {\bibfnamefont {M.}~\bibnamefont {Gregoratti}},\ }\href {https://doi.org/10.1007/978-3-642-01298-3} {\emph {\bibinfo {title} {Quantum trajectories and measurements in continuous time}}}\ (\bibinfo  {publisher} {Springer-Verlag},\ \bibinfo {address} {Berlin, Heidelberg},\ \bibinfo {year} {2009})\BibitemShut {NoStop}%
\bibitem [{\citenamefont {Jacobs}(2014)}]{BookJacobs}%
  \BibitemOpen
  \bibfield  {author} {\bibinfo {author} {\bibfnamefont {K.}~\bibnamefont {Jacobs}},\ }\href {https://doi.org/10.1017/CBO9781139179027} {\emph {\bibinfo {title} {Quantum Measurement Theory and its Applications}}}\ (\bibinfo  {publisher} {Cambridge University Press},\ \bibinfo {year} {2014})\BibitemShut {NoStop}%
\bibitem [{\citenamefont {Jordan}\ and\ \citenamefont {Siddiqi}(2024)}]{BookJordan}%
  \BibitemOpen
  \bibfield  {author} {\bibinfo {author} {\bibfnamefont {A.~N.}\ \bibnamefont {Jordan}}\ and\ \bibinfo {author} {\bibfnamefont {I.~A.}\ \bibnamefont {Siddiqi}},\ }\href@noop {} {\emph {\bibinfo {title} {Quantum Measurement: Theory and Practice}}}\ (\bibinfo  {publisher} {Cambridge University Press},\ \bibinfo {year} {2024})\BibitemShut {NoStop}%
\bibitem [{\citenamefont {Gambetta}\ \emph {et~al.}(2008)\citenamefont {Gambetta}, \citenamefont {Blais}, \citenamefont {Boissonneault}, \citenamefont {Houck}, \citenamefont {Schuster},\ and\ \citenamefont {Girvin}}]{Gambetta2008}%
  \BibitemOpen
  \bibfield  {author} {\bibinfo {author} {\bibfnamefont {J.}~\bibnamefont {Gambetta}}, \bibinfo {author} {\bibfnamefont {A.}~\bibnamefont {Blais}}, \bibinfo {author} {\bibfnamefont {M.}~\bibnamefont {Boissonneault}}, \bibinfo {author} {\bibfnamefont {A.~A.}\ \bibnamefont {Houck}}, \bibinfo {author} {\bibfnamefont {D.~I.}\ \bibnamefont {Schuster}},\ and\ \bibinfo {author} {\bibfnamefont {S.~M.}\ \bibnamefont {Girvin}},\ }\bibfield  {title} {\bibinfo {title} {{Quantum trajectory approach to circuit QED: Quantum jumps and the Zeno effect}},\ }\href {https://doi.org/10.1103/PhysRevA.77.012112} {\bibfield  {journal} {\bibinfo  {journal} {Phys. Rev. A}\ }\textbf {\bibinfo {volume} {77}},\ \bibinfo {pages} {012112} (\bibinfo {year} {2008})}\BibitemShut {NoStop}%
\bibitem [{\citenamefont {Murch}\ \emph {et~al.}(2013)\citenamefont {Murch}, \citenamefont {Weber}, \citenamefont {Macklin},\ and\ \citenamefont {Siddiqi}}]{Murch2013}%
  \BibitemOpen
  \bibfield  {author} {\bibinfo {author} {\bibfnamefont {K.~W.}\ \bibnamefont {Murch}}, \bibinfo {author} {\bibfnamefont {S.~J.}\ \bibnamefont {Weber}}, \bibinfo {author} {\bibfnamefont {C.}~\bibnamefont {Macklin}},\ and\ \bibinfo {author} {\bibfnamefont {I.}~\bibnamefont {Siddiqi}},\ }\bibfield  {title} {\bibinfo {title} {{Observing single quantum trajectories of a superconducting quantum bit}},\ }\href {https://doi.org/10.1038/nature12539} {\bibfield  {journal} {\bibinfo  {journal} {Nature}\ }\textbf {\bibinfo {volume} {502}},\ \bibinfo {pages} {211} (\bibinfo {year} {2013})}\BibitemShut {NoStop}%
\bibitem [{\citenamefont {Hacohen-Gourgy}\ and\ \citenamefont {Martin}(2020)}]{LeighShay2020}%
  \BibitemOpen
  \bibfield  {author} {\bibinfo {author} {\bibfnamefont {S.}~\bibnamefont {Hacohen-Gourgy}}\ and\ \bibinfo {author} {\bibfnamefont {L.~S.}\ \bibnamefont {Martin}},\ }\bibfield  {title} {\bibinfo {title} {Continuous measurements for control of superconducting quantum circuits},\ }\href {https://doi.org/10.1080/23746149.2020.1813626} {\bibfield  {journal} {\bibinfo  {journal} {Advances in Physics: X}\ }\textbf {\bibinfo {volume} {5}},\ \bibinfo {pages} {1813626} (\bibinfo {year} {2020})}\BibitemShut {NoStop}%
\bibitem [{\citenamefont {Blais}\ \emph {et~al.}(2021)\citenamefont {Blais}, \citenamefont {Grimsmo}, \citenamefont {Girvin},\ and\ \citenamefont {Wallraff}}]{Blais_CQED}%
  \BibitemOpen
  \bibfield  {author} {\bibinfo {author} {\bibfnamefont {A.}~\bibnamefont {Blais}}, \bibinfo {author} {\bibfnamefont {A.~L.}\ \bibnamefont {Grimsmo}}, \bibinfo {author} {\bibfnamefont {S.~M.}\ \bibnamefont {Girvin}},\ and\ \bibinfo {author} {\bibfnamefont {A.}~\bibnamefont {Wallraff}},\ }\bibfield  {title} {\bibinfo {title} {Circuit quantum electrodynamics},\ }\href {https://doi.org/10.1103/RevModPhys.93.025005} {\bibfield  {journal} {\bibinfo  {journal} {Rev. Mod. Phys.}\ }\textbf {\bibinfo {volume} {93}},\ \bibinfo {pages} {025005} (\bibinfo {year} {2021})}\BibitemShut {NoStop}%
\bibitem [{\citenamefont {Hacohen-Gourgy}\ \emph {et~al.}(2016)\citenamefont {Hacohen-Gourgy}, \citenamefont {Martin}, \citenamefont {Flurin}, \citenamefont {Ramasesh}, \citenamefont {Whaley},\ and\ \citenamefont {Siddiqi}}]{ShayLeigh2016}%
  \BibitemOpen
  \bibfield  {author} {\bibinfo {author} {\bibfnamefont {S.}~\bibnamefont {Hacohen-Gourgy}}, \bibinfo {author} {\bibfnamefont {L.~S.}\ \bibnamefont {Martin}}, \bibinfo {author} {\bibfnamefont {E.}~\bibnamefont {Flurin}}, \bibinfo {author} {\bibfnamefont {V.~V.}\ \bibnamefont {Ramasesh}}, \bibinfo {author} {\bibfnamefont {K.~B.}\ \bibnamefont {Whaley}},\ and\ \bibinfo {author} {\bibfnamefont {I.}~\bibnamefont {Siddiqi}},\ }\bibfield  {title} {\bibinfo {title} {{Dynamics of simultaneously measured non-commuting observables}},\ }\href {https://doi.org/10.1038/nature19762} {\bibfield  {journal} {\bibinfo  {journal} {Nature}\ }\textbf {\bibinfo {volume} {538}},\ \bibinfo {pages} {491} (\bibinfo {year} {2016})}\BibitemShut {NoStop}%
\bibitem [{\citenamefont {Chantasri}\ \emph {et~al.}(2018)\citenamefont {Chantasri}, \citenamefont {Atalaya}, \citenamefont {Hacohen-Gourgy}, \citenamefont {Martin}, \citenamefont {Siddiqi},\ and\ \citenamefont {Jordan}}]{Chantasri_2017}%
  \BibitemOpen
  \bibfield  {author} {\bibinfo {author} {\bibfnamefont {A.}~\bibnamefont {Chantasri}}, \bibinfo {author} {\bibfnamefont {J.}~\bibnamefont {Atalaya}}, \bibinfo {author} {\bibfnamefont {S.}~\bibnamefont {Hacohen-Gourgy}}, \bibinfo {author} {\bibfnamefont {L.~S.}\ \bibnamefont {Martin}}, \bibinfo {author} {\bibfnamefont {I.}~\bibnamefont {Siddiqi}},\ and\ \bibinfo {author} {\bibfnamefont {A.~N.}\ \bibnamefont {Jordan}},\ }\bibfield  {title} {\bibinfo {title} {Simultaneous continuous measurement of noncommuting observables: Quantum state correlations},\ }\href {https://doi.org/10.1103/PhysRevA.97.012118} {\bibfield  {journal} {\bibinfo  {journal} {Phys. Rev. A}\ }\textbf {\bibinfo {volume} {97}},\ \bibinfo {pages} {012118} (\bibinfo {year} {2018})}\BibitemShut {NoStop}%
\bibitem [{\citenamefont {Lewalle}\ \emph {et~al.}(2017)\citenamefont {Lewalle}, \citenamefont {Chantasri},\ and\ \citenamefont {Jordan}}]{Lewalle_Caustic-thy}%
  \BibitemOpen
  \bibfield  {author} {\bibinfo {author} {\bibfnamefont {P.}~\bibnamefont {Lewalle}}, \bibinfo {author} {\bibfnamefont {A.}~\bibnamefont {Chantasri}},\ and\ \bibinfo {author} {\bibfnamefont {A.~N.}\ \bibnamefont {Jordan}},\ }\bibfield  {title} {\bibinfo {title} {Prediction and characterization of multiple extremal paths in continuously monitored qubits},\ }\href {https://doi.org/10.1103/PhysRevA.95.042126} {\bibfield  {journal} {\bibinfo  {journal} {Phys. Rev. A}\ }\textbf {\bibinfo {volume} {95}},\ \bibinfo {pages} {042126} (\bibinfo {year} {2017})}\BibitemShut {NoStop}%
\bibitem [{\citenamefont {Ficheux}\ \emph {et~al.}(2018)\citenamefont {Ficheux}, \citenamefont {Jezouin}, \citenamefont {Leghtas},\ and\ \citenamefont {Huard}}]{Ficheux2018}%
  \BibitemOpen
  \bibfield  {author} {\bibinfo {author} {\bibfnamefont {Q.}~\bibnamefont {Ficheux}}, \bibinfo {author} {\bibfnamefont {S.}~\bibnamefont {Jezouin}}, \bibinfo {author} {\bibfnamefont {Z.}~\bibnamefont {Leghtas}},\ and\ \bibinfo {author} {\bibfnamefont {B.}~\bibnamefont {Huard}},\ }\bibfield  {title} {\bibinfo {title} {Dynamics of a qubit while simultaneously monitoring its relaxation and dephasing},\ }\href {https://doi.org/10.1038/s41467-018-04372-9} {\bibfield  {journal} {\bibinfo  {journal} {Nat. Comm.}\ }\textbf {\bibinfo {volume} {9}},\ \bibinfo {pages} {1926} (\bibinfo {year} {2018})}\BibitemShut {NoStop}%
\bibitem [{\citenamefont {Lewalle}\ \emph {et~al.}(2018)\citenamefont {Lewalle}, \citenamefont {Steinmetz},\ and\ \citenamefont {Jordan}}]{Lewalle_chaos}%
  \BibitemOpen
  \bibfield  {author} {\bibinfo {author} {\bibfnamefont {P.}~\bibnamefont {Lewalle}}, \bibinfo {author} {\bibfnamefont {J.}~\bibnamefont {Steinmetz}},\ and\ \bibinfo {author} {\bibfnamefont {A.~N.}\ \bibnamefont {Jordan}},\ }\bibfield  {title} {\bibinfo {title} {Chaos in continuously monitored quantum systems: An optimal-path approach},\ }\href {https://doi.org/10.1103/PhysRevA.98.012141} {\bibfield  {journal} {\bibinfo  {journal} {Phys. Rev. A}\ }\textbf {\bibinfo {volume} {98}},\ \bibinfo {pages} {012141} (\bibinfo {year} {2018})}\BibitemShut {NoStop}%
\bibitem [{\citenamefont {Atalaya}\ \emph {et~al.}(2018{\natexlab{a}})\citenamefont {Atalaya}, \citenamefont {Hacohen-Gourgy}, \citenamefont {Martin}, \citenamefont {Siddiqi},\ and\ \citenamefont {Korotkov}}]{Atalaya2018}%
  \BibitemOpen
  \bibfield  {author} {\bibinfo {author} {\bibfnamefont {J.}~\bibnamefont {Atalaya}}, \bibinfo {author} {\bibfnamefont {S.}~\bibnamefont {Hacohen-Gourgy}}, \bibinfo {author} {\bibfnamefont {L.~S.}\ \bibnamefont {Martin}}, \bibinfo {author} {\bibfnamefont {I.}~\bibnamefont {Siddiqi}},\ and\ \bibinfo {author} {\bibfnamefont {A.~N.}\ \bibnamefont {Korotkov}},\ }\bibfield  {title} {\bibinfo {title} {Multitime correlators in continuous measurement of qubit observables},\ }\href {https://doi.org/10.1103/PhysRevA.97.020104} {\bibfield  {journal} {\bibinfo  {journal} {Phys. Rev. A}\ }\textbf {\bibinfo {volume} {97}},\ \bibinfo {pages} {020104} (\bibinfo {year} {2018}{\natexlab{a}})}\BibitemShut {NoStop}%
\bibitem [{\citenamefont {Atalaya}\ \emph {et~al.}(2018{\natexlab{b}})\citenamefont {Atalaya}, \citenamefont {Hacohen-Gourgy}, \citenamefont {Martin}, \citenamefont {Siddiqi},\ and\ \citenamefont {Korotkov}}]{Atalaya2018-2}%
  \BibitemOpen
  \bibfield  {author} {\bibinfo {author} {\bibfnamefont {J.}~\bibnamefont {Atalaya}}, \bibinfo {author} {\bibfnamefont {S.}~\bibnamefont {Hacohen-Gourgy}}, \bibinfo {author} {\bibfnamefont {L.~S.}\ \bibnamefont {Martin}}, \bibinfo {author} {\bibfnamefont {I.}~\bibnamefont {Siddiqi}},\ and\ \bibinfo {author} {\bibfnamefont {A.~N.}\ \bibnamefont {Korotkov}},\ }\bibfield  {title} {\bibinfo {title} {Correlators in simultaneous measurement of non-commuting qubit observables},\ }\href {https://doi.org/10.1038/s41534-018-0091-1} {\bibfield  {journal} {\bibinfo  {journal} {npj Quantum Information}\ }\textbf {\bibinfo {volume} {4}},\ \bibinfo {pages} {41} (\bibinfo {year} {2018}{\natexlab{b}})}\BibitemShut {NoStop}%
\bibitem [{\citenamefont {Atalaya}\ \emph {et~al.}(2019)\citenamefont {Atalaya}, \citenamefont {Hacohen-Gourgy}, \citenamefont {Siddiqi},\ and\ \citenamefont {Korotkov}}]{Atalaya2019}%
  \BibitemOpen
  \bibfield  {author} {\bibinfo {author} {\bibfnamefont {J.}~\bibnamefont {Atalaya}}, \bibinfo {author} {\bibfnamefont {S.}~\bibnamefont {Hacohen-Gourgy}}, \bibinfo {author} {\bibfnamefont {I.}~\bibnamefont {Siddiqi}},\ and\ \bibinfo {author} {\bibfnamefont {A.~N.}\ \bibnamefont {Korotkov}},\ }\bibfield  {title} {\bibinfo {title} {Correlators exceeding one in continuous measurements of superconducting qubits},\ }\href {https://doi.org/10.1103/PhysRevLett.122.223603} {\bibfield  {journal} {\bibinfo  {journal} {Phys. Rev. Lett.}\ }\textbf {\bibinfo {volume} {122}},\ \bibinfo {pages} {223603} (\bibinfo {year} {2019})}\BibitemShut {NoStop}%
\bibitem [{\citenamefont {Jacobs}\ and\ \citenamefont {Shabani}(2008)}]{Jacobs_Shabani_2008}%
  \BibitemOpen
  \bibfield  {author} {\bibinfo {author} {\bibfnamefont {K.}~\bibnamefont {Jacobs}}\ and\ \bibinfo {author} {\bibfnamefont {A.}~\bibnamefont {Shabani}},\ }\bibfield  {title} {\bibinfo {title} {Quantum feedback control: how to use verification theorems and viscosity solutions to find optimal protocols},\ }\href {https://doi.org/10.1080/00107510802601781} {\bibfield  {journal} {\bibinfo  {journal} {Contemporary Physics}\ }\textbf {\bibinfo {volume} {49}},\ \bibinfo {pages} {435} (\bibinfo {year} {2008})}\BibitemShut {NoStop}%
\bibitem [{\citenamefont {Gough}(2012)}]{Gough_2012}%
  \BibitemOpen
  \bibfield  {author} {\bibinfo {author} {\bibfnamefont {J.~E.}\ \bibnamefont {Gough}},\ }\bibfield  {title} {\bibinfo {title} {Principles and applications of quantum control engineering},\ }\href {https://doi.org/10.1098/rsta.2012.0370} {\bibfield  {journal} {\bibinfo  {journal} {Philosophical Transactions of the Royal Society A: Mathematical, Physical and Engineering Sciences}\ }\textbf {\bibinfo {volume} {370}},\ \bibinfo {pages} {5241} (\bibinfo {year} {2012})}\BibitemShut {NoStop}%
\bibitem [{\citenamefont {Zhang}\ \emph {et~al.}(2017)\citenamefont {Zhang}, \citenamefont {Liu}, \citenamefont {Wu}, \citenamefont {Jacobs},\ and\ \citenamefont {Nori}}]{ZhangFeedback2017}%
  \BibitemOpen
  \bibfield  {author} {\bibinfo {author} {\bibfnamefont {J.}~\bibnamefont {Zhang}}, \bibinfo {author} {\bibfnamefont {Y.}~\bibnamefont {Liu}}, \bibinfo {author} {\bibfnamefont {R.-B.}\ \bibnamefont {Wu}}, \bibinfo {author} {\bibfnamefont {K.}~\bibnamefont {Jacobs}},\ and\ \bibinfo {author} {\bibfnamefont {F.}~\bibnamefont {Nori}},\ }\bibfield  {title} {\bibinfo {title} {Quantum feedback: Theory, experiments, and applications},\ }\href {https://doi.org/http://doi.org/10.1016/j.physrep.2017.02.003} {\bibfield  {journal} {\bibinfo  {journal} {Physics Reports}\ }\textbf {\bibinfo {volume} {679}},\ \bibinfo {pages} {1} (\bibinfo {year} {2017})}\BibitemShut {NoStop}%
\bibitem [{\citenamefont {Minev}\ \emph {et~al.}(2019)\citenamefont {Minev}, \citenamefont {Mundhada}, \citenamefont {Shankar}, \citenamefont {Reinhold}, \citenamefont {Gutierrez-Jauregui}, \citenamefont {Schoelkopf}, \citenamefont {Mirrahimi}, \citenamefont {Carmichael},\ and\ \citenamefont {Devoret}}]{Minev2019}%
  \BibitemOpen
  \bibfield  {author} {\bibinfo {author} {\bibfnamefont {Z.~K.}\ \bibnamefont {Minev}}, \bibinfo {author} {\bibfnamefont {S.~O.}\ \bibnamefont {Mundhada}}, \bibinfo {author} {\bibfnamefont {S.}~\bibnamefont {Shankar}}, \bibinfo {author} {\bibfnamefont {P.}~\bibnamefont {Reinhold}}, \bibinfo {author} {\bibfnamefont {R.}~\bibnamefont {Gutierrez-Jauregui}}, \bibinfo {author} {\bibfnamefont {R.~J.}\ \bibnamefont {Schoelkopf}}, \bibinfo {author} {\bibfnamefont {M.}~\bibnamefont {Mirrahimi}}, \bibinfo {author} {\bibfnamefont {H.~J.}\ \bibnamefont {Carmichael}},\ and\ \bibinfo {author} {\bibfnamefont {M.~H.}\ \bibnamefont {Devoret}},\ }\bibfield  {title} {\bibinfo {title} {To catch and reverse a quantum jump mid-flight},\ }\href {https://www.nature.com/articles/s41586-019-1287-z} {\bibfield  {journal} {\bibinfo  {journal} {Nature}\ }\textbf {\bibinfo {volume} {570}},\ \bibinfo {pages} {200} (\bibinfo {year} {2019})}\BibitemShut {NoStop}%
\bibitem [{\citenamefont {Martin}\ \emph {et~al.}(2015)\citenamefont {Martin}, \citenamefont {Motzoi}, \citenamefont {Li}, \citenamefont {Sarovar},\ and\ \citenamefont {Whaley}}]{martin2015deterministic}%
  \BibitemOpen
  \bibfield  {author} {\bibinfo {author} {\bibfnamefont {L.}~\bibnamefont {Martin}}, \bibinfo {author} {\bibfnamefont {F.}~\bibnamefont {Motzoi}}, \bibinfo {author} {\bibfnamefont {H.}~\bibnamefont {Li}}, \bibinfo {author} {\bibfnamefont {M.}~\bibnamefont {Sarovar}},\ and\ \bibinfo {author} {\bibfnamefont {K.~B.}\ \bibnamefont {Whaley}},\ }\bibfield  {title} {\bibinfo {title} {Deterministic generation of remote entanglement with active quantum feedback},\ }\href {https://doi.org/10.1103/PhysRevA.92.062321} {\bibfield  {journal} {\bibinfo  {journal} {Physical Review A}\ }\textbf {\bibinfo {volume} {92}},\ \bibinfo {pages} {062321} (\bibinfo {year} {2015})}\BibitemShut {NoStop}%
\bibitem [{\citenamefont {Martin}\ \emph {et~al.}(2017)\citenamefont {Martin}, \citenamefont {Sayrafi},\ and\ \citenamefont {Whaley}}]{martin2017optimal}%
  \BibitemOpen
  \bibfield  {author} {\bibinfo {author} {\bibfnamefont {L.}~\bibnamefont {Martin}}, \bibinfo {author} {\bibfnamefont {M.}~\bibnamefont {Sayrafi}},\ and\ \bibinfo {author} {\bibfnamefont {K.~B.}\ \bibnamefont {Whaley}},\ }\bibfield  {title} {\bibinfo {title} {What is the optimal way to prepare a bell state using measurement and feedback?},\ }\href {https://doi.org/10.1088/2058-9565/aa804c} {\bibfield  {journal} {\bibinfo  {journal} {Quantum Science and Technology}\ }\textbf {\bibinfo {volume} {2}},\ \bibinfo {pages} {044006} (\bibinfo {year} {2017})}\BibitemShut {NoStop}%
\bibitem [{\citenamefont {Zhang}\ \emph {et~al.}(2020)\citenamefont {Zhang}, \citenamefont {Martin},\ and\ \citenamefont {Whaley}}]{zhang2020locally}%
  \BibitemOpen
  \bibfield  {author} {\bibinfo {author} {\bibfnamefont {S.}~\bibnamefont {Zhang}}, \bibinfo {author} {\bibfnamefont {L.~S.}\ \bibnamefont {Martin}},\ and\ \bibinfo {author} {\bibfnamefont {K.~B.}\ \bibnamefont {Whaley}},\ }\bibfield  {title} {\bibinfo {title} {Locally optimal measurement-based quantum feedback with application to multiqubit entanglement generation},\ }\href {https://doi.org/10.1103/PhysRevA.102.062418} {\bibfield  {journal} {\bibinfo  {journal} {Physical Review A}\ }\textbf {\bibinfo {volume} {102}},\ \bibinfo {pages} {062418} (\bibinfo {year} {2020})}\BibitemShut {NoStop}%
\bibitem [{\citenamefont {Lewalle}\ \emph {et~al.}(2021)\citenamefont {Lewalle}, \citenamefont {Elouard}, \citenamefont {Manikandan}, \citenamefont {Qian}, \citenamefont {Eberly},\ and\ \citenamefont {Jordan}}]{Lewalle:21}%
  \BibitemOpen
  \bibfield  {author} {\bibinfo {author} {\bibfnamefont {P.}~\bibnamefont {Lewalle}}, \bibinfo {author} {\bibfnamefont {C.}~\bibnamefont {Elouard}}, \bibinfo {author} {\bibfnamefont {S.~K.}\ \bibnamefont {Manikandan}}, \bibinfo {author} {\bibfnamefont {X.-F.}\ \bibnamefont {Qian}}, \bibinfo {author} {\bibfnamefont {J.~H.}\ \bibnamefont {Eberly}},\ and\ \bibinfo {author} {\bibfnamefont {A.~N.}\ \bibnamefont {Jordan}},\ }\bibfield  {title} {\bibinfo {title} {Entanglement of a pair of quantum emitters via continuous fluorescence measurements: a tutorial},\ }\href {https://doi.org/10.1364/AOP.399081} {\bibfield  {journal} {\bibinfo  {journal} {Adv. Opt. Photon.}\ }\textbf {\bibinfo {volume} {13}},\ \bibinfo {pages} {517} (\bibinfo {year} {2021})}\BibitemShut {NoStop}%
\bibitem [{\citenamefont {Martin}\ and\ \citenamefont {Whaley}(2019)}]{martin2019single}%
  \BibitemOpen
  \bibfield  {author} {\bibinfo {author} {\bibfnamefont {L.~S.}\ \bibnamefont {Martin}}\ and\ \bibinfo {author} {\bibfnamefont {K.~B.}\ \bibnamefont {Whaley}},\ }\href@noop {} {\bibinfo {title} {Single-shot deterministic entanglement between non-interacting systems with linear optics}} (\bibinfo {year} {2019}),\ \Eprint {https://arxiv.org/abs/1912.00067} {arXiv:1912.00067 [quant-ph]} \BibitemShut {NoStop}%
\bibitem [{\citenamefont {Lewalle}\ \emph {et~al.}(2020)\citenamefont {Lewalle}, \citenamefont {Elouard},\ and\ \citenamefont {Jordan}}]{Lewalle_2020_cycle}%
  \BibitemOpen
  \bibfield  {author} {\bibinfo {author} {\bibfnamefont {P.}~\bibnamefont {Lewalle}}, \bibinfo {author} {\bibfnamefont {C.}~\bibnamefont {Elouard}},\ and\ \bibinfo {author} {\bibfnamefont {A.~N.}\ \bibnamefont {Jordan}},\ }\bibfield  {title} {\bibinfo {title} {Entanglement-preserving limit cycles from sequential quantum measurements and feedback},\ }\href {https://doi.org/10.1103/PhysRevA.102.062219} {\bibfield  {journal} {\bibinfo  {journal} {Phys. Rev. A}\ }\textbf {\bibinfo {volume} {102}},\ \bibinfo {pages} {062219} (\bibinfo {year} {2020})}\BibitemShut {NoStop}%
\bibitem [{\citenamefont {Facchi}\ and\ \citenamefont {Pascazio}(2002)}]{Facchi_2002}%
  \BibitemOpen
  \bibfield  {author} {\bibinfo {author} {\bibfnamefont {P.}~\bibnamefont {Facchi}}\ and\ \bibinfo {author} {\bibfnamefont {S.}~\bibnamefont {Pascazio}},\ }\bibfield  {title} {\bibinfo {title} {Quantum zeno subspaces},\ }\href {https://doi.org/10.1103/PhysRevLett.89.080401} {\bibfield  {journal} {\bibinfo  {journal} {Phys. Rev. Lett.}\ }\textbf {\bibinfo {volume} {89}},\ \bibinfo {pages} {080401} (\bibinfo {year} {2002})}\BibitemShut {NoStop}%
\bibitem [{\citenamefont {Mirrahimi}\ and\ \citenamefont {van Handel}(2007)}]{Mirrahimi_2007}%
  \BibitemOpen
  \bibfield  {author} {\bibinfo {author} {\bibfnamefont {M.}~\bibnamefont {Mirrahimi}}\ and\ \bibinfo {author} {\bibfnamefont {R.}~\bibnamefont {van Handel}},\ }\bibfield  {title} {\bibinfo {title} {Stabilizing feedback controls for quantum systems},\ }\href {https://doi.org/10.1137/050644793} {\bibfield  {journal} {\bibinfo  {journal} {SIAM Journal on Control and Optimization}\ }\textbf {\bibinfo {volume} {46}},\ \bibinfo {pages} {445} (\bibinfo {year} {2007})}\BibitemShut {NoStop}%
\bibitem [{\citenamefont {Ticozzi}\ and\ \citenamefont {Viola}(2008)}]{Ticozzi_2008}%
  \BibitemOpen
  \bibfield  {author} {\bibinfo {author} {\bibfnamefont {F.}~\bibnamefont {Ticozzi}}\ and\ \bibinfo {author} {\bibfnamefont {L.}~\bibnamefont {Viola}},\ }\bibfield  {title} {\bibinfo {title} {Quantum markovian subsystems: Invariance, attractivity, and control},\ }\href {https://doi.org/10.1109/TAC.2008.929399} {\bibfield  {journal} {\bibinfo  {journal} {IEEE Transactions on Automatic Control}\ }\textbf {\bibinfo {volume} {53}},\ \bibinfo {pages} {2048} (\bibinfo {year} {2008})}\BibitemShut {NoStop}%
\bibitem [{\citenamefont {Amini}\ \emph {et~al.}(2011)\citenamefont {Amini}, \citenamefont {Mirrahimi},\ and\ \citenamefont {Rouchon}}]{amini2011stability}%
  \BibitemOpen
  \bibfield  {author} {\bibinfo {author} {\bibfnamefont {H.}~\bibnamefont {Amini}}, \bibinfo {author} {\bibfnamefont {M.}~\bibnamefont {Mirrahimi}},\ and\ \bibinfo {author} {\bibfnamefont {P.}~\bibnamefont {Rouchon}},\ }\href@noop {} {\bibinfo {title} {On stability of continuous-time quantum-filters}} (\bibinfo {year} {2011}),\ \Eprint {https://arxiv.org/abs/1103.2706} {arXiv:1103.2706 [math.OC]} \BibitemShut {NoStop}%
\bibitem [{\citenamefont {Ticozzi}\ \emph {et~al.}(2013)\citenamefont {Ticozzi}, \citenamefont {Nishio},\ and\ \citenamefont {Altafini}}]{Ticozzi_2013}%
  \BibitemOpen
  \bibfield  {author} {\bibinfo {author} {\bibfnamefont {F.}~\bibnamefont {Ticozzi}}, \bibinfo {author} {\bibfnamefont {K.}~\bibnamefont {Nishio}},\ and\ \bibinfo {author} {\bibfnamefont {C.}~\bibnamefont {Altafini}},\ }\bibfield  {title} {\bibinfo {title} {Stabilization of stochastic quantum dynamics via open- and closed-loop control},\ }\href {https://doi.org/10.1109/TAC.2012.2206713} {\bibfield  {journal} {\bibinfo  {journal} {IEEE Transactions on Automatic Control}\ }\textbf {\bibinfo {volume} {58}},\ \bibinfo {pages} {74} (\bibinfo {year} {2013})}\BibitemShut {NoStop}%
\bibitem [{\citenamefont {Benoist}\ \emph {et~al.}(2017)\citenamefont {Benoist}, \citenamefont {Pellegrini},\ and\ \citenamefont {Ticozzi}}]{Benoist_2017}%
  \BibitemOpen
  \bibfield  {author} {\bibinfo {author} {\bibfnamefont {T.}~\bibnamefont {Benoist}}, \bibinfo {author} {\bibfnamefont {C.}~\bibnamefont {Pellegrini}},\ and\ \bibinfo {author} {\bibfnamefont {F.}~\bibnamefont {Ticozzi}},\ }\bibfield  {title} {\bibinfo {title} {Exponential stability of subspaces for quantum stochastic master equations},\ }\href {https://doi.org/https://doi.org/10.1007/s00023-017-0556-3} {\bibfield  {journal} {\bibinfo  {journal} {Ann. Henri Poincaré}\ }\textbf {\bibinfo {volume} {18}},\ \bibinfo {pages} {2045–2074} (\bibinfo {year} {2017})}\BibitemShut {NoStop}%
\bibitem [{\citenamefont {Cardona}\ \emph {et~al.}(2018)\citenamefont {Cardona}, \citenamefont {Sarlette},\ and\ \citenamefont {Rouchon}}]{Cardona_2018}%
  \BibitemOpen
  \bibfield  {author} {\bibinfo {author} {\bibfnamefont {G.}~\bibnamefont {Cardona}}, \bibinfo {author} {\bibfnamefont {A.}~\bibnamefont {Sarlette}},\ and\ \bibinfo {author} {\bibfnamefont {P.}~\bibnamefont {Rouchon}},\ }\bibfield  {title} {\bibinfo {title} {Exponential stochastic stabilization of a two-level quantum system via strict lyapunov control},\ }in\ \href {https://doi.org/10.1109/CDC.2018.8618681} {\emph {\bibinfo {booktitle} {2018 IEEE Conference on Decision and Control (CDC)}}}\ (\bibinfo {year} {2018})\ pp.\ \bibinfo {pages} {6591--6596}\BibitemShut {NoStop}%
\bibitem [{\citenamefont {Liang}\ \emph {et~al.}(2019)\citenamefont {Liang}, \citenamefont {Amini},\ and\ \citenamefont {Mason}}]{liang2019exponential}%
  \BibitemOpen
  \bibfield  {author} {\bibinfo {author} {\bibfnamefont {W.}~\bibnamefont {Liang}}, \bibinfo {author} {\bibfnamefont {N.~H.}\ \bibnamefont {Amini}},\ and\ \bibinfo {author} {\bibfnamefont {P.}~\bibnamefont {Mason}},\ }\href@noop {} {\bibinfo {title} {On exponential stabilization of spin-1/2 systems}} (\bibinfo {year} {2019}),\ \Eprint {https://arxiv.org/abs/1803.10632} {arXiv:1803.10632 [quant-ph]} \BibitemShut {NoStop}%
\bibitem [{\citenamefont {Cardona}\ \emph {et~al.}(2020)\citenamefont {Cardona}, \citenamefont {Sarlette},\ and\ \citenamefont {Rouchon}}]{Cardona_2020}%
  \BibitemOpen
  \bibfield  {author} {\bibinfo {author} {\bibfnamefont {G.}~\bibnamefont {Cardona}}, \bibinfo {author} {\bibfnamefont {A.}~\bibnamefont {Sarlette}},\ and\ \bibinfo {author} {\bibfnamefont {P.}~\bibnamefont {Rouchon}},\ }\bibfield  {title} {\bibinfo {title} {Exponential stabilization of quantum systems under continuous non-demolition measurements},\ }\href {https://doi.org/10.1016/j.automatica.2019.108719} {\bibfield  {journal} {\bibinfo  {journal} {Automatica}\ }\textbf {\bibinfo {volume} {112}},\ \bibinfo {pages} {108719} (\bibinfo {year} {2020})}\BibitemShut {NoStop}%
\bibitem [{\citenamefont {Amini}\ \emph {et~al.}(2023)\citenamefont {Amini}, \citenamefont {Bompais},\ and\ \citenamefont {Pellegrini}}]{amini2023exponential}%
  \BibitemOpen
  \bibfield  {author} {\bibinfo {author} {\bibfnamefont {N.~H.}\ \bibnamefont {Amini}}, \bibinfo {author} {\bibfnamefont {M.}~\bibnamefont {Bompais}},\ and\ \bibinfo {author} {\bibfnamefont {C.}~\bibnamefont {Pellegrini}},\ }\href@noop {} {\bibinfo {title} {Exponential selection and feedback stabilization of invariant subspaces of quantum trajectories}} (\bibinfo {year} {2023}),\ \Eprint {https://arxiv.org/abs/2310.04599} {arXiv:2310.04599 [quant-ph]} \BibitemShut {NoStop}%
\bibitem [{\citenamefont {Liang}\ \emph {et~al.}(2023)\citenamefont {Liang}, \citenamefont {Ohki},\ and\ \citenamefont {Ticozzi}}]{liang2023exploring}%
  \BibitemOpen
  \bibfield  {author} {\bibinfo {author} {\bibfnamefont {W.}~\bibnamefont {Liang}}, \bibinfo {author} {\bibfnamefont {K.}~\bibnamefont {Ohki}},\ and\ \bibinfo {author} {\bibfnamefont {F.}~\bibnamefont {Ticozzi}},\ }\href@noop {} {\bibinfo {title} {Exploring the robustness of stabilizing controls for stochastic quantum evolutions}} (\bibinfo {year} {2023}),\ \Eprint {https://arxiv.org/abs/2311.04428} {arXiv:2311.04428 [quant-ph]} \BibitemShut {NoStop}%
\bibitem [{\citenamefont {Guillaud}\ and\ \citenamefont {Mirrahimi}(2019)}]{Guillaud_2019}%
  \BibitemOpen
  \bibfield  {author} {\bibinfo {author} {\bibfnamefont {J.}~\bibnamefont {Guillaud}}\ and\ \bibinfo {author} {\bibfnamefont {M.}~\bibnamefont {Mirrahimi}},\ }\bibfield  {title} {\bibinfo {title} {Repetition cat qubits for fault-tolerant quantum computation},\ }\href {https://doi.org/10.1103/PhysRevX.9.041053} {\bibfield  {journal} {\bibinfo  {journal} {Phys. Rev. X}\ }\textbf {\bibinfo {volume} {9}},\ \bibinfo {pages} {041053} (\bibinfo {year} {2019})}\BibitemShut {NoStop}%
\bibitem [{\citenamefont {Ahn}\ \emph {et~al.}(2002)\citenamefont {Ahn}, \citenamefont {Doherty},\ and\ \citenamefont {Landahl}}]{Ahn_2002}%
  \BibitemOpen
  \bibfield  {author} {\bibinfo {author} {\bibfnamefont {C.}~\bibnamefont {Ahn}}, \bibinfo {author} {\bibfnamefont {A.~C.}\ \bibnamefont {Doherty}},\ and\ \bibinfo {author} {\bibfnamefont {A.~J.}\ \bibnamefont {Landahl}},\ }\bibfield  {title} {\bibinfo {title} {Continuous quantum error correction via quantum feedback control},\ }\href {https://doi.org/10.1103/PhysRevA.65.042301} {\bibfield  {journal} {\bibinfo  {journal} {Phys. Rev. A}\ }\textbf {\bibinfo {volume} {65}},\ \bibinfo {pages} {042301} (\bibinfo {year} {2002})}\BibitemShut {NoStop}%
\bibitem [{\citenamefont {Ahn}\ \emph {et~al.}(2003)\citenamefont {Ahn}, \citenamefont {Wiseman},\ and\ \citenamefont {Milburn}}]{Ahn_2003}%
  \BibitemOpen
  \bibfield  {author} {\bibinfo {author} {\bibfnamefont {C.}~\bibnamefont {Ahn}}, \bibinfo {author} {\bibfnamefont {H.~M.}\ \bibnamefont {Wiseman}},\ and\ \bibinfo {author} {\bibfnamefont {G.~J.}\ \bibnamefont {Milburn}},\ }\bibfield  {title} {\bibinfo {title} {Quantum error correction for continuously detected errors},\ }\href {https://doi.org/10.1103/PhysRevA.67.052310} {\bibfield  {journal} {\bibinfo  {journal} {Phys. Rev. A}\ }\textbf {\bibinfo {volume} {67}},\ \bibinfo {pages} {052310} (\bibinfo {year} {2003})}\BibitemShut {NoStop}%
\bibitem [{\citenamefont {Ahn}\ \emph {et~al.}(2004)\citenamefont {Ahn}, \citenamefont {Wiseman},\ and\ \citenamefont {Jacobs}}]{Ahn_2004}%
  \BibitemOpen
  \bibfield  {author} {\bibinfo {author} {\bibfnamefont {C.}~\bibnamefont {Ahn}}, \bibinfo {author} {\bibfnamefont {H.}~\bibnamefont {Wiseman}},\ and\ \bibinfo {author} {\bibfnamefont {K.}~\bibnamefont {Jacobs}},\ }\bibfield  {title} {\bibinfo {title} {Quantum error correction for continuously detected errors with any number of error channels per qubit},\ }\href {https://doi.org/10.1103/PhysRevA.70.024302} {\bibfield  {journal} {\bibinfo  {journal} {Phys. Rev. A}\ }\textbf {\bibinfo {volume} {70}},\ \bibinfo {pages} {024302} (\bibinfo {year} {2004})}\BibitemShut {NoStop}%
\bibitem [{\citenamefont {Sarovar}\ \emph {et~al.}(2004)\citenamefont {Sarovar}, \citenamefont {Ahn}, \citenamefont {Jacobs},\ and\ \citenamefont {Milburn}}]{Sarovar_2004}%
  \BibitemOpen
  \bibfield  {author} {\bibinfo {author} {\bibfnamefont {M.}~\bibnamefont {Sarovar}}, \bibinfo {author} {\bibfnamefont {C.}~\bibnamefont {Ahn}}, \bibinfo {author} {\bibfnamefont {K.}~\bibnamefont {Jacobs}},\ and\ \bibinfo {author} {\bibfnamefont {G.~J.}\ \bibnamefont {Milburn}},\ }\bibfield  {title} {\bibinfo {title} {Practical scheme for error control using feedback},\ }\href {https://doi.org/10.1103/PhysRevA.69.052324} {\bibfield  {journal} {\bibinfo  {journal} {Phys. Rev. A}\ }\textbf {\bibinfo {volume} {69}},\ \bibinfo {pages} {052324} (\bibinfo {year} {2004})}\BibitemShut {NoStop}%
\bibitem [{\citenamefont {van Handel}\ and\ \citenamefont {Mabuchi}(2005)}]{vanhandel2005optimal}%
  \BibitemOpen
  \bibfield  {author} {\bibinfo {author} {\bibfnamefont {R.}~\bibnamefont {van Handel}}\ and\ \bibinfo {author} {\bibfnamefont {H.}~\bibnamefont {Mabuchi}},\ }\href@noop {} {\bibinfo {title} {Optimal error tracking via quantum coding and continuous syndrome measurement}} (\bibinfo {year} {2005}),\ \Eprint {https://arxiv.org/abs/quant-ph/0511221} {arXiv:quant-ph/0511221 [quant-ph]} \BibitemShut {NoStop}%
\bibitem [{\citenamefont {Oreshkov}\ and\ \citenamefont {Brun}(2007)}]{Oreshkov_2007}%
  \BibitemOpen
  \bibfield  {author} {\bibinfo {author} {\bibfnamefont {O.}~\bibnamefont {Oreshkov}}\ and\ \bibinfo {author} {\bibfnamefont {T.~A.}\ \bibnamefont {Brun}},\ }\bibfield  {title} {\bibinfo {title} {Continuous quantum error correction for non-markovian decoherence},\ }\href {https://doi.org/10.1103/PhysRevA.76.022318} {\bibfield  {journal} {\bibinfo  {journal} {Phys. Rev. A}\ }\textbf {\bibinfo {volume} {76}},\ \bibinfo {pages} {022318} (\bibinfo {year} {2007})}\BibitemShut {NoStop}%
\bibitem [{\citenamefont {Mascarenhas}\ \emph {et~al.}(2010)\citenamefont {Mascarenhas}, \citenamefont {Marques}, \citenamefont {Cunha},\ and\ \citenamefont {Santos}}]{Mascarenhas_2010}%
  \BibitemOpen
  \bibfield  {author} {\bibinfo {author} {\bibfnamefont {E.}~\bibnamefont {Mascarenhas}}, \bibinfo {author} {\bibfnamefont {B.}~\bibnamefont {Marques}}, \bibinfo {author} {\bibfnamefont {M.~T.}\ \bibnamefont {Cunha}},\ and\ \bibinfo {author} {\bibfnamefont {M.~F.}\ \bibnamefont {Santos}},\ }\bibfield  {title} {\bibinfo {title} {Continuous quantum error correction through local operations},\ }\href {https://doi.org/10.1103/PhysRevA.82.032327} {\bibfield  {journal} {\bibinfo  {journal} {Phys. Rev. A}\ }\textbf {\bibinfo {volume} {82}},\ \bibinfo {pages} {032327} (\bibinfo {year} {2010})}\BibitemShut {NoStop}%
\bibitem [{\citenamefont {Atalaya}\ \emph {et~al.}(2020)\citenamefont {Atalaya}, \citenamefont {Korotkov},\ and\ \citenamefont {Whaley}}]{Atalaya_CQEC}%
  \BibitemOpen
  \bibfield  {author} {\bibinfo {author} {\bibfnamefont {J.}~\bibnamefont {Atalaya}}, \bibinfo {author} {\bibfnamefont {A.~N.}\ \bibnamefont {Korotkov}},\ and\ \bibinfo {author} {\bibfnamefont {K.~B.}\ \bibnamefont {Whaley}},\ }\bibfield  {title} {\bibinfo {title} {Error-correcting bacon-shor code with continuous measurement of noncommuting operators},\ }\href {https://doi.org/10.1103/PhysRevA.102.022415} {\bibfield  {journal} {\bibinfo  {journal} {Phys. Rev. A}\ }\textbf {\bibinfo {volume} {102}},\ \bibinfo {pages} {022415} (\bibinfo {year} {2020})}\BibitemShut {NoStop}%
\bibitem [{\citenamefont {Mohseninia}\ \emph {et~al.}(2020)\citenamefont {Mohseninia}, \citenamefont {Yang}, \citenamefont {Siddiqi}, \citenamefont {Jordan},\ and\ \citenamefont {Dressel}}]{Mohseninia2020alwaysquantumerror}%
  \BibitemOpen
  \bibfield  {author} {\bibinfo {author} {\bibfnamefont {R.}~\bibnamefont {Mohseninia}}, \bibinfo {author} {\bibfnamefont {J.}~\bibnamefont {Yang}}, \bibinfo {author} {\bibfnamefont {I.}~\bibnamefont {Siddiqi}}, \bibinfo {author} {\bibfnamefont {A.~N.}\ \bibnamefont {Jordan}},\ and\ \bibinfo {author} {\bibfnamefont {J.}~\bibnamefont {Dressel}},\ }\bibfield  {title} {\bibinfo {title} {Always-{O}n {Q}uantum {E}rror {T}racking with {C}ontinuous {P}arity {M}easurements},\ }\href {https://doi.org/10.22331/q-2020-11-04-358} {\bibfield  {journal} {\bibinfo  {journal} {{Quantum}}\ }\textbf {\bibinfo {volume} {4}},\ \bibinfo {pages} {358} (\bibinfo {year} {2020})}\BibitemShut {NoStop}%
\bibitem [{\citenamefont {Atalaya}\ \emph {et~al.}(2021)\citenamefont {Atalaya}, \citenamefont {Zhang}, \citenamefont {Niu}, \citenamefont {Babakhani}, \citenamefont {Chan}, \citenamefont {Epstein},\ and\ \citenamefont {Whaley}}]{Atalaya_2021_CQEC}%
  \BibitemOpen
  \bibfield  {author} {\bibinfo {author} {\bibfnamefont {J.}~\bibnamefont {Atalaya}}, \bibinfo {author} {\bibfnamefont {S.}~\bibnamefont {Zhang}}, \bibinfo {author} {\bibfnamefont {M.~Y.}\ \bibnamefont {Niu}}, \bibinfo {author} {\bibfnamefont {A.}~\bibnamefont {Babakhani}}, \bibinfo {author} {\bibfnamefont {H.~C.~H.}\ \bibnamefont {Chan}}, \bibinfo {author} {\bibfnamefont {J.~M.}\ \bibnamefont {Epstein}},\ and\ \bibinfo {author} {\bibfnamefont {K.~B.}\ \bibnamefont {Whaley}},\ }\bibfield  {title} {\bibinfo {title} {Continuous quantum error correction for evolution under time-dependent hamiltonians},\ }\href {https://doi.org/10.1103/PhysRevA.103.042406} {\bibfield  {journal} {\bibinfo  {journal} {Phys. Rev. A}\ }\textbf {\bibinfo {volume} {103}},\ \bibinfo {pages} {042406} (\bibinfo {year} {2021})}\BibitemShut {NoStop}%
\bibitem [{\citenamefont {Convy}\ and\ \citenamefont {Whaley}(2022)}]{Convy2022logarithmicbayesian}%
  \BibitemOpen
  \bibfield  {author} {\bibinfo {author} {\bibfnamefont {I.}~\bibnamefont {Convy}}\ and\ \bibinfo {author} {\bibfnamefont {K.~B.}\ \bibnamefont {Whaley}},\ }\bibfield  {title} {\bibinfo {title} {A {L}ogarithmic {B}ayesian {A}pproach to {Q}uantum {E}rror {D}etection},\ }\href {https://doi.org/10.22331/q-2022-04-04-680} {\bibfield  {journal} {\bibinfo  {journal} {{Quantum}}\ }\textbf {\bibinfo {volume} {6}},\ \bibinfo {pages} {680} (\bibinfo {year} {2022})}\BibitemShut {NoStop}%
\bibitem [{\citenamefont {Livingston}\ \emph {et~al.}(2022)\citenamefont {Livingston}, \citenamefont {Blok}, \citenamefont {Flurin}, \citenamefont {Dressel}, \citenamefont {Jordan},\ and\ \citenamefont {Siddiqi}}]{Livingston_2022}%
  \BibitemOpen
  \bibfield  {author} {\bibinfo {author} {\bibfnamefont {W.~P.}\ \bibnamefont {Livingston}}, \bibinfo {author} {\bibfnamefont {M.~S.}\ \bibnamefont {Blok}}, \bibinfo {author} {\bibfnamefont {E.}~\bibnamefont {Flurin}}, \bibinfo {author} {\bibfnamefont {J.}~\bibnamefont {Dressel}}, \bibinfo {author} {\bibfnamefont {A.~N.}\ \bibnamefont {Jordan}},\ and\ \bibinfo {author} {\bibfnamefont {I.}~\bibnamefont {Siddiqi}},\ }\bibfield  {title} {\bibinfo {title} {Experimental demonstration of continuous quantum error correction},\ }\href {https://doi.org/10.1038/s41467-022-29906-0} {\bibfield  {journal} {\bibinfo  {journal} {Nat. Commun.}\ }\textbf {\bibinfo {volume} {13}},\ \bibinfo {pages} {2307} (\bibinfo {year} {2022})}\BibitemShut {NoStop}%
\bibitem [{\citenamefont {Convy}\ \emph {et~al.}(2022)\citenamefont {Convy}, \citenamefont {Liao}, \citenamefont {Zhang}, \citenamefont {Patel}, \citenamefont {Livingston}, \citenamefont {Nguyen}, \citenamefont {Siddiqi},\ and\ \citenamefont {Whaley}}]{Convy_2022}%
  \BibitemOpen
  \bibfield  {author} {\bibinfo {author} {\bibfnamefont {I.}~\bibnamefont {Convy}}, \bibinfo {author} {\bibfnamefont {H.}~\bibnamefont {Liao}}, \bibinfo {author} {\bibfnamefont {S.}~\bibnamefont {Zhang}}, \bibinfo {author} {\bibfnamefont {S.}~\bibnamefont {Patel}}, \bibinfo {author} {\bibfnamefont {W.~P.}\ \bibnamefont {Livingston}}, \bibinfo {author} {\bibfnamefont {H.~N.}\ \bibnamefont {Nguyen}}, \bibinfo {author} {\bibfnamefont {I.}~\bibnamefont {Siddiqi}},\ and\ \bibinfo {author} {\bibfnamefont {K.~B.}\ \bibnamefont {Whaley}},\ }\bibfield  {title} {\bibinfo {title} {Machine learning for continuous quantum error correction on superconducting qubits},\ }\href {https://doi.org/10.1088/1367-2630/ac66f9} {\bibfield  {journal} {\bibinfo  {journal} {New Journal of Physics}\ }\textbf {\bibinfo {volume} {24}},\ \bibinfo {pages} {063019} (\bibinfo {year} {2022})}\BibitemShut {NoStop}%
\bibitem [{\citenamefont {Misra}\ and\ \citenamefont {Sudarshan}(1977)}]{misra1977zeno}%
  \BibitemOpen
  \bibfield  {author} {\bibinfo {author} {\bibfnamefont {B.}~\bibnamefont {Misra}}\ and\ \bibinfo {author} {\bibfnamefont {E.~C.~G.}\ \bibnamefont {Sudarshan}},\ }\bibfield  {title} {\bibinfo {title} {The {Zeno’s} paradox in quantum theory},\ }\href {https://doi.org/10.1063/1.523304} {\bibfield  {journal} {\bibinfo  {journal} {Journal of Mathematical Physics}\ }\textbf {\bibinfo {volume} {18}},\ \bibinfo {pages} {756} (\bibinfo {year} {1977})}\BibitemShut {NoStop}%
\bibitem [{\citenamefont {Presilla}\ \emph {et~al.}(1996)\citenamefont {Presilla}, \citenamefont {Onofrio},\ and\ \citenamefont {Tambini}}]{Presilla_1996}%
  \BibitemOpen
  \bibfield  {author} {\bibinfo {author} {\bibfnamefont {C.}~\bibnamefont {Presilla}}, \bibinfo {author} {\bibfnamefont {R.}~\bibnamefont {Onofrio}},\ and\ \bibinfo {author} {\bibfnamefont {U.}~\bibnamefont {Tambini}},\ }\bibfield  {title} {\bibinfo {title} {Measurement quantum mechanics and experiments on quantum zeno effect},\ }\href {https://doi.org/10.1006/aphy.1996.0052} {\bibfield  {journal} {\bibinfo  {journal} {Annals of Physics}\ }\textbf {\bibinfo {volume} {248}},\ \bibinfo {pages} {95} (\bibinfo {year} {1996})}\BibitemShut {NoStop}%
\bibitem [{\citenamefont {Sarandy}\ and\ \citenamefont {Lidar}(2005)}]{Sarandy2005adiabaticapprox}%
  \BibitemOpen
  \bibfield  {author} {\bibinfo {author} {\bibfnamefont {M.~S.}\ \bibnamefont {Sarandy}}\ and\ \bibinfo {author} {\bibfnamefont {D.~A.}\ \bibnamefont {Lidar}},\ }\bibfield  {title} {\bibinfo {title} {Adiabatic approximation in open quantum systems},\ }\href {https://doi.org/10.1103/PhysRevA.71.012331} {\bibfield  {journal} {\bibinfo  {journal} {Phys. Rev. A}\ }\textbf {\bibinfo {volume} {71}},\ \bibinfo {pages} {012331} (\bibinfo {year} {2005})}\BibitemShut {NoStop}%
\bibitem [{\citenamefont {Facchi}\ and\ \citenamefont {Pascazio}(2008)}]{facchi2008quantum}%
  \BibitemOpen
  \bibfield  {author} {\bibinfo {author} {\bibfnamefont {P.}~\bibnamefont {Facchi}}\ and\ \bibinfo {author} {\bibfnamefont {S.}~\bibnamefont {Pascazio}},\ }\bibfield  {title} {\bibinfo {title} {Quantum {Z}eno dynamics: mathematical and physical aspects},\ }\href {https://doi.org/10.1088/1751-8113/41/49/493001} {\bibfield  {journal} {\bibinfo  {journal} {Journal of Physics A: Mathematical and Theoretical}\ }\textbf {\bibinfo {volume} {41}},\ \bibinfo {pages} {493001} (\bibinfo {year} {2008})}\BibitemShut {NoStop}%
\bibitem [{\citenamefont {Venuti}\ \emph {et~al.}(2016)\citenamefont {Venuti}, \citenamefont {Albash}, \citenamefont {Lidar},\ and\ \citenamefont {Zanardi}}]{Venuti2016adiabaticinopen}%
  \BibitemOpen
  \bibfield  {author} {\bibinfo {author} {\bibfnamefont {L.~C.}\ \bibnamefont {Venuti}}, \bibinfo {author} {\bibfnamefont {T.}~\bibnamefont {Albash}}, \bibinfo {author} {\bibfnamefont {D.~A.}\ \bibnamefont {Lidar}},\ and\ \bibinfo {author} {\bibfnamefont {P.}~\bibnamefont {Zanardi}},\ }\bibfield  {title} {\bibinfo {title} {Adiabaticity in open quantum systems},\ }\href {https://doi.org/10.1103/PhysRevA.93.032118} {\bibfield  {journal} {\bibinfo  {journal} {Phys. Rev. A}\ }\textbf {\bibinfo {volume} {93}},\ \bibinfo {pages} {032118} (\bibinfo {year} {2016})}\BibitemShut {NoStop}%
\bibitem [{\citenamefont {Burgarth}\ \emph {et~al.}(2020)\citenamefont {Burgarth}, \citenamefont {Facchi}, \citenamefont {Nakazato}, \citenamefont {Pascazio},\ and\ \citenamefont {Yuasa}}]{Burgarth2020quantumzenodynamics}%
  \BibitemOpen
  \bibfield  {author} {\bibinfo {author} {\bibfnamefont {D.}~\bibnamefont {Burgarth}}, \bibinfo {author} {\bibfnamefont {P.}~\bibnamefont {Facchi}}, \bibinfo {author} {\bibfnamefont {H.}~\bibnamefont {Nakazato}}, \bibinfo {author} {\bibfnamefont {S.}~\bibnamefont {Pascazio}},\ and\ \bibinfo {author} {\bibfnamefont {K.}~\bibnamefont {Yuasa}},\ }\bibfield  {title} {\bibinfo {title} {Quantum {Z}eno {D}ynamics from {G}eneral {Q}uantum {O}perations},\ }\href {https://doi.org/10.22331/q-2020-07-06-289} {\bibfield  {journal} {\bibinfo  {journal} {{Quantum}}\ }\textbf {\bibinfo {volume} {4}},\ \bibinfo {pages} {289} (\bibinfo {year} {2020})}\BibitemShut {NoStop}%
\bibitem [{\citenamefont {Kumar}\ \emph {et~al.}(2020)\citenamefont {Kumar}, \citenamefont {Romito},\ and\ \citenamefont {Snizhko}}]{Kumar2020Zeno}%
  \BibitemOpen
  \bibfield  {author} {\bibinfo {author} {\bibfnamefont {P.}~\bibnamefont {Kumar}}, \bibinfo {author} {\bibfnamefont {A.}~\bibnamefont {Romito}},\ and\ \bibinfo {author} {\bibfnamefont {K.}~\bibnamefont {Snizhko}},\ }\bibfield  {title} {\bibinfo {title} {Quantum {Z}eno effect with partial measurement and noisy dynamics},\ }\href {https://doi.org/10.1103/PhysRevResearch.2.043420} {\bibfield  {journal} {\bibinfo  {journal} {Phys. Rev. Research}\ }\textbf {\bibinfo {volume} {2}},\ \bibinfo {pages} {043420} (\bibinfo {year} {2020})}\BibitemShut {NoStop}%
\bibitem [{\citenamefont {Snizhko}\ \emph {et~al.}(2020)\citenamefont {Snizhko}, \citenamefont {Kumar},\ and\ \citenamefont {Romito}}]{Snizhko2020Zeno}%
  \BibitemOpen
  \bibfield  {author} {\bibinfo {author} {\bibfnamefont {K.}~\bibnamefont {Snizhko}}, \bibinfo {author} {\bibfnamefont {P.}~\bibnamefont {Kumar}},\ and\ \bibinfo {author} {\bibfnamefont {A.}~\bibnamefont {Romito}},\ }\bibfield  {title} {\bibinfo {title} {Quantum {Z}eno effect appears in stages},\ }\href {https://doi.org/10.1103/PhysRevResearch.2.033512} {\bibfield  {journal} {\bibinfo  {journal} {Phys. Rev. Research}\ }\textbf {\bibinfo {volume} {2}},\ \bibinfo {pages} {033512} (\bibinfo {year} {2020})}\BibitemShut {NoStop}%
\bibitem [{\citenamefont {Burgarth}\ \emph {et~al.}(2022)\citenamefont {Burgarth}, \citenamefont {Facchi}, \citenamefont {Gramegna},\ and\ \citenamefont {Yuasa}}]{Burgarth2022oneboundtorulethem}%
  \BibitemOpen
  \bibfield  {author} {\bibinfo {author} {\bibfnamefont {D.}~\bibnamefont {Burgarth}}, \bibinfo {author} {\bibfnamefont {P.}~\bibnamefont {Facchi}}, \bibinfo {author} {\bibfnamefont {G.}~\bibnamefont {Gramegna}},\ and\ \bibinfo {author} {\bibfnamefont {K.}~\bibnamefont {Yuasa}},\ }\bibfield  {title} {\bibinfo {title} {One bound to rule them all: from {A}diabatic to {Z}eno},\ }\href {https://doi.org/10.22331/q-2022-06-14-737} {\bibfield  {journal} {\bibinfo  {journal} {{Quantum}}\ }\textbf {\bibinfo {volume} {6}},\ \bibinfo {pages} {737} (\bibinfo {year} {2022})}\BibitemShut {NoStop}%
\bibitem [{\citenamefont {Albert}\ \emph {et~al.}(2016)\citenamefont {Albert}, \citenamefont {Bradlyn}, \citenamefont {Fraas},\ and\ \citenamefont {Jiang}}]{Albert_2016}%
  \BibitemOpen
  \bibfield  {author} {\bibinfo {author} {\bibfnamefont {V.~V.}\ \bibnamefont {Albert}}, \bibinfo {author} {\bibfnamefont {B.}~\bibnamefont {Bradlyn}}, \bibinfo {author} {\bibfnamefont {M.}~\bibnamefont {Fraas}},\ and\ \bibinfo {author} {\bibfnamefont {L.}~\bibnamefont {Jiang}},\ }\bibfield  {title} {\bibinfo {title} {Geometry and response of lindbladians},\ }\href {https://doi.org/10.1103/PhysRevX.6.041031} {\bibfield  {journal} {\bibinfo  {journal} {Phys. Rev. X}\ }\textbf {\bibinfo {volume} {6}},\ \bibinfo {pages} {041031} (\bibinfo {year} {2016})}\BibitemShut {NoStop}%
\bibitem [{\citenamefont {Harrington}\ \emph {et~al.}(2022)\citenamefont {Harrington}, \citenamefont {Mueller},\ and\ \citenamefont {Murch}}]{Harrington_2022}%
  \BibitemOpen
  \bibfield  {author} {\bibinfo {author} {\bibfnamefont {P.}~\bibnamefont {Harrington}}, \bibinfo {author} {\bibfnamefont {E.}~\bibnamefont {Mueller}},\ and\ \bibinfo {author} {\bibfnamefont {K.}~\bibnamefont {Murch}},\ }\bibfield  {title} {\bibinfo {title} {Engineered dissipation for quantum information science},\ }\href {https://doi.org/10.1038/s42254-022-00494-8} {\bibfield  {journal} {\bibinfo  {journal} {Nat. Rev. Phys.}\ }\textbf {\bibinfo {volume} {4}},\ \bibinfo {pages} {660–671} (\bibinfo {year} {2022})}\BibitemShut {NoStop}%
\bibitem [{\citenamefont {Blumenthal}\ \emph {et~al.}(2022)\citenamefont {Blumenthal}, \citenamefont {Mor}, \citenamefont {Diringer}, \citenamefont {Martin}, \citenamefont {Lewalle}, \citenamefont {Burgarth}, \citenamefont {Whaley},\ and\ \citenamefont {Hacohen-Gourgy}}]{blumenthal2021}%
  \BibitemOpen
  \bibfield  {author} {\bibinfo {author} {\bibfnamefont {E.}~\bibnamefont {Blumenthal}}, \bibinfo {author} {\bibfnamefont {C.}~\bibnamefont {Mor}}, \bibinfo {author} {\bibfnamefont {A.~A.}\ \bibnamefont {Diringer}}, \bibinfo {author} {\bibfnamefont {L.~S.}\ \bibnamefont {Martin}}, \bibinfo {author} {\bibfnamefont {P.}~\bibnamefont {Lewalle}}, \bibinfo {author} {\bibfnamefont {D.}~\bibnamefont {Burgarth}}, \bibinfo {author} {\bibfnamefont {K.~B.}\ \bibnamefont {Whaley}},\ and\ \bibinfo {author} {\bibfnamefont {S.}~\bibnamefont {Hacohen-Gourgy}},\ }\bibfield  {title} {\bibinfo {title} {Demonstration of universal control between non-interacting qubits using the quantum {Z}eno effect},\ }\bibfield  {journal} {\bibinfo  {journal} {npj Quantum Information~}\ }\textbf {\bibinfo {volume} {8}},\ \href {https://doi.org/10.1038/s41534-022-00594-4} {10.1038/s41534-022-00594-4} (\bibinfo {year} {2022})\BibitemShut {NoStop}%
\bibitem [{\citenamefont {Lewalle}\ \emph {et~al.}(2023{\natexlab{b}})\citenamefont {Lewalle}, \citenamefont {Martin}, \citenamefont {Flurin}, \citenamefont {Zhang}, \citenamefont {Blumenthal}, \citenamefont {Hacohen-Gourgy}, \citenamefont {Burgarth},\ and\ \citenamefont {Whaley}}]{ZenoGateTheory}%
  \BibitemOpen
  \bibfield  {author} {\bibinfo {author} {\bibfnamefont {P.}~\bibnamefont {Lewalle}}, \bibinfo {author} {\bibfnamefont {L.~S.}\ \bibnamefont {Martin}}, \bibinfo {author} {\bibfnamefont {E.}~\bibnamefont {Flurin}}, \bibinfo {author} {\bibfnamefont {S.}~\bibnamefont {Zhang}}, \bibinfo {author} {\bibfnamefont {E.}~\bibnamefont {Blumenthal}}, \bibinfo {author} {\bibfnamefont {S.}~\bibnamefont {Hacohen-Gourgy}}, \bibinfo {author} {\bibfnamefont {D.}~\bibnamefont {Burgarth}},\ and\ \bibinfo {author} {\bibfnamefont {K.~B.}\ \bibnamefont {Whaley}},\ }\bibfield  {title} {\bibinfo {title} {A {M}ulti-{Q}ubit {Q}uantum {G}ate {U}sing the {Z}eno {E}ffect},\ }\href {https://doi.org/10.22331/q-2023-09-07-1100} {\bibfield  {journal} {\bibinfo  {journal} {{Quantum}}\ }\textbf {\bibinfo {volume} {7}},\ \bibinfo {pages} {1100} (\bibinfo {year} {2023}{\natexlab{b}})}\BibitemShut {NoStop}%
\bibitem [{\citenamefont {Gautier}\ \emph {et~al.}(2023)\citenamefont {Gautier}, \citenamefont {Mirrahimi},\ and\ \citenamefont {Sarlette}}]{gautier2023designing}%
  \BibitemOpen
  \bibfield  {author} {\bibinfo {author} {\bibfnamefont {R.}~\bibnamefont {Gautier}}, \bibinfo {author} {\bibfnamefont {M.}~\bibnamefont {Mirrahimi}},\ and\ \bibinfo {author} {\bibfnamefont {A.}~\bibnamefont {Sarlette}},\ }\bibfield  {title} {\bibinfo {title} {Designing high-fidelity zeno gates for dissipative cat qubits},\ }\href {https://doi.org/10.1103/PRXQuantum.4.040316} {\bibfield  {journal} {\bibinfo  {journal} {PRX Quantum}\ }\textbf {\bibinfo {volume} {4}},\ \bibinfo {pages} {040316} (\bibinfo {year} {2023})}\BibitemShut {NoStop}%
\bibitem [{\citenamefont {Paz-Silva}\ \emph {et~al.}(2012)\citenamefont {Paz-Silva}, \citenamefont {Rezakhani}, \citenamefont {Dominy},\ and\ \citenamefont {Lidar}}]{Paz-Silva_2012}%
  \BibitemOpen
  \bibfield  {author} {\bibinfo {author} {\bibfnamefont {G.~A.}\ \bibnamefont {Paz-Silva}}, \bibinfo {author} {\bibfnamefont {A.~T.}\ \bibnamefont {Rezakhani}}, \bibinfo {author} {\bibfnamefont {J.~M.}\ \bibnamefont {Dominy}},\ and\ \bibinfo {author} {\bibfnamefont {D.~A.}\ \bibnamefont {Lidar}},\ }\bibfield  {title} {\bibinfo {title} {Zeno effect for quantum computation and control},\ }\href {https://doi.org/10.1103/PhysRevLett.108.080501} {\bibfield  {journal} {\bibinfo  {journal} {Phys. Rev. Lett.}\ }\textbf {\bibinfo {volume} {108}},\ \bibinfo {pages} {080501} (\bibinfo {year} {2012})}\BibitemShut {NoStop}%
\bibitem [{\citenamefont {Dominy}\ \emph {et~al.}(2013)\citenamefont {Dominy}, \citenamefont {Paz-Silva}, \citenamefont {Rezakhani},\ and\ \citenamefont {Lidar}}]{Dominy_2013}%
  \BibitemOpen
  \bibfield  {author} {\bibinfo {author} {\bibfnamefont {J.~M.}\ \bibnamefont {Dominy}}, \bibinfo {author} {\bibfnamefont {G.~A.}\ \bibnamefont {Paz-Silva}}, \bibinfo {author} {\bibfnamefont {A.~T.}\ \bibnamefont {Rezakhani}},\ and\ \bibinfo {author} {\bibfnamefont {D.~A.}\ \bibnamefont {Lidar}},\ }\bibfield  {title} {\bibinfo {title} {Analysis of the quantum zeno effect for quantum control and computation},\ }\href {https://doi.org/10.1088/1751-8113/46/7/075306} {\bibfield  {journal} {\bibinfo  {journal} {Journal of Physics A: Mathematical and Theoretical}\ }\textbf {\bibinfo {volume} {46}},\ \bibinfo {pages} {075306} (\bibinfo {year} {2013})}\BibitemShut {NoStop}%
\bibitem [{\citenamefont {Cohen}\ and\ \citenamefont {Mirrahimi}(2014)}]{Cohen_2014}%
  \BibitemOpen
  \bibfield  {author} {\bibinfo {author} {\bibfnamefont {J.}~\bibnamefont {Cohen}}\ and\ \bibinfo {author} {\bibfnamefont {M.}~\bibnamefont {Mirrahimi}},\ }\bibfield  {title} {\bibinfo {title} {Dissipation-induced continuous quantum error correction for superconducting circuits},\ }\href {https://doi.org/10.1103/PhysRevA.90.062344} {\bibfield  {journal} {\bibinfo  {journal} {Phys. Rev. A}\ }\textbf {\bibinfo {volume} {90}},\ \bibinfo {pages} {062344} (\bibinfo {year} {2014})}\BibitemShut {NoStop}%
\bibitem [{\citenamefont {Mirrahimi}\ \emph {et~al.}(2014)\citenamefont {Mirrahimi}, \citenamefont {Leghtas}, \citenamefont {Albert}, \citenamefont {Touzard}, \citenamefont {Schoelkopf}, \citenamefont {Jiang},\ and\ \citenamefont {Devoret}}]{Mirrahimi_2014}%
  \BibitemOpen
  \bibfield  {author} {\bibinfo {author} {\bibfnamefont {M.}~\bibnamefont {Mirrahimi}}, \bibinfo {author} {\bibfnamefont {Z.}~\bibnamefont {Leghtas}}, \bibinfo {author} {\bibfnamefont {V.~V.}\ \bibnamefont {Albert}}, \bibinfo {author} {\bibfnamefont {S.}~\bibnamefont {Touzard}}, \bibinfo {author} {\bibfnamefont {R.~J.}\ \bibnamefont {Schoelkopf}}, \bibinfo {author} {\bibfnamefont {L.}~\bibnamefont {Jiang}},\ and\ \bibinfo {author} {\bibfnamefont {M.~H.}\ \bibnamefont {Devoret}},\ }\bibfield  {title} {\bibinfo {title} {Dynamically protected cat-qubits: a new paradigm for universal quantum computation},\ }\href {https://doi.org/10.1088/1367-2630/16/4/045014} {\bibfield  {journal} {\bibinfo  {journal} {New Journal of Physics}\ }\textbf {\bibinfo {volume} {16}},\ \bibinfo {pages} {045014} (\bibinfo {year} {2014})}\BibitemShut {NoStop}%
\bibitem [{\citenamefont {Lihm}\ \emph {et~al.}(2018)\citenamefont {Lihm}, \citenamefont {Noh},\ and\ \citenamefont {Fischer}}]{Lihm_2018}%
  \BibitemOpen
  \bibfield  {author} {\bibinfo {author} {\bibfnamefont {J.-M.}\ \bibnamefont {Lihm}}, \bibinfo {author} {\bibfnamefont {K.}~\bibnamefont {Noh}},\ and\ \bibinfo {author} {\bibfnamefont {U.~R.}\ \bibnamefont {Fischer}},\ }\bibfield  {title} {\bibinfo {title} {Implementation-independent sufficient condition of the knill-laflamme type for the autonomous protection of logical qudits by strong engineered dissipation},\ }\href {https://doi.org/10.1103/PhysRevA.98.012317} {\bibfield  {journal} {\bibinfo  {journal} {Phys. Rev. A}\ }\textbf {\bibinfo {volume} {98}},\ \bibinfo {pages} {012317} (\bibinfo {year} {2018})}\BibitemShut {NoStop}%
\bibitem [{\citenamefont {Lebreuilly}\ \emph {et~al.}(2021)\citenamefont {Lebreuilly}, \citenamefont {Noh}, \citenamefont {Wang}, \citenamefont {Girvin},\ and\ \citenamefont {Jiang}}]{lebreuilly2021autonomous}%
  \BibitemOpen
  \bibfield  {author} {\bibinfo {author} {\bibfnamefont {J.}~\bibnamefont {Lebreuilly}}, \bibinfo {author} {\bibfnamefont {K.}~\bibnamefont {Noh}}, \bibinfo {author} {\bibfnamefont {C.-H.}\ \bibnamefont {Wang}}, \bibinfo {author} {\bibfnamefont {S.~M.}\ \bibnamefont {Girvin}},\ and\ \bibinfo {author} {\bibfnamefont {L.}~\bibnamefont {Jiang}},\ }\href@noop {} {\bibinfo {title} {Autonomous quantum error correction and quantum computation}} (\bibinfo {year} {2021}),\ \Eprint {https://arxiv.org/abs/2103.05007} {arXiv:2103.05007 [quant-ph]} \BibitemShut {NoStop}%
\bibitem [{\citenamefont {Gertler}\ \emph {et~al.}(2021)\citenamefont {Gertler}, \citenamefont {Baker}, \citenamefont {Li}, \citenamefont {Shirol}, \citenamefont {Koch},\ and\ \citenamefont {Wang}}]{Gertler_2021}%
  \BibitemOpen
  \bibfield  {author} {\bibinfo {author} {\bibfnamefont {J.~M.}\ \bibnamefont {Gertler}}, \bibinfo {author} {\bibfnamefont {B.}~\bibnamefont {Baker}}, \bibinfo {author} {\bibfnamefont {J.}~\bibnamefont {Li}}, \bibinfo {author} {\bibfnamefont {S.}~\bibnamefont {Shirol}}, \bibinfo {author} {\bibfnamefont {J.}~\bibnamefont {Koch}},\ and\ \bibinfo {author} {\bibfnamefont {C.}~\bibnamefont {Wang}},\ }\bibfield  {title} {\bibinfo {title} {Protecting a bosonic qubit with autonomous quantum error correction},\ }\href {https://doi.org/10.1038/s41586-021-03257-0} {\bibfield  {journal} {\bibinfo  {journal} {Nature}\ }\textbf {\bibinfo {volume} {590}},\ \bibinfo {pages} {243} (\bibinfo {year} {2021})}\BibitemShut {NoStop}%
\bibitem [{\citenamefont {Shtanko}\ \emph {et~al.}(2023)\citenamefont {Shtanko}, \citenamefont {Liu}, \citenamefont {Lieu}, \citenamefont {Gorshkov},\ and\ \citenamefont {Albert}}]{shtanko2023bounds}%
  \BibitemOpen
  \bibfield  {author} {\bibinfo {author} {\bibfnamefont {O.}~\bibnamefont {Shtanko}}, \bibinfo {author} {\bibfnamefont {Y.-J.}\ \bibnamefont {Liu}}, \bibinfo {author} {\bibfnamefont {S.}~\bibnamefont {Lieu}}, \bibinfo {author} {\bibfnamefont {A.~V.}\ \bibnamefont {Gorshkov}},\ and\ \bibinfo {author} {\bibfnamefont {V.~V.}\ \bibnamefont {Albert}},\ }\href@noop {} {\bibinfo {title} {Bounds on autonomous quantum error correction}} (\bibinfo {year} {2023}),\ \Eprint {https://arxiv.org/abs/2308.16233} {arXiv:2308.16233 [quant-ph]} \BibitemShut {NoStop}%
\bibitem [{\citenamefont {Krom}(1967)}]{krom1967decision}%
  \BibitemOpen
  \bibfield  {author} {\bibinfo {author} {\bibfnamefont {M.~R.}\ \bibnamefont {Krom}},\ }\bibfield  {title} {\bibinfo {title} {The decision problem for a class of first-order formulas in which all disjunctions are binary},\ }\href@noop {} {\bibfield  {journal} {\bibinfo  {journal} {Mathematical Logic Quarterly}\ }\textbf {\bibinfo {volume} {13}},\ \bibinfo {pages} {15} (\bibinfo {year} {1967})}\BibitemShut {NoStop}%
\bibitem [{\citenamefont {Gent}\ and\ \citenamefont {Walsh}(1994)}]{gent1994easy}%
  \BibitemOpen
  \bibfield  {author} {\bibinfo {author} {\bibfnamefont {I.~P.}\ \bibnamefont {Gent}}\ and\ \bibinfo {author} {\bibfnamefont {T.}~\bibnamefont {Walsh}},\ }\bibfield  {title} {\bibinfo {title} {Easy problems are sometimes hard},\ }\href@noop {} {\bibfield  {journal} {\bibinfo  {journal} {Artificial Intelligence}\ }\textbf {\bibinfo {volume} {70}},\ \bibinfo {pages} {335} (\bibinfo {year} {1994})}\BibitemShut {NoStop}%
\bibitem [{\citenamefont {Korotkov}(1999)}]{Korotkov1}%
  \BibitemOpen
  \bibfield  {author} {\bibinfo {author} {\bibfnamefont {A.~N.}\ \bibnamefont {Korotkov}},\ }\bibfield  {title} {\bibinfo {title} {Continuous quantum measurement of a double dot},\ }\href {https://doi.org/10.1103/PhysRevB.60.5737} {\bibfield  {journal} {\bibinfo  {journal} {Phys. Rev. B}\ }\textbf {\bibinfo {volume} {60}},\ \bibinfo {pages} {5737} (\bibinfo {year} {1999})}\BibitemShut {NoStop}%
\bibitem [{\citenamefont {Korotkov}(2001)}]{Korotkov2}%
  \BibitemOpen
  \bibfield  {author} {\bibinfo {author} {\bibfnamefont {A.~N.}\ \bibnamefont {Korotkov}},\ }\bibfield  {title} {\bibinfo {title} {Selective quantum evolution of a qubit state due to continuous measurement},\ }\href {https://doi.org/10.1103/PhysRevB.63.115403} {\bibfield  {journal} {\bibinfo  {journal} {Phys. Rev. B}\ }\textbf {\bibinfo {volume} {63}},\ \bibinfo {pages} {115403} (\bibinfo {year} {2001})}\BibitemShut {NoStop}%
\bibitem [{\citenamefont {Korotkov}(2011)}]{Korotkov3}%
  \BibitemOpen
  \bibfield  {author} {\bibinfo {author} {\bibfnamefont {A.~N.}\ \bibnamefont {Korotkov}},\ }\bibfield  {title} {\bibinfo {title} {Quantum {B}ayesian approach to circuit {QED} measurement},\ }in\ \href {https://oxford.universitypressscholarship.com/view/10.1093/acprof:oso/9780199681181.001.0001/acprof-9780199681181-chapter-17} {\emph {\bibinfo {booktitle} {Quantum Machines: Measurement and Control of Engineered Quantum Systems: Lecture Notes of the Les Houches Summer School: Volume 96}}},\ \bibinfo {editor} {edited by\ \bibinfo {editor} {\bibfnamefont {M.}~\bibnamefont {Devoret}}, \bibinfo {editor} {\bibfnamefont {B.}~\bibnamefont {Huard}}, \bibinfo {editor} {\bibfnamefont {R.}~\bibnamefont {Schoelkopf}},\ and\ \bibinfo {editor} {\bibfnamefont {L.~F.}\ \bibnamefont {Cugliandolo}}}\ (\bibinfo  {publisher} {Oxford University Press},\ \bibinfo {year} {2011})\BibitemShut {NoStop}%
\bibitem [{\citenamefont {Korotkov}(2016)}]{Korotkov4}%
  \BibitemOpen
  \bibfield  {author} {\bibinfo {author} {\bibfnamefont {A.~N.}\ \bibnamefont {Korotkov}},\ }\bibfield  {title} {\bibinfo {title} {Quantum {B}ayesian approach to circuit {QED} measurement with moderate bandwidth},\ }\href {https://doi.org/10.1103/PhysRevA.94.042326} {\bibfield  {journal} {\bibinfo  {journal} {Phys. Rev. A}\ }\textbf {\bibinfo {volume} {94}},\ \bibinfo {pages} {042326} (\bibinfo {year} {2016})}\BibitemShut {NoStop}%
\bibitem [{\citenamefont {Steinmetz}\ \emph {et~al.}(2022)\citenamefont {Steinmetz}, \citenamefont {Das}, \citenamefont {Siddiqi},\ and\ \citenamefont {Jordan}}]{Steinmetz_2022}%
  \BibitemOpen
  \bibfield  {author} {\bibinfo {author} {\bibfnamefont {J.}~\bibnamefont {Steinmetz}}, \bibinfo {author} {\bibfnamefont {D.}~\bibnamefont {Das}}, \bibinfo {author} {\bibfnamefont {I.}~\bibnamefont {Siddiqi}},\ and\ \bibinfo {author} {\bibfnamefont {A.~N.}\ \bibnamefont {Jordan}},\ }\bibfield  {title} {\bibinfo {title} {Continuous measurement of a qudit using dispersively coupled radiation},\ }\href {https://doi.org/10.1103/PhysRevA.105.052229} {\bibfield  {journal} {\bibinfo  {journal} {Phys. Rev. A}\ }\textbf {\bibinfo {volume} {105}},\ \bibinfo {pages} {052229} (\bibinfo {year} {2022})}\BibitemShut {NoStop}%
\bibitem [{\citenamefont {Lindblad}(1976)}]{Lindblad}%
  \BibitemOpen
  \bibfield  {author} {\bibinfo {author} {\bibfnamefont {G.}~\bibnamefont {Lindblad}},\ }\bibfield  {title} {\bibinfo {title} {On the generators of quantum dynamical semigroups},\ }\href {https://doi.org/10.1007/BF01608499} {\bibfield  {journal} {\bibinfo  {journal} {Commun. Math. Phys.}\ }\textbf {\bibinfo {volume} {48}},\ \bibinfo {pages} {119–130} (\bibinfo {year} {1976})}\BibitemShut {NoStop}%
\bibitem [{\citenamefont {Gorini}\ \emph {et~al.}(1976)\citenamefont {Gorini}, \citenamefont {Kossakowski},\ and\ \citenamefont {Sudarshan}}]{GKS}%
  \BibitemOpen
  \bibfield  {author} {\bibinfo {author} {\bibfnamefont {V.}~\bibnamefont {Gorini}}, \bibinfo {author} {\bibfnamefont {A.}~\bibnamefont {Kossakowski}},\ and\ \bibinfo {author} {\bibfnamefont {E.~C.~G.}\ \bibnamefont {Sudarshan}},\ }\bibfield  {title} {\bibinfo {title} {{Completely positive dynamical semigroups of N‐level systems}},\ }\href {https://doi.org/10.1063/1.522979} {\bibfield  {journal} {\bibinfo  {journal} {Journal of Mathematical Physics}\ }\textbf {\bibinfo {volume} {17}},\ \bibinfo {pages} {821} (\bibinfo {year} {1976})}\BibitemShut {NoStop}%
\bibitem [{\citenamefont {R{\o}nnow}\ \emph {et~al.}(2014)\citenamefont {R{\o}nnow}, \citenamefont {Wang}, \citenamefont {Job}, \citenamefont {Boixo}, \citenamefont {Isakov}, \citenamefont {Wecker}, \citenamefont {Martinis}, \citenamefont {Lidar},\ and\ \citenamefont {Troyer}}]{ronnow2014defining}%
  \BibitemOpen
  \bibfield  {author} {\bibinfo {author} {\bibfnamefont {T.~F.}\ \bibnamefont {R{\o}nnow}}, \bibinfo {author} {\bibfnamefont {Z.}~\bibnamefont {Wang}}, \bibinfo {author} {\bibfnamefont {J.}~\bibnamefont {Job}}, \bibinfo {author} {\bibfnamefont {S.}~\bibnamefont {Boixo}}, \bibinfo {author} {\bibfnamefont {S.~V.}\ \bibnamefont {Isakov}}, \bibinfo {author} {\bibfnamefont {D.}~\bibnamefont {Wecker}}, \bibinfo {author} {\bibfnamefont {J.~M.}\ \bibnamefont {Martinis}}, \bibinfo {author} {\bibfnamefont {D.~A.}\ \bibnamefont {Lidar}},\ and\ \bibinfo {author} {\bibfnamefont {M.}~\bibnamefont {Troyer}},\ }\bibfield  {title} {\bibinfo {title} {Defining and detecting quantum speedup},\ }\href@noop {} {\bibfield  {journal} {\bibinfo  {journal} {science}\ }\textbf {\bibinfo {volume} {345}},\ \bibinfo {pages} {420} (\bibinfo {year} {2014})}\BibitemShut {NoStop}%
\bibitem [{\citenamefont {Wootters}(1998)}]{Wooters_1998}%
  \BibitemOpen
  \bibfield  {author} {\bibinfo {author} {\bibfnamefont {W.~K.}\ \bibnamefont {Wootters}},\ }\bibfield  {title} {\bibinfo {title} {Entanglement of formation of an arbitrary state of two qubits},\ }\href {https://doi.org/10.1103/PhysRevLett.80.2245} {\bibfield  {journal} {\bibinfo  {journal} {Phys. Rev. Lett.}\ }\textbf {\bibinfo {volume} {80}},\ \bibinfo {pages} {2245} (\bibinfo {year} {1998})}\BibitemShut {NoStop}%
\bibitem [{\citenamefont {Goerdt}(1996)}]{GOERDT1996469}%
  \BibitemOpen
  \bibfield  {author} {\bibinfo {author} {\bibfnamefont {A.}~\bibnamefont {Goerdt}},\ }\bibfield  {title} {\bibinfo {title} {A threshold for unsatisfiability},\ }\href {https://doi.org/https://doi.org/10.1006/jcss.1996.0081} {\bibfield  {journal} {\bibinfo  {journal} {Journal of Computer and System Sciences}\ }\textbf {\bibinfo {volume} {53}},\ \bibinfo {pages} {469} (\bibinfo {year} {1996})}\BibitemShut {NoStop}%
\bibitem [{\citenamefont {Leyton-Brown}\ \emph {et~al.}(2014)\citenamefont {Leyton-Brown}, \citenamefont {Hoos}, \citenamefont {Hutter},\ and\ \citenamefont {Xu}}]{10.1145/2594413.2594424}%
  \BibitemOpen
  \bibfield  {author} {\bibinfo {author} {\bibfnamefont {K.}~\bibnamefont {Leyton-Brown}}, \bibinfo {author} {\bibfnamefont {H.~H.}\ \bibnamefont {Hoos}}, \bibinfo {author} {\bibfnamefont {F.}~\bibnamefont {Hutter}},\ and\ \bibinfo {author} {\bibfnamefont {L.}~\bibnamefont {Xu}},\ }\bibfield  {title} {\bibinfo {title} {Understanding the empirical hardness of np-complete problems},\ }\href {https://doi.org/10.1145/2594413.2594424} {\bibfield  {journal} {\bibinfo  {journal} {Commun. ACM}\ }\textbf {\bibinfo {volume} {57}},\ \bibinfo {pages} {98–107} (\bibinfo {year} {2014})}\BibitemShut {NoStop}%
\bibitem [{\citenamefont {Crawford}\ and\ \citenamefont {Auton}(1996)}]{CRAWFORD199631}%
  \BibitemOpen
  \bibfield  {author} {\bibinfo {author} {\bibfnamefont {J.~M.}\ \bibnamefont {Crawford}}\ and\ \bibinfo {author} {\bibfnamefont {L.~D.}\ \bibnamefont {Auton}},\ }\bibfield  {title} {\bibinfo {title} {Experimental results on the crossover point in random 3-sat},\ }\href {https://doi.org/https://doi.org/10.1016/0004-3702(95)00046-1} {\bibfield  {journal} {\bibinfo  {journal} {Artificial Intelligence}\ }\textbf {\bibinfo {volume} {81}},\ \bibinfo {pages} {31} (\bibinfo {year} {1996})},\ \bibinfo {note} {frontiers in Problem Solving: Phase Transitions and Complexity}\BibitemShut {NoStop}%
\bibitem [{\citenamefont {Mitchell}\ \emph {et~al.}(1992)\citenamefont {Mitchell}, \citenamefont {Selman},\ and\ \citenamefont {Levesque}}]{10.5555/1867135.1867206}%
  \BibitemOpen
  \bibfield  {author} {\bibinfo {author} {\bibfnamefont {D.}~\bibnamefont {Mitchell}}, \bibinfo {author} {\bibfnamefont {B.}~\bibnamefont {Selman}},\ and\ \bibinfo {author} {\bibfnamefont {H.}~\bibnamefont {Levesque}},\ }\bibfield  {title} {\bibinfo {title} {Hard and easy distributions of sat problems},\ }in\ \href@noop {} {\emph {\bibinfo {booktitle} {Proceedings of the Tenth National Conference on Artificial Intelligence}}},\ \bibinfo {series and number} {AAAI'92}\ (\bibinfo  {publisher} {AAAI Press},\ \bibinfo {year} {1992})\ p.\ \bibinfo {pages} {459–465}\BibitemShut {NoStop}%
\bibitem [{\citenamefont {Akshay}\ \emph {et~al.}(2020)\citenamefont {Akshay}, \citenamefont {Philathong}, \citenamefont {Morales},\ and\ \citenamefont {Biamonte}}]{PhysRevLett.124.090504}%
  \BibitemOpen
  \bibfield  {author} {\bibinfo {author} {\bibfnamefont {V.}~\bibnamefont {Akshay}}, \bibinfo {author} {\bibfnamefont {H.}~\bibnamefont {Philathong}}, \bibinfo {author} {\bibfnamefont {M.~E.~S.}\ \bibnamefont {Morales}},\ and\ \bibinfo {author} {\bibfnamefont {J.~D.}\ \bibnamefont {Biamonte}},\ }\bibfield  {title} {\bibinfo {title} {Reachability deficits in quantum approximate optimization},\ }\href {https://doi.org/10.1103/PhysRevLett.124.090504} {\bibfield  {journal} {\bibinfo  {journal} {Phys. Rev. Lett.}\ }\textbf {\bibinfo {volume} {124}},\ \bibinfo {pages} {090504} (\bibinfo {year} {2020})}\BibitemShut {NoStop}%
\bibitem [{\citenamefont {Zhang}\ \emph {et~al.}(2022)\citenamefont {Zhang}, \citenamefont {Sone},\ and\ \citenamefont {Zhuang}}]{zhang2022quantum}%
  \BibitemOpen
  \bibfield  {author} {\bibinfo {author} {\bibfnamefont {B.}~\bibnamefont {Zhang}}, \bibinfo {author} {\bibfnamefont {A.}~\bibnamefont {Sone}},\ and\ \bibinfo {author} {\bibfnamefont {Q.}~\bibnamefont {Zhuang}},\ }\bibfield  {title} {\bibinfo {title} {Quantum computational phase transition in combinatorial problems},\ }\href {https://doi.org/https://doi.org/10.1038/s41534-022-00596-2} {\bibfield  {journal} {\bibinfo  {journal} {npj Quantum Information}\ }\textbf {\bibinfo {volume} {8}},\ \bibinfo {pages} {87} (\bibinfo {year} {2022})}\BibitemShut {NoStop}%
\bibitem [{\citenamefont {Albash}\ and\ \citenamefont {Lidar}(2018)}]{albash2018demonstration}%
  \BibitemOpen
  \bibfield  {author} {\bibinfo {author} {\bibfnamefont {T.}~\bibnamefont {Albash}}\ and\ \bibinfo {author} {\bibfnamefont {D.~A.}\ \bibnamefont {Lidar}},\ }\bibfield  {title} {\bibinfo {title} {Demonstration of a scaling advantage for a quantum annealer over simulated annealing},\ }\href@noop {} {\bibfield  {journal} {\bibinfo  {journal} {Physical Review X}\ }\textbf {\bibinfo {volume} {8}},\ \bibinfo {pages} {031016} (\bibinfo {year} {2018})}\BibitemShut {NoStop}%
\bibitem [{\citenamefont {Touzard}\ \emph {et~al.}(2018)\citenamefont {Touzard}, \citenamefont {Grimm}, \citenamefont {Leghtas}, \citenamefont {Mundhada}, \citenamefont {Reinhold}, \citenamefont {Axline}, \citenamefont {Reagor}, \citenamefont {Chou}, \citenamefont {Blumoff}, \citenamefont {Sliwa}, \citenamefont {Shankar}, \citenamefont {Frunzio}, \citenamefont {Schoelkopf}, \citenamefont {Mirrahimi},\ and\ \citenamefont {Devoret}}]{Touzard_2018}%
  \BibitemOpen
  \bibfield  {author} {\bibinfo {author} {\bibfnamefont {S.}~\bibnamefont {Touzard}}, \bibinfo {author} {\bibfnamefont {A.}~\bibnamefont {Grimm}}, \bibinfo {author} {\bibfnamefont {Z.}~\bibnamefont {Leghtas}}, \bibinfo {author} {\bibfnamefont {S.~O.}\ \bibnamefont {Mundhada}}, \bibinfo {author} {\bibfnamefont {P.}~\bibnamefont {Reinhold}}, \bibinfo {author} {\bibfnamefont {C.}~\bibnamefont {Axline}}, \bibinfo {author} {\bibfnamefont {M.}~\bibnamefont {Reagor}}, \bibinfo {author} {\bibfnamefont {K.}~\bibnamefont {Chou}}, \bibinfo {author} {\bibfnamefont {J.}~\bibnamefont {Blumoff}}, \bibinfo {author} {\bibfnamefont {K.~M.}\ \bibnamefont {Sliwa}}, \bibinfo {author} {\bibfnamefont {S.}~\bibnamefont {Shankar}}, \bibinfo {author} {\bibfnamefont {L.}~\bibnamefont {Frunzio}}, \bibinfo {author} {\bibfnamefont {R.~J.}\ \bibnamefont {Schoelkopf}}, \bibinfo {author} {\bibfnamefont {M.}~\bibnamefont {Mirrahimi}},\ and\ \bibinfo {author} {\bibfnamefont {M.~H.}\ \bibnamefont {Devoret}},\ }\bibfield  {title} {\bibinfo
  {title} {Coherent oscillations inside a quantum manifold stabilized by dissipation},\ }\href {https://doi.org/10.1103/PhysRevX.8.021005} {\bibfield  {journal} {\bibinfo  {journal} {Phys. Rev. X}\ }\textbf {\bibinfo {volume} {8}},\ \bibinfo {pages} {021005} (\bibinfo {year} {2018})}\BibitemShut {NoStop}%
\bibitem [{\citenamefont {Martin}\ \emph {et~al.}(2020)\citenamefont {Martin}, \citenamefont {Livingston}, \citenamefont {Hacohen-Gourgy}, \citenamefont {Wiseman},\ and\ \citenamefont {Siddiqi}}]{Leigh_Phase}%
  \BibitemOpen
  \bibfield  {author} {\bibinfo {author} {\bibfnamefont {L.~S.}\ \bibnamefont {Martin}}, \bibinfo {author} {\bibfnamefont {W.~P.}\ \bibnamefont {Livingston}}, \bibinfo {author} {\bibfnamefont {S.}~\bibnamefont {Hacohen-Gourgy}}, \bibinfo {author} {\bibfnamefont {H.~M.}\ \bibnamefont {Wiseman}},\ and\ \bibinfo {author} {\bibfnamefont {I.}~\bibnamefont {Siddiqi}},\ }\bibfield  {title} {\bibinfo {title} {Implementation of a canonical phase measurement with quantum feedback},\ }\href {https://doi.org/10.1038/s41567-020-0939-0} {\bibfield  {journal} {\bibinfo  {journal} {Nat. Phys.}\ }\textbf {\bibinfo {volume} {16}},\ \bibinfo {pages} {1046–1049} (\bibinfo {year} {2020})}\BibitemShut {NoStop}%
\bibitem [{\citenamefont {Chantasri}\ \emph {et~al.}(2013)\citenamefont {Chantasri}, \citenamefont {Dressel},\ and\ \citenamefont {Jordan}}]{Chantasri2013}%
  \BibitemOpen
  \bibfield  {author} {\bibinfo {author} {\bibfnamefont {A.}~\bibnamefont {Chantasri}}, \bibinfo {author} {\bibfnamefont {J.}~\bibnamefont {Dressel}},\ and\ \bibinfo {author} {\bibfnamefont {A.~N.}\ \bibnamefont {Jordan}},\ }\bibfield  {title} {\bibinfo {title} {Action principle for continuous quantum measurement},\ }\href {https://doi.org/10.1103/PhysRevA.88.042110} {\bibfield  {journal} {\bibinfo  {journal} {Phys. Rev. A}\ }\textbf {\bibinfo {volume} {88}},\ \bibinfo {pages} {042110} (\bibinfo {year} {2013})}\BibitemShut {NoStop}%
\bibitem [{\citenamefont {Chantasri}\ and\ \citenamefont {Jordan}(2015)}]{Chantasri2015}%
  \BibitemOpen
  \bibfield  {author} {\bibinfo {author} {\bibfnamefont {A.}~\bibnamefont {Chantasri}}\ and\ \bibinfo {author} {\bibfnamefont {A.~N.}\ \bibnamefont {Jordan}},\ }\bibfield  {title} {\bibinfo {title} {Stochastic path-integral formalism for continuous quantum measurement},\ }\href {https://doi.org/10.1103/PhysRevA.92.032125} {\bibfield  {journal} {\bibinfo  {journal} {Phys. Rev. A}\ }\textbf {\bibinfo {volume} {92}},\ \bibinfo {pages} {032125} (\bibinfo {year} {2015})}\BibitemShut {NoStop}%
\bibitem [{\citenamefont {Bertlmann}\ and\ \citenamefont {Krammer}(2008)}]{Bertlmann2008}%
  \BibitemOpen
  \bibfield  {author} {\bibinfo {author} {\bibfnamefont {R.~A.}\ \bibnamefont {Bertlmann}}\ and\ \bibinfo {author} {\bibfnamefont {P.}~\bibnamefont {Krammer}},\ }\bibfield  {title} {\bibinfo {title} {Bloch vectors for qudits},\ }\href {https://doi.org/10.1088/1751-8113/41/23/235303} {\bibfield  {journal} {\bibinfo  {journal} {Journal of Physics A: Mathematical and Theoretical}\ }\textbf {\bibinfo {volume} {41}},\ \bibinfo {pages} {235303} (\bibinfo {year} {2008})}\BibitemShut {NoStop}%
\bibitem [{\citenamefont {Cylke}\ \emph {et~al.}(2024)\citenamefont {Cylke}, \citenamefont {Lewalle}, \citenamefont {Noungneaw}, \citenamefont {Wiseman}, \citenamefont {Jordan},\ and\ \citenamefont {Chantasri}}]{Cylke_2022inprep}%
  \BibitemOpen
  \bibfield  {author} {\bibinfo {author} {\bibfnamefont {A.~C.}\ \bibnamefont {Cylke}}, \bibinfo {author} {\bibfnamefont {P.}~\bibnamefont {Lewalle}}, \bibinfo {author} {\bibfnamefont {T.}~\bibnamefont {Noungneaw}}, \bibinfo {author} {\bibfnamefont {H.~M.}\ \bibnamefont {Wiseman}}, \bibinfo {author} {\bibfnamefont {A.~N.}\ \bibnamefont {Jordan}},\ and\ \bibinfo {author} {\bibfnamefont {A.}~\bibnamefont {Chantasri}},\ }\href@noop {} {\bibinfo {title} {Stochastic action functionals for diffusive quantum trajectories (in preparation)}} (\bibinfo {year} {2024})\BibitemShut {NoStop}%
\bibitem [{\citenamefont {Karmakar}\ \emph {et~al.}(2022)\citenamefont {Karmakar}, \citenamefont {Lewalle},\ and\ \citenamefont {Jordan}}]{Karmakar_2022}%
  \BibitemOpen
  \bibfield  {author} {\bibinfo {author} {\bibfnamefont {T.}~\bibnamefont {Karmakar}}, \bibinfo {author} {\bibfnamefont {P.}~\bibnamefont {Lewalle}},\ and\ \bibinfo {author} {\bibfnamefont {A.~N.}\ \bibnamefont {Jordan}},\ }\bibfield  {title} {\bibinfo {title} {Stochastic path-integral analysis of the continuously monitored quantum harmonic oscillator},\ }\href {https://doi.org/10.1103/PRXQuantum.3.010327} {\bibfield  {journal} {\bibinfo  {journal} {PRX Quantum}\ }\textbf {\bibinfo {volume} {3}},\ \bibinfo {pages} {010327} (\bibinfo {year} {2022})}\BibitemShut {NoStop}%
\bibitem [{\citenamefont {Lewalle}(2021)}]{Philippe_Thesis}%
  \BibitemOpen
  \bibfield  {author} {\bibinfo {author} {\bibfnamefont {P.}~\bibnamefont {Lewalle}},\ }\href {http://hdl.handle.net/1802/36379} {\bibinfo {title} {Quantum trajectories and their extremal--probability paths: New phenomena and applications}},\ \bibinfo {howpublished} {{PhD Dissertation, University of Rochester}} (\bibinfo {year} {2021})\BibitemShut {NoStop}%
\bibitem [{\citenamefont {Doucet}\ \emph {et~al.}(2020)\citenamefont {Doucet}, \citenamefont {Reiter}, \citenamefont {Ranzani},\ and\ \citenamefont {Kamal}}]{Doucet_2020}%
  \BibitemOpen
  \bibfield  {author} {\bibinfo {author} {\bibfnamefont {E.}~\bibnamefont {Doucet}}, \bibinfo {author} {\bibfnamefont {F.}~\bibnamefont {Reiter}}, \bibinfo {author} {\bibfnamefont {L.}~\bibnamefont {Ranzani}},\ and\ \bibinfo {author} {\bibfnamefont {A.}~\bibnamefont {Kamal}},\ }\bibfield  {title} {\bibinfo {title} {High fidelity dissipation engineering using parametric interactions},\ }\href {https://doi.org/10.1103/PhysRevResearch.2.023370} {\bibfield  {journal} {\bibinfo  {journal} {Phys. Rev. Res.}\ }\textbf {\bibinfo {volume} {2}},\ \bibinfo {pages} {023370} (\bibinfo {year} {2020})}\BibitemShut {NoStop}%
\bibitem [{\citenamefont {Doucet}\ \emph {et~al.}(2023)\citenamefont {Doucet}, \citenamefont {Govia},\ and\ \citenamefont {Kamal}}]{doucet2023scalable}%
  \BibitemOpen
  \bibfield  {author} {\bibinfo {author} {\bibfnamefont {E.}~\bibnamefont {Doucet}}, \bibinfo {author} {\bibfnamefont {L.~C.~G.}\ \bibnamefont {Govia}},\ and\ \bibinfo {author} {\bibfnamefont {A.}~\bibnamefont {Kamal}},\ }\href@noop {} {\bibinfo {title} {Scalable entanglement stabilization with modular reservoir engineering}} (\bibinfo {year} {2023}),\ \Eprint {https://arxiv.org/abs/2301.05725} {arXiv:2301.05725 [quant-ph]} \BibitemShut {NoStop}%
\bibitem [{\citenamefont {Greenfield}\ \emph {et~al.}(2024)\citenamefont {Greenfield}, \citenamefont {Martin}, \citenamefont {Motzoi}, \citenamefont {Whaley}, \citenamefont {Dressel},\ and\ \citenamefont {Levenson-Falk}}]{Greenfield_2024}%
  \BibitemOpen
  \bibfield  {author} {\bibinfo {author} {\bibfnamefont {S.}~\bibnamefont {Greenfield}}, \bibinfo {author} {\bibfnamefont {L.}~\bibnamefont {Martin}}, \bibinfo {author} {\bibfnamefont {F.}~\bibnamefont {Motzoi}}, \bibinfo {author} {\bibfnamefont {K.~B.}\ \bibnamefont {Whaley}}, \bibinfo {author} {\bibfnamefont {J.}~\bibnamefont {Dressel}},\ and\ \bibinfo {author} {\bibfnamefont {E.~M.}\ \bibnamefont {Levenson-Falk}},\ }\bibfield  {title} {\bibinfo {title} {Stabilizing two-qubit entanglement with dynamically decoupled active feedback},\ }\href {https://doi.org/10.1103/PhysRevApplied.21.024022} {\bibfield  {journal} {\bibinfo  {journal} {Phys. Rev. Appl.}\ }\textbf {\bibinfo {volume} {21}},\ \bibinfo {pages} {024022} (\bibinfo {year} {2024})}\BibitemShut {NoStop}%
\end{thebibliography}%
\end{document}